\documentclass[10pt]{article}
\usepackage{amsmath,amssymb,amsthm,graphicx,bbm,geometry,setspace}
\usepackage[longnamesfirst]{natbib}
\newtheorem{lemma}{Lemma}
\newtheorem{theorem}{Theorem}
\newtheorem{corollary}{Corollary}
\newtheorem{proposition}{Proposition}
\newtheorem{assumption}{Assumption}
\newcommand{\mf}{\boldsymbol{g}}
\newcommand{\nf}{\boldsymbol{m}}
\begin{document}
\title{Estimation and Inference for Policy Relevant Treatment Effects\thanks{First arXiv date: May 29, 2018. 
We benefited from very useful comments by 
Serena Ng and Elie Tamer (Editors), an anonymous Associate Editor, anonymous referees, other numerous researchers, and
participants at 
UC Davis, 
2018 CEME Conference: ``Inference on Nonstandard Problems,'' 
the 5th Annual Seattle-Vancouver Econometrics Conference, and 
2018 California Econometrics Conference. All remaining errors are ours.}}
\author{Yuya Sasaki\thanks{Y. Sasaki: Department of Economics, Vanderbilt University, VU Station B \#351819, 2301 Vanderbilt Place, Nashville, TN 37235-1819. Email: yuya.sasaki@vanderbilt.edu}\\ Department of Economics\\ Vanderbilt University
\and 
Takuya Ura\thanks{T. Ura: Department of Economics, University of California, Davis, One Shields Avenue, Davis, CA 95616. Email: takura@ucdavis.edu}\\ Department of Economics\\ University of California, Davis}
\date{} 
\maketitle 
\begin{abstract}
The policy relevant treatment effect (PRTE) measures the average effect of switching from a status-quo policy to a counterfactual policy.
Estimation of the PRTE involves estimation of multiple preliminary parameters, including propensity scores, conditional expectation functions of the outcome and covariates given the propensity score, and marginal treatment effects.
These preliminary estimators can affect the asymptotic distribution of the PRTE estimator in complicated and intractable manners.
In this light, we propose an orthogonal score for double debiased estimation of the PRTE, whereby the asymptotic distribution of the PRTE estimator is obtained without any influence of preliminary parameter estimators as far as they satisfy mild requirements of convergence rates.
To our knowledge, this paper is the first to develop limit distribution theories for inference about the PRTE.
\begin{description}
\item {\bf Keywords:} double debiased estimation, orthogonal score, policy relevant treatment effects
\item {\bf JEL Codes:} C14, C21
\end{description}
\end{abstract}

\section{Introduction}
The policy relevant treatment effect \citep[PRTE,][]{heckman/vytlacil:2001,heckman/vytlacil:2005,heckman/vytlacil:2007} measures the average effect of switching from a status-quo policy to a counterfactual policy.\footnote{Also see \citet{stock:1989} and \citet{ichimura/taber:2000} for parameters related to the PRTE.}
Among a number of measures for treatment effects, the PRTE has the advantage of directly evaluating alternative policy scenarios under consideration.
A rich set of identification results have been established in the literature for this treatment effect parameter \citep{heckman/vytlacil:2001,heckman/vytlacil:2005,heckman/vytlacil:2007} based on the marginal treatment effects \citep{bjorklund/moffitt:1987}.

Estimation of the PRTE involves nonparametric estimation for multiple preliminary parameters.
First, we nonparametrically estimate the propensity score, the conditional probability of being treated given an instrumental variable.
Next, when the outcome model involves additive controls as in \citet[][Section 4.1]{carneiro/lee:2009} and \citet[][Section 8]{carneiro/heckman/vytlacil:2010}, we estimate the conditional expectation functions of the outcome and covariates given the propensity score through the partially linear regression estimation procedure of \citet{robinson1988root}.
Third, with the residual from the partially linear regression, we estimate the marginal treatment effect via a nonparametric derivative estimation.  

These first-, second-, and third-stage estimates can affect the asymptotic distribution of the PRTE estimator in complicated and intractable manners, especially when their convergence rates are slower than the parametric rate of $n^{-1/2}$.
Estimation of the PRTE based on an orthogonal score can mitigate and asymptotically vanish the estimation errors of these first, second, and third preliminary parameters.
In this light, we propose an orthogonal score for double debiased estimation of the PRTE, whereby these first-, second-, and third-stage estimators will not affect the first-order asymptotic distribution of the PRTE estimator.
Consequently, we characterize the asymptotic distribution for the PRTE estimator based on the proposed orthogonal score.
Even conventionally in the absence of orthogonal scores, there has been no semi- or non-parametric inference method for the PRTE available in the existing literature, perhaps because of the complicated dependence of this parameter on multiple preliminary functions.
Therefore, to our knowledge, our work is the first to develop limit distribution theories for inference about the PRTE.

Orthogonal scores for double debiased and doubly robust estimation appear in a few contexts in the literature of econometrics and statistics. 
One instance is about partial linear models,\footnote{Estimation of partial linear models is studied by \citet{robinson1988root} under exogeneity, and is studied by \citet{okui2012doubly} under endogeneity. For a robust minimum-distance approach under endogeneity, see \citet{ai2007estimation}. See \citet{belloni2013honest}, \citet{belloni2014high}, and \citet{farrell2015robust} for machine learning approaches.} and another is about weighted averages\footnote{Weighted estimation is studied by \citet{rosenbaum1987model}, \citet{robins1992estimating}, \citet{imbens1992efficient}, \citet{robins1994estimation}, \citet{robins1995semiparametric}, \citet{lawless1999}, \citet{wooldridge1999,wooldridge2001asymptotic}, \citet{lee/okui/whang:2017} and \citet{sloczynski_wooldridge_2018} under parametric weight function, is studied by \citet{Hahn:1998}, \citet{hirano2003efficient}, \citet{suzukawa2004unbiased}, \citet{NeweyRuud2005}, \citet{Firpo:2007}, \citet{lewbel2007simple}, \citet{magnac2007identification}, and \citet{chen2008semiparametric}, \citet{graham2011efficiency}, \citet{graham2012inverse}, \citet{graham2016efficient}, \citet{sant2018doubly}, and \citet{rothe2019properties} under nonparametric weight function, and is studied by \citet{BCFH:2017} and \citet{Wager/Athey:2018} with machine learning. Particularly, \citet{rothe2019properties} analyzes the orthogonal scores for the policy effect parameter defined in \citet{stock:1989}. 
} such as inverse propensity score weighting.
Estimation of the PRTE consists of both of these two types of estimation procedures: (i) weighted estimation with the propensity score, estimation of a partial linear model to account for additive covariates as in \citet[][Section 4.1]{carneiro/lee:2009} and \citet[][Section 8]{carneiro/heckman/vytlacil:2010}, and (ii) weighted estimation with a counterfactual policy.
As such, it is a natural idea to develop an orthogonal score for estimation and inference for the PRTE by taking advantage of and combining these techniques.
With this said, the existing literature has not proposed an orthogonal score for the PRTE, and hence we aim to contribute to this literature by proposing one in this paper.

There is a large literature on orthogonal scores, and it characterizes desired properties of orthogonal scores such as double robustness against misspecification, local robustness in semiparametric estimation, and even semiparametric efficiency for certain orthogonal scores -- \citet{hitomi2008puzzling} discuss these features of  orthogonal scores in general from geometric perspectives.
Among these characteristics, our purpose of developing an orthogonal score is, as mentioned earlier, to remove the influence of the first, second, and third-stage estimators on the asymptotic distribution of the PRTE estimator, which concerns the local robustness.
There are a number of general procedures for estimation and inference based on orthogonal scores which we rely on in the present paper.
The seminal paper by \citet{newey1994asymptotic} proposes orthogonal scores in semiparametric methods and provides forms of adjustment terms to to obtain orthogonal scores from moment functions.
The way in which we derive our orthogonal score is based on his prescription \citep[cf.][Proposition 4]{newey1994asymptotic}.
\citet{belloni2014uniform} and \citet{belloni2018uniformly} propose a generic Z-estimation framework based on orthogonal scores.
\citet{chernozhukov2016locally} propose a general procedure for construction of orthogonal scores from moment restriction models.
\citet{chernozhukov2018double} combine the use of orthogonal scores and cross-fitting as a generic semiparametric strategy.
Our proposed method of estimation and inference takes advantages of this existing body of knowledge.

To the best of our knowledge, no limit distribution has been established for the PRTE in the existing literature.
Hence, this work is the first to develop asymptotic theories for inference about the PRTE.
The closest is \citet{carneiro/lee:2009} who develop point-wise limit distributions for the MTE, although they do not seem to imply or lead to a limit distribution for the PRTE.
Our orthogonal score does not only pave the way for a limit distribution for the PRTE for the first time in the literature, but also allows for a flexibility in types of preliminary estimators, e.g., kernel, sieve, and lasso.
In order to make our method and theory more accessible in practice, we also provide specific estimation procedures and lower-level sufficient conditions.\footnote{We consider a generalized linear model for the propensity score function allowing for possibly high-dimensional covariates and propose primitive conditions based on nonparametric estimation of other preliminary functional parameters. This framework involves both shrinkage and kernel estimation, and is related to a branch of the recent literature including \citet{kennedy:2017}, \citet{fan:2019}, \citet{su/ura/zhang:2019}, \citet{zimmert/lechner:2019} and \citet{colangelo/lee:2020}.}

Our proposed method is applicable to empirical studies that identify and estimate marginal treatment effects and/or policy relevant treatment effects.
Examples include, but are not limited to, 
\citet{auld:2005}, 
\citet{basu/heckman/navarro/urzua:2007}, 
\citet{doyle:2008}, 
\citet{moffitt:2008}, 
\citet{chuang/lai:2010}, 
\citet{carneiro/heckman/vytlacil:2011}, 
\citet{galasso/schankerman/serrano:2013}, 
\citet{basu/jena/goldman/philipson/dubois:2014}, 
\citet{belskaya/peter/posso:2014}, 
\citet{johar/maruyama:2014}, 
\citet{lindquist/santavirta:2014}, 
\citet{moffitt:2014}, 
\citet{dobbie/song:2015}, 
\citet{jensen/nielsen:2016}, 
\citet{kasahara/liang/rodrigue:2016}, 
\citet{carneiro/lokshin/umapathi:2017}, 
\citet{cornelissen/dustmann/raute/schonberg:2018}, 
\citet{felfe/lalive:2018}, 
and
\citet{kamhofer/schmitz/westphal:2018}. 

The rest of this paper is organized as follows.
Section \ref{sec:model} introduces a model based on \citet{carneiro/lee:2009}.
Section \ref{sec:overview} presents an informal overview of our proposed method.
Section \ref{sec:orthogonal_score} presents an orthogonal score for double debiased estimation of the PRTE.
Section \ref{sec:large_sample} discusses large sample properties of the double debiased estimator for inference.
Section \ref{sec:simulations} presents Monte Carlo simulation studies.
Section \ref{sec:empirical} presents an empirical illustration.
The paper is summarized in Section \ref{sec:summary}.
The appendix contains proofs and additional details that are important but relegated there due to their lengths.

\section{Model}\label{sec:model}

Following the model and notations in \citet[][Section 2]{carneiro/lee:2009}, consider the structure
\begin{align*}
Y=SY_1+\left(1-S\right)Y_0
\end{align*}
of outcome production, where $Y_1$ and $Y_0$ denote the potential outcomes under treatment ($S=1$) and no treatment ($S=0$), respectively.
With covariates $X$, the potential outcomes are modeled by $Y_1=\mu_1\left(X,U_1\right)$ and $Y_0=\mu_0\left(X,U_0\right)$ with unobserved variables $\left(U_0,U_1\right)$.
The binary treatment assignment status $S$ is in turn determined by the threshold crossing model
$S=1\{\mu_S\left(Z\right)-U_S>0\}$, where $Z$ is a vector of exogenous variables and $U_S$ is an error term.  
$Z$ can contain a sub-vector of $X$ as included exogenous variables, and the remaining excluded exogenous variables serve as instruments.
In this model, the dependence between $U_S$ and $\left(U_0,U_1\right)$ is the source of endogeneity in the treatment selection.
Researchers observe the random vector $W = \left(Y,S,X',Z'\right)'$, but do not observe $Y_1$, $Y_0$, $U_1$, $U_0$ or $U_S$.

Following \citet[][Section 3]{carneiro/lee:2009}, we further introduce the following notations for convenience.
Let $V=F_{U_S}\left(U_S\right)$ denote unobserved innate propensity to select into treatment, and let  
$P=F_{U_S}\left(\mu_S\left(Z\right)\right)$ denote the treatment selection probability or the propensity score.
Under these setup and notations, we adopt the following conventional assumption.

\begin{assumption}[\citeauthor{carneiro/lee:2009}, \citeyear{carneiro/lee:2009}, Assumptions 1 and 2]\label{assn1}
(1)
$\mu_S\left(Z\right)$ is non-degenerate conditional on $X$. 
(2) 
$\left(U_1,U_S\right)$ and $\left(U_0,U_S\right)$ are independent of $\left(Z,X\right)$.\footnote{We note that Assumption 1 of \citet{carneiro/lee:2009} only assumes that $\left(U_1,U_S\right)$ and $\left(U_0,U_S\right)$ are independent of $Z$ conditionally on $X$ for the purpose of identification, while their Assumption 2 assumes that $\left(U_1,U_S\right)$ and $\left(U_0,U_S\right)$ are independent of $\left(Z,X\right)$ similarly to our assumption for the purpose of semiparametric estimation.}
(3) 
The distribution of $U_S$ and the conditional distribution of $\mu_S\left(Z\right)$ given $X$ are absolutely continuous with respect to the Lebesgue measure. 
(4) 
$Y_1$ and $Y_0$ have finite first moments. 
(5) 
$0<Pr\left(S=1\mid Z\right)<1$.
(6) 
$p \mapsto E[U_1\mid P=p,S=1]$, $p \mapsto E[U_0\mid P=p,S=0]$, $\left(u_1,p\right) \mapsto f_{U_1\mid P,S=1}\left(u_1\mid p\right)$ and $\left(u_0,p\right) \mapsto f_{U_0\mid P,S=0}\left(u_0\mid p\right)$ are continuously differentiable with respect to $p$. 
\end{assumption}

We can write the treatment selection model in this setting by 
\begin{align*}
S=1\{P>V\}.
\end{align*}
This representation provides the interpretation that those individuals with $V=P$ are at the margin of indifference between the two treatment statuses, and motivates the marginal treatment effect \citep[][]{bjorklund/moffitt:1987} defined by
\begin{align*}
MTE\left(x,p\right)=E[Y_1-Y_0\mid X=x,V=p].
\end{align*}
This treatment parameter serves as a building block for many treatment parameters \citep{heckman/vytlacil:1999,heckman/vytlacil:2001,heckman/vytlacil:2005}.
We consider a counterfactual propensity score $P^\ast=P^\ast\left(P,Z\right)$ with a known function $P^\ast\left(\cdot,\cdot\right)$ satisfying Assumption \ref{assn:pstar} to be stated ahead.\footnote{\citet[][Section 3.2]{carneiro/heckman/vytlacil:2010} consider alternative specifications for counterfactual policies $P^\ast$. One scenario takes the form of $P^\ast=P^\ast\left(P,Z\right)$, in which a policy maker directly affects the probability of being in the treatment group. Another scenario takes the form of $P^\ast=P\left(Z^\ast\left(Z\right)\right)$, in which the counterfactual policy affects the distribution of $Z$. We consider the former scenario in the main text, whereas we consider the latter scenario in Appendix \ref{sec:alternative_P_star}.}
Under a counterfactual propensity score $P^\ast$, the counterfactual treatment and outcome are 
\begin{align*}
S^\ast&=1\{P^\ast>V\}
\qquad\text{and}\\
Y^\ast&=S^\ast Y_1+\left(1-S^\ast\right)Y_0.
\end{align*}
The parameter of our interest is the policy relevant treatment effect \citep[PRTE;][]{heckman/vytlacil:1999,heckman/vytlacil:2001,heckman/vytlacil:2005}: 
\begin{equation}
PRTE=\frac{E[Y^\ast]-E[Y]}{E[S^\ast]-E[S]}.
\end{equation}
This treatment parameter is of policy interest because it allows to compare alternative virtual policies $P^\ast$ under consideration.
\cite{heckman/vytlacil:1999,heckman/vytlacil:2001,heckman/vytlacil:2005} has demonstrated that $PRTE$ can be represented as the weighted average of $MTE\left(x,p\right)$: 
$$
PRTE=\int\int_0^1MTE\left(x,p\right)\frac{F_{P\mid X}\left(p\mid x\right)-F_{P^\ast\mid X}\left(p\mid x\right)}{E[P^\ast]-E[P]}dpf_X\left(x\right)dx. 
$$
Note that $MTE\left(x,p\right)$ measures the average treatment effects for the subpopulation of those individuals with $V=p$ and $F_{P|X}\left(p|x\right)-F_{P^\ast|X}\left(p|x\right)$ measures the probability of compliance with the counterfactual policy for this subpopulation.
Therefore, setting aside the denominator $E[P^\ast]-E[P]$, this integral measures the average treatment effect of the counterfactual policy for the population.
A division of quantity by $E[P^\ast]-E[P]$ in turn yields the average treatment effect among those who complied with the policy.

In the rest of this paper, we propose a method of double debiased estimation and inference for the PRTE.
Since the identification of the PRTE hinges on that of the marginal treatment effect, the theories of our estimation and inference methods rely on the prior result of the identification of the marginal treatment effect, which is formally stated in the following theorem.

\begin{theorem}[\citeauthor{heckman/vytlacil:1999}, \citeyear{heckman/vytlacil:1999,heckman/vytlacil:2001,heckman/vytlacil:2005}; and \citeauthor{carneiro/lee:2009}, \citeyear{carneiro/lee:2009}]\label{theorem:MTE_LIV}
If Assumption \ref{assn1} is satisfied, then
\begin{align*}
MTE\left(x,p\right)=\frac{\partial E[Y\mid X=x,P=p]}{\partial p}
\end{align*}
holds provided that $p \mapsto E[Y\mid X,P=p,S=1]$ and $p \mapsto E[Y\mid X,P=p,S=0]$ are continuously differentiable with respect to $p$ almost surely (with respect to $X$). 
\end{theorem}

In actual empirical research, researchers often use observed controls $X$.
Without imposing additional structural restrictions, they would suffer from the curse of dimensionality of $X$ in estimation of the marginal treatment effect. 
The existing literature proposes alternative suggestions to address this issue.
Following \citet[][Section 4.1]{carneiro/lee:2009} and \citet[][Section 8]{carneiro/heckman/vytlacil:2010}, we employ the following structural restriction on $\mu_1$ and $\mu_0$.

\begin{assumption}[\citeauthor{carneiro/lee:2009}, \citeyear{carneiro/lee:2009}, Section 4.1]\label{assn2}
$Y_1=\mu_1\left(X\right)'\beta_1+U_1$ and $Y_0=\mu_0\left(X\right)'\beta_0+U_0$
for $d_X$-dimensional unknown parameters $\beta_1$ and $\beta_0$ and known functions $\mu_1$ and $\mu_0$.
\end{assumption}
 
Under Assumptions \ref{assn1} and \ref{assn2}, we can write the conditional mean of $Y$ given $\left(X,P\right)$ as 
$$
E[Y\mid X,P]=P\mu_1\left(X\right)'\beta_1+\left(1-P\right)\mu_0\left(X\right)' \beta_0+E[U\mid P],
$$
where $U=SU_1+\left(1-S\right)U_0$. 
Thus the marginal treatment effect takes the partial linear form of
\begin{align}\label{eq:partial_linear_MTE}
MTE\left(x,p\right) = \mu_1\left(x\right)' \beta_1 - \mu_0\left(x\right)' \beta_0 + \Delta_{U\mid P}\left(p\right),
\end{align}
where $\Delta_{U\mid P}\left(p\right) = \frac{d}{dp} E[U|P=p]$.
In the rest of the paper, we shall use this form (\ref{eq:partial_linear_MTE}) of the MTE expression for analysis of the PRTE: 
\begin{align}
PRTE
=&
\frac{E[\left(P^\ast\left(\mf_{S\mid Z}\left(Z\right),Z\right)-\mf_{S\mid Z}\left(Z\right)\right)\mu_1\left(X\right)]'}{E[P^\ast\left(\mf_{S\mid Z}\left(Z\right),Z\right)-\mf_{S\mid Z}\left(Z\right)]}\beta_1-\frac{E[\left(P^\ast\left(\mf_{S\mid Z}\left(Z\right),Z\right)-\mf_{S\mid Z}\left(Z\right)\right)\mu_0\left(X\right)]}{E[P^\ast\left(\mf_{S\mid Z}\left(Z\right),Z\right)-\mf_{S\mid Z}\left(Z\right)]}'\beta_0
\notag\\&+\frac{E\left[\Delta_{U\mid P}\left(\mf_{S\mid Z}\left(Z\right)\right)\frac{F_{P}\left(\mf_{S\mid Z}\left(Z\right)\right)-F_{P^\ast}\left(\mf_{S\mid Z}\left(Z\right)\right)}{f_{P}\left(\mf_{S\mid Z}\left(Z\right)\right)}\right]}{E[P^\ast\left(\mf_{S\mid Z}\left(Z\right),Z\right)-\mf_{S\mid Z}\left(Z\right)]},\label{eq:PRTE_expression}
\end{align}
where $\mf_{S\mid Z}\left(z\right) = E[S|Z=z]$.
Appendix \ref{sec:eq:partial_linear_PRTE_rewritten} provides a derivation of \eqref{eq:PRTE_expression}.

We close this section by discussing feasible and infeasible extensions and variants of our parameter, $PRTE$, of interest.
First, the PRTE may be defined as $E[\omega\left(X\right)PRTE\left(X\right)]$ with a known weight function $\omega$ for heterogeneous policy designs.
Our method and theory presented ahead extends to this variant of the PRTE by replacing $Y$ by $\omega\left(X\right)Y$.
Second, our framework also extends to analysis of $PRTE\left(x\right)$ when $X$ is discrete by focusing on the subpopulation with $X=x$. 
When $X$ is continuous, however, it is generally infeasible to estimate $PRTE\left(x\right)$ at the root-$n$ convergence rate.

\section{An Overview of the Method}\label{sec:overview}

While we present a full-fledged framework and formal asymptotic theories in Sections \ref{sec:orthogonal_score} and \ref{sec:large_sample}, respectively, we first provide an informal overview in this section focusing on a simplified model without covariates.

Define $\theta = \left(\theta_N,\theta_D\right)$ by
\begingroup
\allowdisplaybreaks
\begin{align*}
\theta_N =& 
E\left[
\Delta_{U\mid P}\left(\mf_{S\mid Z}\left(Z\right)\right)\frac{{F}_{P}\left(\mf_{S\mid Z}\left(Z\right)\right)-{F}_{P^\ast}\left(\mf_{S\mid Z}\left(Z\right)\right)}{f_{P}\left(\mf_{S\mid Z}\left(Z\right)\right)}
\right]
\\
\theta_D =&
E\left[
P^\ast\left(\mf_{S\mid Z}\left(Z\right),Z\right)-\mf_{S\mid Z}\left(Z\right)
\right],
\end{align*}
\endgroup
where $\Delta_{U\mid P}\left(p\right) = \frac{d}{dp} E[U|P=p]$ and $\mf_{S\mid Z}\left(z\right) = E[S|Z=z]$ from Section \ref{sec:model}.
In the current setting without $X$, the PRTE consists of the last term in \eqref{eq:PRTE_expression} and can be written as
\begin{align}\label{eq:prte_n_over_d}
PRTE = \theta_N / \theta_D.
\end{align}
We collect the possibly infinite-dimensional nuisance parameters as
\begin{align*}
\gamma=
\left(
\frac{f_{P^\ast}\left({\mf}_{S\mid Z}\right)}{f_{P}\left({\mf}_{S\mid Z}\right)},
\mf_{S\mid Z},
\mf_{U\mid P},
\Delta_{U\mid P}
\right)
\qquad
\text{where $\mf_{U\mid Z}\left(z\right) = E[U|Z=z]$.}
\end{align*}

\subsection{A Step-by-Step Procedure}\label{sec:overview_step_by_step}

Let $L>1$ be a predetermined natural number of folds that is fixed as the sample size increases.
Randomly split the sample into sub-samples $I_1,...,I_L$ of (approximately) equal size, i.e., $|I_\ell| = \lfloor n/L \rfloor$ or $\lfloor n/L \rfloor+1$.
For each $\ell \in L$, estimate $\gamma$ by using the sub-sample $I_\ell^c$, and denote the estimator by $\hat\gamma_\ell$.\footnote{See Appendix \ref{sec:step_by_step_procedure_application} for concrete estimators along with tuning parameter choice rules used for them in our empirical application.}
We then define our double debiased estimator $\hat\theta=\left(\hat\theta_N,\hat\theta_D\right)$ of $\theta$ as the solution to
\begin{align*}
\frac{1}{L} \sum_{\ell=1}^L \frac{1}{|I_\ell|} \sum_{i \in I_\ell} m_N\left(Y_i,Z_i;\hat\theta_N,\hat\gamma_\ell\right) &= 0
\qquad\text{and}\\
\frac{1}{L} \sum_{\ell=1}^L \frac{1}{|I_\ell|} \sum_{i \in I_\ell} m_D\left(Y_i;\hat\theta_D,\hat\gamma_\ell\right) &= 0,
\end{align*}
where $\left(m_N,m_D\right)$ is the orthogonal score defined by
\begingroup
\allowdisplaybreaks
\begin{align}
m_N\left(Y,Z;\tilde\theta_N,\tilde\gamma\right)
=&
\tilde{\mf}_{U\mid P}\left(P^\ast\left(\tilde{\mf}_{S\mid Z}\left(Z\right),Z\right)\right)-Y-\tilde\theta_N
\label{eq:theta_3_a}\\
&+
\frac{\tilde{f}_{P^\ast}\left(\tilde{\mf}_{S\mid Z}\left(Z\right)\right)}{\tilde{f}_{P}\left(\tilde{\mf}_{S\mid Z}\left(Z\right)\right)}\left(Y-\tilde{\mf}_{U\mid P}\left(\tilde{\mf}_{S\mid Z}\left(Z\right)\right)\right)
\label{eq:theta_3_b}\\&+
\left(
\tilde{\Delta}_{U\mid P}\left(P^\ast\left(\tilde{\mf}_{S\mid Z}\left(Z\right),Z\right)\right)\partial P^\ast\left(\tilde{\mf}_{S\mid Z}\left(Z\right),Z\right)
-
\frac{\tilde{f}_{P^\ast}\left(\tilde{\mf}_{S\mid Z}\left(Z\right)\right)}{\tilde{f}_{P}\left(\tilde{\mf}_{S\mid Z}\left(Z\right)\right)}\tilde{\Delta}_{U\mid P}\left(\tilde{\mf}_{S\mid Z}\left(Z\right)\right)
\right)\nonumber
\\
&\times
\left(S-\tilde{\mf}_{S\mid Z}\left(Z\right)\right),
\label{eq:theta_3_c}
\\
m_D\left(Z;\tilde\theta_D,\tilde\gamma\right)
=&
P^\ast\left(\tilde{\mf}_{S\mid Z}\left(Z\right),Z\right)-\tilde{\mf}_{S\mid Z}\left(Z\right)-\tilde\theta_D
\label{eq:theta_2_a}\\
&+
\left(\partial P^\ast\left(\tilde{\mf}_{S\mid Z}\left(Z\right),Z\right)-1\right)
\left(S-\tilde{\mf}_{S\mid Z}\left(Z\right)\right).
\label{eq:theta_2_b}
\end{align}
\endgroup

Plug $\hat\theta$ into \eqref{eq:prte_n_over_d} to in turn obtain an estimator of the PRTE:
$$
\widehat{PRTE} = \hat\theta_N / \hat\theta_D.
$$
This double debiased estimator for the PRTE is root-$n$ asymptotically normal as
$$
\sqrt{n}\left(\widehat{PRTE}-PRTE\right) \rightarrow_d N\left(0,D' \Omega D\right),
$$
where
$
D = \left(1/\theta_D, \ -\theta_N/\theta_D^2\right)'
$
and
$
\Omega = \text{Var}\left(\left( m_N\left(Y,Z;\theta_N,\gamma\right), \ m_D\left(Y;\theta_d,\gamma\right)\right)'\right).
$

\subsection{Intuitions and Discussions}\label{sec:intuitions_discussions}

Having outlined the step-by-step procedure of our proposed method, we next present intuitions behind this proposed method of double debiased estimation and inference for the PRTE as well as some heuristic explanations of why it works.

{\bf On the orthogonal score:}
If we knew the nuisance parameters $(\mf_{S|Z}, \mf_{U|P})$, then we could estimate $\theta$ by using the moment conditions $E[M_N\left(Y,Z;\theta,\gamma\right)]=E[M_D\left(Y,Z;\theta,\gamma\right)]=0$, where 
\begin{align}
M_N\left(Y,Z;\tilde\theta_N,\tilde\gamma\right)=&
\tilde{\mf}_{U\mid P}\left(P^\ast\left(\tilde{\mf}_{S\mid Z}\left(Z\right),Z\right)\right)-Y-\tilde{\theta}_N
\label{eq:momentMN}
\qquad\text{and}\\
M_D\left(Y;\tilde\theta_D,\tilde\gamma\right)=&
P^\ast\left(\tilde{\mf}_{S\mid Z}\left(Z\right),Z\right)-\tilde{\mf}_{S\mid Z}\left(Z\right)-\tilde{\theta}_D.
\label{eq:momentMD}
\end{align}
Since we estimate $(\mf_{S|Z}, \mf_{U|P})$, however, influence function adjustments should be added as in \eqref{eq:theta_3_b}, \eqref{eq:theta_3_c} and \eqref{eq:theta_2_b}.
Lines \eqref{eq:theta_3_a} and \eqref{eq:theta_2_a} precisely correspond to the na\"ive moment functions \eqref{eq:momentMN} and \eqref{eq:momentMD}, respectively.
An influence function adjustment for the estimation of ${\mf}_{U\mid P}$ appears in \eqref{eq:theta_3_b}. 
An estimation of ${\mf}_{U\mid P}\left(P\right)$ entails an influence function adjustment by $Y-{\mf}_{U\mid P}\left({\mf}_{S\mid Z}\left(Z\right)\right)$.
Since the function ${\mf}_{U\mid P}$ is evaluated at $P^\ast\left({\mf}_{S\mid Z}\left(Z\right),Z\right)$ in line \eqref{eq:theta_3_a}, the adjustment is scaled by the coefficient ${{f}_{P^\ast}\left({\mf}_{S\mid Z}\left(Z\right)\right)}/{{f}_{P}\left({\mf}_{S\mid Z}\left(Z\right)\right)}$ as in \eqref{eq:theta_3_b}.
Likewise, an estimation of ${\mf}_{S\mid Z}\left(Z\right)$ entails an influence function adjustment by $S-{\mf}_{S\mid Z}\left(Z\right)$, and it appears in lines \eqref{eq:theta_3_c} and \eqref{eq:theta_2_b}.
Since the function ${\mf}_{S\mid Z}\left(Z\right)$ shows up inside other functions in lines \eqref{eq:theta_3_a} and \eqref{eq:theta_2_a}, the adjustments are scaled by their derivatives in lines \eqref{eq:theta_3_c} and \eqref{eq:theta_2_b}.
Thanks to these influence function adjustments, these moment functions $m_N$ and $m_D$ are robust against local perturbations in $\gamma$, i.e., $m_N$ and $m_D$ constitute an orthogonal score.
This property, together with the cross fitting discussed below, allows the effects of an estimation of $\gamma$ on an estimation of $\theta$ to be asymptotically negligible.

{\bf On the cross fitting:}
Using the same sample to estimate both $\gamma$ and $\theta$ would result in an over-fitting bias.
To circumvent such a bias, we use the sub-sample $I_\ell^c$ to estimate $\gamma$ by $\hat\gamma_\ell$ and use the complementary sub-sample $I_\ell$ to evaluate the orthogonal score, $|I_\ell|^{-1}\sum_{i\in I_\ell} m_N\left(Y_i,Z_i;\theta_N,\hat\gamma_\ell\right)$ and $|I_\ell|^{-1}\sum_{i\in I_\ell} m_D\left(Y_i;\theta_D,\hat\gamma_\ell\right)$, for each $\ell = 1,...,L$.
These lead to the double debiased estimator $\hat\theta$ introduced in Section \ref{sec:overview_step_by_step}.

{\bf On the root-$n$ asymptotic normality:}
Because of the orthogonality property and the cross fitting that allow the effects of the estimation error of $\gamma$ on an estimation of $\theta$ to be asymptotically negligible, $\hat\theta$ enjoys the root-$n$ asymptotic normality
\begin{align*}
\sqrt{n}\left(\hat\theta-\theta\right) \rightarrow_d N\left(0,\Omega\right)
\end{align*}
as if $\gamma$ were known.
The delta method thus yields the root-$n$ asymptotic normality for the PRTE estimator:
$$
\sqrt{n}\left(\widehat{PRTE}-PRTE\right) \rightarrow_d N\left(0,D' \Omega D\right),
$$
again as if $\gamma$ were known.

{\bf Orthogonal Score and Double Robustness:}
The orthogonality property and double robustness are closely related concepts, but neither implies the other -- see \citet{chernozhukov2016locally} for example.
A natural question is whether our orthogonal score also satisfies the double robustness with respect to the nuisance parameters, $f_{P^\ast}\left({\mf}_{S\mid Z}\right)/f_P\left({\mf}_{S\mid Z}\right)$, ${\mf}_{S\mid Z}$, and $\left({\mf}_{U\mid P},\Delta_{U\mid P}\right)$.
It turns out that our orthogonal score does not possess the double robustness property against the propensity score function ${\mf}_{S\mid Z}$ in general.
This is because ${\mf}_{S\mid Z}$ appears in our score inside possibly nonlinear functions -- we show a concrete case in point in Appendix \ref{sec:orthogonal_score_double_robustness}.
On the other hand, given a fixed propensity score function ${\mf}_{S\mid Z}$, our orthogonal score has double robustness between $f_{P^\ast}\left({\mf}_{S\mid Z}\right)/f_P\left({\mf}_{S\mid Z}\right)$ and $\left({\mf}_{U\mid P},\Delta_{U\mid P}\right)$.
This is because the score takes forms of products of affine functions of these nuisance parameters -- see Appendix \ref{sec:orthogonal_score_double_robustness} for details.

{\bf More intuitions:}
Using this discussion on the relation to the double robustness, we now provide more intuitions behind why the orthogonal score works to our goal.
One may wonder why our PRTE estimator converges at the rate of root-$n$ while possibly all the preliminary estimators converge at slower rates.
As demonstrated in Appendix \ref{sec:orthogonal_score_double_robustness}, our orthogonal score after some rewritings takes the form of a product of two estimation errors as in the right-hand side of 
\begin{eqnarray*}
E[m_N\left(Y,Z;\theta_N,\tilde\gamma\right)]
=
\int\left(
\frac{\tilde{f}_{P^\ast}\left(p\right)}{\tilde{f}_{P}\left(p\right)}-\frac{{f}_{P^\ast}\left(p\right)}{{f}_{P}\left(p\right)}\right)\left({\mf}_{U\mid P}\left(p\right)-\tilde{\mf}_{U\mid P}\left(p\right)\right)f_{P}\left(p\right)dp.
\end{eqnarray*}
See Appendix \ref{sec:orthogonal_score_double_robustness} for its derivation.
This shows that, if each of
$
\tilde{f}_{P^\ast}\left( \cdot \right)/\tilde{f}_{P}\left( \cdot \right)-{f}_{P^\ast}\left( \cdot \right)/{f}_{P}\left( \cdot \right)
$
and
$
{\mf}_{U\mid P}\left( \cdot \right)-\tilde{\mf}_{U\mid P}\left( \cdot \right)
$
converges to zero at a rate faster than $n^{1/4}$ (yet slower than $n^{1/2}$), then the product converges at a rate faster than $n^{1/4} \cdot n^{1/4} = n^{1/2}$.
This property of the orthogonality (taking the form of a product in this case) solves the puzzle that the PRTE estimator of our interest can converge at the rate of root-$n$ while convergence rates of the preliminary estimators are often slower.

\section{Orthogonal Score and Double Debiased Estimation}\label{sec:orthogonal_score}

From Theorem \ref{theorem:MTE_LIV}, we can express the PRTE as a function of estimable moments.  
Consider the vector, $\theta=\left(\theta_1',\theta_2',\theta_3\right)'$, of estimable moments defined by 
\begingroup
\allowdisplaybreaks
\begin{align*}
\theta_1
&=
E\left[\xi_1\left(X,Y,\mf_{S\mid Z}\left(Z\right)\right)\right],
&&\left(2p\left(2p+1\right)\text{-dimensional}\right)
\\
\theta_2
&=
E\left[
\left(\mu_0\left(X\right)',\mu_1\left(X\right)',1\right)'\left(P^\ast\left(\mf_{S\mid Z}\left(Z\right),Z\right)-\mf_{S\mid Z}\left(Z\right)\right)
\right],
&&\left(\left(2p+1\right)\text{-dimensional}\right)
\\
\theta_3
&=
E\left[
\Delta_{U\mid P}\left(\mf_{S\mid Z}\left(Z\right)\right)\frac{{F}_{P}\left(\mf_{S\mid Z}\left(Z\right)\right)-{F}_{P^\ast}\left(\mf_{S\mid Z}\left(Z\right)\right)}{f_{P}\left(\mf_{S\mid Z}\left(Z\right)\right)}
\right],
&&\left(\text{1-dimensional}\right)
\end{align*}
\endgroup
where $\mf_{S\mid Z}\left(z\right) = E[S|Z=z]$, 
$\Delta_{U\mid P}\left(p\right) = \frac{d}{dp} E[U|P=p]$,
\begin{align*}
\xi_1\left(x,y,p\right)=\mathrm{vec}
\left(
\left(\begin{array}{c}
\left(1-p\right)\left(\mu_0\left(x\right)-\mf_{\mu_0\left(X\right)\mid P}\left(p\right)\right)\\
p\left(\mu_1\left(x\right)-\mf_{\mu_1\left(X\right)\mid P}\left(p\right)\right)
\end{array}\right)
\left(\begin{array}{c}
\left(1-p\right)\left(\mu_0\left(x\right)-\mf_{\mu_0\left(X\right)\mid P}\left(p\right)\right)\\
p\left(\mu_1\left(x\right)-\mf_{\mu_1\left(X\right)\mid P}\left(p\right)\right)\\
y-\mf_{Y\mid P}\left(p\right)
\end{array}\right)'
\right),
\end{align*}
$\mf_{\mu_0\left(X\right)\mid P}\left(p\right) = E[\mu_0\left(X\right) | P=p]$,
$\mf_{\mu_1\left(X\right)\mid P}\left(p\right) = E[\mu_1\left(X\right) | P=p]$, and 
$\mf_{Y\mid P}\left(p\right) = E[Y | P=p]$.

To express $PRTE$ in (\ref{eq:PRTE_expression}) as a function of $\theta$, we write 
$$
PRTE
=
\Lambda\left(\theta\right)
\equiv
\frac{\theta_{2,1}' \boldsymbol{d}_1\left(\theta_1\right) - \theta_{2,0}' \boldsymbol{d}_0\left(\theta_1\right)+ \theta_3}{\theta_{2,2}},
$$
where 
$\boldsymbol{d}=\left(\boldsymbol{d}_0',\boldsymbol{d}_1'\right)'$ is the function defined by 
$\boldsymbol{d}\left(\mathrm{vec}\left(\mathbf{B},\mathbf{A}\right)\right)={\mathbf{B}^{-1}} {\mathbf{A}}$ for a $2p \times 1$ vector $\mathbf{A}$ and a $2p \times 2p$ matrix $\mathbf{B}$, cf. Equation \eqref{eq:beta_moments} below for the function $\boldsymbol{d}$.

Note that the definition of $\xi_1$ comes from the moment condition for $\left(\beta_0',\beta_1'\right)'$.
Specifically, as in \citet{robinson1988root}, the parameter vector $\left(\beta_0',\beta_1'\right)'$ can be written as 
\begin{align}
\left(\beta_0',\beta_1'\right)'
&=
\left(\boldsymbol{d}_0\left(E\left[\xi_1\left(X,Y,\mf_{S\mid Z}\left(Z\right)\right)\right]\right)',\boldsymbol{d}_1\left(E\left[\xi_1\left(X,Y,\mf_{S\mid Z}\left(Z\right)\right)\right]\right)'\right)'
\label{eq:beta_moments}
\\
&=
E\left[
\left(\begin{array}{c}
\left(1-P\right)\left(\mu_0\left(X\right)-\mf_{\mu_0\left(X\right)\mid P}\left(P\right)\right)\\
P\left(\mu_1\left(X\right)-\mf_{\mu_1\left(X\right)\mid P}\left(P\right)\right)
\end{array}\right)
\left(\begin{array}{c}
\left(1-P\right)\left(\mu_0\left(X\right)-\mf_{\mu_0\left(X\right)\mid P}\left(P\right)\right)\\
P\left(\mu_1\left(X\right)-\mf_{\mu_1\left(X\right)\mid P}\left(P\right)\right)
\end{array}\right)'\right]^{-1}
\nonumber
\\& \qquad\times
E\left[\left(\begin{array}{c}
\left(1-P\right)\left(\mu_0\left(X\right)-\mf_{\mu_0\left(X\right)\mid P}\left(P\right)\right)\\
P\left(\mu_1\left(X\right)-\mf_{\mu_1\left(X\right)\mid P}\left(P\right)\right)
\end{array}\right)\left(Y-\mf_{Y\mid P}\left(P\right)\right)\right].
\nonumber
\end{align}

Recall the notation $W = \left(Y,S,X',Z'\right)'$ from Section \ref{sec:model}.
With these definitions and notations, we now propose an orthogonal score function of the form
\begin{align*}
m\left(W;\tilde{\theta},\tilde{\gamma}\right)=\left(\begin{array}{ccccc}m_1\left(W;\tilde{\theta},\tilde{\gamma}\right)'& m_2\left(W;\tilde{\theta},\tilde{\gamma}\right)'& m_3\left(W;\tilde{\theta},\tilde{\gamma}\right)'\end{array}\right)'
\end{align*}
with all the preliminary parameters collected in the concise notation
\begin{align*}
\gamma=
\left(
\frac{f_{P^\ast}\left({\mf}_{S\mid Z}\right)}{f_{P}\left({\mf}_{S\mid Z}\right)},
\mf_{S\mid Z},
\xi_1,
\zeta,
\mf_{U\mid P},
\Delta_{U\mid P}
\right),
\end{align*}
where 
$\zeta\left(z\right)=E\left[\left.\left.\frac{\partial}{\partial p}
\xi_1\left(X,Y,p\right)\right|_{p=\mf_{S\mid Z}\left(z\right)}
\right\vert Z=z\right]$.
The components of the orthogonal score function are defined by 
\begingroup
\allowdisplaybreaks
\begin{align}
m_1\left(W;\tilde{\theta},\tilde{\gamma}\right)
=&
\tilde{\xi}_1\left(X,Y,\tilde{\mf}_{S\mid Z}\left(Z\right)\right)
-\tilde{\theta}_1
\label{eq:m_1_a}\\
&+
\tilde{\zeta}\left(Z\right)\left(S-\tilde{\mf}_{S\mid Z}\left(Z\right)\right)
\label{eq:m_1_b}
\\
m_2\left(W;\tilde{\theta},\tilde{\gamma}\right)
=&
\left(\mu_0\left(X\right)',\mu_1\left(X\right)',1\right)'\left(P^\ast\left(\tilde{\mf}_{S\mid Z}\left(Z\right),Z\right)-\tilde{\mf}_{S\mid Z}\left(Z\right)\right)-\tilde{\theta}_2
\label{eq:m_2_a}\\
&+
\left(\mu_0\left(X\right)',\mu_1\left(X\right)',1\right)'\left(\partial P^\ast\left(\tilde{\mf}_{S\mid Z}\left(Z\right),Z\right)-1\right)
\left(S-\tilde{\mf}_{S\mid Z}\left(Z\right)\right)
\label{eq:m_2_b}
\\
m_3\left(W;\tilde{\theta},\tilde{\gamma}\right)
=&
\tilde{\mf}_{U\mid P}\left(P^\ast\left(\tilde{\mf}_{S\mid Z}\left(Z\right),Z\right)\right)-\mathcal{U}\left(W,\tilde{\theta}\right)-\tilde{\theta}_3
\label{eq:m_3_a}\\
&+
\frac{\tilde{f}_{P^\ast}\left(\tilde{\mf}_{S\mid Z}\left(Z\right)\right)}{\tilde{f}_{P}\left(\tilde{\mf}_{S\mid Z}\left(Z\right)\right)}\left(\mathcal{U}\left(W,\tilde{\theta}\right)-\tilde{\mf}_{U\mid P}\left(\tilde{\mf}_{S\mid Z}\left(Z\right)\right)\right)
\label{eq:m_3_b}\\&+
\left(
\tilde{\Delta}_{U\mid P}\left(P^\ast\left(\tilde{\mf}_{S\mid Z}\left(Z\right),Z\right)\right)\partial P^\ast\left(\tilde{\mf}_{S\mid Z}\left(Z\right),Z\right)
-
\frac{\tilde{f}_{P^\ast}\left(\tilde{\mf}_{S\mid Z}\left(Z\right)\right)}{\tilde{f}_{P}\left(\tilde{\mf}_{S\mid Z}\left(Z\right)\right)}\tilde{\Delta}_{U\mid P}\left(\tilde{\mf}_{S\mid Z}\left(Z\right)\right)
\right)\nonumber
\\
&\times
\left(S-\tilde{\mf}_{S\mid Z}\left(Z\right)\right),
\label{eq:m_3_c}
\end{align}
\endgroup
where $\mathcal{U}\left(W,\theta\right)=Y-\left(1-S\right)\mu_0\left(X\right)'\beta_0-S\mu_1\left(X\right)'\beta_1$ and $\partial P^\ast\left(p,z\right)=\frac{\partial}{\partial p}P^\ast\left(p,z\right)$.
See Appendix \ref{sec:decomposition_of_moment_functions} for an alternative representation of $m\left(W;\theta,\gamma\right)$

This score function $m$ is \textit{orthogonal} in the sense that 
\begin{equation}\label{eq:orthogonality}
\frac{\partial}{\partial r}\int m\left(w;\theta,\gamma+r\left(\breve{\gamma}-\gamma\right)\right)F_W\left(dw\right)|_{r=0}=0
\qquad
\text{for all } \breve{\gamma} \in \Gamma.
\end{equation}
See Appendix \ref{sec:eq:orthogonality} for a derivation of this property as well as the definition of $\Gamma$.
To arrive at the orthogonal score, we take advantage of two convenient features of $\theta$ and $\gamma$.
First, all the nuisance parameters $\gamma$ take forms of mean-square projections or densities.
Second, the nuisance parameters $\gamma$ enter the equations for $\theta$ through integrals and derivatives, and therefore the Gateaux derivative operator (and thus the orthogonalization operator) can directly act on $\gamma$.
These two features allow us to obtain the adjustment terms using the result of \citet[][Proposition 4]{newey1994asymptotic}.

The orthogonality \eqref{eq:orthogonality} allows for the score to be insensitive to local perturbations $\breve{\gamma}$ of $\gamma$.
In particular, the first-order effect of the estimation error $\hat\gamma-\gamma$ is zero, and the remaining effects are of a smaller order:
$$
\int m\left(w;\theta,\hat{\gamma}\right)F_W\left(dw\right)=o_p\left(n^{-1/2}\right).
$$
In other words, the effects of the estimation error $\hat{\gamma}_\ell - \gamma$ of preliminary parameters are asymptotically negligible relative to the convergence rate of the empirical mean, which is of order $n^{-1/2}$.
Although we postpone rigorous discussions until Section \ref{sec:large_sample} and proofs of the main theorem therein, we here discuss intuitions behind the orthogonality property.
If we knew the true functions ${\xi}_1$, ${\mf}_{U\mid P}$, and ${\mf}_{S\mid Z}$, then we would use the moment conditions    
$$
E\left[
\left(
\begin{array}{c}
{\xi}_1\left(X,Y,{\mf}_{S\mid Z}\left(Z\right)\right)-\tilde{\theta}_1\\
\left(\mu_0\left(X\right)',\mu_1\left(X\right)',1\right)'\left(P^\ast\left({\mf}_{S\mid Z}\left(Z\right),Z\right)-{\mf}_{S\mid Z}\left(Z\right)\right)-\tilde{\theta}_2\\
{\mf}_{U\mid P}\left(P^\ast\left({\mf}_{S\mid Z}\left(Z\right),Z\right)\right)-\mathcal{U}\left(W,\tilde{\theta}\right)-\tilde{\theta}_3\\
\end{array}
\right)
\right]
=0,
$$ 
which correspond to line \eqref{eq:m_1_a} for $m_1$, line \eqref{eq:m_2_a} for $m_2$, and  line \eqref{eq:m_3_a} for $m_3$.
However, since we estimate ${\xi}_1$, ${\mf}_{U\mid P}$, and ${\mf}_{S\mid Z}$, 
influence function adjustments need to be added for these estimators.

First, the influence function adjustment for the estimated ${\xi}_1$ is zero. 
This is because, as is well known in the literature on partial linear models \citep{robinson1988root}, the moment condition $E[{\xi}_1\left(X,Y,{\mf}_{S\mid Z}\left(Z\right)\right)-\tilde{\theta}_1]=0$ satisfies the orthogonality property with respect to ${\xi}_1$.  
Second, the influence function adjustment to account for an estimation of ${\mf}_{U\mid P}$ appears in \eqref{eq:m_3_b}. 
An estimation of ${\mf}_{U\mid P}\left(P\right)$ entails an influence function adjustment by $\mathcal{U}\left(W,{\theta}\right)-{\mf}_{U\mid P}\left({\mf}_{S\mid Z}\left(Z\right)\right)$. 
Since the function ${\mf}_{U\mid P}$ is evaluated at $P^\ast\left({\mf}_{S\mid Z}\left(Z\right),Z\right)$ in line \eqref{eq:m_3_a} for $m_3$, this adjustment is scaled by the coefficient ${{f}_{P^\ast}\left({\mf}_{S\mid Z}\left(Z\right)\right)}/{{f}_{P}\left({\mf}_{S\mid Z}\left(Z\right)\right)}$ as in \eqref{eq:m_3_b}.
Third, the influence function adjustments for the estimated ${\mf}_{S\mid Z}$ appear in line \eqref{eq:m_1_b} for $m_1$, line \eqref{eq:m_2_b} for $m_2$, and line \eqref{eq:m_3_c} for $m_3$. 
An estimation of ${\mf}_{S\mid Z}\left(Z\right)$ entails an influence function adjustment by $S-{\mf}_{S\mid Z}\left(Z\right)$. 
Since the function ${\mf}_{S\mid Z}$ appears inside other functions in lines \eqref{eq:m_1_a}, \eqref{eq:m_2_a}, \eqref{eq:m_3_a} and \eqref{eq:m_3_b}, these adjustments are scaled by their derivatives of the respective functions, which result in the coefficients in front of  $S-{\mf}_{S\mid Z}\left(Z\right)$ in lines \eqref{eq:m_1_b}, \eqref{eq:m_2_b} and \eqref{eq:m_3_c}. 
Note that all these adjustments take the form prescribed by \citet[][Proposition 4]{newey1994asymptotic}.
This is because, as mentioned earlier, our moment functions share two convenient properties.
First, $\gamma$ takes forms of mean-square projections or densities.
Second, $\gamma$ enter the equations for $\theta$ so that the Gateaux derivative operator (and thus the orthogonalization operator) can directly act on $\gamma$.
Formal mathematical analyses rationalizing these intuitions are found in Appendix \ref{sec:orthogonality_lemma}.

Given a random sample $\{W_i\}_{i=1}^n$ of size $n$, we now use the above orthogonal score to construct a double debiased estimator with cross fitting or sample splitting.
Let $L > 1$ be a natural number of folds, and randomly partition the sample index set $\{1,...,n\}$ into $L > 1$ subsets $I_1,...,I_L$ of approximately equal size.
For every subsample index $\ell \in \{1,...,L\}$, let
$$\hat{\gamma}_\ell=
\left(
\frac{\hat{f}_{P^\ast}\left(\hat{\mf}_{S\mid Z}\right)}{\hat{f}_{P}\left(\hat{\mf}_{S\mid Z}\right)},
\hat{\mf}_{S\mid Z},
\hat{\xi}_1,
\hat{\zeta},
\hat{\mf}_{U\mid P},
\hat{\Delta}_{U\mid P}
\right)$$ 
denote a preliminary parameter estimate obtained by using all observations $i \in \{1,...,n\} \backslash I_\ell$.\footnote{For the components of the preliminary parameter estimate $\hat{\gamma}_\ell$, we omit $\ell$ for the sake of notational simplicity.}
We define our double debiased estimator $\hat\theta$ of $\theta$ as the solution to 
\begin{align}\label{eq:double_debiased_theta}
\frac{1}{L} \sum_{\ell=1}^L \frac{1}{|I_\ell|} \sum_{i \in I_\ell} m\left(W_i; \hat\theta , \hat{\gamma}_\ell\right) = 0.
\end{align}
Accordingly, our double debiased estimator $\widehat{PRTE}$ of the PRTE is defined by 
\begin{align}\label{eq:double_debiased_prte}
\widehat{PRTE} = \Lambda\left( \hat\theta \right).
\end{align}

\section{Large Sample Theory for Inference}\label{sec:large_sample}

In this section, we investigate the large sample theory for the double debiased estimators, $\hat\theta$ and $\widehat{PRTE}$, defined in \eqref{eq:double_debiased_theta} and \eqref{eq:double_debiased_prte}, respectively.
To this end, we formally present and discuss relevant assumptions.
First, we consider a counterfactual policy $P^\ast$ satisfying the following conditions.

\begin{assumption}\label{assn:pstar}
The support of $P^\ast$ is a subset of the support of $P$. 
The cumulative distribution function of $P^\ast$ is absolutely continuous with the Lebesgue measure. 
\end{assumption}

Recall that many treatment parameters can be represented in terms of the MTE in similar manners to the PRTE -- see \citet{heckman/vytlacil:2005}.
Assumption \ref{assn:pstar} rules out some of the parameters such as $ATT$, $ATU$, $ATE$, $LATE\left(x\right)$, and $TUT\left(x\right)$.

Second, assume that all the preliminary parameter components of the orthogonal moment function are identified in the following sense.

\begin{assumption}\label{Assn_identification}
$\gamma$ is identifiable, and the following matrix is invertible:
$$
E\left[
\left(\begin{array}{c}
\left(1-P\right)\left(\mu_0\left(X\right)-\mf_{\mu_0\left(X\right)\mid P}\left(P\right)\right)\\
P\left(\mu_1\left(X\right)-\mf_{\mu_1\left(X\right)\mid P}\left(P\right)\right)
\end{array}\right)
\left(\begin{array}{c}
\left(1-P\right)\left(\mu_0\left(X\right)-\mf_{\mu_0\left(X\right)\mid P}\left(P\right)\right)\\
P\left(\mu_1\left(X\right)-\mf_{\mu_1\left(X\right)\mid P}\left(P\right)\right)
\end{array}\right)'
\right].
$$
\end{assumption}
The invertibility condition serves to identify the parameter vector $\left(\beta_0',\beta_1'\right)'$ in the partial linear model as in \citet{carneiro/lee:2009}.
The requirement for $\gamma$ to be identified essentially boils down to the identification of the density functions and the conditional expectation functions, which is satisfied for a large class of data generating processes, and is usually taken for granted in the nonparametrics literature.

We next make three high-level conditions (Assumptions \ref{Assn_slow_conver}--\ref{Assn:Taylor_Errr}) regarding large sample behaviors of the preliminary parameter estimators.
While we stress that each of these three assumptions is a mild requirement for most common nonparametric estimators, we supplement each of the three non-primitive assumptions by lower-level sufficient conditions in Appendix \ref{sec:lowe_level_sufficient_conditions}.
\begin{assumption}\label{Assn_slow_conver}
For every $\hat{\gamma}_\ell=\hat{\gamma}_1,\ldots,\hat{\gamma}_L$, the following objects are $o_p(1)$.\footnotesize
\begingroup
\allowdisplaybreaks
\begin{align*}
&\int \left(\hat{\zeta}\left(z\right)-{\zeta}\left(z\right)\right)\left(\hat{\mf}_{S\mid Z}\left(z\right)-{\mf}_{S\mid Z}\left(z\right)\right)F_W\left(dw\right).
\\
&\int\frac{\hat{f}_{P^\ast}\left(\hat{\mf}_{S\mid Z}\left(z\right)\right)}{\hat{f}_{P}\left(\hat{\mf}_{S\mid Z}\left(z\right)\right)}\left(\hat{\Delta}_{U\mid P}\left(\hat{\mf}_{S\mid Z}\left(z\right)\right)-{\Delta}_{U\mid P}\left({\mf}_{S\mid Z}\left(z\right)\right)
\left({\mf_{S\mid Z}}\left(z\right)-\hat{\mf}_{S\mid Z}\left(z\right)\right)\right)F_W\left(dw\right)
\\
&\int \left(\hat{\Delta}_{U\mid P}\left(P^\ast\left(\hat{\mf}_{S\mid Z}\left(z\right),z\right)\right)\partial P^\ast\left(\hat{\mf}_{S\mid Z}\left(z\right),z\right)-{\Delta}_{U\mid P}\left(P^\ast\left({\mf}_{S\mid Z}\left(z\right),z\right)\right)\partial P^\ast\left({\mf}_{S\mid Z}\left(z\right),z\right)\right)
\left({\mf_{S\mid Z}}\left(z\right)-\hat{\mf}_{S\mid Z}\left(z\right)\right)F_W\left(dw\right).
\end{align*}
\endgroup\normalsize
\end{assumption}
Assumption \ref{Assn_slow_conver} requires that some preliminary parameter estimators to have the $n^{1/4}$ rate of convergence.
This slow rate of convergence suffices because of the orthogonal score we developed and proposed in Section \ref{sec:orthogonal_score} -- see the intuitions and discussions in Section \ref{sec:intuitions_discussions}.
As far as this condition is satisfied, the asymptotic distribution of our parameters of interest, namely $\hat\theta$ and $\widehat{PRTE}$, will not be affected by estimation errors of these preliminary parameter estimates.
Furthermore, this condition can be satisfied by common nonparametric estimators of density and conditional expectation functions.
For completeness, we show lower-level sufficient conditions for Assumption \ref{Assn_slow_conver} in Appendix \ref{sec:lowe_level_sufficient_conditions}.

\begin{assumption}\label{Assn_CV}
The following objects are $o_p(1)$.\footnotesize
\begingroup
\allowdisplaybreaks
\begin{align*}
&\int 
\left(\hat{\mf}_{U\mid P}\left(P^\ast\left(\hat{\mf}_{S\mid Z}\left(z\right),z\right)\right)-{\mf}_{U\mid P}\left(P^\ast\left(\hat{\mf}_{S\mid Z}\left(z\right),z\right)\right)\right)
F_W\left(dw\right)
-\int
\frac{\hat{f}_{P^\ast}\left(\hat{\mf}_{S\mid Z}\left(z\right)\right)}{\hat{f}_{P}\left(\hat{\mf}_{S\mid Z}\left(z\right)\right)}
\left(\hat{\mf}_{U\mid P}\left(\hat{\mf}_{S\mid Z}\left(z\right)-{\mf}_{U\mid P}\left(\hat{\mf}_{S\mid Z}\left(z\right)\right)\right)\right)
F_W\left(dw\right)
\\
&\int \left(\hat{\xi}_1\left(x,y,\hat{\mf}_{S\mid Z}\left(z\right)\right)-{\xi}_1\left(x,y,\hat{\mf}_{S\mid Z}\left(z\right)\right)\right)F_W\left(dw\right)
\end{align*}
\endgroup\normalsize
\end{assumption}
We show lower-level sufficient conditions for Assumption \ref{Assn_CV} in Appendix \ref{sec:lowe_level_sufficient_conditions}.

The final piece of assumptions is to control higher-order effects of the propensity score estimation on the orthogonal score.
Note that the orthogonality property \eqref{eq:orthogonality} forces the first-order effects of nuisance parameter estimation to be zero.
Since all the nuisance parameters except for the propensity score function $\mf_{S|Z}$ appear in our score in a linear manner, the orthogonality property precisely vanishes their estimation effects.
On the other hand, since the propensity score function $\mf_{S|Z}$ appears in our score nonlinearly in general, we need to make sure that the remaining higher-order effects are small enough.
Specifically, we require the following condition.

\begin{assumption}\label{Assn:Taylor_Errr}
$\int\boldsymbol{R}_k\left(z\right)F_Z\left(dz\right)=o_p\left(n^{-1/2}\right)$ for every $\hat{\gamma}_\ell$ and every $k=1,\ldots,4$, where \footnotesize 
\begingroup
\allowdisplaybreaks
\begin{eqnarray*}
\boldsymbol{R}_1\left(z\right)&=&
\int\left({\xi}_1\left(x,y,\hat{\mf}_{S\mid Z}\left(z\right)\right)-{\xi}_1\left(x,y,{\mf}_{S\mid Z}\left(z\right)\right)\right)F_{\left(Y,X\right)\mid Z=z}\left(dy,dx\right)-{\zeta}\left(z\right)\left(\hat{\mf}_{S\mid Z}\left(z\right)-{\mf}_{S\mid Z}\left(z\right)\right)
\\
\boldsymbol{R}_2\left(z\right)&=&
\mf_{\left(\mu_0\left(X\right)',\mu_1\left(X\right)',1\right)'\mid Z}\left(z\right)\left(P^\ast\left(\hat{\mf}_{S\mid Z}\left(z\right),z\right)-P^\ast\left({\mf_{S\mid Z}}\left(z\right),z\right)-\partial P^\ast\left(\hat{\mf}_{S\mid Z}\left(z\right),z\right)
\left(\hat{\mf}_{S\mid Z}\left(z\right)-\mf_{S\mid Z}\left(z\right)\right)\right)
\\
\boldsymbol{R}_3\left(z\right)&=&
{\mf}_{U\mid P}\left(P^\ast\left(\hat{\mf}_{S\mid Z}\left(z\right),z\right)\right)-{\mf_{U\mid P}}\left(P^\ast\left({\mf_{S\mid Z}}\left(z\right),z\right)\right)
-
{\Delta}_{U\mid P}\left(P^\ast\left({\mf}_{S\mid Z}\left(z\right),z\right)\right)\partial P^\ast\left({\mf}_{S\mid Z}\left(z\right),z\right)
\left(\hat{\mf}_{S\mid Z}\left(z\right)-{\mf}_{S\mid Z}\left(z\right)\right)
\\
\boldsymbol{R}_4\left(z\right)&=&
\frac{\hat{f}_{P^\ast}\left(\hat{\mf}_{S\mid Z}\left(z\right)\right)}{\hat{f}_{P}\left(\hat{\mf}_{S\mid Z}\left(z\right)\right)}
\left({\mf_{U\mid P}}\left({\mf_{S\mid Z}}\left(z\right)\right)-{\mf}_{U\mid P}\left(\hat{\mf}_{S\mid Z}\left(z\right)\right)
-{\Delta}_{U\mid P}\left({\mf}_{S\mid Z}\left(z\right)\right)
\left({\mf_{S\mid Z}}\left(z\right)-\hat{\mf}_{S\mid Z}\left(z\right)\right)\right)
\end{eqnarray*}
\endgroup\normalsize
\end{assumption}

Since higher-order effects are usually of smaller magnitudes, Assumption \ref{Assn:Taylor_Errr} is a quite mild condition. 
We present lower-level sufficient conditions for Assumption \ref{Assn:Taylor_Errr} in Appendix \ref{sec:lowe_level_sufficient_conditions}.

In addition to these conditions, we also present regularity conditions (Assumptions \ref{Assn_compact}, \ref{Assn_bounded_moments} and \ref{Assn_just_consis}) in Appendix \ref{sec:regularity}.
These additional conditions are relegated to the appendix for the sake of readability, as they are even more standard, even milder, and are even easier to verify with common nonparametric estimators than Assumptions \ref{Assn_slow_conver}--\ref{Assn:Taylor_Errr}.

We obtain the following asymptotic normality result for the double debiased estimator $\hat\theta$ of the intermediate parameter vector.
\begin{theorem}\label{theorem:normal}
If Assumptions \ref{assn:pstar}, \ref{Assn_identification}, \ref{Assn_slow_conver}, \ref{Assn:Taylor_Errr}, \ref{Assn_compact}, \ref{Assn_bounded_moments} and \ref{Assn_just_consis} are satisfied, then
$$
\sqrt{n}\left(\hat\theta-\theta\right)\rightarrow_dN\left(0,\left(\mathcal{M}'\mathcal{M}\right)^{-1}\mathcal{M}'E[m\left(W;{\theta},{\gamma}\right)m\left(W;{\theta},{\gamma}\right)']\mathcal{M}\left(\mathcal{M}'\mathcal{M}\right)^{-1}\right),
$$ 
where, with $m_{3,2}$ defined in Appendix and \ref{sec:decomposition_of_moment_functions}, the matrix $\mathcal{M}$ takes the form
$$
\mathcal{M}=\left(
\begin{array}{ccc}
I&0&0\\
0&I&0\\
E\left[m_{3,2}\left(W;{\gamma}\right)\right]\frac{\partial}{\partial\theta_1'}\boldsymbol{d}\left(\theta_1\right)&0&I
\end{array}
\right).
$$\end{theorem}

A proof is provided in Appendix \ref{sec:theorem:normal}.
An immediate consequence of this result through the delta method is the following asymptotic normality result for the double debiased estimator $\widehat{PRTE}$ of the PRTE.

\begin{corollary}\label{cor:clt}
If Assumptions \ref{assn:pstar}, \ref{Assn_identification}, \ref{Assn_slow_conver}, \ref{Assn:Taylor_Errr}, \ref{Assn_compact}, \ref{Assn_bounded_moments} and \ref{Assn_just_consis} are satisfied, then
$$
\sqrt{n}\left(\widehat{PRTE}-PRTE\right)\rightarrow_dN\left(0,\lambda\left(\theta\right)\left(\mathcal{M}'\mathcal{M}\right)^{-1}\mathcal{M}'E[m\left(W;{\theta},{\gamma}\right)m\left(W;{\theta},{\gamma}\right)']\mathcal{M}\left(\mathcal{M}'\mathcal{M}\right)^{-1}\lambda\left(\theta\right)'\right).
$$
where $\lambda$ denotes the derivative of $\Lambda$.
\end{corollary}

As we emphasized throughout this paper, this root-$n$ convergence without any influence of preliminary estimation errors $\hat\gamma-\gamma$ is possible as a result of the orthogonality that implies \eqref{eq:orthogonality}.
This method can accommodate a wide array of preliminary estimation techniques including kernel-smoothing, sieve estimation, and shrinkage methods among others.
This flexibility in the choice of preliminary estimation approaches at the current high-level theory is another advantage of using the orthogonal score, unlike conventional methods in absence of orthogonal scores.
A drawback of this root-$n$ asymptotic normality result is that it is not guaranteed to be efficient.

We can estimate every component of the asymptotic variance for $\widehat{PRTE}$ as follows. 
First, we can estimate $\lambda\left(\theta\right)$ by $\lambda\left(\hat\theta\right)$ since $\lambda$ is a known function. 
Second, estimate $\mathcal{M}$ by 
$$
\hat{\mathcal{M}}=\left(
\begin{array}{ccc}
I&0&0\\
0&I&0\\
 \frac{1}{n}\sum_{i=1}^nm_{3,2}\left(W_i;\hat{\gamma}\right)\frac{\partial}{\partial\theta_1'}\boldsymbol{d}\left(\hat\theta_1\right)&0&I
\end{array}
\right).
$$
Last, estimate $E[m\left(W;{\theta},{\gamma}\right)m\left(W;{\theta},{\gamma}\right)']$ by
\begin{align*}
\hat\Sigma = \frac{1}{n}\sum_{i=1}^n
\left[\begin{array}{c}
\left(\begin{array}{c}
\hat m_1\left(W_i;\hat{\theta},\hat{\gamma}\right)\\\hat m_2\left(W_i;\hat{\theta},\hat{\gamma}\right)\\\hat m_3\left(W_i;\hat{\theta},\hat{\gamma}\right)
\end{array}\right)
\left(\begin{array}{c}
\hat m_1\left(W_i;\hat{\theta},\hat{\gamma}\right)\\\hat m_2\left(W_i;\hat{\theta},\hat{\gamma}\right)\\\hat m_3\left(W_i;\hat{\theta},\hat{\gamma}\right)
\end{array}\right)'
\end{array}\right].
\end{align*}
The following theorem guarantees the asymptotic validity of the resultant variance estimator.

\begin{theorem}\label{theorem:var_est}
If Assumptions \ref{assn:pstar}, \ref{Assn_identification}, \ref{Assn_slow_conver}, \ref{Assn:Taylor_Errr}, \ref{Assn_compact}, \ref{Assn_bounded_moments} and \ref{Assn_just_consis} are satisfied, then
$$
\lambda\left(\hat\theta\right)\left(\hat{\mathcal{M}}'\hat{\mathcal{M}}\right)^{-1}\hat{\mathcal{M}}'\hat\Sigma\hat{\mathcal{M}}\left(\hat{\mathcal{M}}'\hat{\mathcal{M}}\right)^{-1}\lambda\left(\hat\theta\right)'
$$ 
is a consistent estimator of the asymptotic variance for $\widehat{PRTE}$. 
\end{theorem}
A proof is provided in Appendix \ref{sec:theorem:variance_est}.

\section{Monte Carlo Simulations}\label{sec:simulations}
In this section, we use Monte Carlo simulations to evaluate finite sample performance of the proposed method of double debiased estimation and inference.
We first consider a benchmark design from the literature in Section \ref{sec:benchmark_design}, and demonstrate that even completely nonparametric preliminary estimation can produce desirable finite-sample performance thanks to the orthogonal score. 
We second extend the design with higher dimensions of $Z$ in Section \ref{sec:extended_designs}.

\subsection{Nonparametric Preliminary Estimation under a Benchmark Design}\label{sec:benchmark_design}
We generate artificial datasets from the same distribution as in \citet[][supplementary Appendix D]{carneiro/lokshin/umapathi:2017}.
Specifically, we generate 
$$
\left(U_0,U_1,U_S\right)=\left(-0.050 \varepsilon_1 + 0.020 \varepsilon_3, 0.012 \varepsilon_1 + 0.010 \varepsilon_2,-1.000 \varepsilon_1\right),
$$
where $\varepsilon_1$ and $\varepsilon_2$ are independent standard normal random variables.
Then, we generate
\begingroup
\allowdisplaybreaks
\begin{align*}
Y_1 =& 0.240 + 0.800 X_1 + 0.400 X_2  + U_1
\\
Y_0 =& 0.020 + 0.500 X_1 + 0.100 X_2 + U_0,
\qquad\text{and}\\
S =& \mathbbm{1}\{ {0.200 + 0.300 Z_1 + 0.100 Z_2} - U_S > 0 \},
\end{align*}
\endgroup
where $X_1 \sim N\left(-2,2^2\right)$, $X_2 \sim N\left(2,2^2\right)$, $Z_1 \sim N\left(-1,3^2\right)$, and $Z_2 \sim N\left(1,3^2\right)$ are mutually independent and are also independent of $\left(U_1,U_0,U_S\right)'$.
The observed outcome is generated in turn as $Y = SY_1 + \left(1-S\right)Y_0$.
We consider two forms of policy changes.
The first takes the form of $P^\ast = P + a\left(1-P\right)$, where $a \in \{0.1,...,0.9\}$.\footnote{Note that $PRTE \rightarrow MPRTE$ \citep[marginal PRTE,][]{carneiro/heckman/vytlacil:2010} as $a \rightarrow 0$ and $PRTE \rightarrow ATU$ as $a \rightarrow 1$.}
The second takes the form of $Z^\ast = Z + a$, where $a \in \{-0.5,...,-0.1,0.1,...,0.5\}$.
See Appendix \ref{sec:details_simulation_setting} for additional details about this simulation setting, including implied analytic expressions for the counterfactual distribution, marginal treatment effects, and the PRTEs.

For each instance of artificial datasets, we estimate the PRTE and its estimated standard error using our proposed method of double debiased estimation and inference with completely nonparametric preliminary estimation.
See Appendix \ref{sec:details_simulation_estimation_p} and \ref{sec:details_simulation_estimation_z} for additional details about concrete estimation and inference procedures.
We experiment with alternative numbers, $L=5$ and $10$, of folds in cross fitting, where the number of observations in each fold is equal.
We report the simulated bias, root mean square error, and coverage frequencies for the nominal probability of 95\%.
Simulated coverage frequencies are computed based on the standard symmetric confidence interval generated by the PRTE estimate and its estimated standard error according to Corollary \ref{cor:clt}.
The number of Monte Carlo iterations is set to 1000 following \citet[][supplementary Appendix D]{carneiro/lokshin/umapathi:2017}.
Since there is no existing limit distribution theory for inference about the PRTE to our best knowledge, we do not have any benchmark in the literature against which to make comparisons of our results.
We therefore present simulation results only for our proposed method.

Tables \ref{tab:simulation_results_p} and \ref{tab:simulation_results_z} summarize simulation results for policy changes of the forms $P^\ast = P + a\left(1-P\right)$ and $Z^\ast = Z + a$, respectively.
The results concerning the mean and bias demonstrate that the proposed double debiased estimator indeed produces small biases (relative to the root mean square error) even in small samples.
In light of these relatively small biases, the results for the root mean square error demonstrate that the estimator converges approximately at the rate of $n^{-1/2}$, consistently with our theory.
The results concerning the coverage frequencies demonstrate that our limit normal distribution results are useful to construct asymptotically valid confidence intervals for the PRTE, except when $a$ is infinitesimal.
The alternative numbers, $L=5$ and $10$, of folds for sample splitting entail quite similar simulation results in terms of all of the bias, root mean square error, and coverage frequencies.
In summary, the proposed estimator enjoys the main properties suggested in this paper even in small samples; namely, it is debiased, converges at the parametric rate, and is asymptotically normal with the proposed asymptotic variance formula, regardless of the number of folds in sample splitting.

\begin{table}
	\centering
	\scalebox{1.00}{
		\begin{tabular}{cccccccc}
			\hline\hline
			    &      &     & True    & \multicolumn{3}{c}{Estimates}& 95\%\\
			\cline{5-7}
			$a$ & $n$  & $L$ & PRTE    & Mean    & Bias    & RMSE    & Coverage\\
			\hline
			0.1 & 1000 & 5   & 0.243 & 0.324 & 0.081 & 0.889 & 0.828\\
					& 2000 & 5   & 0.243 & 0.298 & 0.054 & 0.540 & 0.828\\
			\hline
			0.2 & 1000 & 5   & 0.237 & 0.309 & 0.073 & 0.516 & 0.863\\
					& 2000 & 5   & 0.237 & 0.285 & 0.048 & 0.301 & 0.919\\
			\hline
			0.3 & 1000 & 5   & 0.230 & 0.320 & 0.090 & 0.420 & 0.901\\
					& 2000 & 5   & 0.230 & 0.278 & 0.048 & 0.241 & 0.957\\
			\hline
			0.4 & 1000 & 5   & 0.225 & 0.298 & 0.073 & 0.376 & 0.912\\
					& 2000 & 5   & 0.225 & 0.272 & 0.047 & 0.218 & 0.947\\
			\hline
			0.5 & 1000 & 5   & 0.219 & 0.287 & 0.068 & 0.347 & 0.922\\
					& 2000 & 5   & 0.219 & 0.256 & 0.037 & 0.205 & 0.957\\
			\hline
			0.6 & 1000 & 5   & 0.213 & 0.273 & 0.060 & 0.323 & 0.928\\
					& 2000 & 5   & 0.213 & 0.247 & 0.034 & 0.200 & 0.951\\
			\hline
			0.7 & 1000 & 5   & 0.207 & 0.259 & 0.052 & 0.312 & 0.923\\
					& 2000 & 5   & 0.207 & 0.232 & 0.025 & 0.195 & 0.949\\
			\hline
			0.8 & 1000 & 5   & 0.201 & 0.245 & 0.044 & 0.306 & 0.908\\
					& 2000 & 5   & 0.201 & 0.219 & 0.018 & 0.195 & 0.935\\
			\hline
			0.9 & 1000 & 5   & 0.194 & 0.223 & 0.028 & 0.311 & 0.872\\
					& 2000 & 5   & 0.194 & 0.207 & 0.013 & 0.206 & 0.892\\
			\hline
			    &      &     & True    & \multicolumn{3}{c}{Estimates}& 95\%\\
			\cline{5-7}
			$a$ & $n$  & $L$ & PRTE    & Mean    & Bias    & RMSE    & Coverage\\
			\hline
			0.1 & 1000 & 10  & 0.243 & 0.321 & 0.078 & 0.876 & 0.829\\
					& 2000 & 10  & 0.243 & 0.299 & 0.055 & 0.503 & 0.874\\
			\hline
			0.2 & 1000 & 10  & 0.237 & 0.295 & 0.058 & 0.513 & 0.878\\
					& 2000 & 10  & 0.237 & 0.288 & 0.051 & 0.299 & 0.916\\
			\hline
			0.3 & 1000 & 10  & 0.230 & 0.308 & 0.077 & 0.405 & 0.904\\
					& 2000 & 10  & 0.230 & 0.286 & 0.056 & 0.246 & 0.948\\
			\hline
			0.4 & 1000 & 10  & 0.225 & 0.293 & 0.068 & 0.362 & 0.917\\
					& 2000 & 10  & 0.225 & 0.280 & 0.055 & 0.220 & 0.956\\
			\hline
			0.5 & 1000 & 10  & 0.219 & 0.285 & 0.066 & 0.340 & 0.929\\
					& 2000 & 10  & 0.219 & 0.270 & 0.051 & 0.208 & 0.954\\
			\hline
			0.6 & 1000 & 10  & 0.213 & 0.271 & 0.058 & 0.321 & 0.929\\
					& 2000 & 10  & 0.213 & 0.258 & 0.044 & 0.196 & 0.954\\
			\hline
			0.7 & 1000 & 10  & 0.207 & 0.259 & 0.052 & 0.307 & 0.924\\
					& 2000 & 10  & 0.207 & 0.243 & 0.036 & 0.190 & 0.943\\
			\hline
			0.8 & 1000 & 10  & 0.201 & 0.240 & 0.039 & 0.308 & 0.908\\
					& 2000 & 10  & 0.201 & 0.230 & 0.029 & 0.190 & 0.932\\
			\hline
			0.9 & 1000 & 10  & 0.194 & 0.218 & 0.023 & 0.307 & 0.894\\ 
					& 2000 & 10  & 0.194 & 0.220 & 0.026 & 0.198 & 0.898\\
			\hline\hline
		\end{tabular}
	}
	\caption{Monte Carlo simulation results for the benchmark design with completely nonparametric preliminary estimation under policy changes of the form $P^\ast = P + a \left(1 - P\right)$ for $a \in \{0.1,...,0.9\}$. $n$ denotes the sample size. $L$ is the number of folds in sample splitting. The true PRTE is numerically evaluated. The displayed statistics include the mean, bias, root mean square error, and coverage frequencies with the nominal probability of 95\%.}
	\label{tab:simulation_results_p}
\end{table}

\begin{table}
	\centering
	\scalebox{1.00}{
		\begin{tabular}{cccccccc}
			\hline\hline
			    &      &     & True    & \multicolumn{3}{c}{Estimates}& 95\%\\
			\cline{5-7}
			$a$ & $n$  & $L$ & PRTE    & Mean    & Bias    & RMSE    & Coverage\\
			\hline
			-0.5& 1000 & 5   & 0.223 & 0.187 &-0.035 & 0.166 & 0.946\\
			    & 2000 & 5   & 0.223 & 0.190 &-0.033 & 0.113 & 0.950\\
			\hline
			-0.4& 1000 & 5   & 0.222 & 0.186 &-0.036 & 0.167 & 0.945\\
			    & 2000 & 5   & 0.222 & 0.188 &-0.035 & 0.113 & 0.947\\
			\hline
			-0.3& 1000 & 5   & 0.222 & 0.184 &-0.038 & 0.169 & 0.942\\
			    & 2000 & 5   & 0.222 & 0.186 &-0.036 & 0.114 & 0.948\\
			\hline
			-0.2& 1000 & 5   & 0.221 & 0.183 &-0.038 & 0.171 & 0.933\\
			    & 2000 & 5   & 0.221 & 0.184 &-0.037 & 0.115 & 0.945\\
			\hline
			-0.1& 1000 & 5   & 0.221 & 0.182 &-0.039 & 0.177 & 0.927\\
			    & 2000 & 5   & 0.221 & 0.182 &-0.039 & 0.116 & 0.940\\
			\hline
			0.1 & 1000 & 5   & 0.220 & 0.202 &-0.018 & 0.178 & 0.924\\
			    & 2000 & 5   & 0.220 & 0.186 &-0.034 & 0.121 & 0.930\\
			\hline
			0.2 & 1000 & 5   & 0.219 & 0.206 &-0.014 & 0.177 & 0.923\\
			    & 2000 & 5   & 0.219 & 0.187 &-0.032 & 0.118 & 0.934\\
			\hline
			0.3 & 1000 & 5   & 0.219 & 0.207 &-0.012 & 0.176 & 0.930\\
			    & 2000 & 5   & 0.219 & 0.186 &-0.032 & 0.117 & 0.940\\
			\hline
			0.4 & 1000 & 5   & 0.218 & 0.207 &-0.012 & 0.176 & 0.927\\
			    & 2000 & 5   & 0.218 & 0.186 &-0.032 & 0.116 & 0.946\\
			\hline
			0.5 & 1000 & 5   & 0.218 & 0.206 &-0.012 & 0.174 & 0.934\\
			    & 2000 & 5   & 0.218 & 0.185 &-0.033 & 0.115 & 0.949\\
			\hline\hline
		\end{tabular}
	}
	\caption{Monte Carlo simulation results for the benchmark design with completely nonparametric preliminary estimation under policy changes of the form $Z^\ast = Z + a$ for $a \in \{-0.5,...,-0.1,0.1,...,0.5\}$. $n$ denotes the sample size. $L$ is the number of folds in sample splitting. The true PRTE is numerically evaluated. The displayed statistics include the mean, bias, root mean square error, and coverage frequencies with the nominal probability of 95\%.}
	\label{tab:simulation_results_z}
\end{table}
\clearpage

\subsection{Extended Designs with High Dimensions}\label{sec:extended_designs}

In the previous subsection, we demonstrate that even completely nonparametric preliminary estimation yields desirable finite-sample performances.
We next consider extended designs with high dimensions, employ parametric propensity score estimation with absolute shrinkage penalization, and present finite-sample performances of our estimation and inference procedures in this setting.

Suppose that the selection model is specified by
\begingroup
\allowdisplaybreaks
\begin{align*}
S =& \mathbbm{1}\left\{ \frac{3}{10} - \sum_{k=1}^{\text{dim}\left(Z\right)-2} \frac{3}{10^{\left(k+2\right)/2}} - \frac{1}{10^{{\text{dim}\left(Z\right)}/2}} \right.
\\
&\ \ 
\left. + \frac{3}{10}Z_1 + \sum_{k=1}^{\text{dim}\left(Z\right)-2} \frac{3}{10^{\left(k+2\right)/2}} Z_{k+1} + \frac{1}{10^{\text{dim}\left(Z\right)/2}} Z_{\text{dim}\left(Z\right)} - U_S > 0 \right\},
\end{align*}
\endgroup
where $Z_1 \sim N\left(-1,3^2\right)$ and $Z_2, ..., Z_{\text{dim}\left(Z\right)} \sim N\left(1,3^2\right)$.
This design includes that of the benchmark by \citet[][supplementary Appendix D]{carneiro/lokshin/umapathi:2017} studied in the previous subsection as a special case where $\text{dim}\left(Z\right) = 2$.
Furthermore, this design preserves the same value of PRTE as that in the benchmark by \citet{carneiro/lokshin/umapathi:2017} for each $a$, regardless of the dimension $\text{dim}\left(Z\right) \ge 2$.
In this manner, we devise this extended design to allow for comparisons of simulation results with those presented in Section \ref{sec:benchmark_design}.
We set the high-dimensional setting with $\text{dim}\left(Z\right) = 100$ in this design.

Estimates for the PRTE and their estimated standard errors are obtained using our proposed method of double debiased estimation and inference with preliminary estimation of propensity scores by lasso logit.
See Appendix \ref{sec:details_simulation_parametric_estimation_p} for additional details.
Table \ref{tab:simulation_results_lasso_logit} summarizes simulation results.
Note that we obtain qualitatively similar results to those presented in Table \ref{tab:simulation_results_p}, and hence similar remarks follow.
It is worth noting that the magnitude of the statistics (such as the bias the RMSE) displayed in this table is similar to that in Table \ref{tab:simulation_results_p} despite the difference in $\text{dim}\left(Z\right)$ and the preliminary estimators.
This observation is also consistent with the property of the orthogonal score that it reduces and asymptotically vanishes effects of preliminary estimation errors.

\begin{table}
	\centering
	\scalebox{0.95}{
		\begin{tabular}{cccccccc}
			\hline\hline
			     &      &     & True    & \multicolumn{3}{c}{Estimates}& 95\%\\
			\cline{5-7}
			 $a$ & $n$  & $L$ & PRTE    & Mean    & Bias    & RMSE    & Coverage\\
			\hline
			 0.1 & 1000 & 5   & 0.243 & 0.306 & 0.063 & 0.783 & 0.907\\
			     & 2000 & 5   & 0.243 & 0.294 & 0.051 & 0.457 & 0.919\\
			\hline
			 0.2 & 1000 & 5   & 0.237 & 0.319 & 0.083 & 0.499 & 0.935\\
			     & 2000 & 5   & 0.237 & 0.289 & 0.052 & 0.306 & 0.942\\
			\hline
			 0.3 & 1000 & 5   & 0.230 & 0.327 & 0.096 & 0.422 & 0.935\\
			     & 2000 & 5   & 0.230 & 0.301 & 0.070 & 0.256 & 0.958\\
			\hline
			 0.4 & 1000 & 5   & 0.225 & 0.314 & 0.090 & 0.385 & 0.940\\
			     & 2000 & 5   & 0.225 & 0.281 & 0.057 & 0.242 & 0.949\\
			\hline
			 0.5 & 1000 & 5   & 0.219 & 0.310 & 0.091 & 0.382 & 0.931\\
			     & 2000 & 5   & 0.219 & 0.277 & 0.059 & 0.226 & 0.945\\
			\hline
			 0.6 & 1000 & 5   & 0.213 & 0.281 & 0.068 & 0.358 & 0.929\\
			     & 2000 & 5   & 0.213 & 0.259 & 0.045 & 0.222 & 0.934\\
			\hline
			 0.7 & 1000 & 5   & 0.207 & 0.268 & 0.061 & 0.355 & 0.911\\
			     & 2000 & 5   & 0.207 & 0.247 & 0.040 & 0.208 & 0.935\\
			\hline
			 0.8 & 1000 & 5   & 0.201 & 0.237 & 0.036 & 0.340 & 0.911\\
			     & 2000 & 5   & 0.201 & 0.223 & 0.022 & 0.212 & 0.921\\
			\hline
			 0.9 & 1000 & 5   & 0.194 & 0.219 & 0.025 & 0.378 & 0.860\\
			     & 2000 & 5   & 0.194 & 0.215 & 0.021 & 0.226 & 0.869\\
			\hline\hline
		\end{tabular}
	}
	\caption{Monte Carlo simulation results for the extended design with $\text{dim}\left(Z\right)=100$ based on lasso logit propensity score estimation under policy changes of the form $P^\ast = P + a\left(1-P\right)$ for $a \in \{0.1,...,0.9\}$. $n$ denotes the sample size. $L$ is the number of folds in sample splitting. The true PRTE is numerically evaluated. The displayed statistics include the mean, bias, root mean square error, and coverage frequencies with the nominal probability of 95\%.}
	\label{tab:simulation_results_lasso_logit}
\end{table}
\clearpage

\clearpage
\section{Empirical Illustration}\label{sec:empirical}
In this section, we apply the proposed method to an analysis of the effects of counterfactual policies that expose various fractions of the population to upper secondary schooling in Indonesia.
Following the study by \citet{carneiro/lokshin/umapathi:2017}, we use the data from the third wave of the Indonesia Family Life Survey (IFLS) fielded from June through
November 2000 -- we refer readers to their supplementary appendix for further details of this data set.
We also use the same subsample as that of \citet{carneiro/lokshin/umapathi:2017}, that consists of employed males aged 25--60, who have reported non-missing wage and schooling information.
This subsample consists of 2,608 individuals.

We set our variables following \citet{carneiro/lokshin/umapathi:2017}.
The outcome variable $Y$ denotes the log of hourly wages constructed from self-reported monthly wages and hours worked per week.
The binary treatment variable $S$ indicates attendance of upper secondary school or higher.
Control variables $X$ include age, age squared, an indicator for whether the individual was living in a village at age 12, indicators for the province of residence, an indicator of rural residence, distance from the office of the head of the community of residence to the nearest community health post, and indicators for the level of schooling by each parent.
The excluded instrument is the distance from the office of the head of the community of residence in kilometers to the nearest secondary school.
\citet[][Section 4.1]{carneiro/lokshin/umapathi:2017} provide detailed analysis to support the validity of this instrument.
In this paper, we take advantage of their analysis and discussions, and directly adopt their empirical approach in our estimation and inference framework.
See Appendix \ref{sec:model_application} for details of this model specification.

We consider the counterfactual policies of the form $P^\ast = P + a \left(1-P\right)$ with various levels of $a \in \{0.05,0.10,...,0.45,0.50\}$.
Note that such a counterfactual treatment probability $P^\ast$ arises from the counterfactual policy that exposes fraction $a$ of the population to upper secondary schooling.
The estimation and inference approaches that we take in this analysis are the same as those used for our Monte Carlo simulation studies except that we use the probit propensity score estimation following the benchmark study by \citet{carneiro/lokshin/umapathi:2017} -- see Appendix \ref{sec:step_by_step_procedure_application} for a step-by-step procedure of estimation and inference.
We use $L=5$ as the number of folds in cross fitting.\footnote{We also ran another set of estimation with $L=10$ but the results are almost the same both in terms of estimates and their standard errors, similarly to what we observed in our Monte Carlo simulation studies.}
Finally, to mitigate the finite-sample randomness in estimates due to sample splitting, we use the robust re-randomization method following \citet[][Section 3.4]{chernozhukov2018double}.

Table \ref{tab:empirical} summarizes the results.
For each $a \in \{0.05,0.10,...,0.45,0.50\}$, displayed in this table are the point estimate, standard error, and the 95\% confidence interval.
Observe that the policy that assigns a fraction $a=0.05$ of the population to upper secondary schooling is expected to increase the log of hourly wages by 0.169 (with the standard error of 0.064 and the 95\% confidence interval of $[0.044,0.294]$) per treated individual on average.
As the fraction $a$ of the population assigned to treatment increases, these per-individual average policy relevant treatment effects tend to increase.
Specifically, the policy that assigns a fraction $a=0.50$ of the population to upper secondary schooling is expected to increase the log of hourly wages by 0.225 (with the standard error of 0.076 and the 95\% confidence interval of $[0.076,0.375]$) per treated individual on average.

\begin{table}
	\centering
		\begin{tabular}{ccccc}
		\hline\hline
		$a$ & $L$ & Estimate & Std. Err. & 95\% CI\\
		\hline
		0.05 & 5 & 0.169 & 0.064 & [0.044, 0.294]\\
		0.10 & 5 & 0.172 & 0.064 & [0.046, 0.298]\\
		0.15 & 5 & 0.176 & 0.064 & [0.049, 0.302]\\
		0.20 & 5 & 0.180 & 0.065 & [0.053, 0.308]\\
		0.25 & 5 & 0.186 & 0.066 & [0.056, 0.316]\\
		0.30 & 5 & 0.192 & 0.068 & [0.059, 0.325]\\
		0.35 & 5 & 0.199 & 0.070 & [0.062, 0.336]\\
		0.40 & 5 & 0.207 & 0.072 & [0.066, 0.347]\\
		0.45 & 5 & 0.216 & 0.074 & [0.071, 0.360]\\
		0.50 & 5 & 0.225 & 0.076 & [0.076, 0.375]\\
		\hline\hline
		\end{tabular}
	\caption{Estimates, standard errors, and 95\% confidence intervals of the PRTE for the counterfactual policies of the form $P^\ast = P + a \left(1-P\right)$, $a \in \{0.05,0.10,...,0.45,0.50\}$.}
	\label{tab:empirical}
\end{table}

\section{Summary}\label{sec:summary}
Estimation of the PRTE involves estimation of multiple preliminary parameters.
These preliminary parameters include propensity scores, conditional expectation functions of the outcome and covariates given the propensity score, and marginal treatment effects.
These preliminary estimators can affect the asymptotic distribution of the PRTE estimator in complicated and intractable manners.
To solve this issue, we propose an orthogonal score for double debiased estimation of the PRTE.
Our proposed orthogonal score allows for the asymptotic distribution of the PRTE estimator to be obtained without any influence of preliminary parameter estimators as far as they satisfy mild convergence rate conditions.
Simulation results confirm our theoretical properties, and demonstrate that the method of estimation and inference works well even in small samples.
Our empirical application demonstrates that the proposed method indeed works with real data.
To our knowledge, our work is the first to develop asymptotic distribution theories for inference about the PRTE.
We hope that our proposed method contributes to empirical analyses of policy relevant treatment effects.

\clearpage
\appendix
\section*{Appendix}


\section{Regularity Conditions}\label{sec:regularity}

The following assumption, requiring the parameter set to be compact, is standard.
\begin{assumption}\label{Assn_compact}
$\theta$ belongs to a known compact space $\Theta$.
\end{assumption}

We assume that random variables have bounded fourth moments as stated below.
\begin{assumption}\label{Assn_bounded_moments}
$E[\|\left(Y,\mu_0\left(X\right)',\mu_1\left(X\right)'\right)'\|^4]<\infty$.
\end{assumption}

Finally, we assume that all the preliminary parameter estimators are consistent, as stated below.
There are 12 pieces of conditions contained in the the following assumption, but each of these 12 pieces is a mild assumption that is satisfied by a wide variety of existing nonparametric estimators.
Furthermore, we its provide lower-level primitive sufficient conditions in Appendix \ref{sec:lowe_level_sufficient_conditions}.
\begin{assumption}\label{Assn_just_consis}
For every $\hat{\gamma}_\ell=\hat{\gamma}_1,\ldots,\hat{\gamma}_L$, 
the following random variables are $o_p\left(1\right)$: 
\begin{center}\small
$
\left(\int\left\|\hat{\xi}_1\left(x,y,\hat{\mf}_{S\mid Z}\left(z\right)\right)-{\xi}_1\left(x,y,{\mf}_{S\mid Z}\left(z\right)\right)\right\|^2F_W\left(dw\right)\right)^{1/2}
$
\\
$
\left(\int\left\|\hat{\zeta}\left(z\right)-{\zeta}\left(z\right)\right\|^2F_W\left(dw\right)\right)^{1/2}
$
\\
$
\left(\int\left\|\hat{\zeta}\left(z\right)\hat{\mf}_{S\mid Z}\left(z\right)-{\zeta}\left(z\right){\mf}_{S\mid Z}\left(z\right)\right\|^2F_W\left(dw\right)\right)^{1/2}.
$
\\
$
\left(\int {\mf}_{\|\left(\mu_0\left(x\right)',\mu_1\left(x\right)',1\right)'\|^2\mid Z}\left(z\right)\left(
P^\ast\left(\hat{\mf}_{S\mid Z}\left(z\right),z\right)-P^\ast\left({\mf}_{S\mid Z}\left(z\right),z\right)
\right)^2F_W\left(dw\right)\right)^{1/2}
$
\\
$
\left(\int {\mf}_{\|\left(\mu_0\left(x\right)',\mu_1\left(x\right)',1\right)'\|^2\mid Z}\left(z\right)\left(
\hat{\mf}_{S\mid Z}\left(z\right)-{\mf}_{S\mid Z}\left(z\right)
\right)^2F_W\left(dw\right)\right)^{1/2}
$
\\
$
\left(\int {\mf}_{\|\left(\mu_0\left(x\right)',\mu_1\left(x\right)',1\right)'\|^2\mid Z}\left(z\right)\left(
\partial P^\ast\left(\hat{\mf}_{S\mid Z}\left(z\right),z\right)-\partial P^\ast\left(\mf_{S\mid Z}\left(z\right),z\right)
\right)^2F_W\left(dw\right)\right)^{1/2}
$
\\
$
\left(
\int {\mf}_{\|\left(\mu_0\left(x\right)',\mu_1\left(x\right)',1\right)'\|^2\mid Z}\left(z\right)
\left(
\hat{\mf}_{S\mid Z}\left(z\right)\partial P^\ast\left(\hat{\mf}_{S\mid Z}\left(z\right),z\right)
-
{\mf}_{S\mid Z}\left(z\right)\partial P^\ast\left(\mf_{S\mid Z}\left(z\right),z\right)
\right)^2
F_W\left(dw\right)\right)^{1/2}
$
\\
$
\left(\int
\left(\frac{\hat{f}_{P^\ast}\left(\hat{\mf}_{S\mid Z}\left(z\right)\right)}{\hat{f}_{P}\left(\hat{\mf}_{S\mid Z}\left(z\right)\right)}\hat{\mf}_{U\mid P}\left(\hat{\mf}_{S\mid Z}\left(z\right)\right)-\frac{{f}_{P^\ast}\left({\mf}_{S\mid Z}\left(z\right)\right)}{{f}_{P}\left({\mf}_{S\mid Z}\left(z\right)\right)}{\mf}_{U\mid P}\left({\mf}_{S\mid Z}\left(z\right)\right)\right)^2
F_Z\left(dz\right)\right)^{1/2}
$
\\
$
\left(\int
\left(\frac{\hat{f}_{P^\ast}\left(\hat{\mf}_{S\mid Z}\left(z\right)\right)}{\hat{f}_{P}\left(\hat{\mf}_{S\mid Z}\left(z\right)\right)}\hat{\Delta}_{U\mid P}\left(\hat{\mf}_{S\mid Z}\left(z\right)\right)-\frac{{f}_{P^\ast}\left({\mf}_{S\mid Z}\left(z\right)\right)}{{f}_{P}\left({\mf}_{S\mid Z}\left(z\right)\right)}{\Delta_{U\mid P}}\left({\mf}_{S\mid Z}\left(z\right)\right)\right)^2
F_Z\left(dz\right)\right)^{1/2}
$
\\
$
\left(\int
\left(\frac{\hat{f}_{P^\ast}\left(\hat{\mf}_{S\mid Z}\left(z\right)\right)}{\hat{f}_{P}\left(\hat{\mf}_{S\mid Z}\left(z\right)\right)}\hat{\Delta}_{U\mid P}\left(\hat{\mf}_{S\mid Z}\left(z\right)\right)\hat{\mf}_{S\mid Z}\left(z\right)-\frac{{f}_{P^\ast}\left({\mf}_{S\mid Z}\left(z\right)\right)}{{f}_{P}\left({\mf}_{S\mid Z}\left(z\right)\right)}{\Delta_{U\mid P}}\left({\mf}_{S\mid Z}\left(z\right)\right){\mf}_{S\mid Z}\left(z\right)\right)^2
F_Z\left(dz\right)\right)^{1/2}
$
\\
$
\left(\int
\left(
\hat{\mf}_{U\mid P}\left(P^\ast\left(\hat{\mf}_{S\mid Z}\left(z\right),z\right)\right)-{\mf}_{U\mid P}\left(P^\ast\left({\mf}_{S\mid Z}\left(z\right),z\right)\right)
\right)^2
F_Z\left(dz\right)\right)^{1/2}
$
\\
$
\left(\int
\left(\hat{\Delta}_{U\mid P}\left(P^\ast\left(\hat{\mf}_{S\mid Z}\left(z\right),z\right)\right)\partial P^\ast\left(\hat{\mf}_{S\mid Z}\left(z\right),z\right)-{\Delta_{U\mid P}}\left(P^\ast\left({\mf}_{S\mid Z}\left(z\right),z\right)\right)\partial P^\ast\left({\mf}_{S\mid Z}\left(z\right),z\right)\right)^2
F_Z\left(dz\right)\right)^{1/2}
$
\\

$
\left(\int
\left(\hat{\Delta}_{U\mid P}\left(P^\ast\left(\hat{\mf}_{S\mid Z}\left(z\right),z\right)\right)\partial P^\ast\left(\hat{\mf}_{S\mid Z}\left(z\right),z\right)\hat{\mf}_{S\mid Z}\left(z\right)-{\Delta_{U\mid P}}\left(P^\ast\left({\mf}_{S\mid Z}\left(z\right),z\right)\right)\partial P^\ast\left({\mf}_{S\mid Z}\left(z\right),z\right){\mf}_{S\mid Z}\left(z\right)\right)^2
F_Z\left(dz\right)\right)^{1/2}.
$\normalsize
\end{center}
\end{assumption}

\section{Lower-Level Sufficient Conditions}\label{sec:lowe_level_sufficient_conditions}

The main results presented in the main text presume a general class of estimation procedures and thus state conditions at a high level to accommodate a wide array of estimators.
In this section, we propose a concrete estimation procedure allowing for high-dimensions and provide lower-level sufficient conditions for the high-level statements in Assumptions \ref{Assn_slow_conver}, \ref{Assn_CV} \ref{Assn:Taylor_Errr}, and \ref{Assn_just_consis}.
The framework that we consider involves both shrinkage and kernel estimation, and is related to a branch of the recent literature including \citet{kennedy:2017}, \citet{fan:2019}, \citet{su/ura/zhang:2019}, \citet{zimmert/lechner:2019} and \citet{colangelo/lee:2020}.

Let $\Lambda: \mathbb{R} \rightarrow \left(0,1\right)$ be a twice differentiable link function with bounded first and second derivatives. 
We estimate $\hat{\pi}$ using the observations in $I_\ell^c$, and construct an estimator $\hat{\mf}_{S\mid Z}\left(z\right)$ for ${\mf}_{S\mid Z}\left(z\right)$ by
$$
\hat{\mf}_{S\mid Z}\left(z\right)=\Lambda\left(z'\hat{\pi}\right).
$$ 
We also use an estimator  $\hat{\mf}_{\boldsymbol{XY}\mid Z}\left(z\right)$ for ${\mf}_{\boldsymbol{XY}\mid Z}\left(z\right)$, where 
$$
\boldsymbol{XY}=vec\left(\mu_0\left(X\right),\mu_1\left(X\right),Y,\mu_0\left(X\right)\mu_0\left(X\right)',\mu_0\left(X\right)\mu_1\left(X\right)',\mu_0\left(X\right)Y,\mu_1\left(X\right)\mu_1\left(X\right)',\mu_1\left(X\right)Y,Y^2\right).
$$
For every $z$, define the following density estimators based on the estimated propensity scores:
\begingroup
\allowdisplaybreaks
\begin{align*}
\hat{f}_{P}\left(\hat{\mf}_{S\mid Z}\left(z\right)\right)=&\frac{1}{|I_\ell^c|}\sum_{j\in I_\ell^c}K_{h}\left(\hat{\mf}_{S\mid Z}\left(Z_j\right)-\hat{\mf}_{S\mid Z}\left(z\right)\right),
\\
\hat{f}_{P}\left(P^\ast\left(\hat{\mf}_{S\mid Z}\left(z\right),z\right)\right)=&\frac{1}{|I_\ell^c|}\sum_{j\in I_\ell^c}K_{h}\left(\hat{\mf}_{S\mid Z}\left(Z_j\right)-P^\ast\left(\hat{\mf}_{S\mid Z}\left(z\right),z\right)\right),
\qquad\text{and}
\\
\hat{f}_{P}^{\left(1\right)}\left(\hat{\mf}_{S\mid Z}\left(z\right)\right)=&-\frac{1}{|I_\ell^c|}\sum_{j\in I_\ell^c}K_{h}^{\left(1\right)}\left(\hat{\mf}_{S\mid Z}\left(Z_j\right)-\hat{\mf}_{S\mid Z}\left(z\right)\right),
\end{align*}
\endgroup
where $K_{h}=K\left( \ \cdot \ / h \right)/h$ for a kernel function $K$ satisfying Assumption \ref{assn:14} to be stated ahead.

Consider sequences of constants $\mathbf{f}_n$ and sets $\mathcal{Z}_n$ such that 
\begin{equation}\label{eq:fn}
\max\left\{|\hat{f}_{P}\left(\hat{\mf}_{S\mid Z}\left(z\right)\right)|^{-1},|\hat{f}_{P}\left(P^\ast\left(\hat{\mf}_{S\mid Z}\left(z\right),z\right)\right)|^{-1},
\left|\frac{\hat{f}_{P}^{\left(1\right)}\left(\hat{\mf}_{S\mid Z}\left(z\right)\right)}{\hat{f}_{P}\left(\hat{\mf}_{S\mid Z}\left(z\right)\right)^2}\right|,
\left|\frac{\hat{f}_{P}^{\left(1\right)}\left(P^\ast\left(\hat{\mf}_{S\mid Z}\left(z\right),z\right)\right)}{\hat{f}_{P}\left(P^\ast\left(\hat{\mf}_{S\mid Z}\left(z\right),z\right)\right)^2}\right|\right\}\leq \mathbf{f}_n^{-1}
\end{equation}
for every $z\in\mathcal{Z}_n$. 
For every component  $\omega$  of $vec\left(\mu_0\left(X\right),\mu_1\left(X\right),Y,U\right)$,
we define the $\mathcal{Z}_n$-trimmed estimators,
$\hat{\mf}_{\omega\mid P}\left(\hat{\mf}_{S\mid Z}\left(z\right)\right)$
$\hat{\mf}_{\omega\mid P}\left(P^\ast\left(\hat{\mf}_{S\mid Z}\left(z\right),z\right)\right)$,
$\hat{f}_{P^\ast}\left(\hat{\mf}_{S\mid Z}\left(z\right)\right)$,
$\hat{\Delta}_{U\mid P}\left(\hat{\mf}_{S\mid Z}\left(z\right)\right)$ and 
$\hat{\Delta}_{U\mid P}\left(P^\ast\left(\hat{\mf}_{S\mid Z}\left(z\right),z\right)\right)$, as follows: 
\begingroup
\allowdisplaybreaks
\begin{align*}
\hat{\mf}_{\omega\mid P}\left(\hat{\mf}_{S\mid Z}\left(z\right)\right)
=&
\frac{\hat{\nf}_{\omega\mid P}\left(\hat{\mf}_{S\mid Z}\left(z\right)\right)}{\hat{f}_{P}\left(\hat{\mf}_{S\mid Z}\left(z\right)\right)}1\{z\in\mathcal{Z}_n\},
\\
\hat{\mf}_{\omega\mid P}\left(P^\ast\left(\hat{\mf}_{S\mid Z}\left(z\right),z\right)\right)
=&
\frac{\hat{\nf}_{\omega\mid P}\left(P^\ast\left(\hat{\mf}_{S\mid Z}\left(z\right),z\right)\right)}{\hat{f}_{P}\left(P^\ast\left(\hat{\mf}_{S\mid Z}\left(z\right),z\right)\right)}1\{z\in\mathcal{Z}_n\},
\\
\hat{f}_{P^\ast}\left(\hat{\mf}_{S\mid Z}\left(z\right)\right)
=&
\left(
\frac{1}{|I_\ell^c|}\sum_{j\in I_\ell^c}K_{h}\left(P^\ast\left(\hat{\mf}_{S\mid Z}\left(Z_j\right),Z_j\right)-\hat{\mf}_{S\mid Z}\left(z\right)\right)
\right)
1\{z\in\mathcal{Z}_n\},
\\
\hat{\Delta}_{U\mid P}\left(\hat{\mf}_{S\mid Z}\left(z\right)\right)
=&
\frac{\hat{\nf}_{U\mid P}^{\left(1\right)}\left(\hat{\mf}_{S\mid Z}\left(z\right)\right)}{\hat{f}_{P}\left(\hat{\mf}_{S\mid Z}\left(z\right)\right)}1\{z\in\mathcal{Z}_n\}
-
\hat{\mf}_{U\mid P}\left(\hat{\mf}_{S\mid Z}\left(z\right)\right)
\frac{\hat{f}_{P}^{\left(1\right)}\left(\hat{\mf}_{S\mid Z}\left(z\right)\right)}{\hat{f}_{P}\left(\hat{\mf}_{S\mid Z}\left(z\right)\right)},
\qquad\text{and}
\\
\hat{\Delta}_{U\mid P}\left(P^\ast\left(\hat{\mf}_{S\mid Z}\left(z\right),z\right)\right)
=&
\frac{\hat{\nf}_{U\mid P}^{\left(1\right)}\left(P^\ast\left(\hat{\mf}_{S\mid Z}\left(z\right),z\right)\right)}{\hat{f}_{P}\left(P^\ast\left(\hat{\mf}_{S\mid Z}\left(z\right),z\right)\right)}1\{z\in\mathcal{Z}_n\}
-
\hat{\mf}_{U\mid P}\left(P^\ast\left(\hat{\mf}_{S\mid Z}\left(z\right),z\right)\right)
\frac{\hat{f}_{P}^{\left(1\right)}\left(P^\ast\left(\hat{\mf}_{S\mid Z}\left(z\right),z\right)\right)}{\hat{f}_{P}\left(P^\ast\left(\hat{\mf}_{S\mid Z}\left(z\right),z\right)\right)},
\end{align*}
\endgroup
where $\nf_{\omega\mid\Lambda\left(Z'\pi_0\right)}\left(p\right)$ and $\hat{\nf}_{\omega\mid P}\left(p\right)$ are in turn defined by
\begin{align*}
\nf_{\omega\mid\Lambda\left(Z'\pi_0\right)}\left(p\right)
=&
\mf_{\omega\mid\Lambda\left(Z'\pi_0\right)}\left(p\right)f_{\Lambda\left(Z'\pi_0\right)}\left(p\right)
\qquad\text{and} 
\\
\hat{\nf}_{\omega\mid P}\left(p\right)
=&
\frac{1}{|I_\ell^c|}\sum_{j\in I_\ell^c}\omega_jK_{h}\left(\hat{\mf}_{S\mid Z}\left(Z_j\right)-p\right).
\end{align*}

Finally, we also define 
$$
\hat{\zeta}\left(z\right)=\hat{\zeta}\left(z;\hat{\mf}_{S\mid Z}\left(z\right)\right),
$$
where $\hat\zeta\left(z;p\right) = \left(\hat\zeta_1\left(z;p\right),...,\hat\zeta_6\left(z;p\right)\right)$ consists of the six components defined by
\begingroup
\allowdisplaybreaks
\begin{align*}
\hat{\zeta}_1\left(z;p\right)
=&
2\left(1-p\right)\left(\hat{\mf}_{\mu_0\left(X\right)\mid Z}\left(z\right)-\hat{\mf}_{\mu_0\left(X\right)\mid P}\left(p\right)\right)\left(\hat{\mf}_{\mu_0\left(X\right)\mid Z}\left(z\right)-\hat{\mf}_{\mu_0\left(X\right)\mid P}\left(p\right)\right)'\\
&+2\left(1-p\right)\left(\hat{\mf}_{\mu_0\left(X\right)\mu_0\left(X\right)'\mid Z}\left(z\right)-\hat{\mf}_{\mu_0\left(X\right)\mid Z}\left(z\right)\hat{\mf}_{\mu_0\left(X\right)\mid Z}\left(z\right)'\right)\\
&-\left(1-p\right)^2\frac{\partial}{\partial p}\hat{\mf}_{\mu_0\left(X\right)\mid P}\left(p\right)\left(\hat{\mf}_{\mu_0\left(X\right)\mid Z}\left(z\right)-\hat{\mf}_{\mu_0\left(X\right)\mid P}\left(p\right)\right)'\\
&-\left(1-p\right)^2\left(\hat{\mf}_{\mu_0\left(X\right)\mid Z}\left(z\right)-\hat{\mf}_{\mu_0\left(X\right)\mid P}\left(p\right)\right)\frac{\partial}{\partial p}\hat{\mf}_{\mu_0\left(X\right)\mid P}\left(p\right)',
\\
\hat{\zeta}_2\left(z;p\right)
=&
\left(1-2p\right)\left(\hat{\mf}_{\mu_1\left(X\right)\mid Z}\left(z\right)-\hat{\mf}_{\mu_1\left(X\right)\mid P}\left(p\right)\right)\left(\hat{\mf}_{\mu_0\left(X\right)\mid Z}\left(z\right)-\hat{\mf}_{\mu_0\left(X\right)\mid P}\left(p\right)\right)'\\
&+\left(1-2p\right)\left(\hat{\mf}_{\mu_1\left(X\right)\mu_0\left(X\right)'\mid Z}\left(z\right)-\hat{\mf}_{\mu_1\left(X\right)\mid Z}\left(z\right)\hat{\mf}_{\mu_0\left(X\right)\mid Z}\left(z\right)'\right)\\
&-\left(1-p\right)p\frac{\partial}{\partial p}\hat{\mf}_{\mu_1\left(X\right)\mid P}\left(p\right)\left(\hat{\mf}_{\mu_0\left(X\right)\mid Z}\left(z\right)-\hat{\mf}_{\mu_0\left(X\right)\mid P}\left(p\right)\right)'\\
&-\left(1-p\right)p\left(\hat{\mf}_{\mu_1\left(X\right)\mid Z}\left(z\right)-\hat{\mf}_{\mu_1\left(X\right)\mid P}\left(p\right)\right)\frac{\partial}{\partial p}\hat{\mf}_{\mu_0\left(X\right)\mid P}\left(p\right)',
\\
\hat{\zeta}_3\left(z;p\right)
=&
\left(1-2p\right)\left(\hat{\mf}_{\mu_0\left(X\right)\mid Z}\left(z\right)-\hat{\mf}_{\mu_0\left(X\right)\mid P}\left(p\right)\right)\left(\hat{\mf}_{\mu_1\left(X\right)\mid Z}\left(z\right)-\hat{\mf}_{\mu_1\left(X\right)\mid P}\left(p\right)\right)'\\
&+\left(1-2p\right)\left(\hat{\mf}_{\mu_0\left(X\right)\mu_1\left(X\right)'\mid Z}\left(z\right)-\hat{\mf}_{\mu_0\left(X\right)\mid Z}\left(z\right)\hat{\mf}_{\mu_1\left(X\right)\mid Z}\left(z\right)'\right)\\
&-\left(1-p\right)p\left(\hat{\mf}_{\mu_0\left(X\right)\mid Z}\left(z\right)-\hat{\mf}_{\mu_0\left(X\right)\mid P}\left(p\right)\right)\frac{\partial}{\partial p}\hat{\mf}_{\mu_1\left(X\right)\mid P}\left(p\right)'\\
&-\left(1-p\right)p\frac{\partial}{\partial p}\hat{\mf}_{\mu_0\left(X\right)\mid P}\left(p\right)\left(\hat{\mf}_{\mu_1\left(X\right)\mid Z}\left(z\right)-\hat{\mf}_{\mu_1\left(X\right)\mid P}\left(p\right)\right)',
\\
\hat{\zeta}_4\left(z;p\right)
=&
2p\left(\hat{\mf}_{\mu_1\left(X\right)\mid Z}\left(z\right)-\hat{\mf}_{\mu_1\left(X\right)\mid P}\left(p\right)\right)\left(\hat{\mf}_{\mu_1\left(X\right)\mid Z}\left(z\right)-\hat{\mf}_{\mu_1\left(X\right)\mid P}\left(p\right)\right)'\\
&+2p\left(\hat{\mf}_{\mu_1\left(X\right)\mu_1\left(X\right)'\mid Z}\left(z\right)-\hat{\mf}_{\mu_1\left(X\right)\mid Z}\left(z\right)\hat{\mf}_{\mu_1\left(X\right)\mid Z}\left(z\right)'\right)\\
&-p^2\frac{\partial}{\partial p}\hat{\mf}_{\mu_1\left(X\right)\mid P}\left(p\right)\left(\hat{\mf}_{\mu_1\left(X\right)\mid Z}\left(z\right)-\hat{\mf}_{\mu_1\left(X\right)\mid P}\left(p\right)\right)'\\
&-p^2\left(\hat{\mf}_{\mu_1\left(X\right)\mid Z}\left(z\right)-\hat{\mf}_{\mu_1\left(X\right)\mid P}\left(p\right)\right)\frac{\partial}{\partial p}\hat{\mf}_{\mu_1\left(X\right)\mid P}\left(p\right)',
\\
\hat{\zeta}_5\left(z;p\right)
=&
-p\left(\hat{\mf}_{\mu_0\left(X\right)\mid Z}\left(z\right)-\hat{\mf}_{\mu_0\left(X\right)\mid P}\left(p\right)\right)\left(\hat{\mf}_{Y\mid Z}\left(z\right)-\hat{\mf}_{Y\mid P}\left(p\right)\right)\\
&-p\left(\hat{\mf}_{\mu_0\left(X\right)Y\mid Z}\left(z\right)-\hat{\mf}_{\mu_0\left(X\right)\mid Z}\left(z\right)\hat{\mf}_{Y\mid Z}\left(z\right)\right)\\
&-\left(1-p\right)\frac{\partial}{\partial p}\hat{\mf}_{\mu_0\left(X\right)\mid P}\left(p\right)\left(\hat{\mf}_{Y\mid Z}\left(z\right)-\hat{\mf}_{Y\mid P}\left(p\right)\right)\\
&-\left(1-p\right)\left(\hat{\mf}_{\mu_0\left(X\right)\mid Z}\left(z\right)-\hat{\mf}_{\mu_0\left(X\right)\mid P}\left(p\right)\right)\frac{\partial}{\partial p}\hat{\mf}_{Y\mid P}\left(p\right),
\\
\hat{\zeta}_6\left(z;p\right)
=&
\left(\hat{\mf}_{\mu_1\left(X\right)\mid Z}\left(z\right)-\hat{\mf}_{\mu_1\left(X\right)\mid P}\left(p\right)\right)\left(\hat{\mf}_{Y\mid Z}\left(z\right)-\hat{\mf}_{Y\mid P}\left(p\right)\right)\\
&+\left(\hat{\mf}_{\mu_1\left(X\right)Y\mid Z}\left(z\right)-\hat{\mf}_{\mu_1\left(X\right)\mid Z}\left(z\right)\hat{\mf}_{Y\mid Z}\left(z\right)\right)\\
&-p\frac{\partial}{\partial p}\hat{\mf}_{\mu_1\left(X\right)\mid P}\left(p\right)\left(\hat{\mf}_{Y\mid Z}\left(z\right)-\hat{\mf}_{Y\mid P}\left(p\right)\right)\\
&-p\left(\hat{\mf}_{\mu_1\left(X\right)\mid Z}\left(z\right)-\hat{\mf}_{\mu_1\left(X\right)\mid P}\left(p\right)\right)\frac{\partial}{\partial p}\hat{\mf}_{Y\mid P}\left(p\right).
\end{align*}
\endgroup

Consider constants $r_1$ and $r_2$ satisfying 
$
r_1+r_2\geq \frac{1}{2}
$
and
$
r_1\geq r_2.
$
For example, we can take $\left(r_1,r_2\right)=\left(0.35,0.15\right)$, which is compatible with the assumptions to be stated below. 
For a $d_Z$-dimensional vector, we use the three different norms: the $L_1$ norm defined by $\|z\|_1=\sum_{j}|z_j|$; the $L_2$ norm defined by $\|z\|_2=\sqrt{\sum_{j}z_j^2}$; and the $L_{\infty}$ norm defined by $\|z\|_{\infty}=\max_{j}|z_j|$. 
We use $\|\cdot\|$ for the $L_2$ norm $\|\cdot\|_2$ when we want to simplify the notations.
Here we state lower-level sufficient conditions for the high-level assumptions.

\begin{assumption}[Parameter Estimation Error and Sparsity for $\pi$]\label{assn:11}
There is a non-stochastic set $\mathbf{P}_n\subset\mathbb{R}^{d_Z}$ such that
$\pi_0\in\mathbf{P}_n$,
$\hat\pi\in\mathbf{P}_n\mbox{ with probability approaching one}$,
$
\sup_{\pi\in\mathbf{P}_n}\|{\pi}-\pi_0\|_1=O\left(\sqrt{\frac{\bar{s}\log\left(d_Z\vee n\right)}{n}}\right),
$
and every element $\pi$ of $\mathbf{P}_n$ has at most $\bar{s}$ non-zero components. 
\end{assumption}

\begin{assumption}[Bound on $Z$]\label{assn:12}
$
\sup_{z\in\mathcal{Z}_n}\|z\|_{\infty}\leq K_n\mbox{ a.s.}
$
and
$
K_n=o\left(\sqrt{\frac{n^{1-2r_1}}{\bar{s}\log\left(d_Z\vee n\right)}}\right).
$
\end{assumption}

\begin{assumption}[Bandwidth and Truncation Threshold]\label{assn:13}
$
h^2=o\left(n^{-r_1}\mathbf{f}_n\right),
$
$
h \le K_n,
$
$
-\log\left(h\right)=O\left(\log\left(d_Z\vee n\right)\right),
$
and
$
{\left(1+\bar{s}\right)}{}\log\left(d_Z\vee n\right)=o\left(\min\{n^{3/4-r_1}h\mathbf{f}_n,
n^{1-2r_2}h^3\mathbf{f}_n^2,
n^{3/4-r_2}h^{2}\mathbf{f}_n\}\right).
$
\end{assumption}

In the aforementioned case with $\left(r_1,r_2\right)=\left(0.35,0.15\right)$, this assumption can be satisfied by choosing $h=Cn^{-1/5}$ for instance. 

\begin{assumption}[Kernel]\label{assn:14}
$K$ is twice differentiable and satisfies
$
\int K\left(u\right)du=1,
$
$
\int uK\left(u\right)du=0,
$
$
\int u^2K\left(u\right)du<\infty,
$
$
\int K^{\left(1\right)}\left(u\right)^2du<\infty,
$
$
\sup_{u}|K\left(u\right)|<\infty,
$
$
\sup_{u}|K^{\left(1\right)}\left(u\right)|<\infty,
$
and
$
\sup_{u}|K^{\left(2\right)}\left(u\right)|<\infty.
$

\end{assumption}

\begin{assumption}[Parameter Estimation Error for $\hat{\mf}_{\boldsymbol{XY}\mid Z}\left(z\right)$]\label{assn:15}
$
\sup_{z\in\mathcal{Z}_n}|\hat{\mf}_{\boldsymbol{XY}\mid Z}\left(z\right)-{\mf}_{\boldsymbol{XY}\mid Z}\left(z\right)|=o_p\left(n^{-r_1}\right).
$
\end{assumption}

\begin{assumption}[Regular Bounded Functions]\label{assn:16}
$P^\ast$ is twice differentiable, $\sup_{\left(p,z\right)} \left\vert \frac{\partial}{\partial p} P^\ast\left(p,z\right) \right\vert < \infty$, and $\sup_{\left(p,z\right)} \left\vert \frac{\partial^2}{\partial p^2} P^\ast\left(p,z\right) \right\vert < \infty$.
For every component $\omega$ of $vec\left(\mu_0\left(X\right),\mu_1\left(X\right),Y,U\right)$,
${\mf}_{\omega\mid Z}$ is twice differentiable, 
${\nf}_{\omega\mid\Lambda\left(Z'{\pi}\right)}$ is three-times differentiable,
we have that 
$
E\left[\omega^4\right] < \infty,
 $
 $
\sup_{z}|{\mf}_{\omega^2\mid Z}\left(z\right)|=O\left(1\right),
 $
 $
\sup_{p\in[0,1]}|{\mf}_{\omega\mid P}^{\left(2\right)}\left(p\right)|=O\left(1\right),
 $
 $
\sup_{\left(p,\pi\right)\in[0,1]\times\mathbf{P}_n}\left|{\nf}_{\omega\mid\Lambda\left(Z'{\pi}\right)}^{\left(3\right)}\left(p\right)\right|=O\left(1\right),
 $
 $
\sup_{\left(p,\pi\right)\in[0,1]\times\mathbf{P}_n}{\nf}_{\omega^2\mid \Lambda\left(Z'\pi\right)}\left(p\right)=O\left(1\right),
 $
 $
 \sup_{(p,\pi)\in[0,1]\times\mathbf{P}_n}\frac{{f}_{P^\ast\left(\Lambda\left(Z'\pi\right),Z\right)}\left(p\right)}{{f}_{\Lambda\left(Z'\pi\right)}\left(p\right)}=O\left(1\right),
 $
and
$
\sup_{p}\left|\frac{\partial}{\partial p}\frac{{f}_{P^\ast}\left(p\right)}{{f}_{P}\left(p\right)} \right|=O\left(1\right).
$
 \end{assumption}

\begin{assumption}[Smoothness]\label{assn:17}
$
\sup_{\pi\in\mathbf{P}_n}\left(\int\left(
\tau_{\Lambda\left(Z'\pi\right)}(z,\pi)-
\tau_{\Lambda\left(Z'\pi_0\right)}(z,\pi)\right)^4F_Z\left(dz\right)\right)^{1/4}=o\left(n^{-r_1}\mathbf{f}_n\right)
$
holds for every $\tau_{\mathrm{RV}}(z,\pi)=
{\nf}_{\omega\mid\mathrm{RV}}^{\left(1\right)}\left(\Lambda\left(z'{\pi}\right)\right)$,
${f}_{\mathrm{RV}}^{\left(1\right)}\left(\Lambda\left(z'{\pi}\right)\right)$,
${f}_{\mathrm{RV}}\left(\Lambda\left(z'{\pi}\right)\right)$,
${\nf}_{\omega\mid\mathrm{RV}}\left(\Lambda\left(z'{\pi}\right)\right)$,
${\nf}_{\omega\mid\mathrm{RV}}^{\left(1\right)}\left(P^\ast\left(\Lambda\left(z'{\pi}\right),z\right)\right)$,
${f}_{\mathrm{RV}}^{\left(1\right)}\left(P^\ast\left(\Lambda\left(z'{\pi}\right),z\right)\right)$,
${f}_{\mathrm{RV}}\left(P^\ast\left(\Lambda\left(z'{\pi}\right),z\right)\right)$,
${\nf}_{\omega\mid\mathrm{RV}}\left(P^\ast\left(\Lambda\left(z'{\pi}\right),z\right)\right)$,
$\mf_{\omega\mid\mathrm{RV}}\left(\Lambda\left(z'{\pi}\right)\right)$,
and
$\frac{{f}_{P^\ast(\mathrm{RV},Z)}\left(\Lambda\left(z'\pi\right)\right)}{{f}_{\mathrm{RV}}\left(\Lambda\left(z'\pi\right)\right)}$, 
where $\omega$ is a component of $vec\left(\mu_0\left(X\right),\mu_1\left(X\right),Y,U\right)$.
\end{assumption}

\begin{assumption}[Approximation Error]\label{assn:18}
$
\sup_{z\in\mathcal{Z}_n}|\Lambda\left(z'{\pi}_0\right)-{\mf}_{S\mid Z}\left(z\right)|=o\left(n^{-r_1}\right)
$
holds.
Furthermore, 
$
\sup_{\pi\in\mathbf{P}_n}\left(\int\left(
\tau_{\Lambda\left(Z'\pi_0\right)}(z,\pi)-
\tau_{P}(z,\pi)\right)^4F_Z\left(dz\right)\right)^{1/4}=o\left(n^{-r_1}\mathbf{f}_n\right)
$
holds for every $\tau_{\mathrm{RV}}(z,\pi)=
{\nf}_{\omega\mid\mathrm{RV}}^{\left(1\right)}\left(\Lambda\left(z'{\pi}\right)\right)$,
${f}_{\mathrm{RV}}^{\left(1\right)}\left(\Lambda\left(z'{\pi}\right)\right)$,
${f}_{\mathrm{RV}}\left(\Lambda\left(z'{\pi}\right)\right)$,
${\nf}_{\omega\mid\mathrm{RV}}\left(\Lambda\left(z'{\pi}\right)\right)$,
${\nf}_{\omega\mid\mathrm{RV}}^{\left(1\right)}\left(P^\ast\left(\Lambda\left(z'{\pi}\right),z\right)\right)$,
${f}_{\mathrm{RV}}^{\left(1\right)}\left(P^\ast\left(\Lambda\left(z'{\pi}\right),z\right)\right)$,
${f}_{\mathrm{RV}}\left(P^\ast\left(\Lambda\left(z'{\pi}\right),z\right)\right)$,
${\nf}_{\omega\mid\mathrm{RV}}\left(P^\ast\left(\Lambda\left(z'{\pi}\right),z\right)\right)$,
$\mf_{\omega\mid\mathrm{RV}}\left(\Lambda\left(z'{\pi}\right)\right)$,
and
$\frac{{f}_{P^\ast(\mathrm{RV},Z)}\left(\Lambda\left(z'\pi\right)\right)}{{f}_{\mathrm{RV}}\left(\Lambda\left(z'\pi\right)\right)}$, 
where $\omega$ is a component of $vec\left(\mu_0\left(X\right),\mu_1\left(X\right),Y,U\right)$.
 \end{assumption}

\begin{assumption}[Negligible Truncation Bias]\label{assn:19}
For every component $\omega$ of $vec\left(\mu_0\left(X\right),\mu_1\left(X\right),Y,U\right)$, we have
$
\sup_{\pi\in\mathbf{P}_n}\int_{z\notin\mathcal{Z}_n}  
{\mf_{U\mid P}}\left(P^\ast\left(\Lambda\left(z'\pi\right),z\right)\right)
F_Z\left(dz\right)=o_p\left(n^{-1/2}\right),
$
$
\int_{z\notin\mathcal{Z}_n}{\zeta}\left(z\right){\mf}_{S\mid Z}\left(z\right)F_Z\left(dz\right)=o_p\left(n^{-1/2}\right),
$
$
\sup_{\pi\in\mathbf{P}_n}\left(\int_{z\notin\mathcal{Z}_n}\mf_{\omega\mid P}\left(\Lambda\left(z'{\pi}\right)\right)^4F_Z\left(dz\right)\right)^{1/4}=o_p\left(n^{-1/2}\right),
$
and
$
\int_{z\notin\mathcal{Z}_n}F_Z\left(dz\right)=o_p\left(1\right).
$
\end{assumption}
 
Our trimming is stochastic, i.e., integrating a function over the region under which the estimated density is above some threshold. 
As such, Assumption \ref{assn:19} includes not only the model primitives but also the estimated density inside. 
\citet{lee:2019} uses a stochastic trimming and obtains a condition for the negligible trimming bias using only the model primitives, e.g., condition (iii) of Theorem 3 of \citet{lee:2019}. 
In a similar manner to \citet{lee:2019}, we can also apply the uniform convergence of the density estimator \citep[cf.][Lemma 4]{lee:2019} and obtain a sufficient condition for  Assumption \ref{assn:19} using only the model primitives.

\begin{proposition}\label{prop:low}
If Assumptions \ref{assn:11}-\ref{assn:19} are satisfied, then Assumptions \ref{Assn_slow_conver}, \ref{Assn_CV}, \ref{Assn:Taylor_Errr} and \ref{Assn_just_consis} are satisfied.
\end{proposition}

\begin{proof}
Assumptions \ref{assn:11}-\ref{assn:19} imply Assumption \ref{Assn_slow_conver} by Lemmas \ref{lemma:low1} and \ref{lemma:low2} in Appendix \ref{sec:lemmas_lower_level}.
Assumptions \ref{assn:11}-\ref{assn:14}, \ref{assn:16}-\ref{assn:19} imply Assumption \ref{Assn_CV} by Lemmas \ref{lemma:low3} and \ref{lemma:low4} in Appendix \ref{sec:lemmas_lower_level}.
Assumptions \ref{assn:11}-\ref{assn:14}, \ref{assn:16}-\ref{assn:19} imply Assumption \ref{Assn:Taylor_Errr} by Lemma \ref{lemma:low5} in Appendix \ref{sec:lemmas_lower_level}.
Assumptions \ref{assn:11}-\ref{assn:19} imply Assumption \ref{Assn_just_consis} by Lemma \ref{lemma:Assn9_verification} in Appendix \ref{sec:lemmas_lower_level}.
\end{proof}

\section{An Alternative Specification for $P^\ast$}\label{sec:alternative_P_star}
As in \citet[][Section 3.2]{carneiro/heckman/vytlacil:2010}, we can consider multiple alternative specifications for $P^\ast$. Other than the specification in the main text, the counterfactual propensity score can take the form of $P^\ast=P\left(Z^\ast\left(Z\right)\right)$ with a known function $Z^\ast\left(\cdot\right)$, where the counterfactual policy affects the distribution of $Z$. 
In this section, we consider this alternative type of counterfactual propensity scores, and propose an orthogonal score for this case.
We redefine the preliminary parameters $\gamma$ as
$$
\gamma=
\left(
\frac{f_{Z^\ast\left(Z\right)}}{f_Z},
\mf_{S\mid Z},
\xi_1,
\zeta,
\mf_{U\mid P},
\kappa
\right),
$$
where $\kappa\left(z\right)=\mathrm{diag}\left({\mf}_{\left(\mu_0\left(X\right)',\mu_1\left(X\right)',1\right)'\mid Z}\left(z\right)\right)^{-1}\mathrm{diag}\left({\mf}_{\left(\mu_0\left(X\right)',\mu_1\left(X\right)',1\right)'\mid Z^\ast\left(Z\right)}\left(z\right)\right)$.
The orthogonal score function $m$ in this case has the same form for $m_1$ as in the main text, but has different formulas for $m_2\left(W;\tilde{\theta},\tilde{\gamma}\right)$ and $m_3\left(W;\tilde{\theta},\tilde{\gamma}\right)$:
\begingroup
\allowdisplaybreaks
\begin{align*}
{m}_2\left(W;\tilde{\theta},\tilde{\gamma}\right)
=&
\left(\mu_0\left(X\right)',\mu_1\left(X\right)',1\right)'\left(\tilde{\mf}_{S\mid Z}\left(Z^\ast\left(Z\right)\right)-\tilde{\mf}_{S\mid Z}\left(Z\right)\right)-\tilde{\theta}_2
\\
&+
\left(\mu_0\left(X\right)',\mu_1\left(X\right)',1\right)'\left(\tilde{\kappa}\left(Z\right)\frac{\tilde{f}_{Z^\ast\left(Z\right)}\left(Z\right)}{\tilde{f}_{Z}\left(Z\right)}-I\right)\left(S-\tilde{\mf}_{S\mid Z}\left(Z\right)\right)
\\
{m}_3\left(W;\tilde{\theta},\tilde{\gamma}\right)
=&
\tilde{\mf}_{U\mid P}\left(\tilde{\mf}_{S\mid Z}\left(Z^\ast\left(Z\right)\right)\right)-\mathcal{U}\left(W,\tilde{\theta}\right)-\tilde{\theta}_3
+
\frac{\tilde{f}_{Z^\ast\left(Z\right)}\left(Z\right)}{\tilde{f}_{Z}\left(Z\right)}\left(\mathcal{U}\left(W,\tilde{\theta}\right)-\tilde{\mf}_{U\mid P}\left(\tilde{\mf}_{S\mid Z}\left(Z\right)\right)\right).
\end{align*}
\endgroup
The asymptotic analysis for this orthogonal score is similar to that for the orthogonal score in the main text. 
A drawback of the current specification for $P^\ast$ is that, unlike the case of policies that directly affect $P$, we cannot allow for arbitrary high dimensions of $Z$ because of the need to estimate the joint density of $Z$.
\section{Orthogonal Score and Double Robustness}\label{sec:orthogonal_score_double_robustness}
Our moment condition is not robust against a misspecification of ${\mf}_{S\mid Z}$ in general. 
A counterexample is provided with $P^\ast\left(p,z\right)=p^2$. 
If ${\mf}_{S\mid Z}$ is misspecified by $\tilde{\mf}_{S\mid Z}$ then the expectation of $m_D$ at the true parameter value $\theta_D$ becomes 
\begingroup
\allowdisplaybreaks
\begin{eqnarray*}
E[m_D\left(Z;\theta_D,\tilde\gamma\right)]
&=&
E[P^\ast\left(\tilde{\mf}_{S\mid Z}\left(Z\right),Z\right)-\tilde{\mf}_{S\mid Z}\left(Z\right)-\theta_D+\left(\partial P^\ast\left(\tilde{\mf}_{S\mid Z}\left(Z\right),Z\right)-1\right)\left(S-\tilde{\mf}_{S\mid Z}\left(Z\right)\right)]\\
&=&
E[P^\ast\left(\tilde{\mf}_{S\mid Z}\left(Z\right),Z\right)-P^\ast\left({\mf}_{S\mid Z}\left(Z\right),Z\right)+\partial P^\ast\left(\tilde{\mf}_{S\mid Z}\left(Z\right),Z\right)\left({\mf}_{S\mid Z}\left(Z\right)-\tilde{\mf}_{S\mid Z}\left(Z\right)\right)]\\
&=&
E[\tilde{\mf}_{S\mid Z}\left(Z\right)^2-{\mf}_{S\mid Z}\left(Z\right)^2+2\tilde{\mf}_{S\mid Z}\left(Z\right)\left({\mf}_{S\mid Z}\left(Z\right)-\tilde{\mf}_{S\mid Z}\left(Z\right)\right)]\\
&=&
-E[\left(\tilde{\mf}_{S\mid Z}\left(Z\right)-{\mf}_{S\mid Z}\left(Z\right)\right)^2],
\end{eqnarray*}
\endgroup
where the second equality uses integration by parts, and the third equality uses $P^\ast\left(p,z\right)=p^2$. 
Therefore, unless $\tilde{\mf}_{S\mid Z}\left(Z\right)={\mf}_{S\mid Z}\left(Z\right)$ almost surely, we have
$$
E[m_D\left(Z;\theta_D,\tilde\gamma\right)]\ne 0.
$$
  
If ${\mf}_{S\mid Z}$ is correctly specified, then
we can show that the orthogonal moment condition is doubly robust between $\mf_{U\mid P}$ and $f_{P^\ast}/f_{P}$. 
In other words, at the true parameter value $\left(\theta_N,\theta_D\right)$, we have  
$$
E[m_N\left(Y,Z;\theta_N,\tilde\gamma\right)]=E[m_D\left(Y,Z;\theta_D,\tilde\gamma\right)]=0 
$$ 
as long as one of the following statements is true: 
\begin{equation}\label{eq:dr_assn1}
\left(\tilde{\mf}_{S\mid Z},\tilde\mf_{U\mid P}\right)=\left({\mf}_{S\mid Z},\mf_{U\mid P}\right)
\end{equation}
or 
\begin{equation}\label{eq:dr_assn2}
\left(\tilde{\mf}_{S\mid Z},{\tilde{f}_{P^\ast}}/{\tilde{f}_{P}}\right)=\left({\mf}_{S\mid Z},{f_{P^\ast}}/{f_{P}}\right).
\end{equation}
The equality $E[m_D\left(Y,Z;\theta_D,\tilde\gamma\right)]=0$ holds because $m_D\left(Y,Z;\theta_D,\tilde\gamma\right)$ depends only on $\tilde{\mf}_{S\mid Z}$ and ${\mf}_{S\mid Z}$ is correctly specified (i.e., $\tilde{\mf}_{S\mid Z}={\mf}_{S\mid Z}$). 
Regarding the $m_N$, we have 
\begingroup
\allowdisplaybreaks
\begin{eqnarray*}
E[m_N\left(Y,Z;\theta_N,\tilde\gamma\right)]
&=&
E\left[\tilde{\mf}_{U\mid P}\left(P^\ast\right)-Y-\theta_N
+
\frac{\tilde{f}_{P^\ast}\left(P\right)}{\tilde{f}_{P}\left(P\right)}\left(Y-\tilde{\mf}_{U\mid P}\left(P\right)\right)\right]
\\
&=&
E\left[\tilde{\mf}_{U\mid P}\left(P^\ast\right)-{\mf}_{U\mid P}\left(P^\ast\right)\right]+E\left[
\frac{\tilde{f}_{P^\ast}\left(P\right)}{\tilde{f}_{P}\left(P\right)}\left({\mf}_{U\mid P}\left(P\right)-\tilde{\mf}_{U\mid P}\left(P\right)\right)\right]
\\
&=&
\int \left(\tilde{\mf}_{U\mid P}\left(p\right)-{\mf}_{U\mid P}\left(p\right)\right)f_{P^\ast}\left(p\right)dp
+\int
\frac{\tilde{f}_{P^\ast}\left(p\right)}{\tilde{f}_{P}\left(p\right)}\left({\mf}_{U\mid P}\left(p\right)-\tilde{\mf}_{U\mid P}\left(p\right)\right)f_{P}\left(p\right)dp
\\
&=&
\int \frac{{f}_{P^\ast}\left(p\right)}{{f}_{P}\left(p\right)}\left(\tilde{\mf}_{U\mid P}\left(p\right)-{\mf}_{U\mid P}\left(p\right)\right)f_{P}\left(p\right)dp+\int
\frac{\tilde{f}_{P^\ast}\left(p\right)}{\tilde{f}_{P}\left(p\right)}\left({\mf}_{U\mid P}\left(p\right)-\tilde{\mf}_{U\mid P}\left(p\right)\right)f_{P}\left(p\right)dp
\\
&=&
\int\left(
\frac{\tilde{f}_{P^\ast}\left(p\right)}{\tilde{f}_{P}\left(p\right)}-\frac{{f}_{P^\ast}\left(p\right)}{{f}_{P}\left(p\right)}\right)\left({\mf}_{U\mid P}\left(p\right)-\tilde{\mf}_{U\mid P}\left(p\right)\right)f_{P}\left(p\right)dp,
\end{eqnarray*}
\endgroup
where the first equality uses $\tilde{\mf}_{S\mid Z}\left(Z\right)={\mf}_{S\mid Z}\left(Z\right)=P$, and the second equality follows from $E[m_N\left(Y,Z;\theta_N,\gamma\right)]=0$. 
Therefore, $E[m_N\left(Y,Z;\theta_N,\tilde\gamma\right)]=0$ as long as either \eqref{eq:dr_assn1} or \eqref{eq:dr_assn2} holds.

\section{Auxiliary Lemmas for the Main Results}

\subsection{Orthogonality}\label{sec:orthogonality_lemma}
The following lemma claims that our proposed score satisfies the orthogonality condition.
\begin{lemma}\label{lemma:CEINR_Assn6}
Suppose that Assumptions \ref{Assn_slow_conver} and \ref{Assn:Taylor_Errr} are satisfied.
For every $\hat{\gamma}_\ell=\hat{\gamma}_1,\ldots,\hat{\gamma}_L$, 
$$
\int m\left(w;\theta,\hat{\gamma}_\ell\right)F_W\left(dw\right)=o_p\left(n^{-1/2}\right).
$$
\end{lemma}

A proof is provided in Appendix \ref{sec:lemma:CEINR_Assn6}.
See Section \ref{sec:orthogonal_score} in the main text for discussions of this main result of the paper with intuitions.

\subsection{Consistency of Nuisance Parameter Estimation}
\begin{lemma}\label{lemma:CEINR_Assn4}
Suppose that Assumptions \ref{Assn_bounded_moments} and \ref{Assn_just_consis} are satisfied.
For every $\hat{\gamma}_\ell=\hat{\gamma}_1,\ldots,\hat{\gamma}_L$, 
\begingroup
\allowdisplaybreaks
\begin{align}
&
\int \left\|
\left(
\begin{array}{c}
m_{1,1}\left(w;\hat{\gamma}_\ell\right)\\
m_{2,1}\left(w;\hat{\gamma}_\ell\right)\\
m_{3,1}\left(w;\hat{\gamma}_\ell\right)\\
m_{3,2}\left(w;\hat{\gamma}_\ell\right)
\end{array}
\right)
-
\left(
\begin{array}{c}
m_{1,1}\left(w;{\gamma}\right)\\
m_{2,1}\left(w;{\gamma}\right)\\
m_{3,1}\left(w;{\gamma}\right)\\
m_{3,2}\left(w;{\gamma}\right)
\end{array}
\right)
\right\|F_W\left(dw\right)
=o_p\left(1\right)
\label{eq:first_eq}
\\
&
\int\left\|
\left(
\begin{array}{c}
m_{1,1}\left(w;\hat{\gamma}_\ell\right)\\
m_{2,1}\left(w;\hat{\gamma}_\ell\right)\\
m_{3,1}\left(w;\hat{\gamma}_\ell\right)\\
m_{3,2}\left(w;\hat{\gamma}_\ell\right)
\end{array}
\right)
-
\left(
\begin{array}{c}
m_{1,1}\left(w;{\gamma}\right)\\
m_{2,1}\left(w;{\gamma}\right)\\
m_{3,1}\left(w;{\gamma}\right)\\
m_{3,2}\left(w;{\gamma}\right)
\end{array}
\right)
\right\|^2F_W\left(dw\right)
=o_p\left(1\right)
\label{eq:second_eq}
\\
&
E_n\left[
\left\|
\left(\begin{array}{c}
m_{1,1}\left(W;\hat{\gamma}\right)\\
m_{2,1}\left(W;\hat{\gamma}\right)\\
m_{3,1}\left(W;\hat{\gamma}\right)\\
m_{3,2}\left(W;\hat{\gamma}\right)
\end{array}\right)
-
\left(\begin{array}{c}
m_{1,1}\left(W;{\gamma}\right)\\
m_{2,1}\left(W;{\gamma}\right)\\
m_{3,1}\left(W;{\gamma}\right)\\
m_{3,2}\left(W;{\gamma}\right)
\end{array}\right)
\right\|
\right]
=o_p\left(1\right).
\label{eq:third_eq}
\end{align}
\endgroup
where $m_{1,1}$, $m_{2,1}$, $m_{3,1}$ and $m_{3,2}$ are defined in Appendix \ref{sec:decomposition_of_moment_functions}.
\end{lemma}

A proof is provided in Appendix \ref{sec:lemma:CEINR_Assn4}.

\subsection{Identification of $\theta$}
\begin{lemma}\label{lemma:identification}
If Assumption \ref{Assn_identification} is satisfied, then
$\theta$ is the unique solution to $E[m\left(W;\tilde{\theta},{\gamma}\right)]=0$.
\end{lemma}

A proof is provided in Appendix \ref{sec:lemma:identification}.

\subsection{Consistent Estimation of $\theta$}
\begin{lemma}\label{lemma:consiste}
If Assumptions \ref{Assn_compact}, \ref{Assn_bounded_moments} and \ref{Assn_just_consis} are satisfied, then
$\hat\theta=\theta+o_p\left(1\right)$.
\end{lemma}

A proof is provided in Appendix \ref{sec:lemma:consiste}.

\section{Proofs for the Main Results}

\subsection{Proof of Lemma \ref{lemma:CEINR_Assn6}}\label{sec:lemma:CEINR_Assn6}
\begin{proof}
First, regarding the integral of $m_1$, the definition of $\theta_1$ implies 
\begingroup
\allowdisplaybreaks
\begin{align*}
\int m_1\left(w;\theta,\hat{\gamma}_\ell\right)F_W\left(dw\right)
=&
\int \left(\hat{\xi}_1\left(x,y,\hat{\mf}_{S\mid Z}\left(z\right)\right)-{\xi}_1\left(x,y,{\mf_{S\mid Z}}\left(z\right)\right)\right)F_W\left(dw\right)
\\
&+\int\hat{\zeta}\left(z\right)\left(s-\hat{\mf}_{S\mid Z}\left(z\right)\right)F_W\left(dw\right).
\end{align*}
\endgroup
By the law of total expectation, we have 
\begingroup
\allowdisplaybreaks
\begin{align*}
\int m_1\left(w;\theta,\hat{\gamma}_\ell\right)F_W\left(dw\right)
=&
\int \left(\hat{\xi}_1\left(x,y,\hat{\mf}_{S\mid Z}\left(z\right)\right)-{\xi}_1\left(x,y,{\mf_{S\mid Z}}\left(z\right)\right)\right)F_Z\left(dz\right)
\\
&+\int\hat{\zeta}\left(z\right)\left({\mf}_{S\mid Z}\left(z\right)-\hat{\mf}_{S\mid Z}\left(z\right)\right)F_Z\left(dz\right).
\end{align*}
We can rearrange the terms so that 
\begin{align*}
\int m_1\left(w;\theta,\hat{\gamma}_\ell\right)F_W\left(dw\right)
=&\int \left({\xi}_1\left(x,y,\hat{\mf}_{S\mid Z}\left(z\right)\right)-{\xi}_1\left(x,y,{\mf}_{S\mid Z}\left(z\right)\right)-{\zeta}\left(z\right)\left(\hat{\mf}_{S\mid Z}\left(z\right)-{\mf}_{S\mid Z}\left(z\right)\right)\right)F_W\left(dw\right).
\\&+\int \left(\hat{\xi}_1\left(x,y,\hat{\mf}_{S\mid Z}\left(z\right)\right)-{\xi}_1\left(x,y,\hat{\mf}_{S\mid Z}\left(z\right)\right)\right)F_W\left(dw\right)
\\&-\int \left(\hat{\zeta}\left(z\right)-{\zeta}\left(z\right)\right)\left(\hat{\mf}_{S\mid Z}\left(z\right)-{\mf}_{S\mid Z}\left(z\right)\right)F_W\left(dw\right).
\end{align*}
\endgroup
By Assumptions \ref{Assn_slow_conver}-\ref{Assn:Taylor_Errr}, all the terms on the right hand side of the above equation is $o_p\left(n^{-1/2}\right)$.  

Second, regarding the integral of $m_2$, the definition of $\theta_2$ implies 
\begingroup
\allowdisplaybreaks
\begin{align*}
\int m_2\left(w;\theta,\hat{\gamma}_\ell\right)F_W\left(dw\right)
=&
\int
\left(\mu_0\left(x\right)',\mu_1\left(x\right)',1\right)'\left(P^\ast\left(\hat{\mf}_{S\mid Z}\left(z\right),z\right)-P^\ast\left({\mf_{S\mid Z}}\left(z\right),z\right)-\hat{\mf}_{S\mid Z}\left(z\right)+{\mf_{S\mid Z}}\left(z\right)\right)
F_W\left(dw\right)\\
&+
\int
\left(\mu_0\left(x\right)',\mu_1\left(x\right)',1\right)'\left(\partial P^\ast\left(\hat{\mf}_{S\mid Z}\left(z\right),z\right)-1\right)
\left(s-\hat{\mf}_{S\mid Z}\left(z\right)\right)F_W\left(dw\right).
\end{align*}
\endgroup
Since 
$$
E[S\mid Z=z,X]
=
E[1\{P\leq V\}\mid Z=z,X]
=
E[1\{P\leq V\}\mid Z=z]
=
E[S\mid Z=z]
=
{\mf}_{S\mid Z}\left(z\right)
$$
and 
$$
E[\left(\mu_0\left(X\right)',\mu_1\left(X\right)',1\right)'\mid Z=z]
=
\mf_{\left(\mu_0\left(X\right)',\mu_1\left(X\right)',1\right)'\mid Z}\left(z\right),
$$
the law of total expectation implies 
\begingroup
\allowdisplaybreaks
\begin{align*}
&
\int m_2\left(w;\theta,\hat{\gamma}_\ell\right)F_W\left(dw\right)
\\
=&
\int
\mf_{\left(\mu_0\left(X\right)',\mu_1\left(X\right)',1\right)'\mid Z}\left(z\right)\left(P^\ast\left(\hat{\mf}_{S\mid Z}\left(z\right),z\right)-P^\ast\left({\mf_{S\mid Z}}\left(z\right),z\right)-\hat{\mf}_{S\mid Z}\left(z\right)+{\mf_{S\mid Z}}\left(z\right)\right)
F_Z\left(dz\right)\\
&+
\int
\mf_{\left(\mu_0\left(X\right)',\mu_1\left(X\right)',1\right)'\mid Z}\left(z\right)\left(\partial P^\ast\left(\hat{\mf}_{S\mid Z}\left(z\right),z\right)-1\right)
\left({\mf}_{S\mid Z}\left(z\right)-\hat{\mf}_{S\mid Z}\left(z\right)\right)F_Z\left(dz\right).
\end{align*}
We can rearrange the terms so that 
\begin{align*}
\int m_2\left(w;\theta,\hat{\gamma}_\ell\right)F_W\left(dw\right)
=&
\int
\boldsymbol{R}_2\left(z\right)
F_W\left(dw\right).
\end{align*}
\endgroup
By Assumption \ref{Assn:Taylor_Errr}, it follows that
\begin{align*}
&
\int m_2\left(w;\theta,\hat{\gamma}_\ell\right)F_W\left(dw\right)
= o_p\left(n^{-1/2}\right).
\end{align*}

Finally, regarding the integral of $m_3$, we can rewrite $\theta_3$ as 
\begin{align*}
\theta_3
=&
E\left[
\Delta_{U\mid P}\left(\mf_{S\mid Z}\left(Z\right)\right)\frac{{F}_{P}\left(\mf_{S\mid Z}\left(Z\right)\right)-{F}_{P^\ast}\left(\mf_{S\mid Z}\left(Z\right)\right)}{f_{P}\left(\mf_{S\mid Z}\left(Z\right)\right)}
\right]\\
=&
\int \Delta_{U\mid P}\left(p\right)\frac{{F}_{P}\left(p\right)-{F}_{P^\ast}\left(p\right)}{f_{P}\left(p\right)}f_{P}\left(p\right)dp\\
=&
\int \Delta_{U\mid P}\left(p\right)\left({F}_{P}\left(p\right)-{F}_{P^\ast}\left(p\right)\right)dp\\
=&
\mf_{U\mid P}\left(1\right)\left({F}_{P}\left(1\right)-{F}_{P^\ast}\left(1\right)\right)-\mf_{U\mid P}\left(0\right)\left({F}_{P}\left(0\right)-{F}_{P^\ast}\left(0\right)\right)-\int \mf_{U\mid P}\left(1\right)\left(f_{P}\left(p\right)-f_{P^\ast}\left(p\right)\right)dp\\
=&
E[\mf_{U\mid P}\left(P^\ast\left({\mf_{S\mid Z}}\left(Z\right),Z\right)\right)]-E[\mf_{U\mid P}\left(\mf_{S\mid Z}\left(Z\right)\right)],
\end{align*}
where the fourth equality uses integration by parts. 
Using the above equation, we have 
\begingroup
\allowdisplaybreaks
\begin{align*}
\int m_3\left(w;\theta,\hat{\gamma}_\ell\right)F_W\left(dw\right)
=&
\int \left(\hat{\mf}_{U\mid P}\left(P^\ast\left(\hat{\mf}_{S\mid Z}\left(z\right),z\right)\right)-{\mf_{U\mid P}}\left(P^\ast\left({\mf_{S\mid Z}}\left(z\right),z\right)\right)\right)F_W\left(dw\right)
\\&+
\int\frac{\hat{f}_{P^\ast}\left(\hat{\mf}_{S\mid Z}\left(z\right)\right)}{\hat{f}_{P}\left(\hat{\mf}_{S\mid Z}\left(z\right)\right)}\left(\mathcal{U}\left(w,\theta\right)-\hat{\mf}_{U\mid P}\left(\hat{\mf}_{S\mid Z}\left(z\right)\right)\right)F_W\left(dw\right)
\\&+
\int
\hat{\Delta}_{U\mid P}\left(P^\ast\left(\hat{\mf}_{S\mid Z}\left(z\right),z\right)\right)\partial P^\ast\left(\hat{\mf}_{S\mid Z}\left(z\right),z\right)
\left(s-\hat{\mf}_{S\mid Z}\left(z\right)\right)F_W\left(dw\right)
\\&-
\int
\frac{\hat{f}_{P^\ast}\left(\hat{\mf}_{S\mid Z}\left(z\right)\right)}{\hat{f}_{P}\left(\hat{\mf}_{S\mid Z}\left(z\right)\right)}\hat{\Delta}_{U\mid P}\left(\hat{\mf}_{S\mid Z}\left(z\right)\right)
\left(s-\hat{\mf}_{S\mid Z}\left(z\right)\right)F_W\left(dw\right).
\end{align*}
\endgroup
Since 
\begingroup
\allowdisplaybreaks
\begin{align*}
{\mf_{U\mid Z}}\left(z\right)
=&
E[U\mid Z=z]\\
=&
E[\left(1-S\right)U_0+SU_1\mid Z=z]\\
=&
\left(1-{\mf_{S\mid Z}}\left(z\right)\right)E[U_0\mid V\geq {\mf_{S\mid Z}}\left(z\right),Z=z]+{\mf_{S\mid Z}}\left(z\right)E[U_1\mid V<{\mf_{S\mid Z}}\left(z\right),Z=z]\\
=&
\left(1-{\mf_{S\mid Z}}\left(z\right)\right)E[U_0\mid V\geq {\mf_{S\mid Z}}\left(z\right)]+{\mf_{S\mid Z}}\left(z\right)E[U_1\mid V<{\mf_{S\mid Z}}\left(z\right)]\\
=&
\left(1-{\mf_{S\mid Z}}\left(z\right)\right)E[U_0\mid V\geq {\mf_{S\mid Z}}\left(z\right),P={\mf_{S\mid Z}}\left(z\right)]\\&\quad+{\mf_{S\mid Z}}\left(z\right)E[U_1\mid V<{\mf_{S\mid Z}}\left(z\right),P={\mf_{S\mid Z}}\left(z\right)]\\
=&
E[U\mid P={\mf_{S\mid Z}}\left(z\right)]\\
=&
{\mf_{U\mid P}}\left({\mf_{S\mid Z}}\left(z\right)\right)
\end{align*}
by Assumption \ref{assn1}, 
\endgroup
the law of total expectation implies
\begingroup
\allowdisplaybreaks
\begin{align*}
\int m_3\left(W;\theta,\hat{\gamma}_\ell\right)F_W\left(dw\right)
=&
\int \left(\hat{\mf}_{U\mid P}\left(P^\ast\left(\hat{\mf}_{S\mid Z}\left(z\right),z\right)\right)-{\mf_{U\mid P}}\left(P^\ast\left({\mf_{S\mid Z}}\left(z\right),z\right)\right)\right)F_Z\left(dz\right)
\\&+
\int\frac{\hat{f}_{P^\ast}\left(\hat{\mf}_{S\mid Z}\left(z\right)\right)}{\hat{f}_{P}\left(\hat{\mf}_{S\mid Z}\left(z\right)\right)}\left({\mf_{U\mid P}}\left({\mf_{S\mid Z}}\left(z\right)\right)-\hat{\mf}_{U\mid P}\left(\hat{\mf}_{S\mid Z}\left(z\right)\right)\right)F_Z\left(dz\right)
\\&+
\int
\hat{\Delta}_{U\mid P}\left(P^\ast\left(\hat{\mf}_{S\mid Z}\left(z\right),z\right)\right)\partial P^\ast\left(\hat{\mf}_{S\mid Z}\left(z\right),z\right)
\left({\mf_{S\mid Z}}\left(z\right)-\hat{\mf}_{S\mid Z}\left(z\right)\right)F_Z\left(dz\right)
\\&-
\int
\frac{\hat{f}_{P^\ast}\left(\hat{\mf}_{S\mid Z}\left(z\right)\right)}{\hat{f}_{P}\left(\hat{\mf}_{S\mid Z}\left(z\right)\right)}\hat{\Delta}_{U\mid P}\left(\hat{\mf}_{S\mid Z}\left(z\right)\right)
\left({\mf_{S\mid Z}}\left(z\right)-\hat{\mf}_{S\mid Z}\left(z\right)\right)F_Z\left(dz\right).
\end{align*}
\endgroup
We can rearrange the terms so that 
\begingroup
\allowdisplaybreaks
\begin{align*}
&
\int m_3\left(W;\theta,\hat{\gamma}_\ell\right)F_W\left(dw\right)
\\=&
-\int\frac{\hat{f}_{P^\ast}\left(\hat{\mf}_{S\mid Z}\left(z\right)\right)}{\hat{f}_{P}\left(\hat{\mf}_{S\mid Z}\left(z\right)\right)}
\left(\hat{\Delta}_{U\mid P}\left(\hat{\mf}_{S\mid Z}\left(z\right)\right)-{\Delta}_{U\mid P}\left({\mf}_{S\mid Z}\left(z\right)\right)\right)
\left({\mf_{S\mid Z}}\left(z\right)-\hat{\mf}_{S\mid Z}\left(z\right)\right)F_Z\left(dz\right)
\\&-
\int \left(
\hat{\Delta}_{U\mid P}\left(P^\ast\left(\hat{\mf}_{S\mid Z}\left(z\right),z\right)\right)\partial P^\ast\left(\hat{\mf}_{S\mid Z}\left(z\right),z\right) \right.
\\
&\qquad \left.
-{\Delta}_{U\mid P}\left(P^\ast\left({\mf}_{S\mid Z}\left(z\right),z\right)\right)\partial P^\ast\left({\mf}_{S\mid Z}\left(z\right),z\right)
\right)
\left(\hat{\mf}_{S\mid Z}\left(z\right)-{\mf}_{S\mid Z}\left(z\right)\right)F_Z\left(dz\right)
\\&+
\int \left(\hat{\mf}_{U\mid P}\left(P^\ast\left(\hat{\mf}_{S\mid Z}\left(z\right),z\right)\right)-{\mf}_{U\mid P}\left(P^\ast\left(\hat{\mf}_{S\mid Z}\left(z\right),z\right)\right)\right)F_Z\left(dz\right)
\\
&
-\int\frac{\hat{f}_{P^\ast}\left(\hat{\mf}_{S\mid Z}\left(z\right)\right)}{\hat{f}_{P}\left(\hat{\mf}_{S\mid Z}\left(z\right)\right)}
\left(\hat{\mf}_{U\mid P}\left(\hat{\mf}_{S\mid Z}\left(z\right)\right)-{\mf}_{U\mid P}\left(\hat{\mf}_{S\mid Z}\left(z\right)\right)\right)
F_Z\left(dz\right)
\\&+
\int \left({\mf}_{U\mid P}\left(P^\ast\left(\hat{\mf}_{S\mid Z}\left(z\right),z\right)\right)-{\mf_{U\mid P}}\left(P^\ast\left({\mf_{S\mid Z}}\left(z\right),z\right)\right) \right.
\\
&\qquad \left.
-
{\Delta}_{U\mid P}\left(P^\ast\left({\mf}_{S\mid Z}\left(z\right),z\right)\right)\partial P^\ast\left({\mf}_{S\mid Z}\left(z\right),z\right)
\left(\hat{\mf}_{S\mid Z}\left(z\right)-{\mf}_{S\mid Z}\left(z\right)\right)\right)F_Z\left(dz\right)
\\&+\int\frac{\hat{f}_{P^\ast}\left(\hat{\mf}_{S\mid Z}\left(z\right)\right)}{\hat{f}_{P}\left(\hat{\mf}_{S\mid Z}\left(z\right)\right)}
\left({\mf_{U\mid P}}\left({\mf_{S\mid Z}}\left(z\right)\right)-{\mf}_{U\mid P}\left(\hat{\mf}_{S\mid Z}\left(z\right)\right)
-{\Delta}_{U\mid P}\left({\mf}_{S\mid Z}\left(z\right)\right)
\left({\mf_{S\mid Z}}\left(z\right)-\hat{\mf}_{S\mid Z}\left(z\right)\right)\right)F_Z\left(dz\right).
\end{align*}
The first two terms appear in Assumption \ref{Assn_slow_conver}, 
the third term appears in Assumption \ref{Assn_CV}, and  
the last two terms appear in Assumption \ref{Assn:Taylor_Errr}. All of them are assumed to be $o_p\left(n^{-1/2}\right)$, and therefore  
\begin{align*}
&
\int m_3\left(w;\theta,\hat{\gamma}_\ell\right)F_W\left(dw\right)
= o_p\left(n^{-1/2}\right).
\end{align*}
\endgroup
This completes a proof.
\end{proof}

\subsection{Proof of Lemma \ref{lemma:CEINR_Assn4}}\label{sec:lemma:CEINR_Assn4}
\begin{proof}
We will first show that \eqref{eq:second_eq} holds.
By the triangle inequality for the Euclidean norm $\|\cdot\|$, we have 
\begingroup
\allowdisplaybreaks
\begin{align*}
\left\|m_{1,1}\left(w;\hat{\gamma}_\ell\right)-m_{1,1}\left(w;\gamma\right)\right\|
\leq&
\left\|\hat{\xi}_1\left(x,y,\hat{\mf}_{S\mid Z}\left(z\right)\right)-{\xi}_1\left(x,y,{\mf}_{S\mid Z}\left(z\right)\right)\right\|
\\&+
|s|\left\|\hat{\zeta}\left(z\right)-{\zeta}\left(z\right)\right\|
\\&+
\left\|\hat{\zeta}\left(z\right)\hat{\mf}_{S\mid Z}\left(z\right)-{\zeta}\left(z\right){\mf}_{S\mid Z}\left(z\right)\right\|.
\end{align*}
\endgroup
By the triangle inequality for the $L^2$ norm, we have
\begingroup
\allowdisplaybreaks
\begin{align*}
&
\left(\int \left\|m_{1,1}\left(w;\hat{\gamma}_\ell\right)-m_{1,1}\left(w;\gamma\right)\right\|^2F_W\left(dw\right)\right)^{1/2}
\nonumber\\
\leq&
\left(\int\left\|\hat{\xi}_1\left(x,y,\hat{\mf}_{S\mid Z}\left(z\right)\right)-{\xi}_1\left(x,y,{\mf}_{S\mid Z}\left(z\right)\right)\right\|^2F_W\left(dw\right)\right)^{1/2}
\\&+
\left(\int\left\|\hat{\zeta}\left(z\right)-{\zeta}\left(z\right)\right\|^2F_W\left(dw\right)\right)^{1/2}
\\&+
\left(\int\left\|\hat{\zeta}\left(z\right)\hat{\mf}_{S\mid Z}\left(z\right)-{\zeta}\left(z\right){\mf}_{S\mid Z}\left(z\right)\right\|^2F_W\left(dw\right)\right)^{1/2}.
\end{align*}
\endgroup
The last three terms are $o_p\left(1\right)$ under Assumption \ref{Assn_just_consis}.

Second, by the triangle inequality and Cauchy-Schwartz inequality for the Euclidean norm $\|\cdot\|$, we have 
\begingroup
\allowdisplaybreaks
\begin{align*}
&
\left\|m_{2,1}\left(w;\hat{\gamma}_\ell\right)-m_{2,1}\left(w;\gamma\right)\right\|
\nonumber\\
\leq&
\|\left(\mu_0\left(x\right)',\mu_1\left(x\right)',1\right)'\|\left|
P^\ast\left(\hat{\mf}_{S\mid Z}\left(z\right),z\right)-P^\ast\left({\mf}_{S\mid Z}\left(z\right),z\right)
\right|
+
\|\left(\mu_0\left(x\right)',\mu_1\left(x\right)',1\right)'\|\left|
\hat{\mf}_{S\mid Z}\left(z\right)-{\mf}_{S\mid Z}\left(z\right)
\right|
\\&+
\|\left(\mu_0\left(x\right)',\mu_1\left(x\right)',1\right)'\|\left|
\partial P^\ast\left(\hat{\mf}_{S\mid Z}\left(z\right),z\right)-\partial P^\ast\left(\mf_{S\mid Z}\left(z\right),z\right)
\right|
\\&+
\|\left(\mu_0\left(x\right)',\mu_1\left(x\right)',1\right)'\|\left|
\hat{\mf}_{S\mid Z}\left(z\right)\partial P^\ast\left(\hat{\mf}_{S\mid Z}\left(z\right),z\right)
-{\mf}_{S\mid Z}\left(z\right)\partial P^\ast\left(\mf_{S\mid Z}\left(z\right),z\right)
\right|.
\end{align*}
\endgroup
By the triangle inequality for the $L^2$ norm,
\begingroup
\allowdisplaybreaks
\begin{align*}
&
\left(\int \left\|m_{2,1}\left(w;\hat{\gamma}_\ell\right)-m_{2,1}\left(w;\gamma\right)\right\|^2F_W\left(dw\right)\right)^{1/2}
\nonumber\\
\leq&
\left(\int \left\|\left(\mu_0\left(x\right)',\mu_1\left(x\right)',1\right)'\right\|^2\left(
P^\ast\left(\hat{\mf}_{S\mid Z}\left(z\right),z\right)-P^\ast\left({\mf}_{S\mid Z}\left(z\right),z\right)
\right)^2F_W\left(dw\right)\right)^{1/2}
\\&+
\left(\int \|\left(\mu_0\left(x\right)',\mu_1\left(x\right)',1\right)'\|^2
\left(
\hat{\mf}_{S\mid Z}\left(z\right)-{\mf}_{S\mid Z}\left(z\right)
\right)^2F_W\left(dw\right)\right)^{1/2}
\\&+
\left(\int \|\left(\mu_0\left(x\right)',\mu_1\left(x\right)',1\right)'\|^2\left(
\partial P^\ast\left(\hat{\mf}_{S\mid Z}\left(z\right),z\right)-\partial P^\ast\left(\mf_{S\mid Z}\left(z\right),z\right)
\right)^2F_W\left(dw\right)\right)^{1/2}
\\&+
\left(\int \|\left(\mu_0\left(x\right)',\mu_1\left(x\right)',1\right)'\|^2\left(
\hat{\mf}_{S\mid Z}\left(z\right)\partial P^\ast\left(\hat{\mf}_{S\mid Z}\left(z\right),z\right)
-{\mf}_{S\mid Z}\left(z\right)\partial P^\ast\left(\mf_{S\mid Z}\left(z\right),z\right)
\right)^2F_W\left(dw\right)\right)^{1/2}
\\
\leq&
\left(\int {\mf}_{\|\left(\mu_0\left(x\right)',\mu_1\left(x\right)',1\right)'\|^2\mid Z}\left(z\right)\left(
P^\ast\left(\hat{\mf}_{S\mid Z}\left(z\right),z\right)-P^\ast\left({\mf}_{S\mid Z}\left(z\right),z\right)
\right)^2F_W\left(dw\right)\right)^{1/2}
\\&+
\left(\int {\mf}_{\|\left(\mu_0\left(x\right)',\mu_1\left(x\right)',1\right)'\|^2\mid Z}\left(z\right)\left(
\hat{\mf}_{S\mid Z}\left(z\right)-{\mf}_{S\mid Z}\left(z\right)
\right)^2F_W\left(dw\right)\right)^{1/2}
\\&+
\left(\int {\mf}_{\|\left(\mu_0\left(x\right)',\mu_1\left(x\right)',1\right)'\|^2\mid Z}\left(z\right)\left(
\partial P^\ast\left(\hat{\mf}_{S\mid Z}\left(z\right),z\right)-\partial P^\ast\left(\mf_{S\mid Z}\left(z\right),z\right)
\right)^2F_W\left(dw\right)\right)^{1/2}
\\&+
\left(\int {\mf}_{\|\left(\mu_0\left(x\right)',\mu_1\left(x\right)',1\right)'\|^2\mid Z}\left(z\right)\left(
\hat{\mf}_{S\mid Z}\left(z\right)\partial P^\ast\left(\hat{\mf}_{S\mid Z}\left(z\right),z\right)
-{\mf}_{S\mid Z}\left(z\right)\partial P^\ast\left(\mf_{S\mid Z}\left(z\right),z\right)
\right)^2F_W\left(dw\right)\right)^{1/2}
\end{align*}
\endgroup
and it is $o_p\left(1\right)$ under Assumptions \ref{Assn_bounded_moments} and \ref{Assn_just_consis}.

Third, by the triangle inequality for the Euclidean norm $\|\cdot\|$, we have 
\begingroup
\allowdisplaybreaks
\begin{align*}
&
\left|m_{3,1}\left(w;\hat{\gamma}_\ell\right)-m_{3,1}\left(w;\gamma\right)\right|
\nonumber\\
\leq&
\left|
\hat{\mf}_{U\mid P}\left(P^\ast\left(\hat{\mf}_{S\mid Z}\left(z\right),z\right)\right)-{\mf}_{U\mid P}\left(P^\ast\left({\mf}_{S\mid Z}\left(z\right),z\right)\right)
\right|
\\&+
|y|\left|\frac{\hat{f}_{P^\ast}\left(\hat{\mf}_{S\mid Z}\left(z\right)\right)}{\hat{f}_{P}\left(\hat{\mf}_{S\mid Z}\left(z\right)\right)}-\frac{{f}_{P^\ast}\left({\mf}_{S\mid Z}\left(z\right)\right)}{{f}_{P}\left({\mf}_{S\mid Z}\left(z\right)\right)}\right|
\\&+
\left|\frac{\hat{f}_{P^\ast}\left(\hat{\mf}_{S\mid Z}\left(z\right)\right)}{\hat{f}_{P}\left(\hat{\mf}_{S\mid Z}\left(z\right)\right)}\hat{\mf}_{U\mid P}\left(\hat{\mf}_{S\mid Z}\left(z\right)\right)-\frac{{f}_{P^\ast}\left({\mf}_{S\mid Z}\left(z\right)\right)}{{f}_{P}\left({\mf}_{S\mid Z}\left(z\right)\right)}{\mf}_{U\mid P}\left({\mf}_{S\mid Z}\left(z\right)\right)\right|
\\&+
\left|\frac{\hat{f}_{P^\ast}\left(\hat{\mf}_{S\mid Z}\left(z\right)\right)}{\hat{f}_{P}\left(\hat{\mf}_{S\mid Z}\left(z\right)\right)}\hat{\Delta}_{U\mid P}\left(\hat{\mf}_{S\mid Z}\left(z\right)\right)-\frac{{f}_{P^\ast}\left({\mf}_{S\mid Z}\left(z\right)\right)}{{f}_{P}\left({\mf}_{S\mid Z}\left(z\right)\right)}{\Delta_{U\mid P}}\left({\mf}_{S\mid Z}\left(z\right)\right)\right|
\\&+
\left|\hat{\Delta}_{U\mid P}\left(P^\ast\left(\hat{\mf}_{S\mid Z}\left(z\right),z\right)\right)\partial P^\ast\left(\hat{\mf}_{S\mid Z}\left(z\right),z\right)-{\Delta_{U\mid P}}\left(P^\ast\left({\mf}_{S\mid Z}\left(z\right),z\right)\right)\partial P^\ast\left({\mf}_{S\mid Z}\left(z\right),z\right)\right|
\\&+
\left|\frac{\hat{f}_{P^\ast}\left(\hat{\mf}_{S\mid Z}\left(z\right)\right)}{\hat{f}_{P}\left(\hat{\mf}_{S\mid Z}\left(z\right)\right)}\hat{\Delta}_{U\mid P}\left(\hat{\mf}_{S\mid Z}\left(z\right)\right)\hat{\mf}_{S\mid Z}\left(z\right)-\frac{{f}_{P^\ast}\left({\mf}_{S\mid Z}\left(z\right)\right)}{{f}_{P}\left({\mf}_{S\mid Z}\left(z\right)\right)}{\Delta_{U\mid P}}\left({\mf}_{S\mid Z}\left(z\right)\right){\mf}_{S\mid Z}\left(z\right)\right|
\\&+
\left|\hat{\Delta}_{U\mid P}\left(P^\ast\left(\hat{\mf}_{S\mid Z}\left(z\right),z\right)\right)\partial P^\ast\left(\hat{\mf}_{S\mid Z}\left(z\right),z\right)\hat{\mf}_{S\mid Z}\left(z\right)-{\Delta_{U\mid P}}\left(P^\ast\left({\mf}_{S\mid Z}\left(z\right),z\right)\right)\partial P^\ast\left({\mf}_{S\mid Z}\left(z\right),z\right){\mf}_{S\mid Z}\left(z\right)\right|.
\end{align*}
\endgroup
Therefore, 
\begingroup
\allowdisplaybreaks
\begin{align*}
&
\left(\int\left(m_{3,1}\left(w;\hat{\gamma}_\ell\right)-m_{3,1}\left(w;\gamma\right)\right)^2F_W\left(dw\right)\right)^{1/2}
\nonumber\\
\leq&
\left(\int
\left(
\hat{\mf}_{U\mid P}\left(P^\ast\left(\hat{\mf}_{S\mid Z}\left(z\right),z\right)\right)-{\mf}_{U\mid P}\left(P^\ast\left({\mf}_{S\mid Z}\left(z\right),z\right)\right)
\right)^2
F_Z\left(dz\right)\right)^{1/2}
\\&+
\left(\int
{\mf}_{Y^2\mid Z}\left(z\right)
\left(\frac{\hat{f}_{P^\ast}\left(\hat{\mf}_{S\mid Z}\left(z\right)\right)}{\hat{f}_{P}\left(\hat{\mf}_{S\mid Z}\left(z\right)\right)}-\frac{{f}_{P^\ast}\left({\mf}_{S\mid Z}\left(z\right)\right)}{{f}_{P}\left({\mf}_{S\mid Z}\left(z\right)\right)}\right)^2
F_Z\left(dz\right)\right)^{1/2}
\\&+
\left(\int
\left(\frac{\hat{f}_{P^\ast}\left(\hat{\mf}_{S\mid Z}\left(z\right)\right)}{\hat{f}_{P}\left(\hat{\mf}_{S\mid Z}\left(z\right)\right)}\hat{\mf}_{U\mid P}\left(\hat{\mf}_{S\mid Z}\left(z\right)\right)-\frac{{f}_{P^\ast}\left({\mf}_{S\mid Z}\left(z\right)\right)}{{f}_{P}\left({\mf}_{S\mid Z}\left(z\right)\right)}{\mf}_{U\mid P}\left({\mf}_{S\mid Z}\left(z\right)\right)\right)^2
F_Z\left(dz\right)\right)^{1/2}
\\&+
\left(\int
\left(\frac{\hat{f}_{P^\ast}\left(\hat{\mf}_{S\mid Z}\left(z\right)\right)}{\hat{f}_{P}\left(\hat{\mf}_{S\mid Z}\left(z\right)\right)}\hat{\Delta}_{U\mid P}\left(\hat{\mf}_{S\mid Z}\left(z\right)\right)-\frac{{f}_{P^\ast}\left({\mf}_{S\mid Z}\left(z\right)\right)}{{f}_{P}\left({\mf}_{S\mid Z}\left(z\right)\right)}{\Delta_{U\mid P}}\left({\mf}_{S\mid Z}\left(z\right)\right)\right)^2
F_Z\left(dz\right)\right)^{1/2}
\\&+
\left(\int
\left(\hat{\Delta}_{U\mid P}\left(P^\ast\left(\hat{\mf}_{S\mid Z}\left(z\right),z\right)\right)\partial P^\ast\left(\hat{\mf}_{S\mid Z}\left(z\right),z\right)-{\Delta_{U\mid P}}\left(P^\ast\left({\mf}_{S\mid Z}\left(z\right),z\right)\right)\partial P^\ast\left({\mf}_{S\mid Z}\left(z\right),z\right)\right)^2
F_Z\left(dz\right)\right)^{1/2}
\\&+
\left(\int
\left(\frac{\hat{f}_{P^\ast}\left(\hat{\mf}_{S\mid Z}\left(z\right)\right)}{\hat{f}_{P}\left(\hat{\mf}_{S\mid Z}\left(z\right)\right)}\hat{\Delta}_{U\mid P}\left(\hat{\mf}_{S\mid Z}\left(z\right)\right)\hat{\mf}_{S\mid Z}\left(z\right)-\frac{{f}_{P^\ast}\left({\mf}_{S\mid Z}\left(z\right)\right)}{{f}_{P}\left({\mf}_{S\mid Z}\left(z\right)\right)}{\Delta_{U\mid P}}\left({\mf}_{S\mid Z}\left(z\right)\right){\mf}_{S\mid Z}\left(z\right)\right)^2
F_Z\left(dz\right)\right)^{1/2}
\\&+
\left(\int
\left(\hat{\Delta}_{U\mid P}\left(P^\ast\left(\hat{\mf}_{S\mid Z}\left(z\right),z\right)\right)\partial P^\ast\left(\hat{\mf}_{S\mid Z}\left(z\right),z\right)\hat{\mf}_{S\mid Z}\left(z\right) \right.\right.
\\
&\qquad\left.\left. -{\Delta_{U\mid P}}\left(P^\ast\left({\mf}_{S\mid Z}\left(z\right),z\right)\right)\partial P^\ast\left({\mf}_{S\mid Z}\left(z\right),z\right){\mf}_{S\mid Z}\left(z\right)\right)^2
F_Z\left(dz\right)\right)^{1/2},
\end{align*}
\endgroup
and it is $o_p\left(1\right)$ under Assumptions \ref{Assn_bounded_moments} and \ref{Assn_just_consis}.

Fourth, by the triangle inequality for the Euclidean norm $\|\cdot\|$, we have 
$$
\left|m_{3,2}\left(w;\hat{\gamma}_\ell\right)-m_{3,2}\left(w;\gamma\right)\right|
\leq
\|\left(\left(1-S\right)\mu_0\left(X\right)',S\mu_1\left(X\right)'\right)'\|\left|\frac{\hat{f}_{P^\ast}\left(\hat{\mf}_{S\mid Z}\left(z\right)\right)}{\hat{f}_{P}\left(\hat{\mf}_{S\mid Z}\left(z\right)\right)}-\frac{{f}_{P^\ast}\left({\mf}_{S\mid Z}\left(z\right)\right)}{{f}_{P}\left({\mf}_{S\mid Z}\left(z\right)\right)}\right|.
$$
Therefore, 
\begingroup
\allowdisplaybreaks
\begin{eqnarray*}
&&
\left(\int \left(m_{3,2}\left(w;\hat{\gamma}_\ell\right)-m_{3,2}\left(w;\gamma\right)\right)^2F_W\left(dw\right)\right)^{1/2}
\\
&&\leq
\left(\int{\mf}_{\|\left(\left(1-S\right)\mu_0\left(X\right)',S\mu_1\left(X\right)'\right)'\|^2\mid Z}\left(z\right) \left(\frac{\hat{f}_{P^\ast}\left(\hat{\mf}_{S\mid Z}\left(z\right)\right)}{\hat{f}_{P}\left(\hat{\mf}_{S\mid Z}\left(z\right)\right)}-\frac{{f}_{P^\ast}\left({\mf}_{S\mid Z}\left(z\right)\right)}{{f}_{P}\left({\mf}_{S\mid Z}\left(z\right)\right)}\right)^2F_Z\left(dz\right)\right)^{1/2},
\end{eqnarray*}
\endgroup
and it is $o_p\left(1\right)$ under Assumptions \ref{Assn_bounded_moments} and \ref{Assn_just_consis}.

The above arguments together show that \eqref{eq:second_eq} holds.
Equation \eqref{eq:first_eq} follows from \eqref{eq:second_eq} via H\"older's inequality. 
Moreover, Equation \eqref{eq:third_eq} follows from \eqref{eq:first_eq} due to the independence between $\{W_i: i\in l_\ell\}$ and $\hat{\gamma}_\ell$.
\end{proof}

\subsection{Proof of Lemma \ref{lemma:identification}}\label{sec:lemma:identification}
\begin{proof}
The values of $\theta_1$ and $\theta_2$ are uniquely determined by $E[m_1\left(W;\tilde{\theta},{\gamma}\right)]=0$ and $E[m_2\left(W;\tilde{\theta},{\gamma}\right)]=0$ by construction. 
Once $\theta_1$ is determined, then $\theta_3$ is uniquely determined by $E[m_3\left(W;\tilde{\theta},{\gamma}\right)]=0$ under Assumption \ref{Assn_identification}. 
\end{proof}

\subsection{Proof of Lemma \ref{lemma:consiste}}\label{sec:lemma:consiste}
\begin{proof}
The consistency follows from \citet[Theorem 2.1]{newey/mcfadden:1994}.
It suffices to show that 
$$
\sup_{\tilde{\theta}\in\Theta}\left|\left\|E_n[m\left(W;\tilde{\theta},\hat{\gamma}\right)]\right\|^2-\left\|E[m\left(W;\tilde{\theta},{\gamma}\right)]\right\|^2\right|=o_p\left(1\right).
$$
We are going to show 
$$
\sup_{\tilde{\theta}\in\Theta}\left\|E_n[m\left(W;\tilde{\theta},\hat{\gamma}\right)]-E[m\left(W;\tilde{\theta},{\gamma}\right)]\right\|=o_p\left(1\right).
$$
By Assumption \ref{Assn_compact}, it suffices to show that 
$$
E_n\left[
\left(\begin{array}{c}
m_{1,1}\left(W;\hat{\gamma}\right)\\
m_{2,1}\left(W;\hat{\gamma}\right)\\
m_{3,1}\left(W;\hat{\gamma}\right)\\
m_{3,2}\left(W;\hat{\gamma}\right)
\end{array}\right)
-
\left(\begin{array}{c}
m_{1,1}\left(W;{\gamma}\right)\\
m_{2,1}\left(W;{\gamma}\right)\\
m_{3,1}\left(W;{\gamma}\right)\\
m_{3,2}\left(W;{\gamma}\right)
\end{array}\right)
\right]
=o_p\left(1\right)
$$
and
$$
\left(E_n-E\right)\left[\left(\begin{array}{c}
m_{1,1}\left(W;{\gamma}\right)\\
m_{2,1}\left(W;{\gamma}\right)\\
m_{3,1}\left(W;{\gamma}\right)\\
m_{3,2}\left(W;{\gamma}\right)
\end{array}\right)\right]=o_p\left(1\right),
$$
where $m_{1,1}$, $m_{2,1}$, $m_{3,1}$ and $m_{3,2}$ are defined in Appendix \ref{sec:decomposition_of_moment_functions}.
The first convergence is the third statement in Lemma \ref{lemma:CEINR_Assn4} that holds under Assumptions \ref{Assn_bounded_moments} and \ref{Assn_just_consis}.
The second convergence follows from the weak law of large numbers because Assumption \ref{Assn_bounded_moments} implies the bounded second moments.  
\end{proof}

\subsection{Proof of Theorem \ref{theorem:normal}}\label{sec:theorem:normal}
\begin{proof}
First, note that the identification and consistency follows from Lemma \ref{lemma:identification} and \ref{lemma:consiste}, respectively.
The claimed statement thus follows from \citet[Theorem 16]{chernozhukov2016locally} and Lemma \ref{lemma:consiste}. 
Assumption 4 of \citet{chernozhukov2016locally} follows from the second statement in our Lemma \ref{lemma:CEINR_Assn4}. 
Assumption 5 of \citet{chernozhukov2016locally} holds automatically here because there is no adjustment term for the moment conditions. 
Assumption 6 of \citet{chernozhukov2016locally} follows from our Lemma \ref{lemma:CEINR_Assn6}.
Assumption 7 of \citet{chernozhukov2016locally} follows from the linearity of the moment condition in $\theta$ and the first statement in our Lemma \ref{lemma:CEINR_Assn4}.
\end{proof}

\subsection{Proof of Theorem \ref{theorem:var_est}}\label
{sec:theorem:variance_est}
\begin{proof}
As in Appendix \ref{sec:theorem:normal}, Assumptions 4 and 7 of \citet{chernozhukov2016locally} hold. Thus the statement of this theorem follows from \citet[Theorem 17]{chernozhukov2016locally} and $\lambda\left(\hat\theta\right)=\lambda\left(\theta\right)+o_p\left(1\right)$. 
\end{proof}

\section{Auxiliary Lemmas for Lower-Level Sufficient Conditions}\label{sec:lemmas_lower_level}

This section collects auxiliary lemmas that are used to prove Proposition \ref{prop:low} (Appendix \ref{sec:lowe_level_sufficient_conditions}) for lower-level sufficient conditions.
Proofs of these auxiliary lemmas are also contained together in this section.

\begin{lemma}[Convergence Rate of the High-Dimensional Propensity Score Estimator]\label{lemma:prop_score_conv}
If Assumptions \ref{assn:11}, \ref{assn:12}, and \ref{assn:18} are satisfied, then we have
$$
\sup_{z\in\mathcal{Z}_n}|\hat{\mf}_{S\mid Z}\left(z\right)-{\mf}_{S\mid Z}\left(z\right)|=o_p\left(n^{-r_1}\right).
$$
\end{lemma}
\begin{proof}
The lemma follows from
\begingroup
\allowdisplaybreaks  
\begin{align*}
\sup_{z\in\mathcal{Z}_n}|\hat{\mf}_{S\mid Z}\left(z\right)-{\mf}_{S\mid Z}\left(z\right)|
\leq&
\sup_{z\in\mathcal{Z}_n}|\Lambda\left(z'\hat{\pi}\right)-\Lambda\left(z'{\pi}_0\right)|
+
\sup_{z\in\mathcal{Z}_n}|\Lambda\left(z'{\pi}_0\right)-{\mf}_{S\mid Z}\left(z\right)|
\\
\leq&
\sup_{u}|\Lambda^{\left(1\right)}\left(u\right)|\sup_{z\in\mathcal{Z}_n}\|z\|_{\infty}\|\hat{\pi}-{\pi}_0\|_1
+
\sup_{z\in\mathcal{Z}_n}|\Lambda\left(z'{\pi}_0\right)-{\mf}_{S\mid Z}\left(z\right)|,
\end{align*}
\endgroup
together with Assumptions \ref{assn:11}, \ref{assn:12}, and \ref{assn:18}.
\end{proof}

\begin{lemma}[Convergence Rate of Kernel-Weighted Estimators]\label{lemma:kernel_L2_1}
If Assumption \ref{assn:11}, \ref{assn:12}, \ref{assn:13}, \ref{assn:14}, and \ref{assn:16} are satisfied, then we have
$$
\sup_{\left(p,\pi\right)\in[0,1]\times\mathbf{P}_n}\left|\frac{1}{|I_\ell^c|}\sum_{j\in I_\ell^c}\omega_jK_{h}\left(\Lambda\left(Z_j'\pi\right)-p\right)-{\nf}_{\omega\mid\Lambda\left(Z'{\pi}\right)}\left(p\right)\right|=o_p\left(n^{-r_1}\mathbf{f}_n\right).
$$
\end{lemma}
\begin{proof}
Since 
\begingroup
\allowdisplaybreaks
\begin{align*}
\left|E[\omega K_{h}\left(\Lambda\left(Z'\pi\right)-p\right)]-{\nf}_{\omega\mid\Lambda\left(Z'{\pi}\right)}\left(p\right)\right|
=&
\left|\int {\nf}_{\omega\mid\Lambda\left(Z'{\pi}\right)}\left(p+uh\right)K\left(u\right)du-{\nf}_{\omega\mid\Lambda\left(Z'{\pi}\right)}\left(p\right)\right|\\
\leq&
\frac{1}{2}h^2\int 
u^2K\left(u\right)du\sup_{\left(p,\pi\right)\in[0,1]\times\mathbf{P}_n}\left|{\nf}_{\omega\mid\Lambda\left(Z'{\pi}\right)}^{\left(2\right)}\left(p\right)\right|
\end{align*}
\endgroup
under Assumptions \ref{assn:14} and \ref{assn:16},
we have 
$$
\sup_{\left(p,\pi\right)\in[0,1]\times\mathbf{P}_n}\left|E\left[\omega K_{h}\left(\Lambda\left(Z'{\pi}\right)-p\right)\right]
-{\nf}_{\omega\mid\Lambda\left(Z'{\pi}\right)}\left(p\right)
\right|=O\left(h^2\right)=o\left(n^{-r_1}\mathbf{f}_n\right)
$$
by Assumptions \ref{assn:13}, \ref{assn:14} and \ref{assn:16}.
Therefore, in the remainder of the proof, it suffices to show that
$$
\sup_{\left(p,\pi\right)\in[0,1]\times\mathbf{P}_n}\left|\frac{1}{|I_\ell^c|}\sum_{j\in I_\ell^c}\omega_jK_{h}\left(\Lambda\left(Z_j'\pi\right)-p\right)-E\left[\omega K_{h}\left(\Lambda\left(Z'\pi\right)-p\right)\right]\right|=o_p\left(n^{-r_1}\mathbf{f}_n\right).
$$

Define the function class 
$$
\mathcal{F}=\left\{W\mapsto\omega K_{h}\left(\Lambda\left(Z'\pi\right)-p\right): \left(p,\pi\right)\in[0,1]\times\mathbf{P}_n\right\}
$$ 
with the envelope $F$ given by 
$$
F\left(W\right)=\sigma+\sup_{\left(p,\pi\right)\in[0,1]\times\mathbf{P}_n}\left(|\omega K_{h}\left(\Lambda\left(Z'\pi\right)-p\right)|+hK_n^{-1}\left\|\frac{\partial}{\partial\left(p,\pi'\right)'}\omega K_{h}\left(\Lambda\left(Z'\pi\right)-p\right)\right\|_{\infty}
\right)
$$
where
$$
\sigma^2=h^{-1}\int K\left(u\right)^2du\sup_{\left(p,\pi\right)\in[0,1]\times\mathbf{P}_n}{\nf}_{\omega^2\mid \Lambda\left(Z'\pi\right)}\left(p\right).
$$

First, we are going to show the function class $\mathcal{F}$ is of VC type with characteristics  
$$
\left(A,v\right)=\left(\frac{1+8{\left(d_Z+1\right)}\sup_{\pi\in\mathbf{P}_{n,\mathcal{K}}}\|\pi\|_1}{hK_n^{-1}},1+\bar{s}\right)
$$  and envelope $F$. 
Note that 
$$
\mathcal{F}=\bigcup_{\mathcal{K}\subset\{1,\ldots,d_Z\}: |\mathcal{K}|\leq \bar{s}}\mathcal{F}_\mathcal{K},
$$
where 
\begingroup
\allowdisplaybreaks
\begin{align*}
\mathbf{P}_{n,\mathcal{K}}=&\{\pi\in\mathbf{P}_{n}: \pi_k=0\mbox{ for every }k\notin \mathcal{K}\}
\qquad\text{and}
\\
\mathcal{F}_\mathcal{K}=&\left\{W\mapsto\omega K_{h}\left(\Lambda\left(Z'\pi\right)-p\right): \left(p,\pi\right)\in[0,1]\times\mathbf{P}_{n,\mathcal{K}}\right\}.
\end{align*}
\endgroup
We can bound the covering number as follows:
\begingroup
\allowdisplaybreaks 
\begin{align*}
N\left(\mathcal{F},\|\cdot\|_{Q,2},\epsilon\|F\|_{Q,2}\right)
\leq&
\sum_{\mathcal{K}\subset\{1,\ldots,d_Z\}: |\mathcal{K}|\leq \bar{s}}N\left(\mathcal{F}_\mathcal{K},\|\cdot\|_{Q,2},\epsilon\|F\|_{Q,2}\right)\\
\leq&
\left(d_Z+1\right)^{\bar{s}}\sup_{\mathcal{K}\subset\{1,\ldots,d_Z\}: |\mathcal{K}|\leq \bar{s}}N\left(\mathcal{F}_\mathcal{K},\|\cdot\|_{Q,2},\epsilon\|F\|_{Q,2}\right),
\end{align*}
\endgroup
where the inequality comes from $|\{\mathcal{K}\subset\{1,\ldots,d_Z\}: |\mathcal{K}|\leq \bar{s}\}|\leq \left(d_Z+1\right)^{\bar{s}}$.
Since $\mathbf{P}_{n,\mathcal{K}}$ is an $\bar{s}$-dimensional space and 
$$
\left\|\frac{\partial}{\partial\left(p,\pi'\right)'}\omega K_{h}\left(\Lambda\left(Z'\pi\right)-p\right)\right\|_2
\leq
\sqrt{d_Z+1}\left\|\frac{\partial}{\partial\left(p,\pi'\right)'}\omega K_{h}\left(\Lambda\left(Z'\pi\right)-p\right)\right\|_{\infty}\leq \frac{\sqrt{d_Z+1}}{hK_n^{-1}}F\left(W\right),
$$
Theorem 2.7.11 of \cite{vanweak} gives 
\begingroup
\allowdisplaybreaks
\begin{align*}
N\left(\mathcal{F}_\mathcal{K},\|\cdot\|_{Q,2},\epsilon\|F\|_{Q,2}\right)
=&
N\left(\mathcal{F}_\mathcal{K},\|\cdot\|_{Q,2},\left(\epsilon\frac{hK_n^{-1}}{\sqrt{d_Z+1}}\right)\frac{\sqrt{d_Z+1}}{hK_n^{-1}}\|F\|_{Q,2}\right)
\\
\leq&
N\left(\mathcal{F}_\mathcal{K},\|\cdot\|_{Q,2},\left(\epsilon\frac{hK_n^{-1}}{\sqrt{d_Z+1}}\right)\left\|\sup_{\left(p,\pi\right)\in[0,1]\times\mathbf{P}_n}\left\|\tilde F_{\left(p,\pi\right)}\right\|_2\right\|_{Q,2}\right)
\\
\leq&
\left(1+\frac{4\sup_{\pi_1,\pi_2\in\mathbf{P}_{n,\mathcal{K}}}\|\pi_1-\pi_2\|_2}{\epsilon hK_n^{-1}/\sqrt{d_Z+1}}\right)^{1+\bar{s}}
\\
\leq&
\left(\frac{A}{\epsilon}\right)^{1+\bar{s}}
\end{align*}
\endgroup
for every $\epsilon\in(0,1]$ by Assumption \ref{assn:13}, where $\tilde F_{\left(p,\pi\right)}$ is defined by
$$
\tilde F_{\left(p,\pi\right)}\left(z\right) = \frac{\partial}{\partial\left(p,\pi'\right)'}\omega K_{h}\left(\Lambda\left(z'\pi\right)-p\right).
$$

Second, we are going to show    
\begin{align*}
E[F\left(W\right)^2]^{1/2}&=O\left(h^{-1}\right).
\\
E\left[\max_{j\in I_\ell^c}F\left(W_j\right)^2\right]^{1/2}&=O\left(n^{1/4}h^{-1}\right).
\end{align*}
These inequalities follow from 
\begingroup
\allowdisplaybreaks
\begin{align*}
F\left(W\right)
\leq&
\sigma+h^{-1}|\omega_j| 
\max\left\{
\sup_{u}|K\left(u\right)|+\sup_{u}|K^{\left(1\right)}\left(u\right)|\left(1+\sup_{u}|\Lambda^{\left(1\right)}\left(u\right)|\right)
\right\},
\\
E\left[\max_{j\in I_\ell^c}F\left(W_j\right)^2\right]^{1/2}
=&
O\left(h^{-1}\right) + E\left[\max_{j\in I_\ell^c}\omega_j^2\right]^{1/2}O\left(h^{-1}\right)
\leq
n^{1/4}E\left[\omega^4\right]^{1/4}O\left(h^{-1}\right),
\end{align*}
\endgroup
and Assumptions \ref{assn:13}, \ref{assn:14} and \ref{assn:16}.

Applying Corollary 5.1 of \cite{chernozhukov2014gaussian} with all these characteristics,  
we obtain 
\begin{align*}
&
E\left[\sup_{\left(p,\pi\right)\in[0,1]\times\mathbf{P}_n}\left|\frac{1}{|I_\ell^c|}\sum_{j\in I_\ell^c}\omega_jK_{h}\left(\Lambda\left(Z_j'\pi\right)-p\right)-E\left[\omega K_{h}\left(\Lambda\left(Z'\pi\right)-p\right)\right]\right|\right]
\\
&
=
n^{-1/2}O\left(\sqrt{v\sigma^2\log\left(\frac{AE[F\left(W\right)^2]^{1/2}}{\sigma}\right)}+\frac{vE[\max_{j\in I_\ell^c}F\left(W_j\right)^2]^{1/2}}{\sqrt{n}}\log\left(\frac{AE[F\left(W\right)^2]^{1/2}}{\sigma}\right)\right)
\\
&
=
n^{-1/2}O\left(\sqrt{\left(1+\bar{s}\right)h^{-1}\log\left(d_Z\vee n\right)}+\frac{\left(1+\bar{s}\right)n^{1/4}h^{-1}}{\sqrt{n}}\log\left(d_Z\vee n\right)\right)
\end{align*}
by Assumptions \ref{assn:13}, \ref{assn:14} and \ref{assn:16}.
Therefore, by Assumption  \ref{assn:13}, the statement of the lemma holds. 
\end{proof}

\begin{lemma}\label{lemma:kernel_L2_2}
If Assumptions \ref{assn:11}, \ref{assn:12}, \ref{assn:13}, \ref{assn:14}, and \ref{assn:16} are satisfied, then we have 
$$
\sup_{\left(p,\pi\right)\in[0,1]\times\mathbf{P}_n}\left|
\frac{1}{|I_\ell^c|}\sum_{j\in I_\ell^c}K_{h}\left(P^\ast\left(\Lambda\left(Z_j'\pi\right),Z_j\right)-p\right)
-{f}_{P^\ast\left(\Lambda\left(Z'\pi\right),Z\right)}\left(p\right)\right|
=o_p\left(n^{-r_1}\mathbf{f}_n\right)
$$
\end{lemma}
\begin{proof}
The statement follows through a similar argument to the proof of Lemma \ref{lemma:kernel_L2_1}.
\end{proof}

\begin{lemma}\label{lemma:kernel_L2_3}
If Assumptions \ref{assn:11}, \ref{assn:12}, \ref{assn:13}, \ref{assn:14},  and \ref{assn:16} are satisfied, then we have
$$
\sup_{\left(p,\pi\right)\in[0,1]\times\mathbf{P}_n}\left|\frac{1}{|I_\ell^c|}\sum_{j\in I_\ell^c}\omega_jK_{h}^{\left(1\right)}\left(\Lambda\left(Z_j'\pi\right)-p\right)+{\nf}_{\omega\mid\Lambda\left(Z'{\pi}\right)}^{\left(1\right)}\left(p\right)\right|=o_p\left(n^{-r_2}\mathbf{f}_n\right).
$$
\end{lemma}
\begin{proof}
Since 
\begin{align*}
\left|E[\omega K_{h}^{\left(1\right)}\left(\Lambda\left(Z'\pi\right)-p\right)]+{\nf}_{\omega\mid\Lambda\left(Z'{\pi}\right)}^{\left(1\right)}\left(p\right)\right|
=&
\left|\int {\nf}_{\omega\mid\Lambda\left(Z'{\pi}\right)}^{\left(1\right)}\left(p+uh\right)K\left(u\right)du-{\nf}_{\omega\mid\Lambda\left(Z'{\pi}\right)}^{\left(1\right)}\left(p\right)\right|\\
\leq&
\frac{1}{2}h^2\int 
u^2K\left(u\right)du\sup_{p}\left|{\nf}_{\omega\mid\Lambda\left(Z'{\pi}\right)}^{\left(3\right)}\left(p\right)\right|
\end{align*}
by Assumptions \ref{assn:14} and \ref{assn:16},
we have 
$$
\sup_{\left(p,\pi\right)\in[0,1]\times\mathbf{P}_n}\left|E[\omega K_{h}^{\left(1\right)}\left(\Lambda\left(Z'\pi\right)-p\right)]+{\nf}_{\omega\mid\Lambda\left(Z'{\pi}\right)}^{\left(1\right)}\left(p\right)\right|=O\left(h^2\right)=o\left(n^{-r_2}\mathbf{f}_n\right)
$$
by Assumptions \ref{assn:13}, \ref{assn:14} and \ref{assn:16}.
Therefore, in the remainder of the proof, it suffices to show that
$$
E\left[\sup_{\left(p,\pi\right)\in[0,1]\times\mathbf{P}_n}\left|\frac{1}{|I_\ell^c|}\sum_{j\in I_\ell^c}\omega_jK_{h}^{\left(1\right)}\left(\Lambda\left(Z_j'\pi\right)-p\right)-E\left[\omega K_{h}\left(\Lambda\left(Z'\pi\right)-p\right)\right]\right|\right]=o\left(n^{-r_2}\mathbf{f}_n\right).
$$

Define the following function class 
$$
\mathcal{F}=\left\{W\mapsto\omega K_{h}^{\left(1\right)}\left(\Lambda\left(Z'\pi\right)-p\right): \left(p,\pi\right)\in[0,1]\times\mathbf{P}_n\right\}.
$$ 
with envelope $F$ given by 
$$
F\left(W\right)=\sigma+\sup_{\left(p,\pi\right)\in[0,1]\times\mathbf{P}_n}\left(|\omega K_{h}^{\left(1\right)}\left(\Lambda\left(Z'\pi\right)-p\right)|+hK_n^{-1}\left\|\frac{\partial}{\partial\left(p,\pi'\right)'}\omega K_{h}^{\left(1\right)}\left(\Lambda\left(Z'\pi\right)-p\right)\right\|_{\infty}
\right),
$$
where 
$$
\sigma^2=h^{-3}\int K^{\left(1\right)}\left(u\right)^2du\sup_{\left(p,\pi\right)\in[0,1]\times\mathbf{P}_n}{\nf}_{\omega^2\mid \Lambda\left(Z'\pi\right)}\left(p\right).
$$ 
In a similar way to the proof of Lemma \ref{lemma:kernel_L2_1}, we can show the function class $\mathcal{F}$ is of VC type with characteristics  
$$
\left(A,v\right)=\left(\frac{1+8{\left(d_Z+1\right)}\sup_{\pi\in\mathbf{P}_{n,\mathcal{K}}}\|\pi\|_1}{hK_n^{-1}},1+\bar{s}\right)
$$ and envelope $F$. 
Moreover, we have $E[F\left(W\right)^2]^{1/2}=O\left(h^{-2}\right)$ and $E\left[\max_{j\in I_\ell^c}F\left(W_j\right)^2\right]^{1/2}=O\left(n^{1/4}h^{-2}\right)$ by Assumptions \ref{assn:13}, \ref{assn:14} and \ref{assn:16}.
Applying Corollary 5.1 of \cite{chernozhukov2014gaussian} with all these characteristics,
we obtain 
\begin{align*}
&
E\left[\sup_{\left(p,\pi\right)\in[0,1]\times\mathbf{P}_n}\left|\frac{1}{|I_\ell^c|}\sum_{j\in I_\ell^c}\omega_jK_{h}\left(\Lambda\left(Z_j'\pi\right)-p\right)-E\left[\omega K_{h}\left(\Lambda\left(Z'\pi\right)-p\right)\right]\right|\right]
\\
&
=
n^{-1/2}O\left(\sqrt{v\sigma^2\log\left(\frac{AE[F\left(W\right)^2]^{1/2}}{\sigma}\right)}+\frac{vE[\max_{j\in I_\ell^c}F\left(W_j\right)^2]^{1/2}}{\sqrt{n}}\log\left(\frac{AE[F\left(W\right)^2]^{1/2}}{\sigma}\right)\right)
\\
&
=
n^{-1/2}O\left(\sqrt{\left(1+\bar{s}\right)h^{-3}\log\left(d_Z\vee n\right)}+\frac{\left(1+\bar{s}\right)n^{1/4}h^{-2}}{\sqrt{n}}\log\left(d_Z\vee n\right)\right)
\end{align*}
by Assumptions \ref{assn:13}, \ref{assn:14} and \ref{assn:16}.
Therefore, by Assumption  \ref{assn:13}, the statement of the lemma holds. 
\end{proof}

\begin{lemma}\label{lemma:denstiy_ratio_conv_process}
If Assumptions \ref{assn:11}, \ref{assn:12}, \ref{assn:13}, \ref{assn:14},  and \ref{assn:16} are satisfied, then
$$
\left(\int_{z\in\mathcal{Z}_n}\left(\frac{
\hat{f}_{P^\ast}\left(\Lambda\left(z'\hat\pi\right)\right)
}{\hat{f}_{P}\left(\Lambda\left(z'\hat\pi\right)\right)}-\mathrm{ratio}\left(z,\hat\pi\right)\right)^4F_{Z}\left(dz\right)\right)^{1/4}=o_p\left(n^{-r_1}\right)
$$
holds,
where 
$$
\mathrm{ratio}\left(z,\pi\right)=\frac{f_{P^\ast\left(\Lambda\left(Z'\pi\right),Z\right)}\left(\Lambda\left(z'\pi\right)\right)}{f_{\Lambda\left(Z'\pi\right)}\left(\Lambda\left(z'\pi\right)\right)}.
$$
\end{lemma}
\begin{proof}
For every $z\in\mathcal{Z}_n$, we can write
\begingroup
\allowdisplaybreaks
\begin{align*}
&
\frac{
\frac{1}{|I_\ell^c|}\sum_{j\in I_\ell^c}K_{h}\left(P^\ast\left(\Lambda\left(Z_j'\pi\right),Z_j\right)-\Lambda\left(z'\pi\right)\right)
}{\frac{1}{|I_\ell^c|}\sum_{j\in I_\ell^c}K_{h}\left(\Lambda\left(Z_j'\pi\right)-\Lambda\left(z'\pi\right)\right)}-\frac{f_{P^\ast\left(\Lambda\left(Z'\pi\right),Z\right)}\left(\Lambda\left(z'\pi\right)\right)}{f_{\Lambda\left(Z'\pi\right)}\left(\Lambda\left(z'\pi\right)\right)}
\\
=&
\frac{1}{\frac{1}{|I_\ell^c|}\sum_{j\in I_\ell^c}K_{h}\left(\Lambda\left(Z_j'\pi\right)-\Lambda\left(z'\pi\right)\right)}\left(
\frac{1}{|I_\ell^c|}\sum_{j\in I_\ell^c}K_{h}\left(P^\ast\left(\Lambda\left(Z_j'\pi\right),Z_j\right)-\Lambda\left(z'\pi\right)\right)
-{f}_{P^\ast\left(\Lambda\left(Z'\pi\right),Z\right)}\left(\Lambda\left(z'\pi\right)\right)\right)
\\
&-\frac{1}{\frac{1}{|I_\ell^c|}\sum_{j\in I_\ell^c}K_{h}\left(\Lambda\left(Z_j'\pi\right)-\Lambda\left(z'\pi\right)\right)}\frac{{f}_{P^\ast\left(\Lambda\left(Z'\pi\right),Z\right)}\left(\Lambda\left(z'\pi\right)\right)}{{f}_{\Lambda\left(Z'\pi\right)}\left(\Lambda\left(z'\pi\right)\right)}
\left(\frac{1}{|I_\ell^c|}\sum_{j\in I_\ell^c}K_{h}\left(\Lambda\left(Z_j'\pi\right)-\Lambda\left(z'\pi\right)\right)-{f}_{\Lambda\left(Z'\pi\right)}\left(\Lambda\left(z'\pi\right)\right)\right).
\end{align*}
\endgroup
Thus, from how $\mathbf{f}_n$ is defined in \eqref{eq:fn}, we have
\begingroup
\allowdisplaybreaks
\begin{align*}
&
\left(\int_{z\in\mathcal{Z}_n} \left(
\frac{\hat{f}_{P^\ast}\left(\Lambda\left(z'\hat\pi\right)\right)}{\hat{f}_{P}\left(\Lambda\left(z'\hat\pi\right)\right)}
-
\mathrm{ratio}\left(z,\hat\pi\right)\right)^4
F_Z\left(dz\right)
\right)^{1/4}
\\&\leq
\mathbf{f}_n^{-1}
\sup_{\left(p,\pi\right)\in[0,1]\times\mathbf{P}_n}\left|
\frac{1}{|I_\ell^c|}\sum_{j\in I_\ell^c}K_{h}\left(P^\ast\left(\Lambda\left(Z_j'\pi\right),Z_j\right)-p\right)
-{f}_{P^\ast\left(\Lambda\left(Z'\pi\right),Z\right)}\left(p\right)\right|
\\&+
\mathbf{f}_n^{-1}\sup_{\left(p,\pi\right)\in[0,1]\times\mathbf{P}_n}
\left|
\frac{1}{|I_\ell^c|}\sum_{j\in I_\ell^c}K_{h}\left(\Lambda\left(Z_j'\pi\right)-\Lambda\left(z'\pi\right)\right)-{f}_{\Lambda\left(Z'\pi\right)}\left(\Lambda\left(z'\pi\right)\right)\right|
\\&\times\sup_{\pi\in\mathbf{P}_n}
\left(\int_{z\in\mathcal{Z}_n}
\left(\frac{{f}_{P^\ast\left(\Lambda\left(Z'\pi\right),Z\right)}\left(\Lambda\left(z'\pi\right)\right)}{{f}_{\Lambda\left(Z'\pi\right)}\left(\Lambda\left(z'\pi\right)\right)}\right)^4F_Z\left(dz\right)
\right)^{1/4}.
\end{align*}
\endgroup
By Assumption \ref{assn:16} and Lemmas \ref{lemma:kernel_L2_1} and \ref{lemma:kernel_L2_2} under Assumptions \ref{assn:11}, \ref{assn:12}, \ref{assn:13}, \ref{assn:14},  and \ref{assn:16}, the statement of the lemma follows. 
\end{proof}

\begin{lemma}\label{lemma:mean_P_conv_process}
Under Assumption \ref{assn:11}, \ref{assn:12}, \ref{assn:13}, \ref{assn:14},  \ref{assn:16}, \ref{assn:17}, and \ref{assn:18}, 
$$
\left(\int_{z\in\mathcal{Z}_n}\left(\hat{\mf}_{\omega\mid P}\left(\Lambda\left(z'\hat{\pi}\right)\right)-{\mf}_{\omega\mid P}\left(\Lambda\left(z'\hat{\pi}\right)\right)\right)^4F_Z\left(dz\right)\right)^{1/4}=o_p\left(n^{-r_1}\right).
$$
\end{lemma}
\begin{proof}
For every $z\in\mathcal{Z}_n$, we can write 
\begingroup
\allowdisplaybreaks
\begin{align*}
&
|\hat{\mf}_{\omega\mid P}\left(\Lambda\left(z'\hat{\pi}\right)\right)-{\mf}_{\omega\mid P}\left(\Lambda\left(z'\hat{\pi}\right)\right)|
\\&\leq
\left|\frac{\hat{\nf}_{\omega\mid P}\left(\Lambda\left(z'\hat{\pi}\right)\right)-{\nf}_{\omega\mid P}\left(\Lambda\left(z'\hat{\pi}\right)\right)}{\hat{f}_{P}\left(\Lambda\left(z'\hat\pi\right)\right)}\right|
+
|{\mf}_{\omega\mid P}\left(\Lambda\left(z'\hat{\pi}\right)\right)|\left|\frac{\hat{f}_{P}\left(\Lambda\left(z'\hat\pi\right)\right)-{f}_{P}\left(\Lambda\left(z'\hat\pi\right)\right)}{\hat{f}_{P}\left(\Lambda\left(z'\hat\pi\right)\right)}\right|
\\&\leq
\mathbf{f}_n^{-1}\left|\hat{\nf}_{\omega\mid P}\left(\Lambda\left(z'\hat{\pi}\right)\right)-{\nf}_{\omega\mid P}\left(\Lambda\left(z'\hat{\pi}\right)\right)\right|
+
\mathbf{f}_n^{-1}|{\mf}_{\omega\mid P}\left(\Lambda\left(z'\hat{\pi}\right)\right)|\left|\hat{f}_{P}\left(\Lambda\left(z'\hat\pi\right)\right)-{f}_{P}\left(\Lambda\left(z'\hat\pi\right)\right)\right|
\end{align*}
\endgroup
from how $\mathbf{f}_n$ is defined in \eqref{eq:fn}.
Thus, 
\begingroup
\allowdisplaybreaks
\begin{align*}
&
\left(\int_{z\in\mathcal{Z}_n}\left(\hat{\mf}_{\omega\mid P}\left(\Lambda\left(z'\hat{\pi}\right)\right)-{\mf}_{\omega\mid P}\left(\Lambda\left(z'\hat{\pi}\right)\right)\right)^4F_Z\left(dz\right)\right)^{1/4}
\\&\leq
\mathbf{f}_n^{-1}\left(\int_{z\in\mathcal{Z}_n}\left(\hat{\nf}_{\omega\mid P}\left(\Lambda\left(z'\hat{\pi}\right)\right)-{\nf}_{\omega\mid P}\left(\Lambda\left(z'\hat{\pi}\right)\right)\right)^4F_Z\left(dz\right)\right)^{1/4}
\\&+
\mathbf{f}_n^{-1}\left(\int_{z\in\mathcal{Z}_n}\left(\hat{f}_{P}\left(\Lambda\left(z'\hat\pi\right)\right)-{f}_{P}\left(\Lambda\left(z'\hat\pi\right)\right)\right)^4F_Z\left(dz\right)\right)^{1/4}\sup_{p\in[0,1]}|{\mf}_{\omega\mid P}\left(p\right)|
\\&\leq
\mathbf{f}_n^{-1}\sup_{\pi\in\mathbf{P}_n}\left(\int_{z\in\mathcal{Z}_n}\left(\frac{1}{|I_\ell^c|}\sum_{j\in I_\ell^c}\omega_jK_{h}\left(\Lambda\left(Z_j'\pi\right)-\Lambda\left(z'\pi\right)\right)-{\nf}_{\omega\mid\Lambda\left(Z'\pi\right)}\left(\Lambda\left(z'{\pi}\right)\right)\right)^4F_Z\left(dz\right)\right)^{1/4}
\\&+
\mathbf{f}_n^{-1}\sup_{\pi\in\mathbf{P}_n}
\left(\int_{z\in\mathcal{Z}_n}\left({\nf}_{\omega\mid\Lambda\left(Z'\pi\right)}\left(\Lambda\left(z'{\pi}\right)\right)-{\nf}_{\omega\mid P}\left(\Lambda\left(z'{\pi}\right)\right)\right)^4F_Z\left(dz\right)\right)^{1/4}
\\&+
\mathbf{f}_n^{-1}\sup_{\pi\in\mathbf{P}_n}\left(\int_{z\in\mathcal{Z}_n}\left(\frac{1}{|I_\ell^c|}\sum_{j\in I_\ell^c}K_{h}\left(\Lambda\left(Z_j'\pi\right)-\Lambda\left(z'\pi\right)\right)-{f}_{\Lambda\left(Z'\pi\right)}\left(\Lambda\left(z'\pi\right)\right)\right)^4F_Z\left(dz\right)\right)^{1/4}\sup_{p\in[0,1]}|{\mf}_{\omega\mid P}\left(p\right)|
\\&+
\mathbf{f}_n^{-1}\sup_{\pi\in\mathbf{P}_n}\left(\int_{z\in\mathcal{Z}_n}\left({f}_{\Lambda\left(Z'\pi\right)}\left(\Lambda\left(z'\pi\right)\right)-{f}_{P}\left(\Lambda\left(z'\pi\right)\right)\right)^4F_Z\left(dz\right)\right)^{1/4}\sup_{p\in[0,1]}|{\mf}_{\omega\mid P}\left(p\right)|.
\end{align*}
\endgroup
Note that 
\begingroup
\allowdisplaybreaks
\begin{align*}
&
\sup_{\pi\in\mathbf{P}_n}
\left(\int_{z\in\mathcal{Z}_n}\left({\nf}_{\omega\mid\Lambda\left(Z'\pi\right)}\left(\Lambda\left(z'{\pi}\right)\right)-{\nf}_{\omega\mid P}\left(\Lambda\left(z'{\pi}\right)\right)\right)^4F_Z\left(dz\right)\right)^{1/4}
\\
\leq&
\sup_{\pi\in\mathbf{P}_n}
\left(\int_{z\in\mathcal{Z}_n}\left({\nf}_{\omega\mid\Lambda\left(Z'\pi\right)}\left(\Lambda\left(z'{\pi}\right)\right)-{\nf}_{\omega\mid\Lambda\left(Z'\pi_0\right)}\left(\Lambda\left(z'{\pi}\right)\right)\right)^4F_Z\left(dz\right)\right)^{1/4}
\\&+
\sup_{\pi\in\mathbf{P}_n}\left(\int_{z\in\mathcal{Z}_n}\left({\nf}_{\omega\mid\Lambda\left(Z'\pi_0\right)}\left(\Lambda\left(z'{\pi}\right)\right)-{\nf}_{\omega\mid P}\left(\Lambda\left(z'{\pi}\right)\right)\right)^4F_Z\left(dz\right)\right)^{1/4}
\\
=&
o\left(n^{-r_1}\mathbf{f}_n\right)
\end{align*}
\endgroup
and
\begingroup
\allowdisplaybreaks
\begin{align*}
&
\sup_{\pi\in\mathbf{P}_n}\left(\int_{z\in\mathcal{Z}_n}\left({f}_{\Lambda\left(Z'\pi\right)}\left(\Lambda\left(z'\pi\right)\right)-{f}_{P}\left(\Lambda\left(z'\pi\right)\right)\right)^4F_Z\left(dz\right)\right)^{1/4}
\\
\leq&
\sup_{\pi\in\mathbf{P}_n}\left(\int_{z\in\mathcal{Z}_n}\left({f}_{\Lambda\left(Z'\pi\right)}\left(\Lambda\left(z'\pi\right)\right)-{f}_{\Lambda\left(Z'\pi_0\right)}\left(\Lambda\left(z'\pi\right)\right)\right)^4F_Z\left(dz\right)\right)^{1/4}
\\&+
\sup_{\pi\in\mathbf{P}_n}\left(\int_{z\in\mathcal{Z}_n}\left({f}_{\Lambda\left(Z'\pi_0\right)}\left(\Lambda\left(z'\pi\right)\right)-{f}_{P}\left(\Lambda\left(z'\pi\right)\right)\right)^4F_Z\left(dz\right)\right)^{1/4}
\\
=&
o\left(n^{-r_1}\mathbf{f}_n\right)
\end{align*}
\endgroup
hold by Assumptions \ref{assn:17} and \ref{assn:18}.
Therefore, the statement of this lemma holds
by Assumption \ref{assn:16} and by Lemma \ref{lemma:kernel_L2_1} under Assumption \ref{assn:11}, \ref{assn:12}, \ref{assn:13}, \ref{assn:14}, and \ref{assn:16}. 
\end{proof}

\begin{lemma}\label{lemma:deriv_conv_process}
If Assumptions \ref{assn:11}, \ref{assn:12}, \ref{assn:13}, \ref{assn:14},  \ref{assn:16}, \ref{assn:17}, and \ref{assn:18} are satisfied, then we have 
$$
\left(\int_{z\in\mathcal{Z}_n}
\left(\hat{\mf}_{\omega\mid P}^{\left(1\right)}\left(\Lambda\left(z'\hat{\pi}\right)\right)-{\mf}_{\omega\mid P}^{\left(1\right)}\left(\Lambda\left(z'\hat{\pi}\right)\right)\right)^4F_Z\left(dz\right)\right)^{1/4}=o_p\left(n^{-r_2}\right).
$$
$$
\left(\int_{z\in\mathcal{Z}_n}
\left(\hat{\mf}_{\omega\mid P}^{\left(1\right)}\left(P^\ast\left(\Lambda\left(z'\hat{\pi}\right),z\right)\right)-{\mf}_{\omega\mid P}^{\left(1\right)}\left(P^\ast\left(\Lambda\left(z'\hat{\pi}\right),z\right)\right)\right)^4F_Z\left(dz\right)\right)^{1/4}=o_p\left(n^{-r_2}\right).
$$\end{lemma}
\begin{proof}
For every $z\in\mathcal{Z}_n$, if $p=\Lambda\left(z'\hat{\pi}\right)$ or $p=P^\ast\left(\Lambda\left(z'\hat{\pi}\right),z\right)$, we can write 
\begin{align*}
\hat{\mf}_{\omega\mid P}\left(p\right)-{\mf}_{\omega\mid P}\left(p\right)
=\frac{\hat{\nf}_{\omega\mid P}\left(p\right)-{\nf}_{\omega\mid P}\left(p\right)}{\hat{f}_{P}\left(p\right)}
-{\mf}_{\omega\mid P}\left(p\right)\frac{\hat{f}_{P}\left(p\right)-{f}_{P}\left(p\right)}{\hat{f}_{P}\left(p\right)}
\end{align*}
and 
\begingroup
\allowdisplaybreaks
\begin{align*}
\hat{\mf}_{\omega\mid P}^{\left(1\right)}\left(p\right)-{\mf}_{\omega\mid P}^{\left(1\right)}\left(p\right)
=&
\frac{\hat{\nf}_{\omega\mid P}^{\left(1\right)}\left(p\right)-{\nf}_{\omega\mid P}^{\left(1\right)}\left(p\right)}{\hat{f}_{P}\left(p\right)}
-\frac{\hat{\nf}_{\omega\mid P}\left(p\right)-{\nf}_{\omega\mid P}\left(p\right)}{\hat{f}_{P}\left(p\right)^2}\hat{f}_{P}^{\left(1\right)}\left(p\right)
\\&-
{\mf}_{\omega\mid P}^{\left(1\right)}\left(p\right)\frac{\hat{f}_{P}\left(p\right)-{f}_{P}\left(p\right)}{\hat{f}_{P}\left(p\right)}
-
{\mf}_{\omega\mid P}\left(p\right)\frac{\hat{f}_{P}^{\left(1\right)}\left(p\right)-{f}_{P}^{\left(1\right)}\left(p\right)}{\hat{f}_{P}\left(p\right)}
+
{\mf}_{\omega\mid P}\left(p\right)\frac{\hat{f}_{P}\left(p\right)-{f}_{P}\left(p\right)}{\hat{f}_{P}\left(p\right)^2}\hat{f}_{P}^{\left(1\right)}\left(p\right).
\end{align*}
\endgroup
Therefore, from how $\mathbf{f}_n$ is defined in \eqref{eq:fn} and by Assumption \ref{assn:16}, we can write
\begingroup
\allowdisplaybreaks
\begin{eqnarray*}
&&
\left(\int_{z\in\mathcal{Z}_n}\left(\hat{\mf}_{\omega\mid P}^{\left(1\right)}\left(\Lambda\left(z'\hat{\pi}\right)\right)-{\mf}_{\omega\mid P}^{\left(1\right)}\left(\Lambda\left(z'\hat{\pi}\right)\right)\right)^4F_Z\left(dz\right)\right)^{1/4}
\\
&\leq&
\frac{1}{\mathbf{f}_n}\left(\int_{z\in\mathcal{Z}_n}\left(\hat{\nf}_{\omega\mid P}^{\left(1\right)}\left(\Lambda\left(z'\hat{\pi}\right)\right)-{\nf}_{\omega\mid P}^{\left(1\right)}\left(\Lambda\left(z'\hat{\pi}\right)\right)
\right)^4F_Z\left(dz\right)\right)^{1/4}
\\&&+\frac{1}{\mathbf{f}_n}\left(\int_{z\in\mathcal{Z}_n}\left(\hat{f}_{P}^{\left(1\right)}\left(\Lambda\left(z'\hat{\pi}\right)\right)-{f}_{P}^{\left(1\right)}\left(\Lambda\left(z'\hat{\pi}\right)\right)\right)^4F_Z\left(dz\right)\right)^{1/4}\sup_{p\in[0,1]}|{\mf}_{\omega\mid P}\left(p\right)|
\\&&+\frac{1}{\mathbf{f}_n}\left(\int_{z\in\mathcal{Z}_n}(\hat{f}_{P}\left(\Lambda\left(z'\hat{\pi}\right)\right)-{f}_{P}\left(\Lambda\left(z'\hat{\pi}\right)\right))^4F_Z\left(dz\right)\right)^{1/4}\sup_{p\in[0,1]}\left(|{\mf}_{\omega\mid P}\left(p\right)|+|{\mf}_{\omega\mid P}^{\left(1\right)}\left(p\right)|\right)
\\&&+
\frac{1}{\mathbf{f}_n}\left(\int_{z\in\mathcal{Z}_n}\left(\hat{\nf}_{\omega\mid P}\left(\Lambda\left(z'\hat{\pi}\right)\right)-{\nf}_{\omega\mid P}\left(\Lambda\left(z'\hat{\pi}\right)\right)\right)^4F_Z\left(dz\right)\right)^{1/4}
\\
&\leq&
\frac{1}{\mathbf{f}_n}\sup_{\pi\in\mathbf{P}_n}\left(\int_{z\in\mathcal{Z}_n}\left(
\frac{1}{|I_\ell^c|}\sum_{j\in I_\ell^c}\omega_jK_{h}^{\left(1\right)}\left(\Lambda\left(Z_j'\pi\right)-\Lambda\left(z'\pi\right)\right)
+{\nf}_{\omega\mid\Lambda\left(Z'{\pi}\right)}^{\left(1\right)}\left(\Lambda\left(z'{\pi}\right)\right)
\right)^4F_Z\left(dz\right)\right)^{1/4}
\\&&+
\frac{1}{\mathbf{f}_n}\sup_{\pi\in\mathbf{P}_n}\left(\int_{z\in\mathcal{Z}_n}\left(
{\nf}_{\omega\mid\Lambda\left(Z'{\pi}\right)}^{\left(1\right)}\left(\Lambda\left(z'{\pi}\right)\right)-
{\nf}_{\omega\mid P}^{\left(1\right)}\left(\Lambda\left(z'{\pi}\right)\right)
\right)^4F_Z\left(dz\right)\right)^{1/4}
\\&&+
\frac{1}{\mathbf{f}_n}\sup_{\pi\in\mathbf{P}_n}\left(\int_{z\in\mathcal{Z}_n}\left(\frac{1}{|I_\ell^c|}\sum_{j\in I_\ell^c}K_{h}^{\left(1\right)}\left(\Lambda\left(Z_j'\pi\right)-\Lambda\left(z'\pi\right)\right)+{f}_{\Lambda\left(Z'\pi\right)}^{\left(1\right)}\left(\Lambda\left(z'{\pi}\right)\right)\right)^4F_Z\left(dz\right)\right)^{1/4}O\left(1\right)
\\&&+
\frac{1}{\mathbf{f}_n}\sup_{\pi\in\mathbf{P}_n}\left(\int_{z\in\mathcal{Z}_n}\left({f}_{\Lambda\left(Z'\pi\right)}^{\left(1\right)}\left(\Lambda\left(z'{\pi}\right)\right)-{f}_{P}^{\left(1\right)}\left(\Lambda\left(z'{\pi}\right)\right)\right)^4F_Z\left(dz\right)\right)^{1/4}O\left(1\right)
\\&&+\frac{1}{\mathbf{f}_n}\sup_{\pi\in\mathbf{P}_n}\left(\int_{z\in\mathcal{Z}_n}\left(\frac{1}{|I_\ell^c|}\sum_{j\in I_\ell^c}K_{h}\left(\Lambda\left(Z_j'\pi\right)-\Lambda\left(z'\pi\right)\right)-{f}_{\Lambda\left(Z'\pi\right)}\left(\Lambda\left(z'{\pi}\right)\right)\right)^4F_Z\left(dz\right)\right)^{1/4}O\left(1\right)
\\&&+\frac{1}{\mathbf{f}_n}\sup_{\pi\in\mathbf{P}_n}\left(\int_{z\in\mathcal{Z}_n}\left({f}_{\Lambda\left(Z'\pi\right)}\left(\Lambda\left(z'{\pi}\right)\right)-{f}_{P}\left(\Lambda\left(z'{\pi}\right)\right)\right)^4F_Z\left(dz\right)\right)^{1/4}O\left(1\right)
\\&&+
\frac{1}{\mathbf{f}_n}\sup_{\pi\in\mathbf{P}_n}\left(\int_{z\in\mathcal{Z}_n}\left(\frac{1}{|I_\ell^c|}\sum_{j\in I_\ell^c}\omega_jK_{h}\left(\Lambda\left(Z_j'\pi\right)-\Lambda\left(z'\pi\right)\right)-{\nf}_{\omega\mid\Lambda\left(Z'\pi\right)}\left(\Lambda\left(z'{\pi}\right)\right)\right)^4F_Z\left(dz\right)\right)^{1/4}
\\&&+
\frac{1}{\mathbf{f}_n}\sup_{\pi\in\mathbf{P}_n}\left(\int_{z\in\mathcal{Z}_n}\left({\nf}_{\omega\mid\Lambda\left(Z'\pi\right)}\left(\Lambda\left(z'{\pi}\right)\right)-{\nf}_{\omega\mid P}\left(\Lambda\left(z'{\pi}\right)\right)\right)^4F_Z\left(dz\right)\right)^{1/4}
\end{eqnarray*}
\endgroup
and, similarly,
\begingroup
\allowdisplaybreaks
\begin{eqnarray*}
&&
\left(\int_{z\in\mathcal{Z}_n}\left(\hat{\mf}_{\omega\mid P}^{\left(1\right)}\left(P^\ast\left(\Lambda\left(z'\hat{\pi}\right),z\right)\right)-{\mf}_{\omega\mid P}^{\left(1\right)}\left(P^\ast\left(\Lambda\left(z'\hat{\pi}\right),z\right)\right)\right)^4F_Z\left(dz\right)\right)^{1/4}
\\
&\leq&
\frac{1}{\mathbf{f}_n}\sup_{\pi\in\mathbf{P}_n}\left(\int_{z\in\mathcal{Z}_n}\left(
\frac{1}{|I_\ell^c|}\sum_{j\in I_\ell^c}\omega_jK_{h}^{\left(1\right)}\left(\Lambda\left(Z_j'\pi\right)-P^\ast\left(\Lambda\left(z'{\pi}\right),z\right)\right)
+{\nf}_{\omega\mid\Lambda\left(Z'{\pi}\right)}^{\left(1\right)}\left(P^\ast\left(\Lambda\left(z'{\pi}\right),z\right)\right)
\right)^4F_Z\left(dz\right)\right)^{1/4}
\\&&+
\frac{1}{\mathbf{f}_n}\sup_{\pi\in\mathbf{P}_n}\left(\int_{z\in\mathcal{Z}_n}\left(
{\nf}_{\omega\mid\Lambda\left(Z'{\pi}\right)}^{\left(1\right)}\left(P^\ast\left(\Lambda\left(z'{\pi}\right),z\right)\right)-
{\nf}_{\omega\mid P}^{\left(1\right)}\left(P^\ast\left(\Lambda\left(z'{\pi}\right),z\right)\right)
\right)^4F_Z\left(dz\right)\right)^{1/4}
\\&&+
\frac{1}{\mathbf{f}_n}\sup_{\pi\in\mathbf{P}_n}\left(\int_{z\in\mathcal{Z}_n}\left(\frac{1}{|I_\ell^c|}\sum_{j\in I_\ell^c}K_{h}^{\left(1\right)}\left(\Lambda\left(Z_j'\pi\right)-P^\ast\left(\Lambda\left(z'{\pi}\right),z\right)\right)+{f}_{\Lambda\left(Z'\pi\right)}^{\left(1\right)}\left(P^\ast\left(\Lambda\left(z'{\pi}\right),z\right)\right)\right)^4F_Z\left(dz\right)\right)^{1/4}O\left(1\right)
\\&&+
\frac{1}{\mathbf{f}_n}\sup_{\pi\in\mathbf{P}_n}\left(\int_{z\in\mathcal{Z}_n}\left({f}_{\Lambda\left(Z'\pi\right)}^{\left(1\right)}\left(P^\ast\left(\Lambda\left(z'{\pi}\right),z\right)\right)-{f}_{P}^{\left(1\right)}\left(P^\ast\left(\Lambda\left(z'{\pi}\right),z\right)\right)\right)^4F_Z\left(dz\right)\right)^{1/4}O\left(1\right)
\\&&+\frac{1}{\mathbf{f}_n}\sup_{\pi\in\mathbf{P}_n}\left(\int_{z\in\mathcal{Z}_n}\left(\frac{1}{|I_\ell^c|}\sum_{j\in I_\ell^c}K_{h}\left(\Lambda\left(Z_j'\pi\right)-P^\ast\left(\Lambda\left(z'{\pi}\right),z\right)\right)-{f}_{\Lambda\left(Z'\pi\right)}\left(P^\ast\left(\Lambda\left(z'{\pi}\right),z\right)\right)\right)^4F_Z\left(dz\right)\right)^{1/4}O\left(1\right)
\\&&+\frac{1}{\mathbf{f}_n}\sup_{\pi\in\mathbf{P}_n}\left(\int_{z\in\mathcal{Z}_n}\left({f}_{\Lambda\left(Z'\pi\right)}\left(P^\ast\left(\Lambda\left(z'{\pi}\right),z\right)\right)-{f}_{P}\left(P^\ast\left(\Lambda\left(z'{\pi}\right),z\right)\right)\right)^4F_Z\left(dz\right)\right)^{1/4}O\left(1\right)
\\&&+
\frac{1}{\mathbf{f}_n}\sup_{\pi\in\mathbf{P}_n}\left(\int_{z\in\mathcal{Z}_n}\left(\frac{1}{|I_\ell^c|}\sum_{j\in I_\ell^c}\omega_jK_{h}\left(\Lambda\left(Z_j'\pi\right)-P^\ast\left(\Lambda\left(z'{\pi}\right),z\right)\right)-{\nf}_{\omega\mid\Lambda\left(Z'\pi\right)}\left(P^\ast\left(\Lambda\left(z'{\pi}\right),z\right)\right)\right)^4F_Z\left(dz\right)\right)^{1/4}
\\&&+
\frac{1}{\mathbf{f}_n}\sup_{\pi\in\mathbf{P}_n}\left(\int_{z\in\mathcal{Z}_n}\left({\nf}_{\omega\mid\Lambda\left(Z'\pi\right)}\left(P^\ast\left(\Lambda\left(z'{\pi}\right),z\right)\right)-{\nf}_{\omega\mid P}\left(P^\ast\left(\Lambda\left(z'{\pi}\right),z\right)\right)\right)^4F_Z\left(dz\right)\right)^{1/4}.
\end{eqnarray*}
\endgroup
Note that, similarly to the inequalities in the proof of Lemma \ref{lemma:mean_P_conv_process}, we have   
\begingroup
\allowdisplaybreaks
\begin{align*}
&
\sup_{\pi\in\mathbf{P}_n}\left(\int_{z\in\mathcal{Z}_n}\left(
{\nf}_{\omega\mid\Lambda\left(Z'{\pi}\right)}^{\left(1\right)}\left(\Lambda\left(z'{\pi}\right)\right)-
{\nf}_{\omega\mid P}^{\left(1\right)}\left(\Lambda\left(z'{\pi}\right)\right)
\right)^4F_Z\left(dz\right)\right)^{1/4}
=o\left(n^{-r_2}\mathbf{f}_n\right)
\\
&
\sup_{\pi\in\mathbf{P}_n}\left(\int_{z\in\mathcal{Z}_n}\left({f}_{\Lambda\left(Z'\pi\right)}^{\left(1\right)}\left(\Lambda\left(z'{\pi}\right)\right)-{f}_{P}^{\left(1\right)}\left(\Lambda\left(z'{\pi}\right)\right)\right)^4F_Z\left(dz\right)\right)^{1/4}
=o\left(n^{-r_2}\mathbf{f}_n\right)
\\
&
\sup_{\pi\in\mathbf{P}_n}\left(\int_{z\in\mathcal{Z}_n}\left({f}_{\Lambda\left(Z'\pi\right)}\left(\Lambda\left(z'{\pi}\right)\right)-{f}_{P}\left(\Lambda\left(z'{\pi}\right)\right)\right)^4F_Z\left(dz\right)\right)^{1/4}
=o\left(n^{-r_2}\mathbf{f}_n\right)
\\
&
\sup_{\pi\in\mathbf{P}_n}\left(\int_{z\in\mathcal{Z}_n}\left({\nf}_{\omega\mid\Lambda\left(Z'\pi\right)}\left(\Lambda\left(z'{\pi}\right)\right)-{\nf}_{\omega\mid P}\left(\Lambda\left(z'{\pi}\right)\right)\right)^4F_Z\left(dz\right)\right)^{1/4}
=o\left(n^{-r_2}\mathbf{f}_n\right)
\\
&
\sup_{\pi\in\mathbf{P}_n}\left(\int_{z\in\mathcal{Z}_n}\left(
{\nf}_{\omega\mid\Lambda\left(Z'{\pi}\right)}^{\left(1\right)}\left(P^\ast\left(\Lambda\left(z'{\pi}\right),z\right)\right)-
{\nf}_{\omega\mid P}^{\left(1\right)}\left(P^\ast\left(\Lambda\left(z'{\pi}\right),z\right)\right)
\right)^4F_Z\left(dz\right)\right)^{1/4}
=o\left(n^{-r_2}\mathbf{f}_n\right)
\\
&
\sup_{\pi\in\mathbf{P}_n}\left(\int_{z\in\mathcal{Z}_n}\left({f}_{\Lambda\left(Z'\pi\right)}^{\left(1\right)}\left(P^\ast\left(\Lambda\left(z'{\pi}\right),z\right)\right)-{f}_{P}^{\left(1\right)}\left(P^\ast\left(\Lambda\left(z'{\pi}\right),z\right)\right)\right)^4F_Z\left(dz\right)\right)^{1/4}
=o\left(n^{-r_2}\mathbf{f}_n\right)
\\
&
\sup_{\pi\in\mathbf{P}_n}\left(\int_{z\in\mathcal{Z}_n}\left({f}_{\Lambda\left(Z'\pi\right)}\left(P^\ast\left(\Lambda\left(z'{\pi}\right),z\right)\right)-{f}_{P}\left(P^\ast\left(\Lambda\left(z'{\pi}\right),z\right)\right)\right)^4F_Z\left(dz\right)\right)^{1/4}
=o\left(n^{-r_2}\mathbf{f}_n\right)
\\
&
\sup_{\pi\in\mathbf{P}_n}\left(\int_{z\in\mathcal{Z}_n}\left({\nf}_{\omega\mid\Lambda\left(Z'\pi\right)}\left(P^\ast\left(\Lambda\left(z'{\pi}\right),z\right)\right)-{\nf}_{\omega\mid P}\left(P^\ast\left(\Lambda\left(z'{\pi}\right),z\right)\right)\right)^4F_Z\left(dz\right)\right)^{1/4}
=o\left(n^{-r_2}\mathbf{f}_n\right)
\end{align*}
\endgroup
under Assumptions \ref{assn:17} and \ref{assn:18}.
Therefore, the statement of this lemma follows by Lemmas \ref{lemma:kernel_L2_1} and \ref{lemma:kernel_L2_3} under Assumption \ref{assn:11}, \ref{assn:12}, \ref{assn:13}, \ref{assn:14}, and \ref{assn:16}.
\end{proof}

\begin{lemma}\label{lemma:low1}
If Assumptions \ref{assn:11}, \ref{assn:12}, \ref{assn:13}, \ref{assn:14},  \ref{assn:15}, \ref{assn:16}, \ref{assn:17}, \ref{assn:18}, and \ref{assn:19} are satisfied, then we have 
$$
\int \left(\hat{\zeta}\left(z\right)-{\zeta}\left(z\right)\right)\left(\hat{\mf}_{S\mid Z}\left(z\right)-{\mf}_{S\mid Z}\left(z\right)\right)F_Z\left(dz\right)=o_p\left(n^{-1/2}\right)
$$
\end{lemma}
\begin{proof}
Note that we can write
\begingroup
\allowdisplaybreaks
\begin{eqnarray*}
&&
\left\|\int_{z\in\mathcal{Z}_n} \left(\hat{\zeta}\left(z\right)-{\zeta}\left(z\right)\right)\left(\hat{\mf}_{S\mid Z}\left(z\right)-{\mf}_{S\mid Z}\left(z\right)\right)F_Z\left(dz\right)\right\|
\\
&\leq&
\left\|\int_{z\in\mathcal{Z}_n} \left(\hat{\zeta}\left(z;\hat{\mf}_{S\mid Z}\left(z\right)\right)-{\zeta}\left(z;\hat{\mf}_{S\mid Z}\left(z\right)\right)\right)\left(\hat{\mf}_{S\mid Z}\left(z\right)-{\mf}_{S\mid Z}\left(z\right)\right)F_Z\left(dz\right)\right\|
\\&&+
\left\|\int_{z\in\mathcal{Z}_n} \left({\zeta}\left(z;\hat{\mf}_{S\mid Z}\left(z\right)\right)-{\zeta}\left(z;{\mf}_{S\mid Z}\left(z\right)\right)\right)\left(\hat{\mf}_{S\mid Z}\left(z\right)-{\mf}_{S\mid Z}\left(z\right)\right)F_Z\left(dz\right)\right\|
\\
&\leq&
\int_{z\in\mathcal{Z}_n} \|\hat{\zeta}\left(z;\hat{\mf}_{S\mid Z}\left(z\right)\right)-{\zeta}\left(z;\hat{\mf}_{S\mid Z}\left(z\right)\right)\|F_Z\left(dz\right)\sup_{z \in \mathcal{Z}_n}|\hat{\mf}_{S\mid Z}\left(z\right)-{\mf}_{S\mid Z}\left(z\right)|
\\&&+
\int_{z\in\mathcal{Z}_n} \sup_{p}\left(\frac{\partial}{\partial p}{\zeta}\left(z;p\right)\right|F_Z\left(dz\right)\sup_{z\in\mathcal{Z}_n}\left(\hat{\mf}_{S\mid Z}\left(z\right)-{\mf}_{S\mid Z}\left(z\right)\right)^2
\end{eqnarray*}
\endgroup
by Assumption \ref{assn:16}, where
$
\int_{z\in\mathcal{Z}_n} \sup_{p}\left|\frac{\partial}{\partial p}{\zeta}\left(z;p\right)\right|F_Z\left(dz\right) = O\left(1\right)
$
by Assumption \ref{assn:16}, and
$
\sup_{z\in\mathcal{Z}_n}|\hat{\mf}_{S\mid Z}\left(z\right)-{\mf}_{S\mid Z}\left(z\right)|=o_p\left(n^{-r_1}\right)
$
by Lemma \ref{lemma:prop_score_conv} under Assumptions \ref{assn:11}, \ref{assn:12}, and \ref{assn:18}.
Thus, it remains to show $\int_{z\in\mathcal{Z}_n} \|\hat{\zeta}\left(z;\hat{\mf}_{S\mid Z}\left(z\right)\right)-{\zeta}\left(z;\hat{\mf}_{S\mid Z}\left(z\right)\right)\|F_Z\left(dz\right)=o_p\left(n^{-1/4}\right)$.

We are going to show $\int_{z\in\mathcal{Z}_n} \|\hat{\zeta}\left(z;\hat{\mf}_{S\mid Z}\left(z\right)\right)-{\zeta}\left(z;\hat{\mf}_{S\mid Z}\left(z\right)\right)\|F_Z\left(dz\right)=o_p\left(n^{-1/4}\right)$ only for the square of the $\mu_0\left(X\right)$ term, and results for the other components will similarly follow. 
Since 
\begingroup
\allowdisplaybreaks
\begin{eqnarray*}
\hat{\zeta}_1\left(z;p\right)
&=&
2\left(1-p\right)\left(\hat{\mf}_{\mu_0\left(X\right)\mid Z}\left(z\right)-\hat{\mf}_{\mu_0\left(X\right)\mid P}\left(p\right)\right)\left(\hat{\mf}_{\mu_0\left(X\right)\mid Z}\left(z\right)-\hat{\mf}_{\mu_0\left(X\right)\mid P}\left(p\right)\right)'\\
&&+2\left(1-p\right)\left(\hat{\mf}_{\mu_0\left(X\right)\mu_0\left(X\right)'\mid Z}\left(z\right)-\hat{\mf}_{\mu_0\left(X\right)\mid Z}\left(z\right)\hat{\mf}_{\mu_0\left(X\right)\mid Z}\left(z\right)'\right)\\
&&-\left(1-p\right)^2\frac{\partial}{\partial p}\hat{\mf}_{\mu_0\left(X\right)\mid P}\left(p\right)\left(\hat{\mf}_{\mu_0\left(X\right)\mid Z}\left(z\right)-\hat{\mf}_{\mu_0\left(X\right)\mid P}\left(p\right)\right)'\\
&&-\left(1-p\right)^2\left(\hat{\mf}_{\mu_0\left(X\right)\mid Z}\left(z\right)-\hat{\mf}_{\mu_0\left(X\right)\mid P}\left(p\right)\right)\hat{\mf}_{\mu_0\left(X\right)\mid P}^{\left(1\right)}\left(p\right)',
\end{eqnarray*}
\endgroup
it follows that that $\|\hat{\zeta}_1\left(z;p\right)-{\zeta}_1\left(z;p\right)\|$ is bounded (multiplied by a constant) by the sum of the following terms:  
\begingroup
\allowdisplaybreaks
\begin{eqnarray*}
&&\|\hat{\mf}_{\mu_0\left(X\right)\mid Z}\left(z\right)-{\mf}_{\mu_0\left(X\right)\mid Z}\left(z\right)\|\|\hat{\mf}_{\mu_0\left(X\right)\mid P}\left(p\right)-{\mf}_{\mu_0\left(X\right)\mid P}\left(p\right)\|,
\\&&\|\hat{\mf}_{\mu_0\left(X\right)\mid P}\left(p\right)-{\mf}_{\mu_0\left(X\right)\mid P}\left(p\right)\|^2,
\quad\|\hat{\mf}_{\mu_0\left(X\right)\mid Z}\left(z\right)-{\mf}_{\mu_0\left(X\right)\mid Z}\left(z\right)\|^2,
\\&&\|\hat{\mf}_{\mu_0\left(X\right)\mid Z}\left(z\right)-{\mf}_{\mu_0\left(X\right)\mid Z}\left(z\right)\|\|\hat{\mf}_{\mu_0\left(X\right)\mid P}^{\left(1\right)}\left(p\right)-{\mf}_{\mu_0\left(X\right)\mid P}^{\left(1\right)}\left(p\right)\|,
\\&&\|\hat{\mf}_{\mu_0\left(X\right)\mid P}^{\left(1\right)}\left(p\right)-{\mf}_{\mu_0\left(X\right)\mid P}^{\left(1\right)}\left(p\right)\|\|\hat{\mf}_{\mu_0\left(X\right)\mid P}\left(p\right)-{\mf}_{\mu_0\left(X\right)\mid P}\left(p\right)\|,
\\&&\|\hat{\mf}_{\mu_0\left(X\right)\mid P}^{\left(1\right)}\left(p\right)-{\mf}_{\mu_0\left(X\right)\mid P}^{\left(1\right)}\left(p\right)\|\|\hat{\mf}_{\mu_0\left(X\right)\mid Z}\left(z\right)-{\mf}_{\mu_0\left(X\right)\mid Z}\left(z\right)\|,
\\&&\|{\mf}_{\mu_0\left(X\right)\mid Z}\left(z\right)\|\|\hat{\mf}_{\mu_0\left(X\right)\mid Z}\left(z\right)-{\mf}_{\mu_0\left(X\right)\mid Z}\left(z\right)\|,
\quad\|{\mf}_{\mu_0\left(X\right)\mid P}\left(p\right)\|\|\hat{\mf}_{\mu_0\left(X\right)\mid Z}\left(z\right)-{\mf}_{\mu_0\left(X\right)\mid Z}\left(z\right)\|,
\\&&\|{\mf}_{\mu_0\left(X\right)\mid P}^{\left(1\right)}\left(p\right)\|\|\hat{\mf}_{\mu_0\left(X\right)\mid Z}\left(z\right)-{\mf}_{\mu_0\left(X\right)\mid Z}\left(z\right)\|,
\quad\|{\mf}_{\mu_0\left(X\right)\mid Z}\left(z\right)\|\|\hat{\mf}_{\mu_0\left(X\right)\mid P}\left(p\right)-{\mf}_{\mu_0\left(X\right)\mid P}\left(p\right)\|,
\\&&\|{\mf}_{\mu_0\left(X\right)\mid P}\left(p\right)\|\|\hat{\mf}_{\mu_0\left(X\right)\mid P}\left(p\right)-{\mf}_{\mu_0\left(X\right)\mid P}\left(p\right)\|,
\quad\|{\mf}_{\mu_0\left(X\right)\mid P}^{\left(1\right)}\left(p\right)\|\|\hat{\mf}_{\mu_0\left(X\right)\mid P}\left(p\right)-{\mf}_{\mu_0\left(X\right)\mid P}\left(p\right)\|,
\\&&\|{\mf}_{\mu_0\left(X\right)\mid Z}\left(z\right)\|\|\hat{\mf}_{\mu_0\left(X\right)\mid P}^{\left(1\right)}\left(p\right)-{\mf}_{\mu_0\left(X\right)\mid P}^{\left(1\right)}\left(p\right)\|,
\quad\|{\mf}_{\mu_0\left(X\right)\mid P}\left(p\right)\|\|\hat{\mf}_{\mu_0\left(X\right)\mid P}^{\left(1\right)}\left(p\right)-{\mf}_{\mu_0\left(X\right)\mid P}^{\left(1\right)}\left(p\right)\|, 
\\&&\text{and } \|\hat{\mf}_{\mu_0\left(X\right)\mu_0\left(X\right)'\mid Z}\left(z\right)-{\mf}_{\mu_0\left(X\right)\mu_0\left(X\right)'\mid Z}\left(z\right)\|.
\end{eqnarray*}
\endgroup
Note that
\begingroup
\allowdisplaybreaks
\begin{eqnarray*}
&&
\sup_{z \in \mathcal{Z}_n} \|\hat{\mf}_{\mu_0\left(X\right)\mid Z}\left(z\right)-{\mf}_{\mu_0\left(X\right)\mid Z}\left(z\right)\|
=o_p\left(n^{-r_1}\right)
\\&&
\sup_{z \in \mathcal{Z}_n} \|\hat{\mf}_{\mu_0\left(X\right)\mu_0\left(X\right)'\mid Z}\left(z\right)-{\mf}_{\mu_0\left(X\right)\mu_0\left(X\right)'\mid Z}\left(z\right)\|=o_p\left(n^{-r_1}\right)
\end{eqnarray*}
\endgroup
by Assumption \ref{assn:15},
and 
\begingroup
\allowdisplaybreaks
\begin{eqnarray*}
&&\left(\int_{z\in\mathcal{Z}_n}\|\hat{\mf}_{\mu_0\left(X\right)\mid P}\left(\hat{\mf}_{S\mid Z}\left(z\right)\right)-{\mf}_{\mu_0\left(X\right)\mid P}\left(\hat{\mf}_{S\mid Z}\left(z\right)\right)\|^2F_Z\left(dz\right)\right)^{1/2}=o_p\left(n^{-r_1}\right)
\\&&\left(\int_{z\in\mathcal{Z}_n}\|\hat{\mf}_{\mu_0\left(X\right)\mid P}^{\left(1\right)}\left(\hat{\mf}_{S\mid Z}\left(z\right)\right)-{\mf}_{\mu_0\left(X\right)\mid P}^{\left(1\right)}\left(\hat{\mf}_{S\mid Z}\left(z\right)\right)\|^2F_Z\left(dz\right)\right)^{1/2}=o_p\left(n^{-r_2}\right)
\end{eqnarray*}
\endgroup
by Lemmas \ref{lemma:mean_P_conv_process} and \ref{lemma:deriv_conv_process} under Assumptions \ref{assn:11}, \ref{assn:12}, \ref{assn:13}, \ref{assn:14},  \ref{assn:16}, \ref{assn:17}, and \ref{assn:18}.
Therefore, Assumption \ref{assn:17}, \ref{assn:18} and \ref{assn:19} imply the statement of this lemma. 
\end{proof}

\begin{lemma}\label{lemma:low2}
If Assumptions \ref{assn:11}, \ref{assn:12}, \ref{assn:13}, \ref{assn:14},  \ref{assn:16}, \ref{assn:17}, \ref{assn:18}, and \ref{assn:19} are satisfied, then the following objects are $o_p\left(n^{-1/2}\right)$:
\begingroup
\allowdisplaybreaks
\begin{align*}
&\int\frac{\hat{f}_{P^\ast}\left(\hat{\mf}_{S\mid Z}\left(z\right)\right)}{\hat{f}_{P}\left(\hat{\mf}_{S\mid Z}\left(z\right)\right)}\left(\hat{\Delta}_{U\mid P}\left(\hat{\mf}_{S\mid Z}\left(z\right)\right)-{\Delta}_{U\mid P}\left({\mf}_{S\mid Z}\left(z\right)\right)
\left({\mf_{S\mid Z}}\left(z\right)-\hat{\mf}_{S\mid Z}\left(z\right)\right)\right)F_Z\left(dz\right)
\qquad\text{and}
\\
&\int \left(\hat{\Delta}_{U\mid P}\left(P^\ast\left(\hat{\mf}_{S\mid Z}\left(z\right),z\right)\right)\partial P^\ast\left(\hat{\mf}_{S\mid Z}\left(z\right),z\right)-{\Delta}_{U\mid P}\left(P^\ast\left({\mf}_{S\mid Z}\left(z\right),z\right)\right)\partial P^\ast\left({\mf}_{S\mid Z}\left(z\right),z\right)\right) \times
\\
&\qquad
\left({\mf_{S\mid Z}}\left(z\right)-\hat{\mf}_{S\mid Z}\left(z\right)\right)F_Z\left(dz\right).
\end{align*}
\endgroup
\end{lemma}
\begin{proof}
By the Cauchy-Schwarz inequality, we can write 
\begingroup
\allowdisplaybreaks
\begin{align*}
&
\left|\int\frac{\hat{f}_{P^\ast}\left(\hat{\mf}_{S\mid Z}\left(z\right)\right)}{\hat{f}_{P}\left(\hat{\mf}_{S\mid Z}\left(z\right)\right)}\left(\hat{\Delta}_{U\mid P}\left(\hat{\mf}_{S\mid Z}\left(z\right)\right)-{\Delta}_{U\mid P}\left({\mf}_{S\mid Z}\left(z\right)\right)\right)
\left({\mf_{S\mid Z}}\left(z\right)-\hat{\mf}_{S\mid Z}\left(z\right)\right)F_Z\left(dz\right)
\right|
\\&\leq
\left(\int\left(\frac{\hat{f}_{P^\ast}\left(\hat{\mf}_{S\mid Z}\left(z\right)\right)}{\hat{f}_{P}\left(\hat{\mf}_{S\mid Z}\left(z\right)\right)}\right)^2F_Z\left(dz\right)
\right)^{1/2}
\left(\int_{z\in\mathcal{Z}_n} 
\left(\hat{\Delta}_{U\mid P}\left(\hat{\mf}_{S\mid Z}\left(z\right)\right)-{\Delta}_{U\mid P}\left({\mf}_{S\mid Z}\left(z\right)\right)\right)^2F_Z\left(dz\right)
\right)^{1/2} \times
\\
&\qquad
\sup_{z\in\mathcal{Z}_n}|\hat{\mf}_{S\mid Z}\left(z\right)-{\mf}_{S\mid Z}\left(z\right)|,
\end{align*}
\endgroup
and 
\begingroup
\allowdisplaybreaks
\begin{align*}
&
\left|\int_{z\in\mathcal{Z}_n} \left(\hat{\Delta}_{U\mid P}\left(P^\ast\left(\hat{\mf}_{S\mid Z}\left(z\right),z\right)\right)\partial P^\ast\left(\hat{\mf}_{S\mid Z}\left(z\right),z\right) \right.\right.
\\
&\qquad
\left.\left.-{\Delta}_{U\mid P}\left(P^\ast\left({\mf}_{S\mid Z}\left(z\right),z\right)\right)\partial P^\ast\left({\mf}_{S\mid Z}\left(z\right),z\right)\right)
\left({\mf_{S\mid Z}}\left(z\right)-\hat{\mf}_{S\mid Z}\left(z\right)\right)F_Z\left(dz\right)
\right|
\\&\leq
\int_{z\in\mathcal{Z}_n} |\hat{\Delta}_{U\mid P}\left(P^\ast\left(\hat{\mf}_{S\mid Z}\left(z\right),z\right)\right)\partial P^\ast\left(\hat{\mf}_{S\mid Z}\left(z\right),z\right) -{\Delta}_{U\mid P}\left(P^\ast\left({\mf}_{S\mid Z}\left(z\right),z\right)\right)\partial P^\ast\left({\mf}_{S\mid Z}\left(z\right),z\right)|F_Z\left(dz\right) \times
\\
&\qquad\sup_{z\in\mathcal{Z}_n}|\hat{\mf}_{S\mid Z}\left(z\right)-{\mf}_{S\mid Z}\left(z\right)|
\\&
\leq
\int_{z\in\mathcal{Z}_n} |\hat{\Delta}_{U\mid P}\left(P^\ast\left(\hat{\mf}_{S\mid Z}\left(z\right),z\right)\right)-{\Delta}_{U\mid P}\left(P^\ast\left({\mf}_{S\mid Z}\left(z\right),z\right)\right)|F_Z\left(dz\right) \sup_{\left(p,z\right)}\left|\frac{\partial}{\partial p}P^\ast\left(p,z\right)\right|\sup_{z\in\mathcal{Z}_n}|\hat{\mf}_{S\mid Z}\left(z\right)-{\mf}_{S\mid Z}\left(z\right)|
\\&+ \int_{z\in\mathcal{Z}_n} |{\Delta}_{U\mid P}\left(P^\ast\left({\mf}_{S\mid Z}\left(z\right),z\right)\right)| F_Z\left(dz\right) \sup_{\left(p,z\right)}\left|\frac{\partial^2}{\partial p^2}P^\ast\left(p,z\right)\right|\sup_{z\in\mathcal{Z}_n}\left(\hat{\mf}_{S\mid Z}\left(z\right)-{\mf}_{S\mid Z}\left(z\right)\right)^2.
\end{align*}
\endgroup
under Assumption \ref{assn:16}.
Since 
\begingroup
\allowdisplaybreaks
\begin{align*}
&
\left(\int_{z\in\mathcal{Z}_n} 
\left(\hat{\Delta}_{U\mid P}\left(\hat{\mf}_{S\mid Z}\left(z\right)\right)-{\Delta}_{U\mid P}\left({\mf}_{S\mid Z}\left(z\right)\right)\right)^2F_Z\left(dz\right)
\right)^{1/2}
\\&\leq
\left(\int_{z\in\mathcal{Z}_n} 
\left(\hat{\Delta}_{U\mid P}\left(\hat{\mf}_{S\mid Z}\left(z\right)\right)-{\Delta}_{U\mid P}\left(\hat{\mf}_{S\mid Z}\left(z\right)\right)\right)^2F_Z\left(dz\right)
\right)^{1/2}
\\&+
\left(\int_{z\in\mathcal{Z}_n} 
\left({\Delta}_{U\mid P}\left(\hat{\mf}_{S\mid Z}\left(z\right)\right)-{\Delta}_{U\mid P}\left({\mf}_{S\mid Z}\left(z\right)\right)\right)^2F_Z\left(dz\right)
\right)^{1/2}
\\&\leq
\left(\int_{z\in\mathcal{Z}_n} 
\left(\hat{\Delta}_{U\mid P}\left(\hat{\mf}_{S\mid Z}\left(z\right)\right)-{\Delta}_{U\mid P}\left(\hat{\mf}_{S\mid Z}\left(z\right)\right)\right)^2F_Z\left(dz\right)
\right)^{1/2}
\\&+
\sup_{p}\left|{\Delta}_{U\mid P}^{\left(1\right)}\left(p\right)\right|\sup_{z\in\mathcal{Z}_n}|\hat{\mf}_{S\mid Z}\left(z\right)-{\mf}_{S\mid Z}\left(z\right)|
\end{align*}
\endgroup
and 
\begingroup
\allowdisplaybreaks
\begin{align*}
&
\int_{z\in\mathcal{Z}_n} |\hat{\Delta}_{U\mid P}\left(P^\ast\left(\hat{\mf}_{S\mid Z}\left(z\right),z\right)\right)-{\Delta}_{U\mid P}\left(P^\ast\left({\mf}_{S\mid Z}\left(z\right),z\right)\right)|F_Z\left(dz\right) 
\\&\leq
\int_{z\in\mathcal{Z}_n} |\hat{\Delta}_{U\mid P}\left(P^\ast\left(\hat{\mf}_{S\mid Z}\left(z\right),z\right)\right)-{\Delta}_{U\mid P}\left(P^\ast\left(\hat{\mf}_{S\mid Z}\left(z\right),z\right)\right)|F_Z\left(dz\right) 
\\&+
\int_{z\in\mathcal{Z}_n} |{\Delta}_{U\mid P}\left(P^\ast\left(\hat{\mf}_{S\mid Z}\left(z\right),z\right)\right)-{\Delta}_{U\mid P}\left(P^\ast\left({\mf}_{S\mid Z}\left(z\right),z\right)\right)|F_Z\left(dz\right) 
\\&\leq
\int_{z\in\mathcal{Z}_n} |\hat{\Delta}_{U\mid P}\left(P^\ast\left(\hat{\mf}_{S\mid Z}\left(z\right),z\right)\right)-{\Delta}_{U\mid P}\left(P^\ast\left(\hat{\mf}_{S\mid Z}\left(z\right),z\right)\right)|F_Z\left(dz\right) 
\\&+
\sup_{p}\left|{\Delta}_{U\mid P}^{\left(1\right)}\left(p\right)\right|
\sup_{\left(p,z\right)}\left|\frac{\partial}{\partial p}P^\ast\left(p,z\right)\right|
\sup_{z\in\mathcal{Z}_n}|\hat{\mf}_{S\mid Z}\left(z\right)-{\mf}_{S\mid Z}\left(z\right)|,
\end{align*}
\endgroup
Assumption \ref{assn:16} and Lemmas \ref{lemma:prop_score_conv}, \ref{lemma:denstiy_ratio_conv_process}, and \ref{lemma:deriv_conv_process} under Assumptions \ref{assn:11}, \ref{assn:12}, \ref{assn:13}, \ref{assn:14}, \ref{assn:16}, \ref{assn:17}, and \ref{assn:18} imply the statement of this lemma. 
\end{proof}

\begin{lemma}\label{lemma:low3}
If Assumptions \ref{assn:11}, \ref{assn:12}, \ref{assn:13}, \ref{assn:14},  \ref{assn:16}, \ref{assn:17}, \ref{assn:18}, and \ref{assn:19} are satisfied, then we have\small
\begin{align*}
\int \left(
\hat{\mf}_{U\mid P}\left(P^\ast\left(\hat{\mf}_{S\mid Z}\left(z\right),z\right)\right)-
{\mf_{U\mid P}}\left(P^\ast\left(\hat{\mf}_{S\mid Z}\left(z\right),z\right)\right)
-\frac{\hat{f}_{P^\ast}\left(\hat{\mf}_{S\mid Z}\left(z\right)\right)}{\hat{f}_{P}\left(\hat{\mf}_{S\mid Z}\left(z\right)\right)}\left(\hat{\mf}_{U\mid P}\left(\hat{\mf}_{S\mid Z}\left(z\right)\right)-{\mf_{U\mid P}}\left(\hat{\mf}_{S\mid Z}\left(z\right)\right)\right)
\right)F_Z\left(dz\right)
\\
=o_p\left(n^{-1/2}\right).
\end{align*}
\end{lemma}\normalsize
\begin{proof}
By Assumption \ref{assn:19}, we have
\begingroup
\allowdisplaybreaks
\begin{eqnarray*}
&&
\int \left(
\hat{\mf}_{U\mid P}\left(P^\ast\left(\hat{\mf}_{S\mid Z}\left(z\right),z\right)\right)-
{\mf_{U\mid P}}\left(P^\ast\left(\hat{\mf}_{S\mid Z}\left(z\right),z\right)\right)
\right)F_Z\left(dz\right)
\\
&&=
\int_{z\in\mathcal{Z}_n} \left(
\hat{\mf}_{U\mid P}\left(P^\ast\left(\Lambda\left(z'\hat\pi\right),z\right)\right)-{\mf_{U\mid P}}\left(P^\ast\left(\Lambda\left(z'\hat\pi\right),z\right)\right)
\right)F_Z\left(dz\right)+o\left(n^{-1/2}\right).
\end{eqnarray*}
\endgroup
Using a change of variables, we have 
\begingroup
\allowdisplaybreaks
\begin{eqnarray*}
&&\int_{z\in\mathcal{Z}_n} \left(
\hat{\mf}_{U\mid P}\left(P^\ast\left(\Lambda\left(z'\pi\right),z\right)\right)-{\mf_{U\mid P}}\left(P^\ast\left(\Lambda\left(z'\pi\right),z\right)\right)
\right)F_Z\left(dz\right)\\
&&=
Pr\left(Z\in\mathcal{Z}_n\right)\int\left(
\hat{\mf}_{U\mid P}\left(p\right)-{\mf_{U\mid P}}\left(p\right)
\right)F_{P^\ast\left(\Lambda\left(Z'\pi\right),Z\right)\mid Z\in\mathcal{Z}_n}\left(dp\right)\\
&&=
Pr\left(Z\in\mathcal{Z}_n\right)\int\frac{f_{P^\ast\left(\Lambda\left(Z'\pi\right),Z\right)\mid Z\in\mathcal{Z}_n}\left(p\right)}{f_{\Lambda\left(Z'\pi\right)\mid Z\in\mathcal{Z}_n}\left(p\right)}\left(
\hat{\mf}_{U\mid P}\left(p\right)-{\mf_{U\mid P}}\left(p\right)
\right)F_{\Lambda\left(Z'\pi\right)\mid Z\in\mathcal{Z}_n}\left(dp\right)
\\
&&=
\int_{z\in\mathcal{Z}_n}\mathrm{ratio}\left(z,\pi\right)\left(
\hat{\mf}_{U\mid P}\left(\Lambda\left(z'\pi\right)\right)-{\mf_{U\mid P}}\left(\Lambda\left(z'\pi\right)\right)
\right)F_{Z}\left(dz\right),
\end{eqnarray*}
\endgroup
where 
$$
\mathrm{ratio}\left(z,\pi\right)=\frac{f_{P^\ast\left(\Lambda\left(Z'\pi\right),Z\right)}\left(\Lambda\left(z'\pi\right)\right)}{f_{\Lambda\left(Z'\pi\right)}\left(\Lambda\left(z'\pi\right)\right)}.
$$
Therefore, we can write\small
\begingroup
\allowdisplaybreaks
\begin{eqnarray*}
&&
\left|\int \left(
\hat{\mf}_{U\mid P}\left(P^\ast\left(\hat{\mf}_{S\mid Z}\left(z\right),z\right)\right)-
{\mf_{U\mid P}}\left(P^\ast\left(\hat{\mf}_{S\mid Z}\left(z\right),z\right)\right)
-\frac{\hat{f}_{P^\ast}\left(\hat{\mf}_{S\mid Z}\left(z\right)\right)}{\hat{f}_{P}\left(\hat{\mf}_{S\mid Z}\left(z\right)\right)}\left(\hat{\mf}_{U\mid P}\left(\hat{\mf}_{S\mid Z}\left(z\right)\right)-{\mf_{U\mid P}}\left(\hat{\mf}_{S\mid Z}\left(z\right)\right)\right)
\right)F_Z\left(dz\right)\right|
\\&&\leq
\left(\int_{z\in\mathcal{Z}_n}\left(\hat{\mf}_{U\mid P}\left(\hat{\mf}_{S\mid Z}\left(z\right)\right)-{\mf_{U\mid P}}\left(\hat{\mf}_{S\mid Z}\left(z\right)\right)
\right)^2F_{Z}\left(dz\right)\right)^{1/2}\\&&\times
\left(\int_{z\in\mathcal{Z}_n}\left(\frac{\hat{f}_{P^\ast}\left(\hat{\mf}_{S\mid Z}\left(z\right)\right)}{\hat{f}_{P}\left(\hat{\mf}_{S\mid Z}\left(z\right)\right)}-\mathrm{ratio}\left(z,\hat\pi\right)\right)^2F_{Z}\left(dz\right)\right)^{1/2}
\\&&+o\left(n^{-1/2}\right).
\end{eqnarray*}
\endgroup\normalsize
Lemmas \ref{lemma:denstiy_ratio_conv_process} and \ref{lemma:mean_P_conv_process} under Assumptions \ref{assn:11}, \ref{assn:12}, \ref{assn:13}, \ref{assn:14}, \ref{assn:16}, \ref{assn:17} and \ref{assn:18} imply the statement of the lemma.
\end{proof}

\begin{lemma}\label{lemma:low4}
If Assumptions \ref{assn:11}, \ref{assn:12}, \ref{assn:13}, \ref{assn:14},  \ref{assn:16}, \ref{assn:17}, \ref{assn:18}, and \ref{assn:19} are satisfied, then we have
$$
\int \left(\hat{\xi}_1\left(x,y,\hat{\mf}_{S\mid Z}\left(z\right)\right)-{\xi}_1\left(x,y,\hat{\mf}_{S\mid Z}\left(z\right)\right)\right)F_W\left(dw\right)=o_p\left(n^{-1/2}\right).
$$
\end{lemma}
\begin{proof}
Our proof focuses on the first term $\hat{\xi}_{1,1}\left(x,y,p\right)$ of $\xi_1\left(x,y,p\right)$, but similar arguments apply for the other terms. 
Note that 
\begingroup
\allowdisplaybreaks
\begin{eqnarray*}
&&
\int\left(\hat{\xi}_{1,1}\left(x,y,\Lambda\left(z'\pi\right)\right)-{\xi}_{1,1}\left(x,y,\Lambda\left(z'\pi\right)\right)\right)F_W\left(dw\right)
\\&&=
\int\int\left(\hat{\xi}_{1,1}\left(x,y,p\right)-{\xi}_{1,1}\left(x,y,p\right)\right)F_{\left(X,Y\right)\mid \Lambda\left(Z'\pi\right)=p}\left(dy,dx\right)F_{\Lambda\left(Z'\pi\right)}\left(dp\right).
\end{eqnarray*}
\endgroup
By the definition of $\hat{\xi}_{1,1}$, we have 
\begingroup
\allowdisplaybreaks
\begin{eqnarray*}
&&
\int\left(\hat{\xi}_{1,1}\left(x,y,p\right)-{\xi}_{1,1}\left(x,y,p\right)\right)F_{\left(X,Y\right)\mid \Lambda\left(Z'\pi\right)=p}\left(dy,dx\right)
\\&&=
\left(1-p\right)^2\left(\hat{\mf}_{\mu_0\left(X\right)\mid P}\left(p\right)\hat{\mf}_{\mu_0\left(X\right)\mid P}\left(p\right)'-\hat{\mf}_{\mu_0\left(X\right)\mid P}\left(p\right)\mf_{\mu_0\left(X\right)\mid  \Lambda\left(Z'\pi\right)}\left(p\right)'-\mf_{\mu_0\left(X\right)\mid \Lambda\left(Z'\pi\right)}\left(p\right)\hat{\mf}_{\mu_0\left(X\right)\mid P}\left(p\right)'\right)
\\&&-
\left(1-p\right)^2\left(\mf_{\mu_0\left(X\right)\mid P}\left(p\right)\mf_{\mu_0\left(X\right)\mid P}\left(p\right)'-\mf_{\mu_0\left(X\right)\mid P}\left(p\right)\mf_{\mu_0\left(X\right)\mid\Lambda\left(Z'\pi\right)}\left(p\right)'-\mf_{\mu_0\left(X\right)\mid \Lambda\left(Z'\pi\right)}\left(p\right)\mf_{\mu_0\left(X\right)\mid P}\left(p\right)'\right)
\\&&=
\left(1-p\right)^2\left(\hat{\mf}_{\mu_0\left(X\right)\mid P}\left(p\right)-\mf_{\mu_0\left(X\right)\mid P}\left(p\right)\right)\left(\hat{\mf}_{\mu_0\left(X\right)\mid P}\left(p\right)-\mf_{\mu_0\left(X\right)\mid P}\left(p\right)\right)'.
\\&&-
\left(1-p\right)^2\left({\mf}_{\mu_0\left(X\right)\mid\Lambda\left(Z'\pi\right)}\left(p\right)-\mf_{\mu_0\left(X\right)\mid P}\left(p\right)\right)\left({\mf}_{\mu_0\left(X\right)\mid\Lambda\left(Z'\pi\right)}\left(p\right)-\mf_{\mu_0\left(X\right)\mid P}\left(p\right)\right)'.
\end{eqnarray*}
\endgroup
By the triangle inequality, we have 
\begingroup
\allowdisplaybreaks
\begin{eqnarray*}
&&
\left\|\int\left(\hat{\xi}_{1,1}\left(x,y,\Lambda\left(z'\hat{\pi}\right)\right)-{\xi}_{1,1}\left(x,y,\Lambda\left(z'\hat{\pi}\right)\right)\right)F_{W}\left(dw\right)\right\|
\\&&\leq
\left(\int\left\|\hat{\mf}_{\mu_0\left(X\right)\mid P}\left(\Lambda\left(z'\hat{\pi}\right)\right)-\mf_{\mu_0\left(X\right)\mid P}\left(\Lambda\left(z'\hat{\pi}\right)\right)\right\|^2F_Z\left(dz\right)\right)^{1/2}\\
&&+
\sup_{\pi\in\mathbf{P}_n}\left(\int\left\|{\mf}_{\mu_0\left(X\right)\mid\Lambda\left(Z'\pi\right)}\left(\Lambda\left(z'{\pi}\right)\right)-\mf_{\mu_0\left(X\right)\mid P}\left(\Lambda\left(z'{\pi}\right)\right)\right\|^2F_Z\left(dz\right)\right)^{1/2}
\\&&\leq
\left(\int_{z\in\mathcal{Z}_n}\left\|\hat{\mf}_{\mu_0\left(X\right)\mid P}\left(\Lambda\left(z'\hat{\pi}\right)\right)-\mf_{\mu_0\left(X\right)\mid P}\left(\Lambda\left(z'\hat{\pi}\right)\right)\right\|^2F_Z\left(dz\right)\right)^{1/2}\\
&&+
\sup_{\pi\in\mathbf{P}_n}\left(\int_{z\notin\mathcal{Z}_n}\left\|\mf_{\mu_0\left(X\right)\mid P}\left(\Lambda\left(z'{\pi}\right)\right)\right\|^2F_Z\left(dz\right)\right)^{1/2}\\
&&+
\sup_{\pi\in\mathbf{P}_n}\left(\int\left\|{\mf}_{\mu_0\left(X\right)\mid\Lambda\left(Z'\pi\right)}\left(\Lambda\left(z'{\pi}\right)\right)-\mf_{\mu_0\left(X\right)\mid P}\left(\Lambda\left(z'{\pi}\right)\right)\right\|^2F_Z\left(dz\right)\right)^{1/2}.
\end{eqnarray*}
\endgroup
Since Assumptions \ref{assn:17} and \ref{assn:18} imply 
\begingroup
\allowdisplaybreaks
\begin{eqnarray*}
&&
\sup_{\pi\in\mathbf{P}_n}\left(\int\left\|{\mf}_{\mu_0\left(X\right)\mid\Lambda\left(Z'\pi\right)}\left(\Lambda\left(z'{\pi}\right)\right)-\mf_{\mu_0\left(X\right)\mid P}\left(\Lambda\left(z'{\pi}\right)\right)\right\|^2F_Z\left(dz\right)\right)^{1/2}
\\&&\leq
\sup_{\pi\in\mathbf{P}_n}\left(\int\left\|{\mf}_{\mu_0\left(X\right)\mid\Lambda\left(Z'\pi\right)}\left(\Lambda\left(z'{\pi}\right)\right)-\mf_{\mu_0\left(X\right)\mid\Lambda\left(Z'\pi_0\right)}\left(\Lambda\left(z'{\pi}\right)\right)\right\|^2F_Z\left(dz\right)\right)^{1/2}
\\&&+
\sup_{\pi\in\mathbf{P}_n}\left(\int\left\|{\mf}_{\mu_0\left(X\right)\mid\Lambda\left(Z'\pi_0\right)}\left(\Lambda\left(z'{\pi}\right)\right)-\mf_{\mu_0\left(X\right)\mid P}\left(\Lambda\left(z'{\pi}\right)\right)\right\|^2F_Z\left(dz\right)\right)^{1/2}
\\&&\leq
o_p\left(n^{-r_1}\right),
\end{eqnarray*}
\endgroup
Assumption \ref{assn:19} and Lemma \ref{lemma:mean_P_conv_process} under Assumptions \ref{assn:11}, \ref{assn:12}, \ref{assn:13}, \ref{assn:14}, \ref{assn:16}, \ref{assn:17}, and \ref{assn:18} conclude the statement of this lemma.
\end{proof}

\begin{lemma}\label{lemma:low5}
If Assumptions \ref{assn:11}, \ref{assn:12}, \ref{assn:13}, \ref{assn:14},  \ref{assn:16}, \ref{assn:17}, \ref{assn:18}, and \ref{assn:19} are satisfied, then we have $\int\boldsymbol{R}_k\left(z\right)F_Z\left(dz\right)=o_p\left(n^{-1/2}\right)$ for every $\hat{\gamma}_\ell$ and every $k=1,\ldots,4$.
\end{lemma}
\begin{proof}
By the second-order Taylor expansion under Assumption \ref{assn:16}, we have 
\begingroup
\allowdisplaybreaks
\begin{align*}
\left\|\boldsymbol{R}_1\left(z\right)
\right\|
=&
\left\|
\int \left({\xi}_1\left(x,y,\hat{\mf}_{S\mid Z}\left(z\right)\right)-{\xi}_1\left(x,y,{\mf}_{S\mid Z}\left(z\right)\right)\right)F_{\left(Y,X\right)\mid Z=z}\left(dy,dx\right)-{\zeta}\left(z\right)\left(\hat{\mf}_{S\mid Z}\left(z\right)-{\mf}_{S\mid Z}\left(z\right)\right)
\right\|\\
\leq&
\frac{1}{2}\sup_{p\in[0,1]}\left\|\frac{\partial^2}{\partial p^2}\int{\xi}_1\left(x,y,p\right)F_{\left(Y,X\right)\mid Z=z}\left(dy,dx\right)\right\|\left(\hat{\mf}_{S\mid Z}\left(z\right)-{\mf}_{S\mid Z}\left(z\right)\right)^2,
\end{align*}
\endgroup
and therefore,
\begingroup
\allowdisplaybreaks
\begin{align*}
\left\|\int\boldsymbol{R}_1\left(z\right)F_Z\left(dz\right)
\right\|
\leq&
\frac{1}{2}\sup_{p\in[0,1]}\left\|\frac{\partial^2}{\partial p^2}\int{\xi}_1\left(x,y,p\right)F_{\left(Y,X\right)\mid Z=z}\left(dy,dx\right)\right\|\sup_{z\in\mathcal{Z}_n}\left(\hat{\mf}_{S\mid Z}\left(z\right)-{\mf}_{S\mid Z}\left(z\right)\right)^2
\\
=&
O_p\left(n^{-2r_1}\right),
\end{align*}
\endgroup
where the last equality follows from Assumption \ref{assn:16} and Lemma \ref{lemma:prop_score_conv} under Assumptions \ref{assn:11}, \ref{assn:12}, and \ref{assn:18}.
Applying similar lines of arguments to $\int\boldsymbol{R}_k\left(z\right)F_Z\left(dz\right)$ for $k=2,3,4$, we have 
\begingroup
\allowdisplaybreaks
\begin{align*}
\|\int\boldsymbol{R}_2\left(z\right)F_Z\left(dz\right)\|
\leq&
\frac{1}{2}\sup_{\left(p,z\right)}
\left|\frac{\partial^2}{\partial p^2}P^\ast\left(p,z\right)\right|
\int\left\|\mf_{\left(\mu_0\left(X\right)',\mu_1\left(X\right)',1\right)'\mid Z}\left(z\right)\right\|
F_Z\left(dz\right)\sup_{z\in\mathcal{Z}_n}\left(\hat{\mf}_{S\mid Z}\left(z\right)-{\mf}_{S\mid Z}\left(z\right)\right)^2
\\
=&
O_p\left(n^{-2r_1}\right)
\\
|\int\boldsymbol{R}_3\left(z\right)F_Z\left(dz\right)|
\leq&
\frac{1}{2}\sup_{p\in[0,1]}\left\|\frac{\partial^2}{\partial p^2}{\mf}_{U\mid P}\left(P^\ast\left(p,z\right)\right)\right\|
\sup_{z\in\mathcal{Z}_n}\left(\hat{\mf}_{S\mid Z}\left(z\right)-{\mf}_{S\mid Z}\left(z\right)\right)^2
\\
=&
O_p\left(n^{-2r_1}\right)
\\
|\int\boldsymbol{R}_4\left(z\right)F_Z\left(dz\right)|
\leq&
\frac{1}{2}\int\left|\frac{\hat{f}_{P^\ast}\left(\hat{\mf}_{S\mid Z}\left(z\right)\right)}{\hat{f}_{P}\left(\hat{\mf}_{S\mid Z}\left(z\right)\right)}\right|F_Z\left(dz\right)\sup_{p\in[0,1]}\left|\frac{\partial^2}{\partial u^2}{\mf}_{U\mid P}\left(p\right)\right|
\sup_{z\in\mathcal{Z}_n}\left(\hat{\mf}_{S\mid Z}\left(z\right)-{\mf}_{S\mid Z}\left(z\right)\right)^2\\
=&
O_p\left(n^{-2r_1}\right).
\end{align*}
\endgroup
These complete a proof of the lemma.
\end{proof}

\begin{lemma}\label{lemma:B2_lemma2}
If Assumptions \ref{assn:11}-\ref{assn:19} are satisfied, then $$
\left(\int_{z\in\mathcal{Z}_n}\left\|\hat{\zeta}\left(z\right)-{\zeta}\left(z;\hat{\mf}_{S\mid Z}\left(z\right)\right)\right\|^2F_W\left(dw\right)\right)^{1/2}=o_p\left(1\right)
$$
\end{lemma}
\begin{proof}
We are going to show this lemma only for the square of the $\mu_0\left(X\right)$ term, and results for the other components will similarly follow. 
Since 
\begingroup
\allowdisplaybreaks
\begin{align*}
\hat{\zeta}_1\left(z;p\right)
=&
2\left(1-p\right)\left(\hat{\mf}_{\mu_0\left(X\right)\mid Z}\left(z\right)-\hat{\mf}_{\mu_0\left(X\right)\mid P}\left(p\right)\right)\left(\hat{\mf}_{\mu_0\left(X\right)\mid Z}\left(z\right)-\hat{\mf}_{\mu_0\left(X\right)\mid P}\left(p\right)\right)'\\
&+2\left(1-p\right)\left(\hat{\mf}_{\mu_0\left(X\right)\mu_0\left(X\right)'\mid Z}\left(z\right)-\hat{\mf}_{\mu_0\left(X\right)\mid Z}\left(z\right)\hat{\mf}_{\mu_0\left(X\right)\mid Z}\left(z\right)'\right)\\
&-\left(1-p\right)^2\hat{\mf}_{\mu_0\left(X\right)\mid P}^{(1)}\left(p\right)\left(\hat{\mf}_{\mu_0\left(X\right)\mid Z}\left(z\right)-\hat{\mf}_{\mu_0\left(X\right)\mid P}\left(p\right)\right)'\\
&-\left(1-p\right)^2\left(\hat{\mf}_{\mu_0\left(X\right)\mid Z}\left(z\right)-\hat{\mf}_{\mu_0\left(X\right)\mid P}\left(p\right)\right)\hat{\mf}_{\mu_0\left(X\right)\mid P}^{\left(1\right)}\left(p\right)',
\end{align*}
\endgroup
it follows that that $\|\hat{\zeta}_1\left(z;p\right)-{\zeta}_1\left(z;p\right)\|$ is bounded (multiplied by a constant) by the sum of the following terms:  
\begingroup
\allowdisplaybreaks
\begin{align*}
&\|\hat{\mf}_{\mu_0\left(X\right)\mid Z}\left(z\right)-{\mf}_{\mu_0\left(X\right)\mid Z}\left(z\right)\|\|\hat{\mf}_{\mu_0\left(X\right)\mid P}\left(p\right)-{\mf}_{\mu_0\left(X\right)\mid P}\left(p\right)\|,
\\&\|\hat{\mf}_{\mu_0\left(X\right)\mid P}\left(p\right)-{\mf}_{\mu_0\left(X\right)\mid P}\left(p\right)\|^2,
\quad\|\hat{\mf}_{\mu_0\left(X\right)\mid Z}\left(z\right)-{\mf}_{\mu_0\left(X\right)\mid Z}\left(z\right)\|^2,
\\&\|\hat{\mf}_{\mu_0\left(X\right)\mid Z}\left(z\right)-{\mf}_{\mu_0\left(X\right)\mid Z}\left(z\right)\|\|\hat{\mf}_{\mu_0\left(X\right)\mid P}^{\left(1\right)}\left(p\right)-{\mf}_{\mu_0\left(X\right)\mid P}^{\left(1\right)}\left(p\right)\|,
\\&\|\hat{\mf}_{\mu_0\left(X\right)\mid P}^{\left(1\right)}\left(p\right)-{\mf}_{\mu_0\left(X\right)\mid P}^{\left(1\right)}\left(p\right)\|\|\hat{\mf}_{\mu_0\left(X\right)\mid P}\left(p\right)-{\mf}_{\mu_0\left(X\right)\mid P}\left(p\right)\|,
\\&\|\hat{\mf}_{\mu_0\left(X\right)\mid P}^{\left(1\right)}\left(p\right)-{\mf}_{\mu_0\left(X\right)\mid P}^{\left(1\right)}\left(p\right)\|\|\hat{\mf}_{\mu_0\left(X\right)\mid Z}\left(z\right)-{\mf}_{\mu_0\left(X\right)\mid Z}\left(z\right)\|,
\\&\|{\mf}_{\mu_0\left(X\right)\mid Z}\left(z\right)\|\|\hat{\mf}_{\mu_0\left(X\right)\mid Z}\left(z\right)-{\mf}_{\mu_0\left(X\right)\mid Z}\left(z\right)\|,
\quad\|{\mf}_{\mu_0\left(X\right)\mid P}\left(p\right)\|\|\hat{\mf}_{\mu_0\left(X\right)\mid Z}\left(z\right)-{\mf}_{\mu_0\left(X\right)\mid Z}\left(z\right)\|,
\\&\|{\mf}_{\mu_0\left(X\right)\mid P}^{\left(1\right)}\left(p\right)\|\|\hat{\mf}_{\mu_0\left(X\right)\mid Z}\left(z\right)-{\mf}_{\mu_0\left(X\right)\mid Z}\left(z\right)\|,
\quad\|{\mf}_{\mu_0\left(X\right)\mid Z}\left(z\right)\|\|\hat{\mf}_{\mu_0\left(X\right)\mid P}\left(p\right)-{\mf}_{\mu_0\left(X\right)\mid P}\left(p\right)\|,
\\&\|{\mf}_{\mu_0\left(X\right)\mid P}\left(p\right)\|\|\hat{\mf}_{\mu_0\left(X\right)\mid P}\left(p\right)-{\mf}_{\mu_0\left(X\right)\mid P}\left(p\right)\|,
\quad\|{\mf}_{\mu_0\left(X\right)\mid P}^{\left(1\right)}\left(p\right)\|\|\hat{\mf}_{\mu_0\left(X\right)\mid P}\left(p\right)-{\mf}_{\mu_0\left(X\right)\mid P}\left(p\right)\|,
\\&\|{\mf}_{\mu_0\left(X\right)\mid Z}\left(z\right)\|\|\hat{\mf}_{\mu_0\left(X\right)\mid P}^{\left(1\right)}\left(p\right)-{\mf}_{\mu_0\left(X\right)\mid P}^{\left(1\right)}\left(p\right)\|,
\quad\|{\mf}_{\mu_0\left(X\right)\mid P}\left(p\right)\|\|\hat{\mf}_{\mu_0\left(X\right)\mid P}^{\left(1\right)}\left(p\right)-{\mf}_{\mu_0\left(X\right)\mid P}^{\left(1\right)}\left(p\right)\|, 
\\&\text{and } \|\hat{\mf}_{\mu_0\left(X\right)\mu_0\left(X\right)'\mid Z}\left(z\right)-{\mf}_{\mu_0\left(X\right)\mu_0\left(X\right)'\mid Z}\left(z\right)\|.
\end{align*}
\endgroup
Note that
\begingroup
\allowdisplaybreaks
\begin{align*}
&
\sup_{z \in \mathcal{Z}_n} \|\hat{\mf}_{\mu_0\left(X\right)\mid Z}\left(z\right)-{\mf}_{\mu_0\left(X\right)\mid Z}\left(z\right)\|
=o_p\left(1\right)
\\&
\sup_{z \in \mathcal{Z}_n} \|\hat{\mf}_{\mu_0\left(X\right)\mu_0\left(X\right)'\mid Z}\left(z\right)-{\mf}_{\mu_0\left(X\right)\mu_0\left(X\right)'\mid Z}\left(z\right)\|=o_p\left(1\right)
\end{align*}
\endgroup
by Assumption \ref{assn:15},
and 
\begingroup
\allowdisplaybreaks
\begin{align*}
&\left(\int_{z\in\mathcal{Z}_n}\|\hat{\mf}_{\mu_0\left(X\right)\mid P}\left(\hat{\mf}_{S\mid Z}\left(z\right)\right)-{\mf}_{\mu_0\left(X\right)\mid P}\left(\hat{\mf}_{S\mid Z}\left(z\right)\right)\|^4F_Z\left(dz\right)\right)^{1/4}=o_p\left(1\right)
\\&\left(\int_{z\in\mathcal{Z}_n}\|\hat{\mf}_{\mu_0\left(X\right)\mid P}^{\left(1\right)}\left(\hat{\mf}_{S\mid Z}\left(z\right)\right)-{\mf}_{\mu_0\left(X\right)\mid P}^{\left(1\right)}\left(\hat{\mf}_{S\mid Z}\left(z\right)\right)\|^4F_Z\left(dz\right)\right)^{1/4}=o_p\left(1\right)
\end{align*}
\endgroup
by Lemmas \ref{lemma:mean_P_conv_process} and \ref{lemma:deriv_conv_process} under Assumptions \ref{assn:11}, \ref{assn:12}, \ref{assn:13}, \ref{assn:14},  \ref{assn:16}, \ref{assn:17}, and \ref{assn:18}.
Therefore, Assumption \ref{assn:17}, \ref{assn:18} and \ref{assn:19} imply the statement of this lemma. 
\end{proof}

\begin{lemma}\label{lemma:B2_lemma3}
If Assumptions \ref{assn:11}, \ref{assn:12}, \ref{assn:13}, \ref{assn:14},  \ref{assn:16}, \ref{assn:17}, and \ref{assn:18}  are satisfied, then
$$
\left(\int_{z\in\mathcal{Z}_n}
\left(\frac{\hat{f}_{P^\ast}\left(\hat{\mf}_{S\mid Z}\left(z\right)\right)}{\hat{f}_{P}\left(\hat{\mf}_{S\mid Z}\left(z\right)\right)}\hat{\mf}_{U\mid P}\left(\hat{\mf}_{S\mid Z}\left(z\right)\right)-\frac{{f}_{P^\ast}\left(\hat{\mf}_{S\mid Z}\left(z\right)\right)}{{f}_{P}\left(\hat{\mf}_{S\mid Z}\left(z\right)\right)}{\mf}_{U\mid P}\left(\hat{\mf}_{S\mid Z}\left(z\right)\right)\right)^2
F_Z\left(dz\right)\right)^{1/2}=o_p\left(1\right)
$$
and 
$$
\left(\int_{z\in\mathcal{Z}_n}
\left(\frac{\hat{f}_{P^\ast}\left(\hat{\mf}_{S\mid Z}\left(z\right)\right)}{\hat{f}_{P}\left(\hat{\mf}_{S\mid Z}\left(z\right)\right)}\hat{\Delta}_{U\mid P}\left(\hat{\mf}_{S\mid Z}\left(z\right)\right)-\frac{{f}_{P^\ast}\left(\hat{\mf}_{S\mid Z}\left(z\right)\right)}{{f}_{P}\left(\hat{\mf}_{S\mid Z}\left(z\right)\right)}{\Delta_{U\mid P}}\left(\hat{\mf}_{S\mid Z}\left(z\right)\right)\right)^2
F_Z\left(dz\right)\right)^{1/2}=o_p\left(1\right).
$$
\end{lemma}
\begin{proof}
Since 
\begingroup
\allowdisplaybreaks
\begin{align*}
&\left|\frac{\hat{f}_{P^\ast}\left(\hat{\mf}_{S\mid Z}\left(z\right)\right)}{\hat{f}_{P}\left(\hat{\mf}_{S\mid Z}\left(z\right)\right)}\hat{\mf}_{U\mid P}\left(\hat{\mf}_{S\mid Z}\left(z\right)\right)-\frac{{f}_{P^\ast}\left(\hat{\mf}_{S\mid Z}\left(z\right)\right)}{{f}_{P}\left(\hat{\mf}_{S\mid Z}\left(z\right)\right)}{\mf}_{U\mid P}\left(\hat{\mf}_{S\mid Z}\left(z\right)\right)\right|
\\
&\leq
\left|\frac{\hat{f}_{P^\ast}\left(\hat{\mf}_{S\mid Z}\left(z\right)\right)}{\hat{f}_{P}\left(\hat{\mf}_{S\mid Z}\left(z\right)\right)}-\frac{{f}_{P^\ast}\left(\hat{\mf}_{S\mid Z}\left(z\right)\right)}{{f}_{P}\left(\hat{\mf}_{S\mid Z}\left(z\right)\right)}\right||\hat{\mf}_{U\mid P}\left(\hat{\mf}_{S\mid Z}\left(z\right)\right)-{\mf}_{U\mid P}\left(\hat{\mf}_{S\mid Z}\left(z\right)\right)|
\\&+
\left|\frac{\hat{f}_{P^\ast}\left(\hat{\mf}_{S\mid Z}\left(z\right)\right)}{\hat{f}_{P}\left(\hat{\mf}_{S\mid Z}\left(z\right)\right)}-\frac{{f}_{P^\ast}\left(\hat{\mf}_{S\mid Z}\left(z\right)\right)}{{f}_{P}\left(\hat{\mf}_{S\mid Z}\left(z\right)\right)}\right||{\mf}_{U\mid P}\left(\hat{\mf}_{S\mid Z}\left(z\right)\right)|
\\&+
\left|\frac{{f}_{P^\ast}\left(\hat{\mf}_{S\mid Z}\left(z\right)\right)}{{f}_{P}\left(\hat{\mf}_{S\mid Z}\left(z\right)\right)}\right||\hat{\mf}_{U\mid P}\left(\hat{\mf}_{S\mid Z}\left(z\right)\right)-{\mf}_{U\mid P}\left(\hat{\mf}_{S\mid Z}\left(z\right)\right)|
\end{align*}
\endgroup
and
\begingroup
\allowdisplaybreaks
\begin{align*}
&\left|\frac{\hat{f}_{P^\ast}\left(\hat{\mf}_{S\mid Z}\left(z\right)\right)}{\hat{f}_{P}\left(\hat{\mf}_{S\mid Z}\left(z\right)\right)}\hat{\Delta}_{U\mid P}\left(\hat{\mf}_{S\mid Z}\left(z\right)\right)-\frac{{f}_{P^\ast}\left(\hat{\mf}_{S\mid Z}\left(z\right)\right)}{{f}_{P}\left(\hat{\mf}_{S\mid Z}\left(z\right)\right)}{\Delta}_{U\mid P}\left(\hat{\mf}_{S\mid Z}\left(z\right)\right)\right|
\\
&\leq
\left|\frac{\hat{f}_{P^\ast}\left(\hat{\mf}_{S\mid Z}\left(z\right)\right)}{\hat{f}_{P}\left(\hat{\mf}_{S\mid Z}\left(z\right)\right)}-\frac{{f}_{P^\ast}\left(\hat{\mf}_{S\mid Z}\left(z\right)\right)}{{f}_{P}\left(\hat{\mf}_{S\mid Z}\left(z\right)\right)}\right||\hat{\Delta}_{U\mid P}\left(\hat{\mf}_{S\mid Z}\left(z\right)\right)-{\Delta}_{U\mid P}\left(\hat{\mf}_{S\mid Z}\left(z\right)\right)|
\\&+
\left|\frac{\hat{f}_{P^\ast}\left(\hat{\mf}_{S\mid Z}\left(z\right)\right)}{\hat{f}_{P}\left(\hat{\mf}_{S\mid Z}\left(z\right)\right)}-\frac{{f}_{P^\ast}\left(\hat{\mf}_{S\mid Z}\left(z\right)\right)}{{f}_{P}\left(\hat{\mf}_{S\mid Z}\left(z\right)\right)}\right||{\Delta}_{U\mid P}\left(\hat{\mf}_{S\mid Z}\left(z\right)\right)|
\\&+
\left|\frac{{f}_{P^\ast}\left(\hat{\mf}_{S\mid Z}\left(z\right)\right)}{{f}_{P}\left(\hat{\mf}_{S\mid Z}\left(z\right)\right)}\right||\hat{\Delta}_{U\mid P}\left(\hat{\mf}_{S\mid Z}\left(z\right)\right)-{\Delta}_{U\mid P}\left(\hat{\mf}_{S\mid Z}\left(z\right)\right)|,
\end{align*}
\endgroup
by Assumption \ref{assn:16}, it suffices to show 
\begingroup
\allowdisplaybreaks
\begin{align}\label{eq:some_goal+temp}
\left(\int_{z\in\mathcal{Z}_n}\left(
\frac{\hat{f}_{P^\ast}\left(\hat{\mf}_{S\mid Z}\left(z\right)\right)}{\hat{f}_{P}\left(\hat{\mf}_{S\mid Z}\left(z\right)\right)}-\frac{{f}_{P^\ast}\left(\hat{\mf}_{S\mid Z}\left(z\right)\right)}{{f}_{P}\left(\hat{\mf}_{S\mid Z}\left(z\right)\right)}\right)^4F_Z\left(dz\right)\right)^{1/4}=&o_p\left(1\right)
\\
\left(\int_{z\in\mathcal{Z}_n}\left(
\hat{\mf}_{U\mid P}\left(\hat{\mf}_{S\mid Z}\left(z\right)\right)-{\mf}_{U\mid P}\left(\hat{\mf}_{S\mid Z}\left(z\right)\right)\right)^4F_Z\left(dz\right)\right)^{1/4}=&o_p\left(1\right)
\notag
\\
\left(\int_{z\in\mathcal{Z}_n}\left(
\hat{\Delta}_{U\mid P}\left(\hat{\mf}_{S\mid Z}\left(z\right)\right)-{\Delta}_{U\mid P}\left(\hat{\mf}_{S\mid Z}\left(z\right)\right)\right)^4F_Z\left(dz\right)\right)^{1/4}=&o_p\left(1\right).
\notag
\end{align}
\endgroup
The second and third of these are shown in Lemma \ref{lemma:mean_P_conv_process} and \ref{lemma:deriv_conv_process} under Assumptions \ref{assn:11}, \ref{assn:12}, \ref{assn:13}, \ref{assn:14}, \ref{assn:16}, \ref{assn:17}, and \ref{assn:18}.
We therefore focus on \eqref{eq:some_goal+temp}.
Note that 
\begingroup
\allowdisplaybreaks
\begin{align*}
&
\left(\int_{z\in\mathcal{Z}_n}\left(
\frac{\hat{f}_{P^\ast}\left(\hat{\mf}_{S\mid Z}\left(z\right)\right)}{\hat{f}_{P}\left(\hat{\mf}_{S\mid Z}\left(z\right)\right)}-\frac{{f}_{P^\ast}\left(\hat{\mf}_{S\mid Z}\left(z\right)\right)}{{f}_{P}\left(\hat{\mf}_{S\mid Z}\left(z\right)\right)}\right)^4F_Z\left(dz\right)\right)^{1/4}
\\&\leq
\left(\int_{z\in\mathcal{Z}_n}\left(\frac{
\hat{f}_{P^\ast}\left(\Lambda\left(z'\hat\pi\right)\right)
}{\hat{f}_{P}\left(\Lambda\left(z'\hat\pi\right)\right)}-\mathrm{ratio}\left(z,\hat\pi\right)\right)^4F_{Z}\left(dz\right)\right)^{1/4}
\\&+
\sup_{\pi\in\mathbf{P}_n}\left(\int_{z\in\mathcal{Z}_n}\left(\frac{f_{P^\ast\left(\Lambda\left(Z'\pi\right),Z\right)}\left(\Lambda\left(z'\pi\right)\right)}{f_{\Lambda\left(Z'\pi\right)}\left(\Lambda\left(z'\pi\right)\right)}-\frac{{f}_{P^\ast\left(\Lambda\left(Z'\pi_0\right),Z\right)}\left(\Lambda\left(z'\pi\right)\right)}{{f}_{\Lambda\left(Z'\pi_0\right)}\left(\Lambda\left(z'\pi\right)\right)}\right)^4F_{Z}\left(dz\right)\right)^{1/4}
\\&+
\sup_{\pi\in\mathbf{P}_n}\left(\int_{z\in\mathcal{Z}_n}\left(\frac{{f}_{P^\ast\left(\Lambda\left(Z'\pi_0\right),Z\right)}\left(\Lambda\left(z'\pi\right)\right)}{{f}_{\Lambda\left(Z'\pi_0\right)}\left(\Lambda\left(z'\pi\right)\right)}-\frac{{f}_{P^\ast}\left(\Lambda\left(z'\pi\right)\right)}{{f}_{P}\left(\Lambda\left(z'\pi\right)\right)}\right)^4F_{Z}\left(dz\right)\right)^{1/4},
\end{align*}
\endgroup
where $\mathrm{ratio}\left(z,\pi\right)$ is defined in Lemma \ref{lemma:denstiy_ratio_conv_process}.
By Assumptions \ref{assn:17}-\ref{assn:18}
 and Lemma \ref{lemma:denstiy_ratio_conv_process}, we have \eqref{eq:some_goal+temp}.  
\end{proof}

\begin{lemma}\label{lemma:B2_lemma4}
If Assumptions \ref{assn:11}, \ref{assn:12}, \ref{assn:16} and \ref{assn:18} are satisfied, 
then the following random variables are $o_p\left(1\right)$: \small
\begin{center}
$
\left(\int\|\mf_{\left(\mu_0\left(X\right)',\mu_1\left(X\right)',Y\right)'\mid P}\left(\hat{\mf}_{S\mid Z}\left(z\right)\right)-\mf_{\left(\mu_0\left(X\right)',\mu_1\left(X\right)',Y\right)'\mid P}\left({\mf}_{S\mid Z}\left(z\right)\right)\|^4F_Z\left(dz\right)\right)^{1/4} 
$
\\
$
\left(\int\left\|{\zeta}\left(z;\hat{\mf}_{S\mid Z}\left(z\right)\right)-{\zeta}\left(z;{\mf}_{S\mid Z}\left(z\right)\right)\right\|^2F_W\left(dw\right)\right)^{1/2}
$
\\
$
\left(\int\left\|{\zeta}\left(z;\hat{\mf}_{S\mid Z}\left(z\right)\right)\hat{\mf}_{S\mid Z}\left(z\right)-{\zeta}\left(z\right){\mf}_{S\mid Z}\left(z\right)\right\|^2F_W\left(dw\right)\right)^{1/2}.
$
\\
$
\left(\int{\mf}_{\|\left(\mu_0\left(x\right)',\mu_1\left(x\right)',1\right)'\|^2\mid Z}\left(z\right)\left(
P^\ast\left(\hat{\mf}_{S\mid Z}\left(z\right),z\right)-P^\ast\left({\mf}_{S\mid Z}\left(z\right),z\right)
\right)^2F_W\left(dw\right)\right)^{1/2}
$
\\
$
\left(\int{\mf}_{\|\left(\mu_0\left(x\right)',\mu_1\left(x\right)',1\right)'\|^2\mid Z}\left(z\right)\left(
\hat{\mf}_{S\mid Z}\left(z\right)-{\mf}_{S\mid Z}\left(z\right)
\right)^2F_W\left(dw\right)\right)^{1/2}
$
\\
$
\left(\int{\mf}_{\|\left(\mu_0\left(x\right)',\mu_1\left(x\right)',1\right)'\|^2\mid Z}\left(z\right)\left(
\partial P^\ast\left(\hat{\mf}_{S\mid Z}\left(z\right),z\right)-\partial P^\ast\left(\mf_{S\mid Z}\left(z\right),z\right)
\right)^2F_W\left(dw\right)\right)^{1/2}
$
\\
$
\left(\int{\mf}_{\|\left(\mu_0\left(x\right)',\mu_1\left(x\right)',1\right)'\|^2\mid Z}\left(z\right)\left(
\hat{\mf}_{S\mid Z}\left(z\right)\partial P^\ast\left(\hat{\mf}_{S\mid Z}\left(z\right),z\right)
-{\mf}_{S\mid Z}\left(z\right)\partial P^\ast\left(\mf_{S\mid Z}\left(z\right),z\right)
\right)^2F_W\left(dw\right)\right)^{1/2}
$
\\
$
\left(\int\left(
{\mf}_{U\mid P}\left(P^\ast\left(\hat{\mf}_{S\mid Z}\left(z\right),z\right)\right)-{\mf}_{U\mid P}\left(P^\ast\left({\mf}_{S\mid Z}\left(z\right),z\right)\right)
\right)^2
F_Z\left(dz\right)\right)^{1/2}
$
\\
$
\left(\int\left(\frac{{f}_{P^\ast}\left(\hat{\mf}_{S\mid Z}\left(z\right)\right)}{{f}_{P}\left(\hat{\mf}_{S\mid Z}\left(z\right)\right)}{\mf}_{U\mid P}\left(\hat{\mf}_{S\mid Z}\left(z\right)\right)-\frac{{f}_{P^\ast}\left({\mf}_{S\mid Z}\left(z\right)\right)}{{f}_{P}\left({\mf}_{S\mid Z}\left(z\right)\right)}{\mf}_{U\mid P}\left({\mf}_{S\mid Z}\left(z\right)\right)\right)^2
F_Z\left(dz\right)\right)^{1/2}
$
\\
$
\left(\int
\left(\frac{{f}_{P^\ast}\left(\hat{\mf}_{S\mid Z}\left(z\right)\right)}{{f}_{P}\left(\hat{\mf}_{S\mid Z}\left(z\right)\right)}{\Delta}_{U\mid P}\left(\hat{\mf}_{S\mid Z}\left(z\right)\right)-\frac{{f}_{P^\ast}\left({\mf}_{S\mid Z}\left(z\right)\right)}{{f}_{P}\left({\mf}_{S\mid Z}\left(z\right)\right)}{\Delta_{U\mid P}}\left({\mf}_{S\mid Z}\left(z\right)\right)\right)^2
F_Z\left(dz\right)\right)^{1/2}
$
\\
$
\left(\int\left({\Delta}_{U\mid P}\left(P^\ast\left(\hat{\mf}_{S\mid Z}\left(z\right),z\right)\right)\partial P^\ast\left(\hat{\mf}_{S\mid Z}\left(z\right),z\right)-{\Delta_{U\mid P}}\left(P^\ast\left({\mf}_{S\mid Z}\left(z\right),z\right)\right)\partial P^\ast\left({\mf}_{S\mid Z}\left(z\right),z\right)\right)^2
F_Z\left(dz\right)\right)^{1/2}
$
\\
$
\left(\int
\left(\frac{{f}_{P^\ast}\left(\hat{\mf}_{S\mid Z}\left(z\right)\right)}{{f}_{P}\left(\hat{\mf}_{S\mid Z}\left(z\right)\right)}{\Delta}_{U\mid P}\left(\hat{\mf}_{S\mid Z}\left(z\right)\right)\hat{\mf}_{S\mid Z}\left(z\right)-\frac{{f}_{P^\ast}\left({\mf}_{S\mid Z}\left(z\right)\right)}{{f}_{P}\left({\mf}_{S\mid Z}\left(z\right)\right)}{\Delta_{U\mid P}}\left({\mf}_{S\mid Z}\left(z\right)\right){\mf}_{S\mid Z}\left(z\right)\right)^2
F_Z\left(dz\right)\right)^{1/2}
$
\\
$
\left(\int
\left({\Delta}_{U\mid P}\left(P^\ast\left(\hat{\mf}_{S\mid Z}\left(z\right),z\right)\right)\partial P^\ast\left(\hat{\mf}_{S\mid Z}\left(z\right),z\right)\hat{\mf}_{S\mid Z}\left(z\right)-{\Delta_{U\mid P}}\left(P^\ast\left({\mf}_{S\mid Z}\left(z\right),z\right)\right)\partial P^\ast\left({\mf}_{S\mid Z}\left(z\right),z\right){\mf}_{S\mid Z}\left(z\right)\right)^2
F_Z\left(dz\right)\right)^{1/2}.
$
\end{center}\normalsize
\end{lemma}
\begin{proof}
This lemma follows from Assumption \ref{assn:16} and Lemma \ref{lemma:prop_score_conv} under Assumptions \ref{assn:11}, \ref{assn:12}, and \ref{assn:18}.
\end{proof}

\begin{lemma}\label{lemma:B2_lemma5}
Suppose that Assumptions \ref{assn:11}, \ref{assn:12}, \ref{assn:16} and \ref{assn:18} hold. 
If 
\begin{equation}
\label{eq:temp_L4_conv}
\left(\int\|\hat\mf_{\left(\mu_0\left(X\right)',\mu_1\left(X\right)',Y\right)'\mid P}\left(\hat{\mf}_{S\mid Z}\left(z\right)\right)-\mf_{\left(\mu_0\left(X\right)',\mu_1\left(X\right)',Y\right)'\mid P}\left({\mf}_{S\mid Z}\left(z\right)\right)\|^4F_Z\left(dz\right)\right)^{1/4}=o_p\left(1\right),
\end{equation}
we have 
$$
\left(\int\left\|\hat{\xi}_1\left(x,y,\hat{\mf}_{S\mid Z}\left(z\right)\right)-{\xi}_1\left(x,y,{\mf}_{S\mid Z}\left(z\right)\right)\right\|^2F_W\left(dw\right)\right)^{1/2}=o_p\left(1\right).
$$
\end{lemma}
\begin{proof}
We are going to show it only for the square of the $\mu_0\left(X\right)$ term, and results for the other components will similarly follow. 
We can bound $\|\hat{\xi}_{1,1}\left(x,y,\hat{\mf}_{S\mid Z}\left(z\right)\right)-{\xi}_{1,1}\left(x,y,{\mf}_{S\mid Z}\left(z\right)\right)\|$ as 
\begingroup
\allowdisplaybreaks
\begin{align*}
&\|\hat{\xi}_{1,1}\left(x,y,\hat{\mf}_{S\mid Z}\left(z\right)\right)-{\xi}_{1,1}\left(x,y,{\mf}_{S\mid Z}\left(z\right)\right)\| 
\\&\leq
\|\hat\mf_{\mu_0\left(X\right)\mid P}\left(\hat{\mf}_{S\mid Z}\left(z\right)\right)-\mf_{\mu_0\left(X\right)\mid P}\left({\mf}_{S\mid Z}\left(z\right)\right)\|^2
\\&+
2\|\hat\mf_{\mu_0\left(X\right)\mid P}\left(\hat{\mf}_{S\mid Z}\left(z\right)\right)-\mf_{\mu_0\left(X\right)\mid P}\left({\mf}_{S\mid Z}\left(z\right)\right)\|\|\mf_{\mu_0\left(X\right)\mid P}\left({\mf}_{S\mid Z}\left(z\right)\right)-\mu_0\left(x\right)\|
\\&+
|\left(1-\hat{\mf}_{S\mid Z}\left(z\right)\right)^2-\left(1-{\mf}_{S\mid Z}\left(z\right)\right)^2|\|\mf_{\mu_0\left(X\right)\mid P}\left({\mf}_{S\mid Z}\left(z\right)\right)-\mu_0\left(x\right)\|^2.
\end{align*}
\endgroup
Then 
\begingroup
\allowdisplaybreaks
\begin{align*}
&\left(\int\left\|\hat{\xi}_{1,1}\left(x,y,\hat{\mf}_{S\mid Z}\left(z\right)\right)-{\xi}_{1,1}\left(x,y,{\mf}_{S\mid Z}\left(z\right)\right)\right\|^2F_W\left(dw\right)\right)^{1/2}
\\&\leq
\left(\int\|\hat\mf_{\mu_0\left(X\right)\mid P}\left(\hat{\mf}_{S\mid Z}\left(z\right)\right)-\mf_{\mu_0\left(X\right)\mid P}\left({\mf}_{S\mid Z}\left(z\right)\right)\|^4F_Z\left(dz\right)\right)^{1/2}
\\&+
2\left(\int\|\hat\mf_{\mu_0\left(X\right)\mid P}\left(\hat{\mf}_{S\mid Z}\left(z\right)\right)-\mf_{\mu_0\left(X\right)\mid P}\left({\mf}_{S\mid Z}\left(z\right)\right)\|^2 \times \right.
\\
&\qquad\left.\left(\iint\|\mf_{\mu_0\left(X\right)\mid P}\left({\mf}_{S\mid Z}\left(z\right)\right)-\mu_0\left(x\right)\|^2F_{\left(Y,X\right)\mid Z=z}\left(dy,dx\right)\right)F_Z\left(dz\right)\right)^{1/2}
\\&+
\left(\int\left(\left(1-\hat{\mf}_{S\mid Z}\left(z\right)\right)^2-\left(1-{\mf}_{S\mid Z}\left(z\right)\right)^2\right)^2 \times \right.
\\
&\qquad\left.\left(\iint\|\mf_{\mu_0\left(X\right)\mid P}\left({\mf}_{S\mid Z}\left(z\right)\right)-\mu_0\left(x\right)\|^4F_{\left(Y,X\right)\mid Z=z}\left(dy,dx\right)\right)F_Z\left(dz\right)\right)^{1/2}.
\end{align*}
\endgroup
By Assumption \ref{assn:16}, \eqref{eq:temp_L4_conv}, Lemma \ref{lemma:prop_score_conv} under Assumptions \ref{assn:11}, \ref{assn:12}, and \ref{assn:18}, 
we can bound  
$$\left(\int\left\|\hat{\xi}_{1,1}\left(x,y,\hat{\mf}_{S\mid Z}\left(z\right)\right)-{\xi}_{1,1}\left(x,y,{\mf}_{S\mid Z}\left(z\right)\right)\right\|^2F_W\left(dw\right)\right)^{1/2}$$
 by $o_p\left(1\right).$
\end{proof}

\begin{lemma}\label{lemma:Assn9_verification}
Assumptions \ref{assn:11}-\ref{assn:19}  imply Assumption \ref{Assn_just_consis}. 
\end{lemma}
\begin{proof}
We are going to see each term in Assumption \ref{Assn_just_consis}. 
Regarding the first term in Assumption \ref{Assn_just_consis}, by Lemma \ref{lemma:B2_lemma5}, we consider 
$$
\left(\int\left(\hat\mf_{\omega\mid P}\left(\hat{\mf}_{S\mid Z}\left(z\right)\right)-\mf_{\omega\mid P}\left({\mf}_{S\mid Z}\left(z\right)\right)\right)^4F_Z\left(dz\right)\right)^{1/4}
$$ 
for every component $\omega$ of $\left(\mu_0\left(X\right)',\mu_1\left(X\right)',Y\right)'$. 
We can bound it as follows: 
\begingroup
\allowdisplaybreaks
\begin{eqnarray*}
&&
\left(\int\left(\hat\mf_{\omega\mid P}\left(\hat{\mf}_{S\mid Z}\left(z\right)\right)-\mf_{\omega\mid P}\left({\mf}_{S\mid Z}\left(z\right)\right)\right)^4F_Z\left(dz\right)\right)^{1/4}
\\&&\leq
\left(\int_{z\notin\mathcal{Z}_n}\mf_{\omega\mid P}\left({\mf}_{S\mid Z}\left(z\right)\right)^4F_Z\left(dz\right)\right)^{1/4}
\\&&+
\left(\int_{z\in\mathcal{Z}_n}\left(\hat\mf_{\omega\mid P}\left(\hat{\mf}_{S\mid Z}\left(z\right)\right)-\mf_{\omega\mid P}\left(\hat{\mf}_{S\mid Z}\left(z\right)\right)\right)^4F_Z\left(dz\right)\right)^{1/4}
\\&&+
\left(\int_{z\in\mathcal{Z}_n}\left(\mf_{\omega\mid P}\left(\hat{\mf}_{S\mid Z}\left(z\right)\right)-\mf_{\omega\mid P}\left({\mf}_{S\mid Z}\left(z\right)\right)\right)^4F_Z\left(dz\right)\right)^{1/4}.
\end{eqnarray*}
\endgroup
The first term on the right hand side is $o(1)$ by Assumption \ref{assn:19}, the second term on the right hand side is $o_p(1)$ by Lemma \ref{lemma:mean_P_conv_process}, and the third term on the right hand side is $o_p(1)$ by Lemma \ref{lemma:B2_lemma4}. 

Regarding the second and third terms in Assumption \ref{Assn_just_consis},  we can bound them similarly. The second term is bounded as follows: 
\begingroup
\allowdisplaybreaks
\begin{eqnarray*}
&&\left(\int\left\|\hat{\zeta}\left(z\right)-{\zeta}\left(z\right)\right\|^2F_W\left(dw\right)\right)^{1/2}
\\&&\leq
\left(\int_{z\notin\mathcal{Z}_n}\left\|{\zeta}\left(z\right)\right\|^2F_W\left(dw\right)\right)^{1/2}
\\&&+
\left(\int_{z\in\mathcal{Z}_n}\left\|\hat{\zeta}\left(z\right)-{\zeta}\left(z;\hat{\mf}_{S\mid Z}\left(z\right)\right)\right\|^2F_W\left(dw\right)\right)^{1/2}
\\&&+
\left(\int_{z\in\mathcal{Z}_n}\left\|{\zeta}\left(z;\hat{\mf}_{S\mid Z}\left(z\right)\right)-{\zeta}\left(z\right)\right\|^2F_W\left(dw\right)\right)^{1/2}.
\end{eqnarray*}
\endgroup
The first term on the right hand side is $o(1)$ by Assumption \ref{assn:19}, the second term on the right hand side is $o_p(1)$ by Lemma \ref{lemma:B2_lemma2}, and the third term on the right hand side is $o_p(1)$ by Lemma \ref{lemma:B2_lemma4}.

The fourth to seventh terms in Assumption \ref{Assn_just_consis} are $o_p(1)$ by Lemma \ref{lemma:prop_score_conv} and Assumption \ref{assn:16}. 

The eighth to tenth terms in Assumption \ref{Assn_just_consis} are $o_p(1)$ by Assumption \ref{assn:19} and Lemma \ref{lemma:B2_lemma3} and  \ref{lemma:B2_lemma4}.

Regarding the last three terms  in Assumption \ref{Assn_just_consis},  we can bound them in a similar manner. 
The ninth term is bounded as follows: 
\begingroup
\allowdisplaybreaks
\begin{eqnarray*}
&&\left(\int
\left(
\hat{\mf}_{U\mid P}\left(P^\ast\left(\hat{\mf}_{S\mid Z}\left(z\right),z\right)\right)-{\mf}_{U\mid P}\left(P^\ast\left({\mf}_{S\mid Z}\left(z\right),z\right)\right)
\right)^2
F_Z\left(dz\right)\right)^{1/2}
\\&&\leq
\left(\int_{z\notin\mathcal{Z}_n}
\left(
{\mf}_{U\mid P}\left(P^\ast\left({\mf}_{S\mid Z}\left(z\right),z\right)\right)
\right)^2
F_Z\left(dz\right)\right)^{1/2}
\\&&+
\left(\int_{z\in\mathcal{Z}_n}
\left(
\hat{\mf}_{U\mid P}\left(P^\ast\left(\hat{\mf}_{S\mid Z}\left(z\right),z\right)\right)-{\mf}_{U\mid P}\left(P^\ast\left(\hat{\mf}_{S\mid Z}\left(z\right),z\right)\right)
\right)^2
F_Z\left(dz\right)\right)^{1/2}
\\&&+
\left(\int_{z\in\mathcal{Z}_n}
\left(
{\mf}_{U\mid P}\left(P^\ast\left(\hat{\mf}_{S\mid Z}\left(z\right),z\right)\right)-{\mf}_{U\mid P}\left(P^\ast\left({\mf}_{S\mid Z}\left(z\right),z\right)\right)
\right)^2
F_Z\left(dz\right)\right)^{1/2}.
 \end{eqnarray*}
\endgroup
The first term on the right hand side is $o(1)$ by Assumption \ref{assn:19}, the second term on the right hand side is $o_p(1)$ by Lemma \ref{lemma:mean_P_conv_process}, and the third term on the right hand side is $o_p(1)$ by Lemma \ref{lemma:B2_lemma4}. 
\end{proof}

\section{Auxiliary Equations}
\subsection{Derivation of Equation \eqref{eq:partial_linear_MTE}}\label{sec:eq:partial_linear_MTE}
Observe that
\begingroup
\allowdisplaybreaks
\begin{align*}
E[SU_1+\left(1-S\right)U_0\mid X,P=p]
=&
E[1\{p>F_{U_S}[U_S]\}U_1+1\{p\leq F_{U_S}[U_S]\}U_0\mid X,P=p]\\
=&
E[1\{p>F_{U_S}[U_S]\}U_1+1\{p\leq F_{U_S}[U_S]\}U_0\mid P=p]\\
=&
E[SU_1+\left(1-S\right)U_0\mid P=p]\\
=&
E[U\mid P=p]
\end{align*}
\endgroup
holds under Assumption \ref{assn1}.
Therefore, we have
\begingroup
\allowdisplaybreaks
\begin{align}
E[Y\mid X,P=p]
=&
E[S\left(\mu_1\left(X\right)'\beta_1+U_1\right)+\left(1-S\right)\left(\mu_0\left(X\right)'\beta_0+U_0\right)\mid X,P=p]\notag\\
=&
p\mu_1\left(X\right)'\beta_1+\left(1-p\right)\mu_0\left(X\right)'\beta_0+E[SU_1+\left(1-S\right)U_0\mid X,P=p]\notag\\
=&
p\mu_1\left(X\right)'\beta_1+\left(1-p\right)\mu_0\left(X\right)'\beta_0+E[U\mid P=p]\label{eq:decompo}
\end{align}
\endgroup
under Assumption \ref{assn2}.
Applying Theorem \ref{theorem:MTE_LIV} yields
$$
MTE\left(x,p\right)
=\mu_1\left(x\right)'\beta_1-\mu_0\left(x\right)'\beta_0
+\Delta_{U\mid P}\left(p\right)
$$
under Assumption \ref{assn1}.
\qed

\subsection{Derivation of Equation \eqref{eq:PRTE_expression}}\label{sec:eq:partial_linear_PRTE_rewritten}
Based on the relationship between $PRTE$ and $MTE$, we have 
$$
PRTE
=
\int\int_0^1MTE\left(x,p\right)\frac{F_{P\mid X}\left(p\mid x\right)-F_{P^\ast\mid X}\left(p\mid x\right)}{E[P^\ast]-E[P]}dpf_X\left(x\right)dx.
$$
By \eqref{eq:partial_linear_MTE}, we can write
\begin{align*}
\left(E[P^\ast]-E[P]\right)PRTE
=&
\left(\int\mu_1\left(x\right)\left(\int_0^1F_{P\mid X}\left(p\mid x\right)dp-\int_0^1F_{P^\ast\mid X}\left(p\mid x\right)dp\right)f_X\left(x\right)dx\right)'\beta_1
\\&-\left(\int\mu_0\left(x\right)\left(\int_0^1F_{P\mid X}\left(p\mid x\right)dp-\int_0^1F_{P^\ast\mid X}\left(p\mid x\right)dp\right)f_X\left(x\right)dx\right)'\beta_0
\\&+\left(\int\int_0^1\Delta_{U\mid P}\left(p\right)\left(F_{P\mid X}\left(p\mid x\right)-F_{P^\ast\mid X}\left(p\mid x\right)\right)dpf_X\left(x\right)dx\right).
\end{align*}
Using integration by parts, we can obtain the following two equations: 
\begingroup
\allowdisplaybreaks
\begin{align*}
&\int_0^1F_{P\mid X}\left(p\mid x\right)dp
=
1\cdot F_{P\mid X=x}\left(1\right)-0\cdot F_{P\mid X=x}\left(0\right)-\int_0^1pf_{P\mid X}\left(p\mid x\right)dp
=
-E[P\mid X=x],
\\
&\int_0^1F_{P^\ast\mid X}\left(p\mid x\right)dp
=
1\cdot F_{P^\ast\mid X=x}\left(1\right)-0\cdot F_{P^\ast\mid X=x}\left(0\right)-\int_0^1pf_{P^\ast\mid X}\left(p\mid x\right)dp
=
-E[P^\ast\mid X=x].
\end{align*}
\endgroup
Therefore, combining the above equations, we have
\begin{align*}
\left(E[P^\ast]-E[P]\right)PRTE
=&
\left(\int\mu_1\left(x\right)\left(E[P^\ast\mid X=x]-E[P\mid X=x]\right)f_X\left(x\right)dx\right)'\beta_1
\\&-\left(\int\mu_0\left(x\right)\left(E[P^\ast\mid X=x]-E[P\mid X=x]\right)f_X\left(x\right)dx\right)'\beta_0
\\&+\left(\int\int_0^1\Delta_{U\mid P}\left(p\right)\left(F_{P\mid X}\left(p\mid x\right)-F_{P^\ast\mid X}\left(p\mid x\right)\right)dpf_X\left(x\right)dx\right)\\
=&
E[\left(P^\ast\left(\mf_{S\mid Z}\left(Z\right),Z\right)-\mf_{S\mid Z}\left(Z\right)\right)\mu_1\left(X\right)]'\beta_1\\&-E[\left(P^\ast\left(\mf_{S\mid Z}\left(Z\right),Z\right)-\mf_{S\mid Z}\left(Z\right)\right)\mu_0\left(X\right)]'\beta_0
\\&+E\left[\Delta_{U\mid P}\left(\mf_{S\mid Z}\left(Z\right)\right)\frac{F_{P}\left(\mf_{S\mid Z}\left(Z\right)\right)-F_{P^\ast}\left(\mf_{S\mid Z}\left(Z\right)\right)}{f_{P}\left(\mf_{S\mid Z}\left(Z\right)\right)}\right].
\end{align*}
This yields \eqref{eq:PRTE_expression}.

\subsection{Derivation of Equation \eqref{eq:orthogonality}}\label{sec:eq:orthogonality}

Define $\Gamma$ as the set of values $\breve{\gamma}$ of $\gamma$ such that 
\begin{align*}
\breve{\xi}_1\left(x,y,p\right)=\mathrm{vec}
\left(
\left(\begin{array}{c}
\left(1-p\right)\left(\mu_0\left(x\right)-\breve\mf_{\mu_0\left(X\right)\mid P}\left(p\right)\right)\\
p\left(\mu_1\left(x\right)-\breve\mf_{\mu_1\left(X\right)\mid P}\left(p\right)\right)
\end{array}\right)
\left(\begin{array}{c}
\left(1-p\right)\left(\mu_0\left(x\right)-\breve\mf_{\mu_0\left(X\right)\mid P}\left(p\right)\right)\\
p\left(\mu_1\left(x\right)-\breve\mf_{\mu_1\left(X\right)\mid P}\left(p\right)\right)\\
y-\breve\mf_{Y\mid P}\left(p\right)
\end{array}\right)'
\right).
\end{align*}
We write $\tilde{\gamma}=\gamma+r\left(\breve{\gamma}-\gamma\right)$ so that $\tilde{\gamma}$ is a linear function of $r$. Note that, in the following derivations, we are going to frequently use the fact that $\tilde{\gamma}=\gamma$ when $r=0$. 

As in the proof of Lemma \ref{lemma:CEINR_Assn6}, we can obtain the three equalities:
\begingroup
\allowdisplaybreaks
\begin{eqnarray*}
&&
\int m_1\left(w;\theta,\tilde{\gamma}\right)F_W\left(dw\right)
\\
&&\quad=
\int \left({\xi}_1\left(x,y,\tilde{\mf}_{S\mid Z}\left(z\right)\right)-{\xi}_1\left(x,y,{\mf}_{S\mid Z}\left(z\right)\right)-{\zeta}\left(z\right)\left(\tilde{\mf}_{S\mid Z}\left(z\right)-{\mf}_{S\mid Z}\left(z\right)\right)\right)F_W\left(dw\right)
\\&&\qquad
+\int \left(\tilde{\xi}_1\left(x,y,\tilde{\mf}_{S\mid Z}\left(z\right)\right)-{\xi}_1\left(x,y,\tilde{\mf}_{S\mid Z}\left(z\right)\right)\right)F_W\left(dw\right)
\\&&\qquad
-\int \left(\tilde{\zeta}\left(z\right)-{\zeta}\left(z\right)\right)\left(\tilde{\mf}_{S\mid Z}\left(z\right)-{\mf}_{S\mid Z}\left(z\right)\right)F_W\left(dw\right),
\\&&
\int m_2\left(w;\theta,\tilde{\gamma}\right)F_W\left(dw\right)
\\&&
\quad=
\int
\mf_{\left(\mu_0\left(X\right)',\mu_1\left(X\right)',1\right)'\mid Z}\left(z\right)\left(P^\ast\left(\tilde{\mf}_{S\mid Z}\left(z\right),z\right)-P^\ast\left({\mf_{S\mid Z}}\left(z\right),z\right)-\tilde{\mf}_{S\mid Z}\left(z\right)+{\mf_{S\mid Z}}\left(z\right)\right)
F_W\left(dw\right)
\\&&\qquad
+\int
{\mf}_{\left(\mu_0\left(X\right)',\mu_1\left(X\right)',1\right)'\mid Z}\left(z\right)
\left(\partial P^\ast\left(\tilde{\mf}_{S\mid Z}\left(z\right),z\right)-1\right)
\left(\mf_{S\mid Z}\left(z\right)-\tilde{\mf}_{S\mid Z}\left(z\right)\right)F_W\left(dw\right),
\\
&&
\int m_3\left(W;\theta,\tilde{\gamma}\right)F_W\left(dw\right)
\\
&&\quad=
\int \left({\mf}_{U\mid P}\left(P^\ast\left(\tilde{\mf}_{S\mid Z}\left(z\right),z\right)\right)-{\mf_{U\mid P}}\left(P^\ast\left({\mf_{S\mid Z}}\left(z\right),z\right)\right)
-
{\Delta}_{U\mid P}\left(P^\ast\left({\mf}_{S\mid Z}\left(z\right),z\right)\right)\partial P^\ast\left({\mf}_{S\mid Z}\left(z\right),z\right)\right)\\&&\qquad\qquad\times
\left(\tilde{\mf}_{S\mid Z}\left(z\right)-{\mf}_{S\mid Z}\left(z\right)\right)F_W\left(dw\right)
\\&&\qquad
+\int\frac{\tilde{f}_{P^\ast}\left(\tilde{\mf}_{S\mid Z}\left(z\right)\right)}{\tilde{f}_{P}\left(\tilde{\mf}_{S\mid Z}\left(z\right)\right)}
\left({\mf_{U\mid P}}\left({\mf_{S\mid Z}}\left(z\right)\right)-{\mf}_{U\mid P}\left(\tilde{\mf}_{S\mid Z}\left(z\right)\right)
-{\Delta}_{U\mid P}\left({\mf}_{S\mid Z}\left(z\right)\right)
\left({\mf_{S\mid Z}}\left(z\right)-\tilde{\mf}_{S\mid Z}\left(z\right)\right)\right)F_W\left(dw\right)
\\&&\qquad
-
\int \left(
\tilde{\Delta}_{U\mid P}\left(P^\ast\left(\tilde{\mf}_{S\mid Z}\left(z\right),z\right)\right)\partial P^\ast\left(\tilde{\mf}_{S\mid Z}\left(z\right),z\right) \right.
\\
&&\qquad\qquad \left.
-{\Delta}_{U\mid P}\left(P^\ast\left({\mf}_{S\mid Z}\left(z\right),z\right)\right)\partial P^\ast\left({\mf}_{S\mid Z}\left(z\right),z\right)
\right)
\left(\tilde{\mf}_{S\mid Z}\left(z\right)-{\mf}_{S\mid Z}\left(z\right)\right)F_W\left(dw\right)
\\&&\qquad
-\int\frac{\tilde{f}_{P^\ast}\left(\tilde{\mf}_{S\mid Z}\left(z\right)\right)}{\tilde{f}_{P}\left(\tilde{\mf}_{S\mid Z}\left(z\right)\right)}
\left(\tilde{\Delta}_{U\mid P}\left(\tilde{\mf}_{S\mid Z}\left(z\right)\right)-{\Delta}_{U\mid P}\left({\mf}_{S\mid Z}\left(z\right)\right)\right)
\left({\mf_{S\mid Z}}\left(z\right)-\tilde{\mf}_{S\mid Z}\left(z\right)\right)F_W\left(dw\right)
\\&&\qquad
+
\int \left(\tilde{\mf}_{U\mid P}\left(P^\ast\left(\tilde{\mf}_{S\mid Z}\left(z\right),z\right)\right)-{\mf}_{U\mid P}\left(P^\ast\left(\tilde{\mf}_{S\mid Z}\left(z\right),z\right)\right)\right)F_W\left(dw\right)
\\&&\qquad
-\int\frac{\tilde{f}_{P^\ast}\left(\tilde{\mf}_{S\mid Z}\left(z\right)\right)}{\tilde{f}_{P}\left(\tilde{\mf}_{S\mid Z}\left(z\right)\right)}
\left(\tilde{\mf}_{U\mid P}\left(\tilde{\mf}_{S\mid Z}\left(z\right)\right)-{\mf}_{U\mid P}\left(\tilde{\mf}_{S\mid Z}\left(z\right)\right)\right)
F_W\left(dw\right).
\end{eqnarray*}
\endgroup

First, we differentiate $\int m_1\left(w;\theta,\tilde{\gamma}\right)F_W\left(dw\right)$ with respect to $r$:   
\begingroup
\allowdisplaybreaks
\begin{align*}
\frac{\partial}{\partial r}\int m_1\left(w;\theta,\tilde{\gamma}\right)F_W\left(dw\right)
=&
\int \left(\frac{\partial}{\partial p}{\xi}_1\left(x,y,p\right)|_{p=\tilde{\mf}_{S\mid Z}\left(z\right)}\left(\breve{\mf}_{S\mid Z}\left(z\right)-{\mf}_{S\mid Z}\left(z\right)\right)-{\zeta}\left(z\right)\left(\breve{\mf}_{S\mid Z}\left(z\right)-{\mf}_{S\mid Z}\left(z\right)\right)\right)F_W\left(dw\right).
\\&
+\int \left(\frac{\partial}{\partial r}\left(\tilde{\xi}_1\left(x,y,p\right)-{\xi}_1\left(x,y,p\right)\right)\right)|_{p=\tilde{\mf}_{S\mid Z}\left(z\right)}
F_W\left(dw\right)
\\&
+\int \left(\frac{\partial}{\partial p}\left(\tilde{\xi}_1\left(x,y,p\right)-{\xi}_1\left(x,y,p\right)\right)\right)|_{p=\tilde{\mf}_{S\mid Z}\left(z\right)}
\left(\breve{\mf}_{S\mid Z}\left(z\right)-{\mf}_{S\mid Z}\left(z\right)\right)
F_W\left(dw\right)
\\&
+\int \frac{\partial}{\partial p}\left(\tilde{\xi}_1\left(x,y,p\right)-{\xi}_1\left(x,y,p\right)\right)|_{p=\tilde{\mf}_{S\mid Z}\left(z\right)}\left(\breve{\mf}_{S\mid Z}\left(z\right)-{\mf}_{S\mid Z}\left(z\right)\right)F_W\left(dw\right)
\\&-\int \left(\breve{\zeta}\left(z\right)-{\zeta}\left(z\right)\right)\left(\tilde{\mf}_{S\mid Z}\left(z\right)-{\mf}_{S\mid Z}\left(z\right)\right)F_W\left(dw\right)
\\&-\int \left(\tilde{\zeta}\left(z\right)-{\zeta}\left(z\right)\right)\left(\breve{\mf}_{S\mid Z}\left(z\right)-{\mf}_{S\mid Z}\left(z\right)\right)F_W\left(dw\right).
\end{align*}
\endgroup
Since 
\begingroup
\allowdisplaybreaks
\begin{eqnarray*}
&&
\iint \frac{\partial}{\partial r}\tilde{\xi}_1\left(x,y,p\right)|_{r=0}F_{\left(Y,X\right)\mid P=p}\left(dy,dx\right)
\\
&=&
-\mathrm{vec}
\left(
\left(\begin{array}{c}
\left(1-p\right)\left(\breve\mf_{\mu_0\left(X\right)\mid P}\left(p\right)-\mf_{\mu_0\left(X\right)\mid P}\left(p\right)\right)\\
p\left(\breve\mf_{\mu_1\left(X\right)\mid P}\left(p\right)-\mf_{\mu_1\left(X\right)\mid P}\left(p\right)\right)
\end{array}\right)
\iint\left(\begin{array}{c}
\left(1-p\right)\left(\mu_0\left(x\right)-\mf_{\mu_0\left(X\right)\mid P}\left(p\right)\right)\\
p\left(\mu_1\left(x\right)-\mf_{\mu_1\left(X\right)\mid P}\left(p\right)\right)\\
y-\mf_{Y\mid P}\left(p\right)
\end{array}\right)'F_{\left(Y,X\right)\mid P=p}\left(dy,dx\right)
\right)
\\&&-
\mathrm{vec}
\left(\int
\left(\begin{array}{c}
\left(1-p\right)\left(\mu_0\left(x\right)-\mf_{\mu_0\left(X\right)\mid P}\left(p\right)\right)\\
p\left(\mu_1\left(x\right)-\mf_{\mu_1\left(X\right)\mid P}\left(p\right)\right)
\end{array}\right)
F_{X\mid P=p}\left(dx\right)
\left(\begin{array}{c}
\left(1-p\right)\left(\breve\mf_{\mu_0\left(X\right)\mid P}\left(p\right)-\mf_{\mu_0\left(X\right)\mid P}\left(p\right)\right)\\
p\left(\breve\mf_{\mu_1\left(X\right)\mid P}\left(p\right)-\mf_{\mu_1\left(X\right)\mid P}\left(p\right)\right)\\
\left(\breve\mf_{Y\mid P}\left(p\right)-\mf_{Y\mid P}\left(p\right)\right)
\end{array}\right)'
\right)\\
&=&
0,
\end{eqnarray*}
\endgroup
the Gateaux derivative is 
\begingroup
\allowdisplaybreaks
\begin{align*}
&
\frac{\partial}{\partial r}\int m_1\left(w;\theta,\gamma+r\left(\breve{\gamma}-\gamma\right)\right)F_W\left(dw\right)|_{r=0}
=
0.
\end{align*}
\endgroup

Second, we differentiate $\int m_2\left(w;\theta,\tilde{\gamma}\right)F_W\left(dw\right)$ with respect to $r$: 
\begingroup
\allowdisplaybreaks
\begin{align*}
&
\frac{\partial}{\partial r}\int m_2\left(w;\theta,\tilde{\gamma}\right)F_W\left(dw\right)
\\
=&
\int
\mf_{\left(\mu_0\left(X\right)',\mu_1\left(X\right)',1\right)'\mid Z}\left(z\right)\left(\partial P^\ast\left(\tilde{\mf}_{S\mid Z}\left(z\right),z\right)\left(\breve\mf_{S\mid Z}\left(z\right)-{\mf}_{S\mid Z}\left(z\right)\right)-\left(\breve\mf_{S\mid Z}\left(z\right)-{\mf}_{S\mid Z}\left(z\right)\right)\right)
F_W\left(dw\right)\\
&-
\int
{\mf}_{\left(\mu_0\left(X\right)',\mu_1\left(X\right)',1\right)'\mid Z}\left(z\right)
\left(\partial P^\ast\left(\tilde{\mf}_{S\mid Z}\left(z\right),z\right)-1\right)
\left(\breve\mf_{S\mid Z}\left(z\right)-{\mf}_{S\mid Z}\left(z\right)\right)F_W\left(dw\right)\\
&+
\int
{\mf}_{\left(\mu_0\left(X\right)',\mu_1\left(X\right)',1\right)'\mid Z}\left(z\right)
\frac{\partial^2}{\partial p^2} P^\ast\left(p,z\right)|_{p=\tilde{\mf}_{S\mid Z}\left(z\right)}
\left(\mf_{S\mid Z}\left(z\right)-\tilde{\mf}_{S\mid Z}\left(z\right)\right)F_W\left(dw\right)
\\
=&
\int
{\mf}_{\left(\mu_0\left(X\right)',\mu_1\left(X\right)',1\right)'\mid Z}\left(z\right)
\frac{\partial^2}{\partial p^2} P^\ast\left(p,z\right)|_{p=\tilde{\mf}_{S\mid Z}\left(z\right)}
\left(\mf_{S\mid Z}\left(z\right)-\tilde{\mf}_{S\mid Z}\left(z\right)\right)F_W\left(dw\right).
\end{align*}
\endgroup
Thus, the Gateaux derivative is 
\begin{align*}
\frac{\partial}{\partial r}\int m_2\left(w;\theta,\tilde{\gamma}\right)F_W\left(dw\right)|_{r=0}
=0.
\end{align*}
 
Third, we differentiate $\int m_3\left(w;\theta,\tilde{\gamma}\right)F_W\left(dw\right)$ with respect to $r$: 
\begingroup
\allowdisplaybreaks
\begin{align*}
&
\frac{\partial}{\partial r}\int m_3\left(W;\theta,\tilde{\gamma}\right)F_W\left(dw\right)
\\=&
\int 
\left({\Delta}_{U\mid P}\left(P^\ast\left(\tilde{\mf}_{S\mid Z}\left(z\right),z\right)\right)\partial P^\ast\left(\tilde{\mf}_{S\mid Z}\left(z\right),z\right) \right.
\\
&\qquad\left. -{\Delta}_{U\mid P}\left(P^\ast\left({\mf}_{S\mid Z}\left(z\right),z\right)\right)\partial P^\ast\left({\mf}_{S\mid Z}\left(z\right),z\right)\right)
\left(\breve{\mf}_{S\mid Z}\left(z\right)-{\mf}_{S\mid Z}\left(z\right)\right)F_W\left(dw\right)
\\&+\int\left(\frac{\partial}{\partial r}\frac{\tilde{f}_{P^\ast}\left(\tilde{\mf}_{S\mid Z}\left(z\right)\right)}{\tilde{f}_{P}\left(\tilde{\mf}_{S\mid Z}\left(z\right)\right)}\right)
\left({\mf_{U\mid P}}\left({\mf_{S\mid Z}}\left(z\right)\right)-{\mf}_{U\mid P}\left(\tilde{\mf}_{S\mid Z}\left(z\right)\right) \right.
\\
&\qquad\left.
-{\Delta}_{U\mid P}\left({\mf}_{S\mid Z}\left(z\right)\right)
\left({\mf_{S\mid Z}}\left(z\right)-\tilde{\mf}_{S\mid Z}\left(z\right)\right)\right)F_W\left(dw\right)
\\&-\int\frac{\tilde{f}_{P^\ast}\left(\tilde{\mf}_{S\mid Z}\left(z\right)\right)}{\tilde{f}_{P}\left(\tilde{\mf}_{S\mid Z}\left(z\right)\right)}
\left({\Delta}_{U\mid P}\left(\tilde{\mf}_{S\mid Z}\left(z\right)\right)-{\Delta}_{U\mid P}\left({\mf}_{S\mid Z}\left(z\right)\right)\right)
\left(\breve\mf_{S\mid Z}\left(z\right)-{\mf}_{S\mid Z}\left(z\right)\right)F_W\left(dw\right)
\\&-
\int \left(\frac{\partial}{\partial r}
\tilde{\Delta}_{U\mid P}\left(P^\ast\left(\tilde{\mf}_{S\mid Z}\left(z\right),z\right)\right)\right)\partial P^\ast\left(\tilde{\mf}_{S\mid Z}\left(z\right),z\right)
\left(\tilde{\mf}_{S\mid Z}\left(z\right)-{\mf}_{S\mid Z}\left(z\right)\right)F_W\left(dw\right)
\\&-
\int \left(
\tilde{\Delta}_{U\mid P}\left(P^\ast\left(\tilde{\mf}_{S\mid Z}\left(z\right),z\right)\right)\partial P^\ast\left(\tilde{\mf}_{S\mid Z}\left(z\right),z\right) \right.
\\
&\left.
-{\Delta}_{U\mid P}\left(P^\ast\left({\mf}_{S\mid Z}\left(z\right),z\right)\right)\partial P^\ast\left({\mf}_{S\mid Z}\left(z\right),z\right)
\right)
\left(\breve{\mf}_{S\mid Z}\left(z\right)-{\mf}_{S\mid Z}\left(z\right)\right)
F_W\left(dw\right)
\\&-\int\left(\frac{\partial}{\partial r}\frac{\tilde{f}_{P^\ast}\left(\tilde{\mf}_{S\mid Z}\left(z\right)\right)}{\tilde{f}_{P}\left(\tilde{\mf}_{S\mid Z}\left(z\right)\right)}\right)
\left(\tilde{\Delta}_{U\mid P}\left(\tilde{\mf}_{S\mid Z}\left(z\right)\right)-{\Delta}_{U\mid P}\left({\mf}_{S\mid Z}\left(z\right)\right)\right)
\left({\mf_{S\mid Z}}\left(z\right)-\tilde{\mf}_{S\mid Z}\left(z\right)\right)F_W\left(dw\right)
\\&-\int\frac{\tilde{f}_{P^\ast}\left(\tilde{\mf}_{S\mid Z}\left(z\right)\right)}{\tilde{f}_{P}\left(\tilde{\mf}_{S\mid Z}\left(z\right)\right)}
\left(\frac{\partial}{\partial r}\tilde{\Delta}_{U\mid P}\left(\tilde{\mf}_{S\mid Z}\left(z\right)\right)\right)
\left({\mf_{S\mid Z}}\left(z\right)-\tilde{\mf}_{S\mid Z}\left(z\right)\right)F_W\left(dw\right)
\\&+\int\frac{\tilde{f}_{P^\ast}\left(\tilde{\mf}_{S\mid Z}\left(z\right)\right)}{\tilde{f}_{P}\left(\tilde{\mf}_{S\mid Z}\left(z\right)\right)}
\left(\tilde{\Delta}_{U\mid P}\left(\tilde{\mf}_{S\mid Z}\left(z\right)\right)-{\Delta}_{U\mid P}\left({\mf}_{S\mid Z}\left(z\right)\right)\right)
\left(\frac{\partial}{\partial r}\tilde{\mf}_{S\mid Z}\left(z\right)\right)F_W\left(dw\right)
\\&+
\int \left(\frac{\partial}{\partial r}\tilde{\mf}_{U\mid P}\left(P^\ast\left(\tilde{\mf}_{S\mid Z}\left(z\right),z\right)\right)-\frac{\partial}{\partial r}{\mf}_{U\mid P}\left(P^\ast\left(\tilde{\mf}_{S\mid Z}\left(z\right),z\right)\right)\right)F_W\left(dw\right)
\\&
-\int\left(\frac{\partial}{\partial r}\frac{\tilde{f}_{P^\ast}\left(\tilde{\mf}_{S\mid Z}\left(z\right)\right)}{\tilde{f}_{P}\left(\tilde{\mf}_{S\mid Z}\left(z\right)\right)}\right)
\left(\tilde{\mf}_{U\mid P}\left(\tilde{\mf}_{S\mid Z}\left(z\right)\right)-{\mf}_{U\mid P}\left(\tilde{\mf}_{S\mid Z}\left(z\right)\right)\right)
F_W\left(dw\right)
\\&
-\int\frac{\tilde{f}_{P^\ast}\left(\tilde{\mf}_{S\mid Z}\left(z\right)\right)}{\tilde{f}_{P}\left(\tilde{\mf}_{S\mid Z}\left(z\right)\right)}
\left(\frac{\partial}{\partial r}\tilde{\mf}_{U\mid P}\left(\tilde{\mf}_{S\mid Z}\left(z\right)\right)-\frac{\partial}{\partial r}{\mf}_{U\mid P}\left(\tilde{\mf}_{S\mid Z}\left(z\right)\right)\right)
F_W\left(dw\right).
\end{align*}
\endgroup
Thus, the Gateaux derivative is 
\begingroup
\allowdisplaybreaks
\begin{align*}
\frac{\partial}{\partial r}\int m_3\left(W;\theta,\tilde{\gamma}\right)F_W\left(dw\right)|_{r=0}
=&
\int \left(\frac{\partial}{\partial r}\tilde{\mf}_{U\mid P}\left(P^\ast\left(p,z\right)\right)\right)|_{p=\tilde{\mf}_{S\mid Z}\left(z\right)}F_W\left(dw\right)
\\&
-\int\frac{{f}_{P^\ast}\left({\mf}_{S\mid Z}\left(z\right)\right)}{{f}_{P}\left({\mf}_{S\mid Z}\left(z\right)\right)}
\left(\frac{\partial}{\partial r}\tilde{\mf}_{U\mid P}\left(p\right)\right)|_{p=\tilde{\mf}_{S\mid Z}\left(z\right)}
F_W\left(dw\right)
\\=&
\int \left(\breve{\mf}_{U\mid P}\left(P^\ast\left({\mf}_{S\mid Z}\left(z\right),z\right)\right)-{\mf}_{U\mid P}\left(P^\ast\left({\mf}_{S\mid Z}\left(z\right),z\right)\right)\right)F_W\left(dw\right)
\\&
-\int\frac{{f}_{P^\ast}\left({\mf}_{S\mid Z}\left(z\right)\right)}{{f}_{P}\left({\mf}_{S\mid Z}\left(z\right)\right)}
\left(\breve{\mf}_{U\mid P}\left({\mf}_{S\mid Z}\left(z\right)\right)-{\mf}_{U\mid P}\left({\mf}_{S\mid Z}\left(z\right)\right)\right)
F_W\left(dw\right)
\\=&
\int\breve{\mf}_{U\mid P}\left(p\right)-{\mf}_{U\mid P}\left(p\right){f}_{P^\ast}\left(p\right)dp
-\int\frac{{f}_{P^\ast}\left(p\right)}{{f}_{P}\left(p\right)}
\left(\breve{\mf}_{U\mid P}\left(p\right)-{\mf}_{U\mid P}\left(p\right)\right)
F_P\left(dp\right)
\\=&
0.
\end{align*}
These show \eqref{eq:orthogonality}.
\endgroup

\subsection{Decomposition of the Moment Functions}\label{sec:decomposition_of_moment_functions}
We make a decomposition of the moment function $m$ defined in Section \ref{sec:orthogonal_score} for convenience of stating and proving some auxiliary lemmas and to introduce notations to be used in their proofs.
Specifically, we write
\begingroup
\allowdisplaybreaks
\begin{align*}
m_1\left(W;\tilde{\theta},\tilde{\gamma}\right)=&m_{1,1}\left(W;\tilde{\gamma}\right)-\tilde{\theta}_1\\
m_2\left(W;\tilde{\theta},\tilde{\gamma}\right)=&m_{2,1}\left(W;\tilde{\gamma}\right)-\tilde{\theta}_2\\
m_3\left(W;\tilde{\theta},\tilde{\gamma}\right)=&m_{3,1}\left(W;\tilde{\gamma}\right)-\tilde{\theta}_3-m_{3,2}\left(W;\tilde{\gamma}\right)'\boldsymbol{d}\left(\tilde{\theta}_1\right).
\end{align*}
\endgroup
where 
\begingroup
\allowdisplaybreaks
\begin{align*}
m_{1,1}\left(W;\tilde{\gamma}\right)
=& \tilde{\xi}_1\left(X,Y,\tilde{\mf}_{S\mid Z}\left(Z\right)\right)
+
\tilde{\zeta}\left(Z\right)\left(S-\tilde{\mf}_{S\mid Z}\left(Z\right)\right)
\\
m_{2,1}\left(W;\tilde{\gamma}\right)
=& \left(\mu_0\left(X\right)',\mu_1\left(X\right)',1\right)'\left(P^\ast\left(\tilde{\mf}_{S\mid Z}\left(Z\right),Z\right)-\tilde{\mf}_{S\mid Z}\left(Z\right)\right)
\\&\quad+ \left(\mu_0\left(X\right)',\mu_1\left(X\right)',1\right)'\left(\partial P^\ast\left(\tilde{\mf}_{S\mid Z}\left(Z\right),Z\right)-1\right)
\left(S-\tilde{\mf}_{S\mid Z}\left(Z\right)\right)
\\
m_{3,1}\left(W;\tilde{\gamma}\right)
=&
\tilde{\mf}_{U\mid P}\left(P^\ast\left(\tilde{\mf}_{S\mid Z}\left(Z\right),Z\right)\right)-Y
+
\frac{\tilde{f}_{P^\ast}\left(\tilde{\mf}_{S\mid Z}\left(Z\right)\right)}{\tilde{f}_{P}\left(\tilde{\mf}_{S\mid Z}\left(Z\right)\right)}\left(Y-\tilde{\mf}_{U\mid P}\left(\tilde{\mf}_{S\mid Z}\left(Z\right)\right)\right)
\\&+
\tilde{\Delta}_{U\mid P}\left(P^\ast\left(\tilde{\mf}_{S\mid Z}\left(Z\right),Z\right)\right)
\partial P^\ast\left(\tilde{\mf}_{S\mid Z}\left(Z\right),Z\right)\left(S-\tilde{\mf}_{S\mid Z}\left(Z\right)\right)
\\&-
\frac{\tilde{f}_{P^\ast}\left(\tilde{\mf}_{S\mid Z}\left(Z\right)\right)}{\tilde{f}_{P}\left(\tilde{\mf}_{S\mid Z}\left(Z\right)\right)}\tilde{\Delta}_{U\mid P}\left(\tilde{\mf}_{S\mid Z}\left(Z\right)\right)\left(S-\tilde{\mf}_{S\mid Z}\left(Z\right)\right)
\\
m_{3,2}\left(W;\tilde{\gamma}\right)
=&
\left(\frac{\tilde{f}_{P^\ast}\left(\tilde{\mf}_{S\mid Z}\left(Z\right)\right)}{\tilde{f}_{P}\left(\tilde{\mf}_{S\mid Z}\left(Z\right)\right)}-1\right)
\left(\left(1-S\right)\mu_0\left(X\right)',S\mu_1\left(X\right)'\right).
\end{align*}
\endgroup

\section{Additional Details about the Monte Carlo Simulations}\label{sec:details_simulations}

\subsection{Additional Details about the Simulation Setting}\label{sec:details_simulation_setting}
In the simulation setting of \citet[][Appendix D]{carneiro/lokshin/umapathi:2017} presented in Section \ref{sec:simulations}, we have 
\begin{align*}
V =& 1 - \Phi\left(\varepsilon_1\right)
\qquad\text{and}\\
P =& 1 - \Phi\left( \underbrace{- 0.200 - 0.300 Z_1 - 0.100 Z_2}_{-\mu_S\left(Z_1,Z_2\right)} \right),
\end{align*}
where $\Phi$ denotes the cumulative distribution function of standard normal random variables.
The true distribution of the propensity score $P$ is given by
\begin{align*}
F_{P|X_1X_2}\left(p|x_1,x_2\right) = \Phi\left( - \frac{\sqrt{10}}{3} \Phi^{-1}\left( 1-p\right) \right).
\end{align*}
The true distributions of the propensity scores $P^\ast$ under policy changes $P^\ast = P + a\left(1-P\right)$ take the forms of
\begin{align*}
F_{P^\ast|X_1X_2}\left(p|x_1,x_2\right) = \Phi\left( - \frac{\sqrt{10}}{3} \Phi^{-1}\left( 1-\frac{p-a}{1-a} \right) \right)
\end{align*}
for $p \in [a,1]$.
The true distributions of the propensity scores $P^\ast$ under policy changes $Z_1^\ast = Z_1 + a$ take the forms of
\begin{align*}
F_{P^\ast|X_1X_2}\left(p|x_1,x_2\right) = \Phi\left( - \frac{\sqrt{10}}{3} \Phi^{-1}\left( 1-p\right) - \frac{a}{\sqrt{10}}\right).
\end{align*}
The marginal treatment effect takes the form of
\begin{align*}
MTE\left(x_1,x_2,p\right) = 0.220 + 0.300 x_1 + 0.300 x_2 + 0.062 \Phi^{-1}\left( 1-p \right).
\end{align*}
Therefore, the PRTE under policy changes $P^\ast = P + a\left(1-P\right)$ takes the form of
\begin{align*}
PRTE =& 0.220
+
0.062 \cdot \frac{\int_0^1 \Phi^{-1}\left(1-p\right) \cdot \left[ \Phi\left( - \frac{\sqrt{10}}{3} \Phi^{-1}\left( 1-p \right) \right) - \Phi\left( - \frac{\sqrt{10}}{3} \Phi^{-1}\left( 1-\frac{p-a}{1-a} \right) \right) \right] dp}{\int_0^1 \left[ \Phi\left( - \frac{\sqrt{10}}{3} \Phi^{-1}\left( 1-p \right) \right) - \Phi\left( - \frac{\sqrt{10}}{3} \Phi^{-1}\left( 1-\frac{p-a}{1-a} \right) \right) \right] dp}.
\end{align*}
The PRTE under policy changes $Z_1^\ast = Z_1 + a$ takes the form of
\begin{align*}
PRTE =& 0.220
+ 
0.062 \cdot \frac{\int_0^1 \Phi^{-1}\left(1-p\right) \cdot \left[ \Phi\left( - \frac{\sqrt{10}}{3} \Phi^{-1}\left( 1-p \right) \right) - \Phi\left( - \frac{\sqrt{10}}{3} \Phi^{-1}\left( 1-p\right) - \frac{a}{\sqrt{10}}\right) \right] dp}{\int_0^1 \left[ \Phi\left( - \frac{\sqrt{10}}{3} \Phi^{-1}\left( 1-p \right) \right) - \Phi\left( - \frac{\sqrt{10}}{3} \Phi^{-1}\left( 1-p\right) - \frac{a}{\sqrt{10}}\right) \right] dp}.
\end{align*}
We compute the true PRTEs under the setting of \citet[][Appendix D]{carneiro/lokshin/umapathi:2017} based on these formulas.

\subsection{Additional Details about the Estimation Procedure for Simulations}\label{sec:details_simulation_estimation}
\subsubsection{Case: Nonparametric Estimation under Policy Changes $P^\ast = P + a\left(1-P\right)$}\label{sec:details_simulation_estimation_p}
We present the concrete nonparametric estimation procedure used in the Monte Carlo simulations for policy changes of the form $P^\ast = P + a\left(1-P\right)$.

For all $i \in I_\ell^c$, the propensity score $P_i$ is estimated by
\begin{align*}
\hat P_i = \hat{\mf}_{S\mid Z}\left(Z_i\right) = \frac{\sum_{j \in I_\ell^c \backslash \{i\}} S_j K_{h_1}\left(Z_{j1}-Z_{i1}\right) K_{h_2}\left(Z_{j2}-Z_{i2}\right)}{\sum_{j \in I_\ell^c \backslash \{i\}} K_{h_1}\left(Z_{j1}-Z_{i1}\right) K_{h_2}\left(Z_{j2}-Z_{i2}\right)}
\end{align*}
where $K_{h} = K\left( \ \cdot \ / h \right) / h$ for the Epanechnikov kernel function $K$, $h_1= c_1 \hat\sigma_{Z_1} n^{-1/6}$ and $h_1= c_2 \hat\sigma_{Z_2} n^{-1/6}$ with $c_1=c_2=2$.
For all $i \in I_\ell$, the densities $f_{P}\left( {\mf}_{S\mid Z}\left(Z_i\right) \right)$ and $f_{P^\ast}\left( {\mf}_{S\mid Z}\left(Z_i\right) \right)$ are estimated by
\begingroup
\allowdisplaybreaks
\begin{align*}
\hat f_{P}\left( \hat{\mf}_{S\mid Z}\left(Z_i\right) \right) &= \frac{1}{|I_\ell^c|} \sum_{j \in I_\ell^c} K_{h_3}\left(\hat P_j - \hat{\mf}_{S\mid Z}\left(Z_i\right)\right)
\qquad\text{and}\\
\hat f_{P^\ast}\left( \hat{\mf}_{S\mid Z}\left(Z_i\right) \right) &= \frac{1}{|I_\ell^c|} \sum_{j \in I_\ell^c} K_{h_3}\left(\hat P_j + a\left(1 - \hat P_j\right) - \hat{\mf}_{S\mid Z}\left(Z_i\right)\right)
\end{align*}
\endgroup
where $h_3 = c_3 \hat\sigma_{\hat P} n^{-1/5}$ with $c_3 = 1.06$.

For all $i \in I_\ell$, the conditional expectations of $\left(X_1,X_2,Y\right)$ given $P$ evaluated at data point $i$ are  estimated by
\begingroup
\allowdisplaybreaks
\begin{align*}
\hat \mf_{X_1\mid P}\left( \hat{\mf}_{S\mid Z}\left(Z_i\right) \right)
=&
\frac{ \sum_{j \in I_\ell^c} X_{j1} K_{h_3}\left(\hat P_j - \hat{\mf}_{S\mid Z}\left(Z_i\right) \right) }{ \sum_{j \in I_\ell^c} K_{h_3}\left(\hat P_j - \hat{\mf}_{S\mid Z}\left(Z_i\right) \right) }
\\
\hat \mf_{X_2\mid P}\left( \hat{\mf}_{S\mid Z}\left(Z_i\right) \right)
=&
\frac{ \sum_{j \in I_\ell^c} X_{j2} K_{h_3}\left(\hat P_j - \hat{\mf}_{S\mid Z}\left(Z_i\right) \right) }{ \sum_{j \in I_\ell^c} K_{h_3}\left(\hat P_j - \hat{\mf}_{S\mid Z}\left(Z_i\right) \right) }
\\
\hat \mf_{Y\mid P}\left( \hat{\mf}_{S\mid Z}\left(Z_i\right) \right)
=&
\frac{ \sum_{j \in I_\ell^c} Y_j K_{h_3}\left(\hat P_j - \hat{\mf}_{S\mid Z}\left(Z_i\right) \right) }{ \sum_{j \in I_\ell^c} K_{h_3}\left(\hat P_j - \hat{\mf}_{S\mid Z}\left(Z_i\right) \right) }.
\end{align*}
\endgroup

The infinite-dimensional parameter $\xi_1$ evaluated at data point $i$ is estimated by
\begin{align*}
&\hat\xi_1\left(X_i,Y_i,\hat{\mf}_{S\mid Z}\left(Z_i\right)\right)=
\\
&\mathrm{vec}
\left[
\left(\begin{array}{c}
\left(1-\hat{\mf}_{S\mid Z}\left(Z_i\right)\right)\left(X_{i1}-\hat\mf_{X_1\mid P}\left(\hat{\mf}_{S\mid Z}\left(Z_i\right)\right)\right)\\
\left(1-\hat{\mf}_{S\mid Z}\left(Z_i\right)\right)\left(X_{i2}-\hat\mf_{X_2\mid P}\left(\hat{\mf}_{S\mid Z}\left(Z_i\right)\right)\right)\\
\hat{\mf}_{S\mid Z}\left(Z_i\right)\left(X_{i1}-\hat\mf_{X_1\mid P}\left(\hat{\mf}_{S\mid Z}\left(Z_i\right)\right)\right)\\
\hat{\mf}_{S\mid Z}\left(Z_i\right)\left(X_{i2}-\hat\mf_{X_2\mid P}\left(\hat{\mf}_{S\mid Z}\left(Z_i\right)\right)\right)
\end{array}\right)
\left(\begin{array}{c}
\left(1-\hat{\mf}_{S\mid Z}\left(Z_i\right)\right)\left(X_{i1}-\hat\mf_{X_1\mid P}\left(\hat{\mf}_{S\mid Z}\left(Z_i\right)\right)\right)\\
\left(1-\hat{\mf}_{S\mid Z}\left(Z_i\right)\right)\left(X_{i2}-\hat\mf_{X_2\mid P}\left(\hat{\mf}_{S\mid Z}\left(Z_i\right)\right)\right)\\
\hat{\mf}_{S\mid Z}\left(Z_i\right)\left(X_{i1}-\hat\mf_{X_1\mid P}\left(\hat{\mf}_{S\mid Z}\left(Z_i\right)\right)\right)\\
\hat{\mf}_{S\mid Z}\left(Z_i\right)\left(X_{i2}-\hat\mf_{X_2\mid P}\left(\hat{\mf}_{S\mid Z}\left(Z_i\right)\right)\right)\\
Y_{i}-\hat\mf_{Y\mid P}\left(\hat{\mf}_{S\mid Z}\left(Z_i\right)\right)
\end{array}\right)'\right].
\end{align*}
Its partial derivative $\xi_2 = \partial \xi_1 / \partial p$ is in turn obtained by the approximate local derivative estimator through numerical approximation as the difference quotient
\begin{align*}
\hat\xi_2\left(X_i,Y_i,\hat{\mf}_{S\mid Z}\left(Z_i\right)\right)
=
\frac{\hat\xi_1\left(X_i,Y_i,\hat{\mf}_{S\mid Z}\left(Z_i\right) + \delta\right) - \hat\xi_1\left(X_i,Y_i,\hat{\mf}_{S\mid Z}\left(Z_i\right) - \delta\right)}{2\delta}
\end{align*}
where $\delta = 0.01$.
For all $i \in I_\ell$, the projection $\xi_2$ on $Z_i$ is estimated by
\begin{align*}
\hat\zeta\left(Z_i\right)
=
\frac{\sum_{j \in I_\ell^c} \hat\xi_2\left(X_j,Y_j,\hat{\mf}_{S\mid Z}\left(Z_j\right)\right) K_{h_1}\left(Z_{j1}-Z_{i1}\right) K_{h_2}\left(Z_{j2}-Z_{i2}\right)}{\sum_{j \in I_\ell^c} K_{h_1}\left(Z_{j1}-Z_{i1}\right) K_{h_2}\left(Z_{j2}-Z_{i2}\right)}.
\end{align*}

The parameter $\theta_1$ is estimated by
\begin{align*}
\hat\theta_1 = \frac{1}{L}\sum_{\ell = 1}^L
\frac{1}{|I_\ell|} \sum_{i \in I_\ell} \left\{
\hat{\xi}_1\left(X_i,Y_i,\hat{\mf}_{S\mid Z}\left(Z_i\right)\right)
+
\hat\zeta\left(Z_i\right)
\left(S_i-\hat{\mf}_{S\mid Z}\left(Z_i\right)\right)
\right\}.
\end{align*}
With this estimate $\hat\theta_1 = \left(\hat\theta_{1,1},...,\hat\theta_{1,20}\right)'$, the partial linear coefficients $\left(\beta_0',\beta_1'\right)'$ for the sub-sample $I_\ell$ are estimated by
\begin{align*}
\left[\begin{array}{c}
\hat\beta_{0} \\
\hat\beta_{1}
\end{array}\right]
=
\left[\begin{array}{c}
\hat\beta_{01} \\
\hat\beta_{02} \\
\hat\beta_{11} \\
\hat\beta_{12}
\end{array}\right]
=
\boldsymbol{d}\left(\hat\theta_1\right)
=
\left[\begin{array}{cccc}
\hat\theta_{1,1} & \hat\theta_{1,5} & \hat\theta_{1,9} & \hat\theta_{1,13} \\
\hat\theta_{1,2} & \hat\theta_{1,6} & \hat\theta_{1,10} & \hat\theta_{1,14} \\
\hat\theta_{1,3} & \hat\theta_{1,7} & \hat\theta_{1,11} & \hat\theta_{1,15} \\
\hat\theta_{1,4} & \hat\theta_{1,8} & \hat\theta_{1,12} & \hat\theta_{1,16}
\end{array}\right]^{-1}
\left[\begin{array}{c}
\hat\theta_{1,17} \\
\hat\theta_{1,18} \\
\hat\theta_{1,19} \\
\hat\theta_{1,20}
\end{array}\right].
\end{align*}

Given the policy changes $P^\ast = P + a\left(1-P\right)$ under consideration, the parameter $\theta_2$ is therefore estimated by
\begin{align*}
\hat\theta_2 = 
\left[\begin{array}{c}
\hat\theta_{2,0}\\
\hat\theta_{2,2}
\end{array}\right]
=
\frac{1}{L}\sum_{\ell = 1}^L
\frac{1}{|I_\ell|} \sum_{i \in I_\ell}
a
\left[\begin{array}{c}
\left(\begin{array}{c}
X_{i1}\\
X_{i2}
\end{array}\right)
\\
1
\end{array}\right]
\left( 1 - S_i \right).
\end{align*}
Note that $\hat\theta_{2,0} = \hat\theta_{2,1}$ in the current model.

For the parameter $\theta_3$, it is required to estimate  $\mathcal{U}\left(W_i,\theta\right)$, ${\mf}_{U\mid P}\left(P^\ast\left({\mf}_{S\mid Z}\left(Z_i\right),Z_i\right)\right)$, ${\Delta}_{U\mid P}\left({\mf}_{S\mid Z}\left(Z_i\right)\right)$ and ${\Delta}_{U\mid P}\left(P^\ast\left({\mf}_{S\mid Z}\left(Z_i\right),Z_i\right)\right)$. 
$\mathcal{U}\left(W_i,\theta\right)$ is estimated by
\begin{align*}
{\mathcal{U}}_i\left(\hat\theta\right) = Y_i - S_i \left(X_{i1}, X_{i2}\right) \hat \beta_0 - \left(1-S_i\right) \left(X_{i1}, X_{i2}\right) \hat \beta_1.
\end{align*}
${\mf}_{U\mid P}\left({\mf}_{S\mid Z}\left(Z_i\right)\right)$ is estimated by
\begin{align*}
\hat{\mf}_{U\mid P}\left(\hat{\mf}_{S\mid Z}\left(Z_i\right)\right)
=\frac{\sum_{j \in I_\ell^c} {\mathcal{U}}_j\left(\hat\theta\right) K_{h_3}\left(P_j - \hat{\mf}_{S\mid Z}\left(Z_i\right)\right) }{\sum_{j \in I_\ell^c} K_{h_3}\left(P_j - \hat{\mf}_{S\mid Z}\left(Z_i\right)\right)}.
\end{align*}
Similarly, ${\mf}_{U\mid P}\left(P^\ast\left({\mf}_{S\mid Z}\left(Z_i\right),Z_i\right)\right)$ is estimated by
\begin{align*}
\hat{\mf}_{U\mid P}\left(P^\ast\left(\hat{\mf}_{S\mid Z}\left(Z_i\right),Z_i\right)\right)
=&\frac{\sum_{j \in I_\ell^c} {\mathcal{U}}_j\left(\hat\theta\right) K_{h_3}\left(P_j - \left(1-a\right)\hat{\mf}_{S\mid Z}\left(Z_i\right) - a\right) }{\sum_{j \in I_\ell^c} K_{h_3}\left(P_j - \left(1-a\right)\hat{\mf}_{S\mid Z}\left(Z_i\right) - a\right)}.
\end{align*}
The partial derivatives, ${\Delta}_{U\mid P}\left({\mf}_{S\mid Z}\left(Z_i\right)\right)$ and ${\Delta}_{U\mid P}\left(P^\ast\left({\mf}_{S\mid Z}\left(Z_i\right),Z_i\right)\right)$ are estimated by the approximate local derivative estimators through numerical approximation as the difference quotients:
\begingroup
\allowdisplaybreaks
\begin{align*}
\hat{\Delta}_{U\mid P}\left(\hat{\mf}_{S\mid Z}\left(Z_i\right)\right)
=&
\frac{\hat{\mf}_{U\mid P}\left(\hat{\mf}_{S\mid Z}\left(Z_i\right) + \delta\right) - \hat{\mf}_{U\mid P}\left(\hat{\mf}_{S\mid Z}\left(Z_i\right) - \delta\right)}{2\delta}
\qquad\text{and}\\
\hat{\Delta}_{U\mid P}\left(P^\ast\left(\hat{\mf}_{S\mid Z}\left(Z_i\right),Z_i\right)\right)
=&
\frac{\hat{\mf}_{U\mid P}\left(P^\ast\left(\hat{\mf}_{S\mid Z}\left(Z_i\right) + \delta,Z_i\right)\right) - \hat{\mf}_{U\mid P}\left(P^\ast\left(\hat{\mf}_{S\mid Z}\left(Z_i\right) - \delta,Z_i\right)\right)}{2\delta}.
\end{align*}
\endgroup

With these estimates, the parameter $\theta_3$ is estimated by
\begingroup
\allowdisplaybreaks
\begin{align*}
\hat\theta_3
=& \frac{1}{L}\sum_{\ell = 1}^L \frac{1}{|I_\ell|} \sum_{i \in I_\ell} \left\{
\hat{\mf}_{U\mid P}\left(P^\ast\left(\hat{\mf}_{S\mid Z}\left(Z_i\right),Z_i\right)\right)-{\mathcal{U}}_i\left(\hat\theta\right)
+
\rho\left( \frac{\hat f_{P^\ast}\left( \hat{\mf}_{S\mid Z}\left(Z_i\right) \right)}{\hat f_{P}\left( \hat{\mf}_{S\mid Z}\left(Z_i\right) \right)} \right) \left({\mathcal{U}}_i\left(\hat\theta\right)-\hat{\mf}_{U\mid P}\left(\hat{\mf}_{S\mid Z}\left(Z_i\right)\right)\right) \right.
\\ &\left.
+
\left( \left(1-a\right)
\hat{\Delta}_{U\mid P}\left(P^\ast\left(\hat{\mf}_{S\mid Z}\left(Z_i\right),Z_i\right)\right)
-
\rho\left( \frac{\hat{f}_{P^\ast}\left(\hat{\mf}_{S\mid Z}\left(Z_i\right)\right)}{\hat{f}_{P}\left(\hat{\mf}_{S\mid Z}\left(Z_i\right)\right)} \right)\hat{\Delta}_{U\mid P}\left(\hat{\mf}_{S\mid Z}\left(Z_i\right)\right)
\right)
\left(S_i-\hat{\mf}_{S\mid Z}\left(Z_i\right)\right) \right\},
\end{align*}
\endgroup
where we use a regularization function $\rho$ to shrink ${\hat{f}_{P^\ast}\left(\hat{\mf}_{S\mid Z}\left(Z_i\right)\right)} / {\hat{f}_{P}\left(\hat{\mf}_{S\mid Z}\left(Z_i\right)\right)}$ toward one.
We use $\rho\left(x\right) = x^\alpha$ for $\alpha \in \left(0,1\right)$.
Simulation results indicate that $\alpha \approx 0.3$ works well across alternative data generating designs, and we therefore recommend this number. 
Simulation results displayed in the main text are also based on this value of $\alpha$.

Given the parameter estimates $\left(\hat\theta_1',\hat\theta_2',\hat\theta_3\right)'$, the PRTE is estimated by 
\begin{align*}
\widehat{PRTE} = \hat\theta_{2,2}^{-1} \hat\theta_{2,0}' \left(\begin{array}{c}\hat\beta_{11}-\hat\beta_{01}\\\hat\beta_{12}-\hat\beta_{02}\end{array}\right) + \hat\theta_{2,2}^{-1} \hat\theta_3.
\end{align*}
Note that $\hat\theta_{2,0} = \hat\theta_{2,1}$ in the current model, and hence this simplified form is obtained.

Finally, we estimate the variance matrix for $\sqrt{n}\left(\widehat{PRTE}-PRTE\right)$.
We obtain the $4 \times 20$ matrix
\begin{align*}
\frac{\partial}{\partial\theta_1'}
\boldsymbol{d}\left(\hat\theta_1\right)
=
\frac{\partial}{\partial\theta_1'}\left.
\left[\begin{array}{cccc}
\theta_{1,1} & \theta_{1,5} & \theta_{1,9} & \theta_{1,13} \\
\theta_{1,2} & \theta_{1,6} & \theta_{1,10} & \theta_{1,14} \\
\theta_{1,3} & \theta_{1,7} & \theta_{1,11} & \theta_{1,15} \\
\theta_{1,4} & \theta_{1,8} & \theta_{1,12} & \theta_{1,16}
\end{array}\right]^{-1}
\left[\begin{array}{c}
\theta_{1,17} \\
\theta_{1,18} \\
\theta_{1,19} \\
\theta_{1,20}
\end{array}\right]
\right\vert_{\theta_1 = \left(\hat\theta_{1,1},\cdots,\hat\theta_{1,20}\right)'}
\end{align*}
as a numerical derivative.
Using this matrix, in turn, we compute the $1 \times 24$ derivative matrix
\begin{align*}
\lambda\left(\hat\theta\right) = \hat\theta_{2,2}^{-1} \left[
\hat\theta_{2,0}' \left(\begin{array}{cccc}-1&0&1&0\\0&-1&0&1\end{array}\right) \frac{\partial}{\partial\theta_1'}
\boldsymbol{d}\left(\hat\theta_1\right), \ \
\hat\beta_{11}-\hat\beta_{01}, \ \
\hat\beta_{12}-\hat\beta_{02}, \ \
-\widehat{PRTE}, \ \
1
\right]
\end{align*}
and the $24 \times 24$ derivative matrix $\hat{\mathcal{M}}$ by 
$$
\left(
\begin{array}{ccc}
I&0&0\\
0&I&0\\
\frac{1}{n}\sum_{i=1}^n\left[\left(\rho\left(\frac{\hat{f}_{P^\ast}\left(\hat{\mf}_{S\mid Z}\left(Z_i\right)\right)}{\hat{f}_{P}\left(\hat{\mf}_{S\mid Z}\left(Z_i\right)\right)}\right)-1\right)
\left(\left(1-S_i\right)X_{i1},\left(1-S_i\right)X_{i2},S_iX_{i1},S_iX_{i2}\right)
\right]\frac{\partial}{\partial\theta_1'}\boldsymbol{d}\left(\hat\theta_1\right)&0&I
\end{array}
\right).
$$
The $24 \times 24$ variance matrix $E[m\left(W;{\theta},{\gamma}\right)m\left(W;{\theta},{\gamma}\right)']$ is estimated by
\begin{align*}
\hat\Sigma = \frac{1}{n}\sum_{i=1}^n
\left[\begin{array}{c}
\left(\begin{array}{c}
\hat m_1\left(W_i;\hat{\theta},\hat{\gamma}\right)\\\hat m_2\left(W_i;\hat{\theta},\hat{\gamma}\right)\\\hat m_3\left(W_i;\hat{\theta},\hat{\gamma}\right)
\end{array}\right)
\left(\begin{array}{c}
\hat m_1\left(W_i;\hat{\theta},\hat{\gamma}\right)\\\hat m_2\left(W_i;\hat{\theta},\hat{\gamma}\right)\\\hat m_3\left(W_i;\hat{\theta},\hat{\gamma}\right)
\end{array}\right)'
\end{array}\right],
\end{align*}
where 
\begingroup
\allowdisplaybreaks
\begin{align*}
\hat m_1\left(W_i;\hat\theta,\hat\gamma\right) &= \hat{\xi}_1\left(X_i,Y_i,\hat{\mf}_{S\mid Z}\left(Z_i\right)\right) - \hat\theta_1
+
\hat\zeta\left(Z_i\right)
\left(S_i-\hat{\mf}_{S\mid Z}\left(Z_i\right)\right)
\\
\hat m_2\left(W_i;\hat\theta,\hat\gamma\right) &=a \left[\begin{array}{c}
\left(\begin{array}{c}
X_{i1}\\
X_{i2}
\end{array}\right)
\\
1
\end{array}\right]
\left( 1 - S_i \right)-\hat\theta_2\\
\hat m_3\left(W_i;\hat{\theta},\hat{\gamma}\right)&=
\hat{\mf}_{U\mid P}\left(P^\ast\left(\hat{\mf}_{S\mid Z}\left(Z_i\right),Z_i\right)\right)-{\mathcal{U}}_i\left(\hat\theta\right)-\hat{\theta}_3
+
\rho\left(\frac{\hat f_{P^\ast}\left( \hat{\mf}_{S\mid Z}\left(Z_i\right) \right)}{\hat f_{P}\left( \hat{\mf}_{S\mid Z}\left(Z_i\right) \right)}\right)\left({\mathcal{U}}_i\left(\hat\theta\right)-\hat{\mf}_{U\mid P}\left(\hat{\mf}_{S\mid Z}\left(Z_i\right)\right)\right) 
\\&
+
\left( \left(1-a\right)
\hat{\Delta}_{U\mid P}\left(P^\ast\left(\hat{\mf}_{S\mid Z}\left(Z_i\right),Z_i\right)\right)
-
\rho\left(\frac{\hat{f}_{P^\ast}\left(\hat{\mf}_{S\mid Z}\left(Z_i\right)\right)}{\hat{f}_{P}\left(\hat{\mf}_{S\mid Z}\left(Z_i\right)\right)}\right)\hat{\Delta}_{U\mid P}\left(\hat{\mf}_{S\mid Z}\left(Z_i\right)\right)
\right)\\&\qquad\times
\left(S_i-\hat{\mf}_{S\mid Z}\left(Z_i\right)\right).
\end{align*}
\endgroup
The asymptotic variance for $\sqrt{n}\left(\widehat{PRTE}-PRTE\right)$ can be now estimated by
$$
\lambda\left(\hat\theta\right)\left(\hat{\mathcal{M}}'\hat{\mathcal{M}}\right)^{-1}\hat{\mathcal{M}}'\hat\Sigma\hat{\mathcal{M}}\left(\hat{\mathcal{M}}'\hat{\mathcal{M}}\right)^{-1}\lambda\left(\hat\theta\right)'.
$$

\subsubsection{Case: Parametric and High-Dimensional Propensity Score Estimation}\label{sec:details_simulation_parametric_estimation_p}

The nonparametric estimator for the propensity score $P_i$ is replaced by a parametric and lasso estimators as follows.
Let $L$ be a link function.
In parametric estimation, for all $i \in I_\ell^c$,
\begin{align*}
\hat P_i = L\left(\hat b_0 + \sum_{k=1}^{d_Z} \hat b_k Z_k\right)
\end{align*}
where 
\begingroup
\allowdisplaybreaks
\begin{align*}
\left(\hat b_0,...,\hat b_{d_Z}\right)
=&
\arg\max \sum_{j \in I_\ell^c} \left\{ S_j \log L\left(b_0 + \sum_{k=1}^{d_Z} b_k Z_k\right) \right.
\\
&\qquad\qquad\qquad\qquad
\left. 
+ \left(1-S_j\right) \log \left(1 - L\left(b_0 + \sum_{k=1}^{d_Z} b_k Z_k\right) \right) \right\}
\end{align*}
\endgroup
Lasso estimation adds the absolute deviation penalty $- \nu_{1,n} \|b\|_1$ to the objective, where the tuning parameter $\nu_{1,n}$ is chosen by cross validation.

The nonparametric model of $\zeta$ is also replaced by a parametric estimator as follows.
Let $a_k \mapsto R_k\left(z,a_k\right)$ be a parametric model of $\zeta_k\left(z\right)=E\left[\left.\left.\frac{\partial}{\partial p}
\xi_{1,k}\left(x,y,p\right)\right|_{p=\mf_{S\mid Z}\left(z\right)}
\right\vert Z=z\right]$ for each $k \in \{1,...,20\}$.
In parametric estimation, for all $i \in I_\ell$, the projection $\xi_2$ on $Z_i$ is estimated by
\begin{align*}
\hat\zeta\left(Z_i\right) =
\left(\hat\zeta_{1}\left(Z_i\right), ..., \hat\zeta_{20}\left(Z_i\right)\right) = 
\left( R_1\left(Z_i,\hat a_1\right), ..., R_{20}\left(Z_i,\hat a_{20}\right) \right)
\end{align*}
where
\begin{align*}
\hat a_k = \arg\min \sum_{j \in I_\ell^c} \left( \hat\xi_{2,k}\left(X_j,Y_j,\hat{\mf}_{S\mid Z}\left(Z_j\right)\right) - R_k\left(Z_i,\hat a_k\right) \right)^2
\end{align*}
for each $k \in \{1,...,20\}$.
In simulations, we use a linear model $R_k\left(z,a\right) = a_{k,0} + a_{k,1} z_1 + \cdots + a_{k,d_Z} z_{d_Z}$.
Lasso estimation adds the absolute deviation penalty $\nu_{2,n} \|a\|_1$ to the objective, where the tuning parameter $\nu_{2,n}$ is chosen by cross validation.

\subsubsection{Case: Nonparametric Estimation under Policy Changes $Z_1^\ast = Z_1 + a$}\label{sec:details_simulation_estimation_z}
We present the concrete nonparametric estimation procedure used in the Monte Carlo simulations for policy changes of the form $Z_1^\ast = Z_1 + a$.

For all $i \in I_\ell^c$, the propensity score $P_i$ and the counterfactual propensity score are estimated by
\begingroup
\allowdisplaybreaks
\begin{align*}
\hat P_i =& \hat{\mf}_{S\mid Z}\left(Z_i\right) = \frac{\sum_{j \in I_\ell^c \backslash \{i\}} S_j K_{h_1}\left(Z_{j1}-Z_{i1}\right) K_{h_2}\left(Z_{j2}-Z_{i2}\right)}{\sum_{j \in I_\ell^c \backslash \{i\}} K_{h_1}\left(Z_{j1}-Z_{i1}\right) K_{h_2}\left(Z_{j2}-Z_{i2}\right)}
\qquad\text{and}\\
\hat P^\ast_i =& \hat{\mf}_{S\mid Z}\left(Z^\ast\left(Z_i\right)\right) = \frac{\sum_{j \in I_\ell^c \backslash \{i\}} S_j K_{h_1}\left(Z_{j1}-Z_{i1}-a\right) K_{h_2}\left(Z_{j2}-Z_{i2}\right)}{\sum_{j \in I_\ell^c \backslash \{i\}} K_{h_1}\left(Z_{j1}-Z_{i1}-a\right) K_{h_2}\left(Z_{j2}-Z_{i2}\right)},
\end{align*}
\endgroup
where $K_{h} = K\left( \ \cdot \ / h \right) / h$ for the Epanechnikov kernel function $K$, $h_1= c_1 \hat\sigma_{Z_1} n^{-1/6}$ and $h_1= c_2 \hat\sigma_{Z_2} n^{-1/6}$ with $c_1=c_2=2$.

For all $i \in I_\ell$, the conditional expectations of $\left(X_1,X_2,Y\right)$ given $P$ evaluated at data point $i$ are  estimated by
\begingroup
\allowdisplaybreaks
\begin{align*}
\hat \mf_{X_1\mid P}\left( \hat{\mf}_{S\mid Z}\left(Z_i\right) \right)
=&
\frac{ \sum_{j \in I_\ell^c} X_{j1} K_{h_3}\left(\hat P_j - \hat{\mf}_{S\mid Z}\left(Z_i\right) \right) }{ \sum_{j \in I_\ell^c} K_{h_3}\left(\hat P_j - \hat{\mf}_{S\mid Z}\left(Z_i\right) \right) }
\\
\hat \mf_{X_2\mid P}\left( \hat{\mf}_{S\mid Z}\left(Z_i\right) \right)
=&
\frac{ \sum_{j \in I_\ell^c} X_{j2} K_{h_3}\left(\hat P_j - \hat{\mf}_{S\mid Z}\left(Z_i\right) \right) }{ \sum_{j \in I_\ell^c} K_{h_3}\left(\hat P_j - \hat{\mf}_{S\mid Z}\left(Z_i\right) \right) }
\\
\hat \mf_{Y\mid P}\left( \hat{\mf}_{S\mid Z}\left(Z_i\right) \right)
=&
\frac{ \sum_{j \in I_\ell^c} Y_j K_{h_3}\left(\hat P_j - \hat{\mf}_{S\mid Z}\left(Z_i\right) \right) }{ \sum_{j \in I_\ell^c} K_{h_3}\left(\hat P_j - \hat{\mf}_{S\mid Z}\left(Z_i\right) \right) },
\end{align*}
\endgroup
where $h_3 = c_3 \hat\sigma_{\hat P} n^{-1/5}$ with $c_3 = 1.06$

The infinite-dimensional parameter $\xi_1$ evaluated at data point $i$ is estimated by
\begin{align*}
&\hat\xi_1\left(X_i,Y_i,\hat{\mf}_{S\mid Z}\left(Z_i\right)\right)=
\\
&\mathrm{vec}
\left[
\left(\begin{array}{c}
\left(1-\hat{\mf}_{S\mid Z}\left(Z_i\right)\right)\left(X_{i1}-\hat\mf_{X_1\mid P}\left(\hat{\mf}_{S\mid Z}\left(Z_i\right)\right)\right)\\
\left(1-\hat{\mf}_{S\mid Z}\left(Z_i\right)\right)\left(X_{i2}-\hat\mf_{X_2\mid P}\left(\hat{\mf}_{S\mid Z}\left(Z_i\right)\right)\right)\\
\hat{\mf}_{S\mid Z}\left(Z_i\right)\left(X_{i1}-\hat\mf_{X_1\mid P}\left(\hat{\mf}_{S\mid Z}\left(Z_i\right)\right)\right)\\
\hat{\mf}_{S\mid Z}\left(Z_i\right)\left(X_{i2}-\hat\mf_{X_2\mid P}\left(\hat{\mf}_{S\mid Z}\left(Z_i\right)\right)\right)
\end{array}\right)
\left(\begin{array}{c}
\left(1-\hat{\mf}_{S\mid Z}\left(Z_i\right)\right)\left(X_{i1}-\hat\mf_{X_1\mid P}\left(\hat{\mf}_{S\mid Z}\left(Z_i\right)\right)\right)\\
\left(1-\hat{\mf}_{S\mid Z}\left(Z_i\right)\right)\left(X_{i2}-\hat\mf_{X_2\mid P}\left(\hat{\mf}_{S\mid Z}\left(Z_i\right)\right)\right)\\
\hat{\mf}_{S\mid Z}\left(Z_i\right)\left(X_{i1}-\hat\mf_{X_1\mid P}\left(\hat{\mf}_{S\mid Z}\left(Z_i\right)\right)\right)\\
\hat{\mf}_{S\mid Z}\left(Z_i\right)\left(X_{i2}-\hat\mf_{X_2\mid P}\left(\hat{\mf}_{S\mid Z}\left(Z_i\right)\right)\right)\\
Y_{i}-\hat\mf_{Y\mid P}\left(\hat{\mf}_{S\mid Z}\left(Z_i\right)\right)
\end{array}\right)'\right].
\end{align*}
Its partial derivative $\xi_2 = \partial \xi_1 / \partial p$ is in turn obtained by the approximate local derivative estimator through numerical approximation as the difference quotient
\begin{align*}
\hat\xi_2\left(X_i,Y_i,\hat{\mf}_{S\mid Z}\left(Z_i\right)\right)
=
\frac{\hat\xi_1\left(X_i,Y_i,\hat{\mf}_{S\mid Z}\left(Z_i\right) + \delta\right) - \hat\xi_1\left(X_i,Y_i,\hat{\mf}_{S\mid Z}\left(Z_i\right) - \delta\right)}{2\delta}
\end{align*}
where $\delta = 0.01$.
For all $i \in I_\ell$, the projection $\xi_2$ on $Z_i$ is estimated by
\begin{align*}
\hat\zeta\left(Z_i\right)
=
\frac{\sum_{j \in I_\ell^c} \hat\xi_2\left(X_j,Y_j,\hat{\mf}_{S\mid Z}\left(Z_j\right)\right) K_{h_1}\left(Z_{j1}-Z_{i1}\right) K_{h_2}\left(Z_{j2}-Z_{i2}\right)}{\sum_{j \in I_\ell^c} K_{h_1}\left(Z_{j1}-Z_{i1}\right) K_{h_2}\left(Z_{j2}-Z_{i2}\right)}.
\end{align*}

The parameter $\theta_1$ is estimated by
\begin{align*}
\hat\theta_1 = \frac{1}{L}\sum_{\ell = 1}^L
\frac{1}{|I_\ell|} \sum_{i \in I_\ell} \left\{
\hat{\xi}_1\left(X_i,Y_i,\hat{\mf}_{S\mid Z}\left(Z_i\right)\right)
+
\hat\zeta\left(Z_i\right)
\left(S_i-\hat{\mf}_{S\mid Z}\left(Z_i\right)\right)
\right\}.
\end{align*}
With this estimate $\hat\theta_1 = \left(\hat\theta_{1,1},...,\hat\theta_{1,20}\right)'$, the partial linear coefficients $\left(\beta_0',\beta_1'\right)'$ for the sub-sample $I_\ell$ are estimated by
\begin{align*}
\left[\begin{array}{c}
\hat\beta_{0} \\
\hat\beta_{1}
\end{array}\right]
=
\left[\begin{array}{c}
\hat\beta_{01} \\
\hat\beta_{02} \\
\hat\beta_{11} \\
\hat\beta_{12}
\end{array}\right]
=
\boldsymbol{d}\left(\hat\theta_1\right)
=
\left[\begin{array}{cccc}
\hat\theta_{1,1} & \hat\theta_{1,5} & \hat\theta_{1,9} & \hat\theta_{1,13} \\
\hat\theta_{1,2} & \hat\theta_{1,6} & \hat\theta_{1,10} & \hat\theta_{1,14} \\
\hat\theta_{1,3} & \hat\theta_{1,7} & \hat\theta_{1,11} & \hat\theta_{1,15} \\
\hat\theta_{1,4} & \hat\theta_{1,8} & \hat\theta_{1,12} & \hat\theta_{1,16}
\end{array}\right]^{-1}
\left[\begin{array}{c}
\hat\theta_{1,17} \\
\hat\theta_{1,18} \\
\hat\theta_{1,19} \\
\hat\theta_{1,20}
\end{array}\right].
\end{align*}

For all $i \in I_\ell$, the density functions of $Z$ and $Z^\ast$ evaluated at data point $i$ are estimated by 
\begin{align*}
\hat f_Z\left(Z_i\right) =& \frac{1}{|I_\ell^c|} \sum_{j \in I_\ell^c} K_{h_1}\left(Z_{j1} - Z_{i1}\right) K_{h_2}\left(Z_{j2} - Z_{i2}\right)
\\
\hat f_{Z^\ast}\left(Z_i\right) =& \frac{1}{|I_\ell^c|} \sum_{j \in I_\ell^c} K_{h_1}\left(Z_{j1} + a - Z_{i1}\right) K_{h_2}\left(Z_{j2} - Z_{i2}\right),
\end{align*}
the conditional expectations of $\left(X_1,X_2\right)$ given $Z$ evaluated at data point $i$ are estimated by
\begin{align*}
\hat \mf_{X_1\mid Z}\left( Z_i \right)
=&
\frac{ \sum_{j \in I_\ell^c} X_{j1} K_{h_1}\left( Z_{j1} - Z_{i1} \right) K_{h_2}\left( Z_{j2} - Z_{i2} \right) }{ \sum_{j \in I_\ell^c} K_{h_1}\left( Z_{j1} - Z_{i1} \right) K_{h_2}\left( Z_{j2} - Z_{i2} \right) }
\\
\hat \mf_{X_2\mid Z}\left( Z_i \right)
=&
\frac{ \sum_{j \in I_\ell^c} X_{j2} K_{h_1}\left( Z_{j1} - Z_{i1} \right) K_{h_2}\left( Z_{j2} - Z_{i2} \right) }{ \sum_{j \in I_\ell^c} K_{h_1}\left( Z_{j1} - Z_{i1} \right) K_{h_2}\left( Z_{j2} - Z_{i2} \right) },
\end{align*}
and the conditional expectations of $\left(X_1,X_2\right)$ given $Z^\ast$ evaluated at data point $i$ are estimated by
\begin{align*}
\hat \mf_{X_1\mid Z^\ast}\left( Z_i \right)
=&
\frac{ \sum_{j \in I_\ell^c} X_{j1} K_{h_1}\left( Z_{j1} + a - Z_{i1} \right) K_{h_2}\left( Z_{j2} - Z_{i2} \right) }{ \sum_{j \in I_\ell^c} K_{h_1}\left( Z_{j1} - Z_{i1} \right) K_{h_2}\left( Z_{j2} - Z_{i2} \right) }
\\
\hat \mf_{X_2\mid Z^\ast}\left( Z_i \right)
=&
\frac{ \sum_{j \in I_\ell^c} X_{j2} K_{h_1}\left( Z_{j1} + a - Z_{i1} \right) K_{h_2}\left( Z_{j2} - Z_{i2} \right) }{ \sum_{j \in I_\ell^c} K_{h_1}\left( Z_{j1} - Z_{i1} \right) K_{h_2}\left( Z_{j2} - Z_{i2} \right) }.
\end{align*}

Given these estimates, the parameter $\theta_2$ is estimated by
\begin{align*}
\hat\theta_2 = 
\left[\begin{array}{c}
\hat\theta_{2,0}\\
\hat\theta_{2,2}
\end{array}\right]
=&
\frac{1}{L}\sum_{\ell = 1}^L
\frac{1}{|I_\ell|} \sum_{i \in I_\ell}
\left[\begin{array}{c}
\left(\begin{array}{c}
X_{i1}\\
X_{i2}
\end{array}\right)
\\
1
\end{array}\right]
\left( \hat\mf_{S|Z}\left(Z^\ast\left(Z_i\right)\right) - \hat\mf_{S|Z}\left(Z_i\right) \right)
\\
+&
\frac{1}{L}\sum_{\ell = 1}^L
\frac{1}{|I_\ell|} \sum_{i \in I_\ell}
\left[\begin{array}{c}
\left(\begin{array}{c}
X_{i1} \cdot \left( \rho\left(\frac{ \hat\mf_{X_1|Z^\ast}\left(Z_i\right) f_{Z^\ast}\left(Z_i\right)}{ \hat\mf_{X_1|Z}\left(Z_i\right) f_{Z}\left(Z_i\right)}\right) - 1\right)\\
X_{i2} \cdot \left( \rho\left(\frac{ \hat\mf_{X_2|Z^\ast}\left(Z_i\right) f_{Z^\ast}\left(Z_i\right)}{ \hat\mf_{X_2|Z}\left(Z_i\right) f_{Z}\left(Z_i\right)}\right) - 1\right)
\end{array}\right)
\\
\rho\left(\frac{f_{Z^\ast}\left(Z_i\right)}{f_{Z}\left(Z_i\right)}\right) - 1
\end{array}\right]
\left( S_i - \hat\mf_{S|Z}\left(Z_i\right) \right).
\end{align*}
Note that $\hat\theta_{2,0} = \hat\theta_{2,1}$ in the current model.

For the parameter $\theta_3$, it is required to estimate  $\mathcal{U}\left(W_i,\theta\right)$, ${\mf}_{U\mid P}\left(P^\ast\left({\mf}_{S\mid Z}\left(Z_i\right),Z_i\right)\right)$, ${\Delta}_{U\mid P}\left({\mf}_{S\mid Z}\left(Z_i\right)\right)$ and ${\Delta}_{U\mid P}\left(P^\ast\left({\mf}_{S\mid Z}\left(Z_i\right),Z_i\right)\right)$. 
$\mathcal{U}\left(W_i,\theta\right)$ is estimated by
\begin{align*}
{\mathcal{U}}_i\left(\hat\theta\right) = Y_i - S_i \left(X_{i1}, X_{i2}\right) \hat \beta_0 - \left(1-S_i\right) \left(X_{i1}, X_{i2}\right) \hat \beta_1.
\end{align*}
${\mf}_{U\mid P}\left({\mf}_{S\mid Z}\left(Z_i\right)\right)$ is estimated by
\begin{align*}
\hat{\mf}_{U\mid P}\left(\hat{\mf}_{S\mid Z}\left(Z_i\right)\right)
=\frac{\sum_{j \in I_\ell^c} {\mathcal{U}}_j\left(\hat\theta\right) K_{h_3}\left(P_j - \hat{\mf}_{S\mid Z}\left(Z_i\right)\right) }{\sum_{j \in I_\ell^c} K_{h_3}\left(P_j - \hat{\mf}_{S\mid Z}\left(Z_i\right)\right)}.
\end{align*}
Similarly, ${\mf}_{U\mid P}\left({\mf}_{S\mid Z}\left(Z^\ast\left(Z_i\right)\right)\right)$ is estimated by
\begin{align*}
\hat{\mf}_{U\mid P}\left(\hat{\mf}_{S\mid Z}\left(Z^\ast\left(Z_i\right)\right)\right)
=\frac{\sum_{j \in I_\ell^c} {\mathcal{U}}_j\left(\hat\theta\right) K_{h_3}\left(P_j - \hat{\mf}_{S\mid Z}\left(Z^\ast\left(Z_i\right)\right)\right) }{\sum_{j \in I_\ell^c} K_{h_3}\left(P_j - \hat{\mf}_{S\mid Z}\left(Z^\ast\left(Z_i\right)\right)\right)}.
\end{align*}

With these estimates, the parameter $\theta_3$ is estimated by
\begin{align*}
\hat\theta_3
=& \frac{1}{L}\sum_{\ell = 1}^L \frac{1}{|I_\ell|} \sum_{i \in I_\ell} \left\{
\hat{\mf}_{U\mid P}\left(\hat{\mf}_{S\mid Z}\left(Z^\ast\left(Z_i\right)\right)\right)-{\mathcal{U}}_i\left(\hat\theta\right)
+
\rho\left( \frac{\hat f_{Z^\ast}\left( Z_i \right)}{\hat f_{Z}\left( Z_i \right)} \right) \left({\mathcal{U}}_i\left(\hat\theta\right)-\hat{\mf}_{U\mid P}\left(\hat{\mf}_{S\mid Z}\left(Z_i\right)\right)\right) \right\},
\end{align*}

Given the parameter estimates $\left(\hat\theta_1',\hat\theta_2',\hat\theta_3\right)'$, the PRTE is estimated by 
\begin{align*}
\widehat{PRTE} = \hat\theta_{2,2}^{-1} \hat\theta_{2,0}' \left(\begin{array}{c}\hat\beta_{11}-\hat\beta_{01}\\\hat\beta_{12}-\hat\beta_{02}\end{array}\right) + \hat\theta_{2,2}^{-1} \hat\theta_3.
\end{align*}
Note that $\hat\theta_{2,0} = \hat\theta_{2,1}$ in the current model, and hence this simplified form is obtained.

Finally, we estimate the variance matrix for $\sqrt{n}\left(\widehat{PRTE}-PRTE\right)$.
We obtain the $4 \times 20$ matrix
\begin{align*}
\frac{\partial}{\partial\theta_1'}
\boldsymbol{d}\left(\hat\theta_1\right)
=
\frac{\partial}{\partial\theta_1'}\left.
\left[\begin{array}{cccc}
\theta_{1,1} & \theta_{1,5} & \theta_{1,9} & \theta_{1,13} \\
\theta_{1,2} & \theta_{1,6} & \theta_{1,10} & \theta_{1,14} \\
\theta_{1,3} & \theta_{1,7} & \theta_{1,11} & \theta_{1,15} \\
\theta_{1,4} & \theta_{1,8} & \theta_{1,12} & \theta_{1,16}
\end{array}\right]^{-1}
\left[\begin{array}{c}
\theta_{1,17} \\
\theta_{1,18} \\
\theta_{1,19} \\
\theta_{1,20}
\end{array}\right]
\right\vert_{\theta_1 = \left(\hat\theta_{1,1},\cdots,\hat\theta_{1,20}\right)'}
\end{align*}
as a numerical derivative.
Using this matrix, in turn, we compute the $1 \times 24$ derivative matrix
\begin{align*}
\lambda\left(\hat\theta\right) = \hat\theta_{2,2}^{-1} \left[
\hat\theta_{2,0}' \left(\begin{array}{cccc}-1&0&1&0\\0&-1&0&1\end{array}\right) \frac{\partial}{\partial\theta_1'}
\boldsymbol{d}\left(\hat\theta_1\right), \ \
\hat\beta_{11}-\hat\beta_{01}, \ \
\hat\beta_{12}-\hat\beta_{02}, \ \
-\widehat{PRTE}, \ \
1
\right]
\end{align*}
and the $24 \times 24$ derivative matrix $\hat{\mathcal{M}}$ by 
$$
\left(
\begin{array}{ccc}
I&0&0\\
0&I&0\\
\frac{1}{n}\sum_{i=1}^n\left[\left(\rho\left(\frac{\hat{f}_{Z^\ast}\left(Z_i\right)}{\hat{f}_{Z}\left(Z_i\right)}\right)-1\right)
\left(\left(1-S_i\right)X_{i1},\left(1-S_i\right)X_{i2},S_iX_{i1},S_iX_{i2}\right)
\right]\frac{\partial}{\partial\theta_1'}\boldsymbol{d}\left(\hat\theta_1\right)&0&I
\end{array}
\right).
$$
The $24 \times 24$ variance matrix $E[m\left(W;{\theta},{\gamma}\right)m\left(W;{\theta},{\gamma}\right)']$ is estimated by
\begin{align*}
\hat\Sigma = \frac{1}{n}\sum_{i=1}^n
\left[\begin{array}{c}
\left(\begin{array}{c}
\hat m_1\left(W_i;\hat{\theta},\hat{\gamma}\right)\\\hat m_2\left(W_i;\hat{\theta},\hat{\gamma}\right)\\\hat m_3\left(W_i;\hat{\theta},\hat{\gamma}\right)
\end{array}\right)
\left(\begin{array}{c}
\hat m_1\left(W_i;\hat{\theta},\hat{\gamma}\right)\\\hat m_2\left(W_i;\hat{\theta},\hat{\gamma}\right)\\\hat m_3\left(W_i;\hat{\theta},\hat{\gamma}\right)
\end{array}\right)'
\end{array}\right],
\end{align*}
where 
\begingroup
\allowdisplaybreaks
\begin{align*}
\hat m_1\left(W_i;\hat\theta,\hat\gamma\right) &= \hat{\xi}_1\left(X_i,Y_i,\hat{\mf}_{S\mid Z}\left(Z_i\right)\right) - \hat\theta_1
+
\hat\zeta\left(Z_i\right)
\left(S_i-\hat{\mf}_{S\mid Z}\left(Z_i\right)\right)
\\
\hat m_2\left(W_i;\hat\theta,\hat\gamma\right) &=
\left[\begin{array}{c}
\left(\begin{array}{c}
X_{i1}\\
X_{i2}
\end{array}\right)
\\
1
\end{array}\right]
\left( \hat\mf_{S|Z}\left(Z^\ast\left(Z_i\right)\right) - \hat\mf_{S|Z}\left(Z_i\right) \right)
-
\hat\theta_2\\
&+
\left[\begin{array}{c}
\left(\begin{array}{c}
X_{i1} \cdot \left( \frac{ \hat\mf_{X_1|Z^\ast}\left(Z_i\right) f_{Z^\ast}\left(Z_i\right)}{ \hat\mf_{X_1|Z}\left(Z_i\right) f_{Z}\left(Z_i\right)} - 1\right)\\
X_{i2} \cdot \left(\frac{ \hat\mf_{X_2|Z^\ast}\left(Z_i\right) f_{Z^\ast}\left(Z_i\right)}{ \hat\mf_{X_2|Z}\left(Z_i\right) f_{Z}\left(Z_i\right)} - 1\right)
\end{array}\right)
\\
\frac{f_{Z^\ast}\left(Z_i\right)}{f_{Z}\left(Z_i\right)} - 1
\end{array}\right]
\left( S_i - \hat\mf_{S|Z}\left(Z_i\right) \right)
\\
\end{align*}
\endgroup
The asymptotic variance for $\sqrt{n}\left(\widehat{PRTE}-PRTE\right)$ can be now estimated by
$$
\lambda\left(\hat\theta\right)\left(\hat{\mathcal{M}}'\hat{\mathcal{M}}\right)^{-1}\hat{\mathcal{M}}'\hat\Sigma\hat{\mathcal{M}}\left(\hat{\mathcal{M}}'\hat{\mathcal{M}}\right)^{-1}\lambda\left(\hat\theta\right)'.
$$

\section{Details about the Application}\label{sec:details_application}

\subsection{Model Specification for the Application}\label{sec:model_application}

The outcome variable $Y$ denotes the log of hourly wages.
The binary treatment variable $S$ indicates attendance of upper secondary school or higher.
The outcome is produced by 
\begin{align*}
Y = SY_1 + \left(1-S\right)Y_0
\end{align*}
where the treatment variable is selected by
\begingroup
\allowdisplaybreaks
\begin{align*}
S =& \{
b_{0} + b_{1} age + b_{2} age^2 + b_{3} rural + b_{4} kmsd + b_{5} protest + b_{6} cathol + b_{7} other
\\
&\ + b_{8}elem\_f + b_{9}jsec\_f + b_{10}edumiss\_f + b_{11}elem\_m + b_{12}jsec\_m + b_{13}edumiss\_m
\\
&\ +
b_{14}prov\_NSUM + b_{15}prov\_WSUM  + b_{16}prov\_SSUM  + b_{17}prov\_LAMP 
\\
&\ + b_{18}prov\_JAKA  + b_{19}prov\_CJAV  + b_{20}prov\_YOGI  + b_{21}prov\_EJAV 
\\
&\ + b_{22}prov\_BALI  + b_{23}prov\_WNUSA  + b_{24}prov\_SKALI + b_{25}prov\_SSUL
\\
&\ + b_{26} kmsmp + b_{27} kmsmp \cdot age + b_{28} kmsmp \cdot age^2 + b_{29} kmsmp \cdot rural + b_{30} kmsmp \cdot kmsd 
\\
&\ + b_{31} kmsmp \cdot protest + b_{32} kmsmp \cdot cathol + b_{33} kmsmp \cdot other + b_{34} kmsmp \cdot elem\_f 
\\
&\ + b_{35} kmsmp \cdot jsec\_f + b_{36} kmsmp \cdot edumiss\_f + b_{37} kmsmp \cdot elem\_m + b_{38} kmsmp \cdot jsec\_m 
\\
&\ + b_{39} kmsmp \cdot edumiss\_m + b_{40} kmsmp \cdot prov\_NSUM + b_{41} kmsmp \cdot prov\_WSUM 
\\
&\ + b_{42} kmsmp \cdot prov\_SSUM  + b_{43} kmsmp \cdot prov\_LAMP + b_{44} kmsmp \cdot prov\_JAKA 
\\
&\ + b_{45} kmsmp \cdot prov\_CJAV  + b_{46} kmsmp \cdot prov\_YOGI  + b_{47} kmsmp \cdot prov\_EJAV 
\\
&\ + b_{48} kmsmp \cdot prov\_BALI  + b_{49} kmsmp \cdot prov\_WNUSA + b_{50} kmsmp \cdot prov\_SKALI
\\
&\ + b_{51} kmsmp \cdot prov\_SSUL
>
V\}
\end{align*}
\endgroup
and the potential outcomes are produced by
\begingroup
\allowdisplaybreaks
\begin{align*}
Y_1 =&
\beta_{1,0} + \beta_{1,1} age + \beta_{1,2} age^2 + \beta_{1,3} rural + \beta_{1,4} kmsd + \beta_{1,5} protest + \beta_{1,6} cathol + \beta_{1,7} other
\\
& + \beta_{8}elem\_f + \beta_{9}jsec\_f + \beta_{10}edumiss\_f + \beta_{11}elem\_m + \beta_{12}jsec\_m + \beta_{13}edumiss\_m
\\
& +
\beta_{1,14}prov\_NSUM + \beta_{1,15}prov\_WSUM  + \beta_{1,16}prov\_SSUM  + \beta_{1,17}prov\_LAMP 
\\
& + \beta_{1,18}prov\_JAKA  + \beta_{1,19}prov\_CJAV  + \beta_{1,20}prov\_YOGI  + \beta_{1,21}prov\_EJAV 
\\
& + \beta_{1,22}prov\_BALI  + \beta_{1,23}prov\_WNUSA  + \beta_{1,24}prov\_SKALI + \beta_{1,25}prov\_SSUL
\end{align*}
\endgroup
and
\begingroup
\allowdisplaybreaks
\begin{align*}
Y_0 =&
\beta_{0,0} + \beta_{0,1} age + \beta_{0,2} age^2 + \beta_{0,3} rural + \beta_{0,4} kmsd + \beta_{0,5} protest + \beta_{0,6} cathol + \beta_{0,7} other
\\
& + \beta_{8}elem\_f + \beta_{9}jsec\_f + \beta_{10}edumiss\_f + \beta_{11}elem\_m + \beta_{12}jsec\_m + \beta_{13}edumiss\_m
\\
& +
\beta_{0,14}prov\_NSUM + \beta_{0,15}prov\_WSUM  + \beta_{0,16}prov\_SSUM  + \beta_{0,17}prov\_LAMP 
\\
& + \beta_{0,18}prov\_JAKA  + \beta_{0,19}prov\_CJAV  + \beta_{0,20}prov\_YOGI  + \beta_{0,21}prov\_EJAV 
\\
& + \beta_{0,22}prov\_BALI  + \beta_{0,23}prov\_WNUSA  + \beta_{0,24}prov\_SKALI + \beta_{0,25}prov\_SSUL.
\end{align*}
\endgroup
Here
$age$ denotes the age,
$rural$ indicates that the individual was living in a village at age 12,
$kmsd$ denotes the distance from the office of the head of the community of residence to the nearest community health post,
$protest$ indicates that the individual is protestant,
$cathol$ indicates that the individual is catholic,
$other$ indicates other religions,
$elem\_f$, $jsec\_f$ and $edumiss\_f$ indicate the level of schooling (respectively, elementary education, secondary education, and an indicator for unreported parental education) by father,
$elem\_m$, $jsec\_m$ and $edumiss\_m$ indicate the level of schooling by mother,
$prov\_NSUM$, $prov\_WSUM$, $prov\_SSUM$, $prov\_LAMP$, $prov\_JAKA$, $prov\_CJAV$, $prov\_YOGI$, $prov\_EJAV$, $prov\_BALI$, $prov\_WNUSA$, $prov\_SKALI$, and $prov\_SSUL$ indicate province of residence, and
$kmsmp$ denotes the distance from the office of the head of the community of residence in kilometers to the nearest secondary school.
Notice that $kmsmp$ is excluded from the outcome equation, and thus serves as an instrument.

\subsection{Step-by-Step Procedure for the Application}\label{sec:step_by_step_procedure_application}

In this section, we present a step-by-step estimation and inference procedure that we actually use in the application.
This procedure is an extension to the procedure that we use for simulations (see Section \ref{sec:details_simulation_estimation}), with the treatment of higher dimensionality of $X$ as the sole difference. 

For all $i \in I_\ell^c$, the propensity score $P_i$ is estimated by
\begin{align*}
\hat P_i = \hat{\mf}_{S|Z}\left(Z_i\right) = L\left(\hat b_0 + \sum_{k=1}^{d_Z} \hat b_k Z_k\right)
\end{align*}
where 
\begin{align*}
\left(\hat b_0,...,\hat b_{d_Z}\right)
=&
\arg\max \sum_{j \in I_\ell^c} \left\{ S_j \log L\left(b_0 + \sum_{k=1}^{d_Z} b_k Z_k\right) \right.
\\
&\qquad\qquad\qquad\qquad
\left. 
+ \left(1-S_j\right) \log \left(1 - L\left(b_0 + \sum_{k=1}^{d_Z} b_k Z_k\right) \right) \right\}
\end{align*}
Lasso estimation adds the absolute deviation penalty $- \nu_{1,n} \|b\|_1$ to the objective, where the tuning parameter $\nu_{1,n}$ is chosen by cross validation.

For all $i \in I_\ell$, the densities $f_{P}\left( {\mf}_{S\mid Z}\left(Z_i\right) \right)$ and $f_{P^\ast}\left( {\mf}_{S\mid Z}\left(Z_i\right) \right)$ are estimated by
\begingroup
\allowdisplaybreaks
\begin{align*}
\hat f_{P}\left( \hat{\mf}_{S\mid Z}\left(Z_i\right) \right) &= \frac{1}{|I_\ell^c|} \sum_{j \in I_\ell^c} K_{h}\left(\hat P_j - \hat{\mf}_{S\mid Z}\left(Z_i\right)\right)
\qquad\text{and}\\
\hat f_{P^\ast}\left( \hat{\mf}_{S\mid Z}\left(Z_i\right) \right) &= \frac{1}{|I_\ell^c|} \sum_{j \in I_\ell^c} K_{h}\left(\hat P_j + a\left(1 - \hat P_j\right) - \hat{\mf}_{S\mid Z}\left(Z_i\right)\right)
\end{align*}
\endgroup
where $h = c_3 \hat\sigma_{\hat P} n^{-1/5}$ with $c_3 = 1.06$.

For all $i \in I_\ell$, the conditional expectations of $\left(X_1,...,X_{d_X},Y\right)$ given $P$ evaluated at data point $i$ are  estimated by
\begingroup
\allowdisplaybreaks
\begin{align*}
\hat \mf_{X_1\mid P}\left( \hat{\mf}_{S\mid Z}\left(Z_i\right) \right)
=&
\frac{ \sum_{j \in I_\ell^c} X_{j1} K_{h}\left(\hat P_j - \hat{\mf}_{S\mid Z}\left(Z_i\right) \right) }{ \sum_{j \in I_\ell^c} K_{h}\left(\hat P_j - \hat{\mf}_{S\mid Z}\left(Z_i\right) \right) }
\\
& \qquad\vdots
\\
\hat \mf_{X_{d_X}\mid P}\left( \hat{\mf}_{S\mid Z}\left(Z_i\right) \right)
=&
\frac{ \sum_{j \in I_\ell^c} X_{j d_X} K_{h}\left(\hat P_j - \hat{\mf}_{S\mid Z}\left(Z_i\right) \right) }{ \sum_{j \in I_\ell^c} K_{h}\left(\hat P_j - \hat{\mf}_{S\mid Z}\left(Z_i\right) \right) }
\\
\hat \mf_{Y\mid P}\left( \hat{\mf}_{S\mid Z}\left(Z_i\right) \right)
=&
\frac{ \sum_{j \in I_\ell^c} Y_j K_{h}\left(\hat P_j - \hat{\mf}_{S\mid Z}\left(Z_i\right) \right) }{ \sum_{j \in I_\ell^c} K_{h}\left(\hat P_j - \hat{\mf}_{S\mid Z}\left(Z_i\right) \right) }.
\end{align*}
\endgroup

The infinite-dimensional parameter $\xi_1$ evaluated at data point $i$ is estimated by
\begingroup
\allowdisplaybreaks
\begin{align*}
&\hat\xi_1\left(X_i,Y_i,\hat{\mf}_{S\mid Z}\left(Z_i\right)\right)=
\\
&\mathrm{vec}
\left[
\left(\begin{array}{c}
\left(1-\hat{\mf}_{S\mid Z}\left(Z_i\right)\right)\left(X_{i1}-\hat\mf_{X_1\mid P}\left(\hat{\mf}_{S\mid Z}\left(Z_i\right)\right)\right)\\
\vdots\\
\left(1-\hat{\mf}_{S\mid Z}\left(Z_i\right)\right)\left(X_{id_X}-\hat\mf_{X_{d_X}\mid P}\left(\hat{\mf}_{S\mid Z}\left(Z_i\right)\right)\right)\\
\hat{\mf}_{S\mid Z}\left(Z_i\right)\left(X_{i1}-\hat\mf_{X_1\mid P}\left(\hat{\mf}_{S\mid Z}\left(Z_i\right)\right)\right)\\
\vdots\\
\hat{\mf}_{S\mid Z}\left(Z_i\right)\left(X_{id_X}-\hat\mf_{X_{d_X}\mid P}\left(\hat{\mf}_{S\mid Z}\left(Z_i\right)\right)\right)
\end{array}\right)
\left(\begin{array}{c}
\left(1-\hat{\mf}_{S\mid Z}\left(Z_i\right)\right)\left(X_{i1}-\hat\mf_{X_1\mid P}\left(\hat{\mf}_{S\mid Z}\left(Z_i\right)\right)\right)\\
\vdots\\
\left(1-\hat{\mf}_{S\mid Z}\left(Z_i\right)\right)\left(X_{id_X}-\hat\mf_{X_{d_X}\mid P}\left(\hat{\mf}_{S\mid Z}\left(Z_i\right)\right)\right)\\
\hat{\mf}_{S\mid Z}\left(Z_i\right)\left(X_{i1}-\hat\mf_{X_1\mid P}\left(\hat{\mf}_{S\mid Z}\left(Z_i\right)\right)\right)\\
\vdots\\
\hat{\mf}_{S\mid Z}\left(Z_i\right)\left(X_{id_X}-\hat\mf_{X_{d_X}\mid P}\left(\hat{\mf}_{S\mid Z}\left(Z_i\right)\right)\right)\\
Y_{i}-\hat\mf_{Y\mid P}\left(\hat{\mf}_{S\mid Z}\left(Z_i\right)\right)
\end{array}\right)'\right].
\end{align*}
\endgroup
Its partial derivative $\xi_2 = \partial \xi_1 / \partial p$ is in turn obtained by the approximate local derivative estimator through numerical approximation as the difference quotient
\begin{align*}
\hat\xi_2\left(X_i,Y_i,\hat{\mf}_{S\mid Z}\left(Z_i\right)\right)
=
\frac{\hat\xi_1\left(X_i,Y_i,\hat{\mf}_{S\mid Z}\left(Z_i\right) + \delta\right) - \hat\xi_1\left(X_i,Y_i,\hat{\mf}_{S\mid Z}\left(Z_i\right) - \delta\right)}{2\delta}
\end{align*}
where $\delta = 0.01$.

Let the linear model $R_k\left(z,a\right) = a_{k,0} + a_{k,1} z_1 + \cdots + a_{k,d_Z} z_{d_Z}$ for each $k \in \{1,...,2d_X\left(2d_X+1\right)\}$.
For all $i \in I_\ell$, the projection $\xi_2$ on $Z_i$ is estimated by
\begin{align*}
\hat\zeta\left(Z_i\right) =
\left(\hat\zeta_{1}\left(Z_i\right), ..., \hat\zeta_{2d_X\left(2d_X+1\right)}\left(Z_i\right)\right) = 
\left( R_1\left(Z_i,\hat a_1\right), ..., R_{2d_X\left(2d_X+1\right)}\left(Z_i,\hat a_{2d_X\left(2d_X+1\right)}\right) \right)
\end{align*}
where
\begin{align*}
\hat a_k = \arg\min \sum_{j \in I_\ell^c} \left( \hat\xi_{2,k}\left(X_j,Y_j,\hat{\mf}_{S\mid Z}\left(Z_j\right)\right) - R_k\left(Z_i,\hat a_k\right) \right)^2
\end{align*}
Lasso estimation adds the absolute deviation penalty $\nu_{2,n} \|a\|_1$ to the objective, where the tuning parameter $\nu_{2,n}$ is chosen by cross validation.

The parameter $\theta_1$ is estimated by
\begin{align*}
\hat\theta_1 = \frac{1}{L}\sum_{\ell = 1}^L
\frac{1}{|I_\ell|} \sum_{i \in I_\ell} \left\{
\hat{\xi}_1\left(X_i,Y_i,\hat{\mf}_{S\mid Z}\left(Z_i\right)\right)
+
\hat\zeta\left(Z_i\right)
\left(S_i-\hat{\mf}_{S\mid Z}\left(Z_i\right)\right)
\right\}.
\end{align*}
With this estimate $\hat\theta_1 = \left(\hat\theta_{1,1},...,\hat\theta_{1,2d_X\left(2d_X+1\right)}\right)'$, the partial linear coefficients $\left(\beta_0',\beta_1'\right)'$ for the sub-sample $I_\ell$ are estimated by
\begin{align*}
\left[\begin{array}{c}
\hat\beta_{0} \\
\hat\beta_{1}
\end{array}\right]
=
\left[\begin{array}{c}
\hat\beta_{01} \\
\vdots\\
\hat\beta_{0d_X} \\
\hat\beta_{11} \\
\vdots\\
\hat\beta_{1d_X}
\end{array}\right]
=
\boldsymbol{d}\left(\hat\theta_1\right)
=
\left[\begin{array}{ccccc}
\hat\theta_{1,1} & \hat\theta_{1,2d_X+1} & \cdots & \hat\theta_{1,4d_X^2-2d_X+1} \\
\vdots & \vdots && \vdots\\
\hat\theta_{1,d_X} & \hat\theta_{1,3d_X} &        & \hat\theta_{1,4d_X^2-d_X} \\
\hat\theta_{1,d_X+1} & \hat\theta_{1,3d_X+1} &        & \hat\theta_{1,4d_X^2-d_X+1} \\
\vdots & \vdots &&  \vdots\\
\hat\theta_{1,2d_X} & \hat\theta_{1,4d_X} & \cdots &  \hat\theta_{1,4d_X^2}
\end{array}\right]^{-1}
\left[\begin{array}{c}
\hat\theta_{1,4d_X^2+1} \\
\hat\theta_{1,4d_X^2+2} \\
\vdots\\
\hat\theta_2d_X\left(2d_X+1-1\right) \\
\hat\theta_{1,2d_X\left(2d_X+1\right)}
\end{array}\right].
\end{align*}

Given the policy changes $P^\ast = P + a\left(1-P\right)$ under consideration, the parameter $\theta_2$ is therefore estimated by
\begin{align*}
\hat\theta_2 = 
\left[\begin{array}{c}
\hat\theta_{2,0}\\
\hat\theta_{2,2}
\end{array}\right]
=
\frac{1}{L}\sum_{\ell = 1}^L
\frac{1}{|I_\ell|} \sum_{i \in I_\ell}
a
\left[\begin{array}{c}
\left(\begin{array}{c}
X_{i1}\\
\vdots\\
X_{id_X}
\end{array}\right)
\\
1
\end{array}\right]
\left( 1 - S_i \right).
\end{align*}
Note that $\hat\theta_{2,0} = \hat\theta_{2,1}$ in the current model.

Next, the parameter $\theta_3$ requires to estimate  $\mathcal{U}\left(W_i,\theta\right)$, ${\mf}_{U\mid P}\left(P^\ast\left({\mf}_{S\mid Z}\left(Z_i\right),Z_i\right)\right)$, ${\Delta}_{U\mid P}\left({\mf}_{S\mid Z}\left(Z_i\right)\right)$ and ${\Delta}_{U\mid P}\left(P^\ast\left({\mf}_{S\mid Z}\left(Z_i\right),Z_i\right)\right)$. 
$\mathcal{U}\left(W_i,\theta\right)$ is estimated by
\begin{align*}
{\mathcal{U}}_i\left(\hat\theta\right) = Y_i - S_i \left(X_{i1}, X_{i2}\right) \hat \beta_0 - \left(1-S_i\right) \left(X_{i1}, X_{i2}\right) \hat \beta_1.
\end{align*}
${\mf}_{U\mid P}\left({\mf}_{S\mid Z}\left(Z_i\right)\right)$ is estimated by
\begin{align*}
\hat{\mf}_{U\mid P}\left(\hat{\mf}_{S\mid Z}\left(Z_i\right)\right)
=\frac{\sum_{j \in I_\ell^c} {\mathcal{U}}_j\left(\hat\theta\right) K_{h}\left(P_j - \hat{\mf}_{S\mid Z}\left(Z_i\right)\right) }{\sum_{j \in I_\ell^c} K_{h}\left(P_j - \hat{\mf}_{S\mid Z}\left(Z_i\right)\right)}.
\end{align*}
Similarly, ${\mf}_{U\mid P}\left(P^\ast\left({\mf}_{S\mid Z}\left(Z_i\right),Z_i\right)\right)$ is estimated by
\begin{align*}
\hat{\mf}_{U\mid P}\left(P^\ast\left(\hat{\mf}_{S\mid Z}\left(Z_i\right),Z_i\right)\right)
=&\frac{\sum_{j \in I_\ell^c} {\mathcal{U}}_j\left(\hat\theta\right) K_{h}\left(P_j - \left(1-a\right)\hat{\mf}_{S\mid Z}\left(Z_i\right) - a\right) }{\sum_{j \in I_\ell^c} K_{h}\left(P_j - \left(1-a\right)\hat{\mf}_{S\mid Z}\left(Z_i\right) - a\right)}.
\end{align*}
The partial derivatives, ${\Delta}_{U\mid P}\left({\mf}_{S\mid Z}\left(Z_i\right)\right)$ and ${\Delta}_{U\mid P}\left(P^\ast\left({\mf}_{S\mid Z}\left(Z_i\right),Z_i\right)\right)$ are estimated by the approximate local derivative estimators through numerical approximation as the difference quotients:
\begingroup
\allowdisplaybreaks
\begin{align*}
\hat{\Delta}_{U\mid P}\left(\hat{\mf}_{S\mid Z}\left(Z_i\right)\right)
=&
\frac{\hat{\mf}_{U\mid P}\left(\hat{\mf}_{S\mid Z}\left(Z_i\right) + \delta\right) - \hat{\mf}_{U\mid P}\left(\hat{\mf}_{S\mid Z}\left(Z_i\right) - \delta\right)}{2\delta}
\qquad\text{and}\\
\hat{\Delta}_{U\mid P}\left(P^\ast\left(\hat{\mf}_{S\mid Z}\left(Z_i\right),Z_i\right)\right)
=&
\frac{\hat{\mf}_{U\mid P}\left(P^\ast\left(\hat{\mf}_{S\mid Z}\left(Z_i\right) + \delta,Z_i\right)\right) - \hat{\mf}_{U\mid P}\left(P^\ast\left(\hat{\mf}_{S\mid Z}\left(Z_i\right) - \delta,Z_i\right)\right)}{2\delta}.
\end{align*}
\endgroup

With these estimates, the parameter $\theta_3$ is estimated by
\begin{align*}
\hat\theta_3
=& 
\frac{1}{L}\sum_{\ell = 1}^L \frac{1}{|I_\ell|} \sum_{i \in I_\ell} \left\{\hat{\mf}_{U\mid P}\left(P^\ast\left(\hat{\mf}_{S\mid Z}\left(Z_i\right),Z_i\right)\right)-{\mathcal{U}}_i\left(\hat\theta\right)+\rho\left( \frac{\hat f_{P^\ast}\left( \hat{\mf}_{S\mid Z}\left(Z_i\right) \right)}{\hat f_{P}\left( \hat{\mf}_{S\mid Z}\left(Z_i\right) \right)} \right) \left({\mathcal{U}}_i\left(\hat\theta\right)-\hat{\mf}_{U\mid P}\left(\hat{\mf}_{S\mid Z}\left(Z_i\right)\right)\right) \right.\\ &\left.+\left( \left(1-a\right)\hat{\Delta}_{U\mid P}\left(P^\ast\left(\hat{\mf}_{S\mid Z}\left(Z_i\right),Z_i\right)\right)-\rho\left( \frac{\hat{f}_{P^\ast}\left(\hat{\mf}_{S\mid Z}\left(Z_i\right)\right)}{\hat{f}_{P}\left(\hat{\mf}_{S\mid Z}\left(Z_i\right)\right)} \right)\hat{\Delta}_{U\mid P}\left(\hat{\mf}_{S\mid Z}\left(Z_i\right)\right)\right)\left(S_i-\hat{\mf}_{S\mid Z}\left(Z_i\right)\right) \right\},
\end{align*}
where we use a regularization function $\rho$ to shrink ${\hat{f}_{P^\ast}\left(\hat{\mf}_{S\mid Z}\left(Z_i\right)\right)} / {\hat{f}_{P}\left(\hat{\mf}_{S\mid Z}\left(Z_i\right)\right)}$ toward one.
We use $\rho\left(x\right) = x^\alpha$ for $\alpha \in \left(0,1\right)$.
Simulation results indicate that $\alpha \approx 1/3$ works well across alternative data generating designs, and we therefore recommend this number. 
Simulation results displayed in the main text are also based on this value of $\alpha$.

Given the parameter estimates $\left(\hat\theta_1',\hat\theta_2',\hat\theta_3\right)'$, the PRTE is estimated by 
\begin{align*}
\widehat{PRTE} = \hat\theta_{2,2}^{-1} \hat\theta_{2,0}' \left(\begin{array}{c}\hat\beta_{11}-\hat\beta_{01}\\\vdots\\\hat\beta_{1d_X}-\hat\beta_{0d_X}\end{array}\right) + \hat\theta_{2,2}^{-1} \hat\theta_3.
\end{align*}
Note that $\hat\theta_{2,0} = \hat\theta_{2,1}$ in the current model, and hence this simplified form is obtained.

Finally, we estimate the variance matrix for $\sqrt{n}\left(\widehat{PRTE}-PRTE\right)$.
We obtain the $2d_X \times 2d_X\left(2d_X+1\right)$ matrix 
\begin{align*}
\frac{\partial}{\partial\theta_1'}
\boldsymbol{d}\left(\hat\theta_1\right)
=
\frac{\partial}{\partial\theta_1'}\left.
\left[\begin{array}{ccccc}
\hat\theta_{1,1} & \hat\theta_{1,2d_X+1} & \cdots & \hat\theta_{1,4d_X^2-2d_X+1} \\
\vdots & \vdots && \vdots\\
\hat\theta_{1,d_X} & \hat\theta_{1,3d_X} &        & \hat\theta_{1,4d_X^2-d_X} \\
\hat\theta_{1,d_X+1} & \hat\theta_{1,3d_X+1} &        & \hat\theta_{1,4d_X^2-d_X+1} \\
\vdots & \vdots &&  \vdots\\
\hat\theta_{1,2d_X} & \hat\theta_{1,4d_X} & \cdots &  \hat\theta_{1,4d_X^2}
\end{array}\right]^{-1}
\left[\begin{array}{c}
\hat\theta_{1,4d_X^2+1} \\
\hat\theta_{1,4d_X^2+2} \\
\vdots\\
\hat\theta_2d_X\left(2d_X+1-1\right) \\
\hat\theta_{1,2d_X\left(2d_X+1\right)}
\end{array}\right]
\right\vert_{\theta_1 = \left(\hat\theta_{1,1},\cdots,\hat\theta_{1,2d_X\left(2d_X+1\right)}\right)'}
\end{align*}
as a numerical derivative.
Using this matrix, in turn, we compute the $1 \times \left(2d_X\left(2d_X+1\right) + d_X + 2\right)$ derivative matrix
\begin{align*}
\lambda\left(\hat\theta\right) = \hat\theta_{2,2}^{-1} \left[
\hat\theta_{2,0}' \left(\begin{array}{cc}-I & I\end{array}\right) \frac{\partial}{\partial\theta_1'}
\boldsymbol{d}\left(\hat\theta_1\right), \ \
\hat\beta_{11}-\hat\beta_{01}, \ \
\hat\beta_{12}-\hat\beta_{02}, \ \
-\widehat{PRTE}, \ \
1
\right]
\end{align*}
and the $\left(2d_X\left(2d_X+1\right) + d_X + 2\right) \times \left(2d_X\left(2d_X+1\right) + d_X + 2\right)$ derivative matrix $\hat{\mathcal{M}}$ by 
$$
\left(
\begin{array}{ccc}
I&0&0\\
0&I&0\\
\frac{1}{n}\sum_{i=1}^n\left[\left(\rho\left(\frac{\hat{f}_{P^\ast}\left(\hat{\mf}_{S\mid Z}\left(Z_i\right)\right)}{\hat{f}_{P}\left(\hat{\mf}_{S\mid Z}\left(Z_i\right)\right)}\right)-1\right)
\left(\left(1-S_i\right)X_{i1},\left(1-S_i\right)X_{i2},S_iX_{i1},S_iX_{i2}\right)
\right]\frac{\partial}{\partial\theta_1'}\boldsymbol{d}\left(\hat\theta_1\right)&0&I
\end{array}
\right).
$$
The $\left(2d_X\left(2d_X+1\right) + d_X + 2\right) \times \left(2d_X\left(2d_X+1\right) + d_X + 2\right)$ variance matrix $E[m\left(W;{\theta},{\gamma}\right)m\left(W;{\theta},{\gamma}\right)']$ is estimated by
\begin{align*}
\hat\Sigma = \frac{1}{n}\sum_{i=1}^n
\left[\begin{array}{c}
\left(\begin{array}{c}
\hat m_1\left(W_i;\hat{\theta},\hat{\gamma}\right)\\\hat m_2\left(W_i;\hat{\theta},\hat{\gamma}\right)\\\hat m_3\left(W_i;\hat{\theta},\hat{\gamma}\right)
\end{array}\right)
\left(\begin{array}{c}
\hat m_1\left(W_i;\hat{\theta},\hat{\gamma}\right)\\\hat m_2\left(W_i;\hat{\theta},\hat{\gamma}\right)\\\hat m_3\left(W_i;\hat{\theta},\hat{\gamma}\right)
\end{array}\right)'
\end{array}\right],
\end{align*}
where 
\begingroup
\allowdisplaybreaks
\begin{align*}
\hat m_1\left(W_i;\hat\theta,\hat\gamma\right) 
&= 
\hat{\xi}_1\left(X_i,Y_i,\hat{\mf}_{S\mid Z}\left(Z_i\right)\right) - \hat\theta_1+\hat\zeta\left(Z_i\right)\left(S_i-\hat{\mf}_{S\mid Z}\left(Z_i\right)\right)
\\
\hat m_2\left(W_i;\hat\theta,\hat\gamma\right) 
&=
a \left[\begin{array}{c}\left(\begin{array}{c}X_{i1}\\ \vdots\\ X_{id_X}\end{array}\right)\\ 1\end{array}\right]\left( 1 - S_i \right)-\hat\theta_2\\
\hat m_3\left(W_i;\hat{\theta},\hat{\gamma}\right)
&=
\hat{\mf}_{U\mid P}\left(P^\ast\left(\hat{\mf}_{S\mid Z}\left(Z_i\right),Z_i\right)\right)-{\mathcal{U}}_i\left(\hat\theta\right)-\hat{\theta}_3+\rho\left(\frac{\hat f_{P^\ast}\left( \hat{\mf}_{S\mid Z}\left(Z_i\right) \right)}{\hat f_{P}\left( \hat{\mf}_{S\mid Z}\left(Z_i\right) \right)}\right)\left({\mathcal{U}}_i\left(\hat\theta\right)-\hat{\mf}_{U\mid P}\left(\hat{\mf}_{S\mid Z}\left(Z_i\right)\right)\right) 
\\&
+\left( \left(1-a\right)\hat{\Delta}_{U\mid P}\left(P^\ast\left(\hat{\mf}_{S\mid Z}\left(Z_i\right),Z_i\right)\right)-\rho\left(\frac{\hat{f}_{P^\ast}\left(\hat{\mf}_{S\mid Z}\left(Z_i\right)\right)}{\hat{f}_{P}\left(\hat{\mf}_{S\mid Z}\left(Z_i\right)\right)}\right)\hat{\Delta}_{U\mid P}\left(\hat{\mf}_{S\mid Z}\left(Z_i\right)\right)\right)\\&\qquad\times\left(S_i-\hat{\mf}_{S\mid Z}\left(Z_i\right)\right).
\end{align*}
\endgroup
The asymptotic variance for $\sqrt{n}\left(\widehat{PRTE}-PRTE\right)$ can be now estimated by
$$
\lambda\left(\hat\theta\right)\left(\hat{\mathcal{M}}'\hat{\mathcal{M}}\right)^{-1}\hat{\mathcal{M}}'\hat\Sigma\hat{\mathcal{M}}\left(\hat{\mathcal{M}}'\hat{\mathcal{M}}\right)^{-1}\lambda\left(\hat\theta\right)'.
$$ 

\bibliography{mybib}

\begin{thebibliography}{74}
\newcommand{\enquote}[1]{``#1''}
\expandafter\ifx\csname natexlab\endcsname\relax\def\natexlab#1{#1}\fi

\bibitem[\protect\citeauthoryear{Ai and Chen}{Ai and
  Chen}{2007}]{ai2007estimation}
\textsc{Ai, C. and X.~Chen} (2007): \enquote{Estimation of possibly
  misspecified semiparametric conditional moment restriction models with
  different conditioning variables,} \emph{Journal of Econometrics}, 141,
  5--43.

\bibitem[\protect\citeauthoryear{Auld}{Auld}{2005}]{auld:2005}
\textsc{Auld, M.~C.} (2005): \enquote{Causal Effect of Early Initiation on
  Adolescent Smoking Patterns,} \emph{Canadian Journal of Economics/Revue
  canadienne d'{\'e}conomique}, 38, 709--734.

\bibitem[\protect\citeauthoryear{Basu, Heckman, Navarro-Lozano, and Urzua}{Basu
  et~al.}{2007}]{basu/heckman/navarro/urzua:2007}
\textsc{Basu, A., J.~J. Heckman, S.~Navarro-Lozano, and S.~Urzua} (2007):
  \enquote{Use of Instrumental Variables in the Presence of Heterogeneity and
  Self-Selection: an Application to Treatments of Breast Cancer Patients,}
  \emph{Health Economics}, 16, 1133--1157.

\bibitem[\protect\citeauthoryear{Basu, Jena, Goldman, Philipson, and
  Dubois}{Basu et~al.}{2014}]{basu/jena/goldman/philipson/dubois:2014}
\textsc{Basu, A., A.~B. Jena, D.~P. Goldman, T.~J. Philipson, and R.~Dubois}
  (2014): \enquote{Heterogeneity in Action: the Role of Passive Personalization
  in Comparative Effectiveness Research,} \emph{Health Economics}, 23,
  359--373.

\bibitem[\protect\citeauthoryear{Belloni, Chernozhukov, Chetverikov, and
  Wei}{Belloni et~al.}{2018}]{belloni2018uniformly}
\textsc{Belloni, A., V.~Chernozhukov, D.~Chetverikov, and Y.~Wei} (2018):
  \enquote{Uniformly valid post-regularization confidence regions for many
  functional parameters in z-estimation framework,} \emph{Annals of
  statistics}, 46, 3643.

\bibitem[\protect\citeauthoryear{Belloni, Chernozhukov, Fern\'andez-Val, and
  Hansen}{Belloni et~al.}{2017}]{BCFH:2017}
\textsc{Belloni, A., V.~Chernozhukov, I.~Fern\'andez-Val, and C.~Hansen}
  (2017): \enquote{Program Evaluation with High-dimensional Data,}
  \emph{Econometrica}, 85, 233--298.

\bibitem[\protect\citeauthoryear{Belloni, Chernozhukov, and Hansen}{Belloni
  et~al.}{2014{\natexlab{a}}}]{belloni2014high}
\textsc{Belloni, A., V.~Chernozhukov, and C.~Hansen} (2014{\natexlab{a}}):
  \enquote{High-dimensional methods and inference on structural and treatment
  effects,} \emph{Journal of Economic Perspectives}, 28, 29--50.

\bibitem[\protect\citeauthoryear{Belloni, Chernozhukov, and Kato}{Belloni
  et~al.}{2014{\natexlab{b}}}]{belloni2014uniform}
\textsc{Belloni, A., V.~Chernozhukov, and K.~Kato} (2014{\natexlab{b}}):
  \enquote{Uniform post-selection inference for least absolute deviation
  regression and other Z-estimation problems,} \emph{Biometrika}, 102, 77--94.

\bibitem[\protect\citeauthoryear{Belloni, Chernozhukov, and Wei}{Belloni
  et~al.}{2013}]{belloni2013honest}
\textsc{Belloni, A., V.~Chernozhukov, and Y.~Wei} (2013): \enquote{Honest
  confidence regions for a regression parameter in logistic regression with a
  large number of controls,} Tech. rep., cemmap working paper, Centre for
  Microdata Methods and Practice.

\bibitem[\protect\citeauthoryear{Belskaya, Peter, and Posso}{Belskaya
  et~al.}{2014}]{belskaya/peter/posso:2014}
\textsc{Belskaya, O., K.~S. Peter, and C.~Posso} (2014): \enquote{College
  Expansion and the Marginal Returns to Education: Evidence from Russia,} IZA
  Discussion Paper No. 8735.

\bibitem[\protect\citeauthoryear{Bj\"orklund and Moffitt}{Bj\"orklund and
  Moffitt}{1987}]{bjorklund/moffitt:1987}
\textsc{Bj\"orklund, A. and R.~Moffitt} (1987): \enquote{The Estimation of Wage
  Gains and Welfare Gains in Self-Selection Models,} \emph{The Review of
  Economics and Statistics}, 69, 42--49.

\bibitem[\protect\citeauthoryear{Carneiro, Heckman, and Vytlacil}{Carneiro
  et~al.}{2010}]{carneiro/heckman/vytlacil:2010}
\textsc{Carneiro, P., J.~J. Heckman, and E.~Vytlacil} (2010):
  \enquote{Evaluating Marginal Policy Changes and the Average Effect of
  Treatment for Individuals at the Margin,} \emph{Econometrica}, 78, 377--394.

\bibitem[\protect\citeauthoryear{Carneiro, Heckman, and Vytlacil}{Carneiro
  et~al.}{2011}]{carneiro/heckman/vytlacil:2011}
---\hspace{-.1pt}---\hspace{-.1pt}--- (2011): \enquote{Estimating Marginal
  Returns to Education,} \emph{American Economic Review}, 101, 2754--81.

\bibitem[\protect\citeauthoryear{Carneiro and Lee}{Carneiro and
  Lee}{2009}]{carneiro/lee:2009}
\textsc{Carneiro, P. and S.~Lee} (2009): \enquote{Estimating Distributions of
  Potential Outcomes Using Local Instrumental Variables with an Application to
  Changes in College Enrollment and Wage Inequality,} \emph{Journal of
  Econometrics}, 149, 191--208.

\bibitem[\protect\citeauthoryear{Carneiro, Lokshin, and Umapathi}{Carneiro
  et~al.}{2017}]{carneiro/lokshin/umapathi:2017}
\textsc{Carneiro, P., M.~Lokshin, and N.~Umapathi} (2017): \enquote{Average and
  Marginal Returns to Upper Secondary Schooling in Indonesia,} \emph{Journal of
  Applied Econometrics}, 32, 16--36.

\bibitem[\protect\citeauthoryear{Chen, Hong, and Tarozzi}{Chen
  et~al.}{2008}]{chen2008semiparametric}
\textsc{Chen, X., H.~Hong, and A.~Tarozzi} (2008): \enquote{Semiparametric
  efficiency in GMM models with auxiliary data,} \emph{The Annals of
  Statistics}, 36, 808--843.

\bibitem[\protect\citeauthoryear{Chernozhukov, Chetverikov, Demirer, Duflo,
  Hansen, Newey, and Robins}{Chernozhukov
  et~al.}{2018{\natexlab{a}}}]{chernozhukov2018double}
\textsc{Chernozhukov, V., D.~Chetverikov, M.~Demirer, E.~Duflo, C.~Hansen,
  W.~Newey, and J.~Robins} (2018{\natexlab{a}}): \enquote{Double/debiased
  machine learning for treatment and structural parameters,} \emph{The
  Econometrics Journal}, 21, C1--C68.

\bibitem[\protect\citeauthoryear{Chernozhukov, Chetverikov, Kato
  et~al.}{Chernozhukov et~al.}{2014}]{chernozhukov2014gaussian}
\textsc{Chernozhukov, V., D.~Chetverikov, K.~Kato, et~al.} (2014):
  \enquote{Gaussian approximation of suprema of empirical processes,} \emph{The
  Annals of Statistics}, 42, 1564--1597.

\bibitem[\protect\citeauthoryear{Chernozhukov, Escanciano, Ichimura, Newey, and
  Robins}{Chernozhukov et~al.}{2018{\natexlab{b}}}]{chernozhukov2016locally}
\textsc{Chernozhukov, V., J.~C. Escanciano, H.~Ichimura, W.~K. Newey, and J.~M.
  Robins} (2018{\natexlab{b}}): \enquote{Locally Robust Semiparametric
  Estimation,} Working paper.

\bibitem[\protect\citeauthoryear{Chuang and Lai}{Chuang and
  Lai}{2010}]{chuang/lai:2010}
\textsc{Chuang, Y.-C. and W.-W. Lai} (2010): \enquote{Heterogeneity,
  comparative advantage, and return to education: The case of Taiwan,}
  \emph{Economics of Education Review}, 29, 804 -- 812.

\bibitem[\protect\citeauthoryear{Colangelo and Lee}{Colangelo and
  Lee}{2020}]{colangelo/lee:2020}
\textsc{Colangelo, K. and Y.-Y. Lee} (2020): \enquote{Double debiased machine
  learning nonparametric inference with continuous treatments,} \emph{arXiv
  preprint arXiv:2004.03036}.

\bibitem[\protect\citeauthoryear{Cornelissen, Dustmann, Raute, and
  Sch\"onberg}{Cornelissen
  et~al.}{2018}]{cornelissen/dustmann/raute/schonberg:2018}
\textsc{Cornelissen, T., C.~Dustmann, A.~Raute, and U.~Sch\"onberg} (2018):
  \enquote{Who Benefits from Universal Child Care? Estimating Marginal Returns
  to Early Child Care Attendance,} CEPR Discussion Paper No. DP13050.

\bibitem[\protect\citeauthoryear{Dobbie and Song}{Dobbie and
  Song}{2015}]{dobbie/song:2015}
\textsc{Dobbie, W. and J.~Song} (2015): \enquote{Debt Relief and Debtor
  Outcomes: Measuring the Effects of Consumer Bankruptcy Protection,}
  \emph{American Economic Review}, 105, 1272--1311.

\bibitem[\protect\citeauthoryear{Doyle~Jr.}{Doyle~Jr.}{2008}]{doyle:2008}
\textsc{Doyle~Jr., J.~J.} (2008): \enquote{Child Protection and Adult Crime:
  Using Investigator Assignment to Estimate Causal Effects of Foster Care,}
  \emph{Journal of Political Economy}, 116, 746--770.

\bibitem[\protect\citeauthoryear{Fan, Hsu, Lieli, and Zhang}{Fan
  et~al.}{2019}]{fan:2019}
\textsc{Fan, Q., Y.-C. Hsu, R.~P. Lieli, and Y.~Zhang} (2019):
  \enquote{Estimation of conditional average treatment effects with
  high-dimensional data,} \emph{arXiv preprint arXiv:1908.02399}.

\bibitem[\protect\citeauthoryear{Farrell}{Farrell}{2015}]{farrell2015robust}
\textsc{Farrell, M.~H.} (2015): \enquote{Robust inference on average treatment
  effects with possibly more covariates than observations,} \emph{Journal of
  Econometrics}, 189, 1--23.

\bibitem[\protect\citeauthoryear{Felfe and Lalive}{Felfe and
  Lalive}{2018}]{felfe/lalive:2018}
\textsc{Felfe, C. and R.~Lalive} (2018): \enquote{Does Early Child Care Affect
  Children's Development?} \emph{Journal of Public Economics}, 159, 33 -- 53.

\bibitem[\protect\citeauthoryear{Firpo}{Firpo}{2007}]{Firpo:2007}
\textsc{Firpo, S.} (2007): \enquote{Efficient semiparametric estimation of
  quantile treatment effects,} \emph{Econometrica}, 75, 259--276.

\bibitem[\protect\citeauthoryear{Galasso, Schankerman, and Serrano}{Galasso
  et~al.}{2013}]{galasso/schankerman/serrano:2013}
\textsc{Galasso, A., M.~Schankerman, and C.~J. Serrano} (2013):
  \enquote{Trading and Enforcing Patent Rights,} \emph{The RAND Journal of
  Economics}, 44, 275--312.

\bibitem[\protect\citeauthoryear{Graham}{Graham}{2011}]{graham2011efficiency}
\textsc{Graham, B.~S.} (2011): \enquote{Efficiency bounds for missing data
  models with semiparametric restrictions,} \emph{Econometrica}, 79, 437--452.

\bibitem[\protect\citeauthoryear{Graham, Pinto, and Egel}{Graham
  et~al.}{2012}]{graham2012inverse}
\textsc{Graham, B.~S., C.~C. d.~X. Pinto, and D.~Egel} (2012): \enquote{Inverse
  probability tilting for moment condition models with missing data,} \emph{The
  Review of Economic Studies}, 79, 1053--1079.

\bibitem[\protect\citeauthoryear{Graham, Pinto, and Egel}{Graham
  et~al.}{2016}]{graham2016efficient}
---\hspace{-.1pt}---\hspace{-.1pt}--- (2016): \enquote{Efficient estimation of
  data combination models by the method of auxiliary-to-study tilting (AST),}
  \emph{Journal of Business \& Economic Statistics}, 34, 288--301.

\bibitem[\protect\citeauthoryear{Hahn}{Hahn}{1998}]{Hahn:1998}
\textsc{Hahn, J.} (1998): \enquote{On the role of the propensity score in
  efficient semiparametric estimation of average treatment effects,}
  \emph{Econometrica}, 315--331.

\bibitem[\protect\citeauthoryear{Heckman and Vytlacil}{Heckman and
  Vytlacil}{2001}]{heckman/vytlacil:2001}
\textsc{Heckman, J.~J. and E.~Vytlacil} (2001): \enquote{Policy-Relevant
  Treatment Effects,} \emph{American Economic Review}, 91, 107--111.

\bibitem[\protect\citeauthoryear{Heckman and Vytlacil}{Heckman and
  Vytlacil}{2005}]{heckman/vytlacil:2005}
---\hspace{-.1pt}---\hspace{-.1pt}--- (2005): \enquote{Structural Equations,
  Treatment Effects, and Econometric Policy Evaluation,} \emph{Econometrica},
  73, 669--738.

\bibitem[\protect\citeauthoryear{Heckman and Vytlacil}{Heckman and
  Vytlacil}{2007}]{heckman/vytlacil:2007}
---\hspace{-.1pt}---\hspace{-.1pt}--- (2007): \enquote{Econometric Evaluation
  of Social Programs, Part II: Using the Marginal Treatment Effect to Organize
  Alternative Econometric Estimators to Evaluate Social Programs, and to
  Forecast their Effects in New Environments,} in \emph{Handbook of
  Econometrics}, ed. by J.~J. Heckman and E.~E. Leamer, Elsevier, vol.~6,
  chap.~71, 4875--5143.

\bibitem[\protect\citeauthoryear{Heckman and Vytlacil}{Heckman and
  Vytlacil}{1999}]{heckman/vytlacil:1999}
\textsc{Heckman, J.~J. and E.~J. Vytlacil} (1999): \enquote{Local instrumental
  variables and latent variable models for identifying and bounding treatment
  effects,} \emph{Proceedings of the national Academy of Sciences}, 96,
  4730--4734.

\bibitem[\protect\citeauthoryear{Hirano, Imbens, and Ridder}{Hirano
  et~al.}{2003}]{hirano2003efficient}
\textsc{Hirano, K., G.~W. Imbens, and G.~Ridder} (2003): \enquote{Efficient
  estimation of average treatment effects using the estimated propensity
  score,} \emph{Econometrica}, 71, 1161--1189.

\bibitem[\protect\citeauthoryear{Hitomi, Nishiyama, and Okui}{Hitomi
  et~al.}{2008}]{hitomi2008puzzling}
\textsc{Hitomi, K., Y.~Nishiyama, and R.~Okui} (2008): \enquote{A puzzling
  phenomenon in semiparametric estimation problems with infinite-dimensional
  nuisance parameters,} \emph{Econometric Theory}, 24, 1717--1728.

\bibitem[\protect\citeauthoryear{Ichimura and Taber}{Ichimura and
  Taber}{2000}]{ichimura/taber:2000}
\textsc{Ichimura, H. and C.~Taber} (2000): \enquote{Direct Estimation of Policy
  Impacts,} NBER Working Paper No. t0254.

\bibitem[\protect\citeauthoryear{Imbens}{Imbens}{1992}]{imbens1992efficient}
\textsc{Imbens, G.~W.} (1992): \enquote{An efficient method of moments
  estimator for discrete choice models with choice-based sampling,}
  \emph{Econometrica}, 1187--1214.

\bibitem[\protect\citeauthoryear{Joensen and Nielsen}{Joensen and
  Nielsen}{2016}]{jensen/nielsen:2016}
\textsc{Joensen, J.~S. and H.~S. Nielsen} (2016): \enquote{Mathematics and
  Gender: Heterogeneity in Causes and Consequences,} \emph{The Economic
  Journal}, 126, 1129--1163.

\bibitem[\protect\citeauthoryear{Johar and Maruyama}{Johar and
  Maruyama}{2014}]{johar/maruyama:2014}
\textsc{Johar, M. and S.~Maruyama} (2014): \enquote{Does Coresidence Improve an
  Elderly Parent's Health?} \emph{Journal of Applied Econometrics}, 29,
  965--983.

\bibitem[\protect\citeauthoryear{Kamh{\"o}fer, Schmitz, and
  Westphal}{Kamh{\"o}fer et~al.}{2018}]{kamhofer/schmitz/westphal:2018}
\textsc{Kamh{\"o}fer, D.~A., H.~Schmitz, and M.~Westphal} (2018):
  \enquote{Heterogeneity in Marginal Non-Monetary Returns to Higher Education,}
  \emph{Journal of the European Economic Association}, jvx058.

\bibitem[\protect\citeauthoryear{Kasahara, Liang, and Rodrigue}{Kasahara
  et~al.}{2016}]{kasahara/liang/rodrigue:2016}
\textsc{Kasahara, H., Y.~Liang, and J.~Rodrigue} (2016): \enquote{Does
  Importing Intermediates Increase the Demand for Skilled Workers? Plant-Level
  Evidence from Indonesia,} \emph{Journal of International Economics}, 102, 242
  -- 261.

\bibitem[\protect\citeauthoryear{Kennedy, Ma, McHugh, and Small}{Kennedy
  et~al.}{2017}]{kennedy:2017}
\textsc{Kennedy, E.~H., Z.~Ma, M.~D. McHugh, and D.~S. Small} (2017):
  \enquote{Nonparametric methods for doubly robust estimation of continuous
  treatment effects,} \emph{Journal of the Royal Statistical Society. Series B,
  Statistical Methodology}, 79, 1229.

\bibitem[\protect\citeauthoryear{Lawless, Kalbfleisch, and Wild}{Lawless
  et~al.}{1999}]{lawless1999}
\textsc{Lawless, J.~F., J.~D. Kalbfleisch, and C.~J. Wild} (1999):
  \enquote{Semiparametric Methods for Response-Selective and Missing Data
  Problems in Regression,} \emph{Journal of the Royal Statistical Society.
  Series B (Statistical Methodology)}, 61, 413--438.

\bibitem[\protect\citeauthoryear{Lee, Okui, and Whang}{Lee
  et~al.}{2017}]{lee/okui/whang:2017}
\textsc{Lee, S., R.~Okui, and Y.-J. Whang} (2017): \enquote{Doubly robust
  uniform confidence band for the conditional average treatment effect
  function,} \emph{Journal of Applied Econometrics}, 32, 1207--1225.

\bibitem[\protect\citeauthoryear{Lee}{Lee}{2019}]{lee:2019}
\textsc{Lee, Y.-Y.} (2019): \enquote{Nonparametric Weighted Average Quantile
  Derivative,} Working paper.

\bibitem[\protect\citeauthoryear{Lewbel and Schennach}{Lewbel and
  Schennach}{2007}]{lewbel2007simple}
\textsc{Lewbel, A. and S.~M. Schennach} (2007): \enquote{A simple ordered data
  estimator for inverse density weighted expectations,} \emph{Journal of
  Econometrics}, 136, 189--211.

\bibitem[\protect\citeauthoryear{Lindquist and Santavirta}{Lindquist and
  Santavirta}{2014}]{lindquist/santavirta:2014}
\textsc{Lindquist, M.~J. and T.~Santavirta} (2014): \enquote{Does Placing
  Children in Foster Care Increase Their Adult Criminality?} \emph{Labour
  Economics}, 31, 72 -- 83.

\bibitem[\protect\citeauthoryear{Magnac and Maurin}{Magnac and
  Maurin}{2007}]{magnac2007identification}
\textsc{Magnac, T. and E.~Maurin} (2007): \enquote{Identification and
  information in monotone binary models,} \emph{Journal of Econometrics}, 139,
  76--104.

\bibitem[\protect\citeauthoryear{Moffitt}{Moffitt}{2008}]{moffitt:2008}
\textsc{Moffitt, R.} (2008): \enquote{Estimating Marginal Treatment Effects in
  Heterogeneous Populations,} \emph{Annales d'Economie et de Statistique},
  239--261.

\bibitem[\protect\citeauthoryear{Moffitt}{Moffitt}{2014}]{moffitt:2014}
---\hspace{-.1pt}---\hspace{-.1pt}--- (2014): \enquote{Estimating Marginal
  Treatment Effects of Transfer Programs on Labor Supply,} Johns Hopkins
  University Working Paper.

\bibitem[\protect\citeauthoryear{Newey and Ruud}{Newey and
  Ruud}{2005}]{NeweyRuud2005}
\textsc{Newey, W. and P.~Ruud} (2005): \enquote{Density weighted linear least
  squares,} in \emph{Identification and Inference for Econometric Models,
  Essays in Honor of Thomas Rothenberg}, ed. by D.~Andrews and J.~Stock,
  Cambridge University Press, 554--573.

\bibitem[\protect\citeauthoryear{Newey}{Newey}{1994}]{newey1994asymptotic}
\textsc{Newey, W.~K.} (1994): \enquote{The asymptotic variance of
  semiparametric estimators,} \emph{Econometrica: Journal of the Econometric
  Society}, 1349--1382.

\bibitem[\protect\citeauthoryear{Newey and McFadden}{Newey and
  McFadden}{1994}]{newey/mcfadden:1994}
\textsc{Newey, W.~K. and D.~McFadden} (1994): \enquote{Large Sample Estimation
  and Hypothesis Testing,} in \emph{Handbook of Econometrics}, ed. by R.~F.
  Engle and D.~L. McFadden, Elsevier, vol.~4, chap.~36, 2111--2245.

\bibitem[\protect\citeauthoryear{Okui, Small, Tan, and Robins}{Okui
  et~al.}{2012}]{okui2012doubly}
\textsc{Okui, R., D.~S. Small, Z.~Tan, and J.~M. Robins} (2012):
  \enquote{Doubly robust instrumental variable regression,} \emph{Statistica
  Sinica}, 173--205.

\bibitem[\protect\citeauthoryear{Robins, Mark, and Newey}{Robins
  et~al.}{1992}]{robins1992estimating}
\textsc{Robins, J.~M., S.~D. Mark, and W.~K. Newey} (1992): \enquote{Estimating
  exposure effects by modelling the expectation of exposure conditional on
  confounders,} \emph{Biometrics}, 479--495.

\bibitem[\protect\citeauthoryear{Robins and Rotnitzky}{Robins and
  Rotnitzky}{1995}]{robins1995semiparametric}
\textsc{Robins, J.~M. and A.~Rotnitzky} (1995): \enquote{Semiparametric
  efficiency in multivariate regression models with missing data,}
  \emph{Journal of the American Statistical Association}, 90, 122--129.

\bibitem[\protect\citeauthoryear{Robins, Rotnitzky, and Zhao}{Robins
  et~al.}{1994}]{robins1994estimation}
\textsc{Robins, J.~M., A.~Rotnitzky, and L.~P. Zhao} (1994):
  \enquote{Estimation of regression coefficients when some regressors are not
  always observed,} \emph{Journal of the American statistical Association}, 89,
  846--866.

\bibitem[\protect\citeauthoryear{Robinson}{Robinson}{1988}]{robinson1988root}
\textsc{Robinson, P.~M.} (1988): \enquote{Root-N-consistent semiparametric
  regression,} \emph{Econometrica: Journal of the Econometric Society},
  931--954.

\bibitem[\protect\citeauthoryear{Rosenbaum}{Rosenbaum}{1987}]{rosenbaum1987model}
\textsc{Rosenbaum, P.~R.} (1987): \enquote{Model-based direct adjustment,}
  \emph{Journal of the American Statistical Association}, 82, 387--394.

\bibitem[\protect\citeauthoryear{Rothe and Firpo}{Rothe and
  Firpo}{2019}]{rothe2019properties}
\textsc{Rothe, C. and S.~Firpo} (2019): \enquote{Properties of doubly robust
  estimators when nuisance functions are estimated nonparametrically,}
  \emph{Econometric Theory}, 35, 1048--1087.

\bibitem[\protect\citeauthoryear{Sant'Anna and Zhao}{Sant'Anna and
  Zhao}{2018}]{sant2018doubly}
\textsc{Sant'Anna, P.~H. and J.~B. Zhao} (2018): \enquote{Doubly Robust
  Difference-in-Differences Estimators,} \emph{arXiv:1812.01723}.

\bibitem[\protect\citeauthoryear{S\l{}oczy\'nski and
  Wooldridge}{S\l{}oczy\'nski and
  Wooldridge}{2018}]{sloczynski_wooldridge_2018}
\textsc{S\l{}oczy\'nski, T. and J.~M. Wooldridge} (2018): \enquote{A general
  double robustness result for estimating average treatment effects,}
  \emph{Econometric Theory}, 34, 112–133.

\bibitem[\protect\citeauthoryear{Stock}{Stock}{1989}]{stock:1989}
\textsc{Stock, J.~H.} (1989): \enquote{Nonparametric Policy Analysis,}
  \emph{Journal of the American Statistical Association}, 84, 567--575.

\bibitem[\protect\citeauthoryear{Su, Ura, and Zhang}{Su
  et~al.}{2019}]{su/ura/zhang:2019}
\textsc{Su, L., T.~Ura, and Y.~Zhang} (2019): \enquote{Non-separable models
  with high-dimensional data,} \emph{Journal of Econometrics}, 212, 646--677.

\bibitem[\protect\citeauthoryear{Suzukawa}{Suzukawa}{2004}]{suzukawa2004unbiased}
\textsc{Suzukawa, A.} (2004): \enquote{Unbiased estimation of functionals under
  random censorship,} \emph{Journal of the Japan Statistical Society}, 34,
  153--172.

\bibitem[\protect\citeauthoryear{{van der Vaart} and Wellner}{{van der Vaart}
  and Wellner}{2001}]{vanweak}
\textsc{{van der Vaart}, A.~W. and J.~A. Wellner} (2001): \emph{Weak
  Convergence and Empirical Processes With Applications to Statistics},
  Springer-Verlag New York.

\bibitem[\protect\citeauthoryear{Wager and Athey}{Wager and
  Athey}{2018}]{Wager/Athey:2018}
\textsc{Wager, S. and S.~Athey} (2018): \enquote{Estimation and inference of
  heterogeneous treatment effects using random forests,} \emph{Journal of the
  American Statistical Association}, 113, 1228--1242.

\bibitem[\protect\citeauthoryear{Wooldridge}{Wooldridge}{1999}]{wooldridge1999}
\textsc{Wooldridge, J.~M.} (1999): \enquote{Asymptotic Properties of Weighted
  M-Estimators for Variable Probability Samples,} \emph{Econometrica}, 67,
  1385--1406.

\bibitem[\protect\citeauthoryear{Wooldridge}{Wooldridge}{2001}]{wooldridge2001asymptotic}
---\hspace{-.1pt}---\hspace{-.1pt}--- (2001): \enquote{Asymptotic properties of
  weighted M-estimators for standard stratified samples,} \emph{Econometric
  Theory}, 17, 451--470.

\bibitem[\protect\citeauthoryear{Zimmert and Lechner}{Zimmert and
  Lechner}{2019}]{zimmert/lechner:2019}
\textsc{Zimmert, M. and M.~Lechner} (2019): \enquote{Nonparametric estimation
  of causal heterogeneity under high-dimensional confounding,} \emph{arXiv
  preprint arXiv:1908.08779}.

\end{thebibliography}
\end{document}